\documentclass[oneside,12pt]{book}

\usepackage{scrextend}

\usepackage[linedheaders=true, dottedtoc=true, eulermath=true, floatperchapter=true]{classicthesis}
\usepackage[margin=2.5cm]{geometry}

\usepackage[toc, page]{appendix}

\usepackage{graphicx}

\usepackage{pgfplots}

\usepackage{array}
\usepackage{supertabular}

\usepackage{setspace}
\usepackage[style=numeric-comp, maxbibnames=5, maxcitenames=5,sorting=none]{biblatex}
\usepackage{amsmath}

\usepackage{svg}

\usepackage{bbold}

\usepackage{bm}

\usepackage{amsthm}

\usepackage{amssymb}

\usepackage{pstricks}

\usepackage{acro}

\newcommand{\inintro}[1]{#1}

\acsetup{barriers/use=true, barriers/reset=true}
\DeclareAcronym{GPS}{short=GPS, long=Gaussian~Process~State}
\DeclareAcronym{qGPS}{short=qGPS, long=quantum~Gaussian~Process~State}
\DeclareAcronym{CPS}{short=CPS, long=Correlator~Product~State}
\DeclareAcronym{MPS}{short=MPS, long=Matrix~Product~State}
\DeclareAcronym{DMRG}{short=DMRG, long=density~matrix~renormalization~group}
\DeclareAcronym{HF}{short=HF, long=Hartree-Fock}
\DeclareAcronym{CCSD}{short=CCSD, long=coupled cluster with single and double excitations}
\DeclareAcronym{CCSD(T)}{short=CCSD(T), long=coupled cluster with single and double excitations and pertubative treatment of triple excitations}
\DeclareAcronym{SD}{short=SD, long=Slater~determinant}
\DeclareAcronym{NQS}{short=NQS, long=Neural~Quantum~State}
\DeclareAcronym{TNS}{short=TNS, long=Tensor~Network~State}
\DeclareAcronym{NN}{short=NN, long=artificial~neural~network}
\DeclareAcronym{GP}{short=GP, long=Gaussian~Process, long-plural=Gaussian~Processes}
\DeclareAcronym{GPR}{short=GPR, long=Gaussian~Process~regression}
\DeclareAcronym{SWO}{short=SWO, long=supervised~wavefunction~optimization}
\DeclareAcronym{VMC}{short=VMC, long=Variational~Monte~Carlo}
\DeclareAcronym{RBM}{short=RBM, long=restricted~Boltzmann~machine}
\DeclareAcronym{RVM}{short=RVM, long=Relevance~Vector~Machine}
\DeclareAcronym{CP}{short=CP, long=CANDECOMP/PARAFAC}
\DeclareAcronym{ML}{short=ML, long=machine~learning}
\DeclareAcronym{t-SNE}{short=t-SNE, long=t-distributed~stochastic~neighbour~embedding}
\DeclareAcronym{ALS}{short=ALS, long=alternating~least~squares}
\DeclareAcronym{SR}{short=SR, long=Stochastic~Reconfiguration}
\DeclareAcronym{MSR}{short=MSR, long=Marshall~Sign~Rule}
\DeclareAcronym{DQG}{short=DQG, long=variational two-electron reduced density matrix}
\DeclareAcronym{1RDM}{short=$1$-RDM, long=one-body reduced density matrix}

\graphicspath{{./figures/03/Random_basis_selection/}{./figures/03/Marginal_likelihood_model_selection/}{./figures/03/Basis_clustering/}{./figures/04/Bootstrapping_schematic/}}

\newtheorem*{ansatz}{Ansatz}

\begin{document}

\thispagestyle{empty}

\quad

\vspace{1cm}

{\noindent \textit{Doctor of Philosophy} dissertation}

\noindent \rule{\textwidth}{1pt}

\vspace{0,5cm}

{\Large \noindent Bayesian Modelling Approaches for Quantum States}

\vspace{0,5cm}

{\large  \noindent The Ultimate Gaussian Process States Handbook}

\noindent \rule{\textwidth}{1pt}

\vspace{0,5cm}

{\large \raggedright \hfill Yannic Rath}

\vspace{14cm}
Department of Physics

King's College London

\vspace{0.5cm}

Supervisor: Dr George Booth

\vspace{0.5cm}
\today

\newpage

\chapter*{Abstract}

Capturing the correlation emerging between constituents of many-body systems is one of the key challenges to describe various quantum systems accurately.
This thesis discusses novel tools and techniques for the numerical modelling of quantum many-body wavefunctions exhibiting non-trivial correlations.
It is outlined how synergies with standard machine learning frameworks can be exploited to design efficient representations enabling an automated representation of the relevant characteristics.
In particular, it is presented how rigorous Bayesian regression techniques, e.g., formalized via Gaussian Processes, can be utilized to introduce compact forms for correlated many-body states.
Based on the probabilistic regression techniques forming the foundation of the resulting ansatz, coined the Gaussian Process State, different compression techniques are discussed to efficiently extract a numerically feasible representation.
By following physically motivated modelling principles, the obtained representations carry a high degree of interpretability and offer an easily applicable tool for the study of challenging many-body systems.

This work discusses different perspectives on the Gaussian Process State representation of many-body quantum systems, and presents practically applicable methods and techniques to utilize the framework in the numerical practice.
A strong focus is to show how rigorous Bayesian modelling principles can be used to find a compact description of intricate quantum states based on (potentially incomplete) wavefunction data.
On the one hand, these schemes can be exploited to extract and uncover physically interpretable characteristics, such as information about the correlation within the state.
On the other hand, these also offer an easily applicable scheme to infer an approximate state spanning across the full Hilbert space only relying on the information from a small subsection of the state space.

Following the Gaussian Process regression framework to extract a probabilistic representation of data points, the definition of Gaussian Process States explicitly relies on the specification of suitable physical configurations.
To improve the compactness of standard Gaussian Process regression models, two approaches are presented to achieve a selection of particularly sparse wavefunction models.
These are based on the extraction of appropriate configurations, either in an explicitly data-driven framework via (potentially iterative) compression of data from presented states, or via a direct variational optimization of parameterized product states.

The practical efficiency of the Gaussian Process State ansatz is demonstrated for ground state approximations with standard Variational Monte Carlo techniques.
Results are presented for prototypical quantum lattice models, Fermi-Hubbard models and $J_1$-$J_2$ models, as well as simple ab-initio quantum chemical systems.
It is demonstrated that competitive accuracies can be achieved practically with the Gaussian Process representation for different challenging systems.

This thesis also aims to identify how the Gaussian Process ansatz fits into broader classifications of compact many-body quantum state representations.
The Gaussian Process State is linked to neural network, as well as tensor network representations of quantum states, and current challenges and limitations of the applied methods are discussed.

\chapter*{Acknowledgements}

I wish to express my sincere gratitude towards those many people who contributed to my PhD efforts.

Most importantly, I want to thank my PhD supervisor, Dr George Booth, for providing the best imaginable platform to study fascinating research questions and offering unlimited assistance regarding all aspects of my research experience at King's College.
Not only am I still immensely grateful that he has offered me the opportunity to pursue a doctorate in physics under his supervision, and hands-on support for theoretical and practical questions, but his endless enthusiasm and optimism for the project has also become one of the main drivers of my own motivation.
Thank you, George!

Secondly, I want to say `thank you' to my amazing wife Maja who offers unconditional support in any situation, makes my life in London an absolute joy, and (rather unexpectedly) also ended up being my main (home) office partner over the course of my PhD following the outbreak of the COVID-19 pandemic.

It was an absolute pleasure for me to contribute to highly collaborative research efforts and expand my horizons through various stimulating discussions.
In particular, I want to thank Dr Aldo Glielmo who, in addition to laying the foundations for the research presented in this work and significantly advancing the project with his wide-reaching technical knowledge, also welcomed me with open arms when I joined the group.
Moreover, I thank Prof. Dr Gábor Csányi for valuable contributions to the research and insightful exchanges.
Furthermore, I also want to thank Massimo Bortone for contributing interesting thoughts and ideas, extending the functionality of the developed code, and significantly pushing the understanding about the developed methodology forward.

Feedback is always a key element to improve the quality of research (and in particular the communication of results).
I am therefore highly thankful to all the people taking the time to offer comments, suggestions and novel perspectives on presented findings.
I especially want to thank Prof. Dr Stephen Clark and Prof. Dr Andrew Green for taking on the examination of my PhD degree.
Furthermore, I thank my co-supervisor Prof. Dr Lev Kantorovich, as well as Dr Joe Bhaseen and Prof. Dr Jean Alexandre for the feedback provided as part of my `upgrade viva'.

I also want to extend this thanks to all the people that made my experience at King's College particularly enjoyable and helped to broaden my interests and expand my knowledge.
In particular, I want to thank the other past and present colleagues of the `Booth group', Rob, Ollie, Max, Sriluckshmy, Edoardo, Chris, James, Charlie, Basil, Zelong and Terence, as well as all the other colleagues I had the pleasure of sharing an office with, for contributing to a stimulating research atmosphere and interesting discussions.
Additionally, I would like to show my gratitude to the broader community at King's College, especially all the people ensuring frictionless operations within the department of Physics to facilitate my doctorate.

The results presented in this work were obtained from various numerical simulations performed on different computing platforms and I acknowledge use of the research computing facilities at King's College London, Rosalind (\url{https://rosalind.kcl.ac.uk/}) and CREATE (\url{https://doi.org/10.18742/rnvf-m076}).
Moreover, I am grateful to the UK Materials and Molecular Modelling Hub for computational resources, which is partially funded by EPSRC (EP/P020194/1 and EP/T022213/1).

Lastly, I want to thank my parents, Mama Mo and Papa Tho, for their support and guidance, as well as my sister Leo, friends, and other family members, for providing the most amazing social network I can always rely on.

\quad

Thank you all!

\tableofcontents

\listoffigures

\printacronyms[template=supertabular]


\chapter{Introduction}
\section{Motivation}
Accurately simulating the behaviour of electrons in a material compound or molecule is the key ingredient to understand various system properties based on the underlying physical principles.
Theoretically, such simulations would, among other things, also make it possible to predict system properties and could therefore directly aid the discovery of new materials and substances for various sought after applications.
However, capturing the electronic behaviour of interest with great accuracy is intrinsically limited by the inherent complexity of the quantum mechanical laws governing the system characteristics.

The description of electrons in molecules and materials is only one manifestation of the key challenges emerging in the context of quantum many-body physics.
These restrict the availability of known exact solutions to the well-defined laws describing multiple interacting system constituents to few specific cases.
This poses significant challenges in understanding the physical phenomena appearing in different areas of quantum many-body physics, such as quantum chemistry, materials science, quantum information, and many more.
With the inherent complexity scaling of many-body systems often severely limiting the accessibility of exact descriptions, to uncover insights and information about these systems, in practice one often has to resort to numerical approaches employing suitable approximations to the problem.
In principle, it is possible to introduce such approximations on different levels of abstraction making it possible to access different physical regimes and accuracy levels.
Practically, this typically means that the overall accuracy of methods decreases with increasing system sizes that are treatable by the approach.
The limitations coming with numerical approaches naturally give rise to a very fundamental question: What is the best approach to study a specific system of interest with the available computational resources to the greatest accuracy possible?

While this work certainly does not provide a general answer to this question, it explores how many-body wavefunctions, which are the core backbone of the exact quantum mechanical description, can be represented efficiently aided by modern computing hardware.
Designing a compressed representation of the many-body wavefunction is particularly appealing as most system properties of interest can directly be extracted from this.
Numerical schemes explicitly relying on a representation of the wavefunction represent a low abstraction level and can often achieve high accuracy approximations to the exact solution in regimes of relatively small system sizes.
This makes it possible to uncover and describe additional physical phenomena emerging from the correlations and interactions between the system components.
These would otherwise be inaccessible with higher level approaches significantly simplifying underlying quantum mechanical problems for large systems.
Examples include Density Functional Theory~\cite{shollDensityFunctionalTheory2011}, or material and object studies on even higher levels of abstraction, e.g., with the Finite Element Analysis~\cite{zienkiewiczFiniteElementMethod2005}.

The key question for wavefunction descriptions, also explored in this work, is how the (often unknown) exact many-body state of the system can be represented most efficiently.
The overall dimension of the space of potential states typically grows exponentially in the number of system constituents.
Nonetheless, many physically meaningful target states can be made numerically treatable by introducing compact representations exploiting some underlying structure emerging from physical principles.
However, exploiting this underlying structure in order to find the best trade-off between accuracy and affordable computational complexity represents quite often a challenging task.
In particular, many-body wavefunctions take various different forms and show very different physical characteristics depending on the specifics of the studied systems.
For example, a ground state of weakly interacting freely moving electrons exhibits a significantly different structure than one of a one-dimensional array of fixed spins only interacting with their nearest neighbours.
The inherent structure extracted by a good, numerically feasible representations of these states should therefore also be very different.

\section{Related work}
Various different schemes to efficiently extract information directly from a many-body state have been introduced over the years.
Many of these approaches are based around the idea of finding a representation explicitly encoding the physical properties that are expected for the system.
As an example, in the \ac{HF} method, the ground state of a many-electron system is approximated as a single anti-symmetrized product of single-particle wavefunctions~\cite{szaboModernQuantumChemistry2012}.
This mean-field approach therefore explicitly neglects many-particle correlations emerging from interactions between the different electrons in the Hamiltonian.
If, however, it is expected that the many-body correlations are important for the description, other representations explicitly incorporating them would be more suitable.
Many approaches have been developed incorporating the correlation properties based on some physical intuition, which are however often very particularly tailored for a specific physical regime.

Although the state described depends on the context, one might ask whether there is a general approach to obtaining an efficient representation of many-body wavefunctions, applicable to different types of systems and different degrees of correlation.
Exploring this question and developing general schemes to efficiently capture many-body effects of interest with a compact functional representation of the wavefunction is one of the central objectives of this work.
Especially by exploiting direct correspondences of this problem with tasks from the field of \ac{ML}, the main goal is to develop a flexible model to represent many-body states.
This should, in particular, incorporate expected physical structure for different scenarios but overcome the limitations of rigid system-specific representations.

Following the same motivations, significant progress has been made towards this goal recently with the introduction of quantum states parametrized by \acp{NN}.
While it was not necessarily the first presentation of approaches using \acp{NN} to parametrize a wavefunction~\cite{lagarisArtificialNeuralNetwork1997}, especially the application of \acp{RBM} for many-body spin systems discussed in Ref.~\cite{Carleo2017}, which was published in 2017, sparked an increased interest in utilizing such approaches for many-body problems, resulting in a plethora of publications building upon such ideas over the last few years~\cite{abrahamsenTamingSignProblem2022, bagrovKineticSamplersNeural2020, barrettAutoregressiveNeuralnetworkWavefunctions2021, battagliaMachineLearningWavefunction2022, Kochkov2018, Borin2019, bukovLearningGroundState2020, caiApproximatingQuantumManybody2018, Carleo2017, Carleo2018, carrasquillaNeuralNetworksQuantum2021, cassellaDiscoveringQuantumPhase2022, cevenNeuralnetworkQuantumStates2022, chenNeuralNetworkEvolution2021, chenSystematicImprovementNeural2022, Choo2019, Choo2019a, chooSymmetriesManybodyExcited2018, Clark2018, czischekDataEnhancedVariationalMonte2022, czischekNEURALNETWORKSIMULATIONSTRONGLY2020, dengQuantumEntanglementNeural2017, donatellaDynamicsAutoregressiveNeural2022, Dornheim2019, duricMachineLearningQuantum2021, fuLatticeConvolutionalNetworks2022, gaoAbInitioPotentialEnergy2021, gerardGoldstandardSolutionsSchr2022, girardinBuildingSeparableApproximations2021, hanNeuralQuantumStates2021, havlicekAmplitudeRatiosNeural2022, Heinen2019, Hendry2019, hendryChebyshevExpansionSpectral2021, Hermann2019, hermannAbinitioQuantumChemistry2022, hibat-allahSupplementingRecurrentNeural2022, hibat-allahVariationalNeuralAnnealing2021, HibatAllah2020, hofmannRoleStochasticNoise2021, Hutter2020, inackNeuralAnnealingVisualization2022, inuiDeterminantfreeFermionicWave2021, irikuraNeuralnetworkQuantumStates2020, jonssonNeuralnetworkStatesClassical2018, Kessler2019, klassertVariationalLearningQuantum2021, kochkovLearningGroundStates2021, Liang2018, liangHybridConvolutionalNeural2021, liFermionicNeuralNetwork2021, linExplicitlyAntisymmetrizedNeural2021, linScalingNeuralnetworkQuantum2021, lovatoHiddennucleonsNeuralnetworkQuantum2022, luoInfiniteNeuralNetwork2021, medvidovicClassicalVariationalSimulation2021, morenoFermionicWaveFunctions2021, Neugebauer2020, Nomura2017, nomuraDiractypeNodalSpin2021, nomuraHelpingRestrictedBoltzmann2020, nomuraInvestigatingNetworkParameters2022, nomuraPurifyingDeepBoltzmann2021, noormandipourRestrictedBoltzmannMachine2022, parkExpressivePowerComplexvalued2021, parkGeometryLearningNeural2020, patrascuAreClassicalNeural2022, peiCompactNeuralnetworkQuantum2021, peiNeuralnetworkQuantumStates2021, pesciaNeuralNetworkQuantumStates2021, Pfau2019, spencerBetterFasterFermionic2020, Puente2020, rehTimedependentVariationalPrinciple2021, rothGroupConvolutionalNeural2021, rzadkowskiArtificialNeuralNetwork2021, scherbelaSolvingElectronicSchr2021, schmittJVMCVersatilePerformant2021, schmittQuantumManybodyDynamics2020, Schuett2019, Sharir_2020, sharirNeuralTensorContractions2021, Shinjo2019, Sinitskiy2019, Stokes2020, stokesContinuousvariableNeuralnetworkQuantum2021, sunEntanglementFeaturesRandom2022, Szabo2020, szabó_2021, vargas-calderonEmpiricalStudyQuantum2022, vicentiniNetKetMachineLearning2021, vicentiniPositivedefiniteParametrizationMixed2022, Vieijra2019, vieijraManyBodyQuantumStates2021, viterittiAccuracyRestrictedBoltzmann2022, Westerhout2019, westerhoutUnveilingGroundState2022, wilsonSimulationsStateoftheartFermionic2021, wilsonWaveFunctionAnsatz2022, wuTensorNetworkQuantum2022, Yang2019, Zen2019, zhangContinuousvariableOptimizationNeural2021, zhangGroundStateSearch2022, zhangHamiltonianReconstructionMetric2021, zhaoScalableNeuralQuantum2022, zhengSpeedingLearningQuantum2021, zouLearningCompassSpin2021, harneyEntanglementClassificationNeural2020, frankLearningNeuralNetwork2021, wangSolvingSchrOdinger2022, humeniukAutoregressiveNeuralSlaterJastrow2022, Ferrari2019, zhaoOvercomingBarriersScalability2021, albashQuantumInspiredTemperingGround2022, Han2018, Luo2019, liInitioCalculationReal2022, chenNeuralNetworkQuantum2022, knitterNeuralNetworkSimulation2022, liNeuralnetworkbasedMultistateSolver2021,mattheakisFirstPrinciplesPhysicsinformed2022, rothHighaccuracyVariationalMonte2022, zhangUnderstandingEliminatingSpurious2022, vonglehnSelfAttentionAnsatzAbinitio2022, yangDeepneuralnetworkSolutionInitio2022, passettiCanNeuralQuantum2022, Huang2017, sharirNeuralVariationalMonte, Sehayek, Shi, Saito2017, Zheng2019, 2007.14282, Chen2018, Czischek2019, Gao2017, Gardas2018, Glasser2018,Hartmann2019, Torlai2019a, Nagy2019, Vicentini2019, Pilati2019, LopezGutierrez2019, Jia2019, Levine2018, Schmitt2018, Inack2018, Zhang2019, Ruggeri2018, Saito2018, Saito2018a, Rocchetto2018, Torlai2018, McBrian2019, Pastori2019, Kaubruegger2018, Yoshioka2019, Sellier2019, Lu2019, Wu2019, martynVariationalNeuralNetworkAnsatz2022, chenSimulating1DLattice2022, foreDiluteNeutronStar2022, Freitas_2018, https://doi.org/10.48550/arxiv.2212.13678, https://doi.org/10.48550/arxiv.2212.13453, doi:10.1142/S0218271817430209, https://doi.org/10.48550/arxiv.2301.02683, https://doi.org/10.48550/arxiv.2301.03755, https://doi.org/10.48550/arxiv.2301.06788, https://doi.org/10.48550/arxiv.2301.09923, Yang2020, doi:10.1063/5.0040785, https://doi.org/10.48550/arxiv.2302.00173, zhao2022, https://doi.org/10.48550/arxiv.2302.02523, https://doi.org/10.48550/arxiv.2302.01941, https://doi.org/10.48550/arxiv.2302.04168, https://doi.org/10.48550/arxiv.2205.14962, https://doi.org/10.48550/arxiv.2302.04919, https://doi.org/10.48550/arxiv.2211.04614, https://doi.org/10.48550/arxiv.2302.08965, https://doi.org/10.48550/arxiv.2302.11588, https://doi.org/10.48550/arxiv.2303.08184, scherbela2023foundation, hibatallah2023investigating, abrahamsen2023convergence, mezera2023neural, chen2023autoregressive, radu2023deep, PhysRevResearch.5.013216, roth2020iterative, joshi2023ground, Bennewitz_2022, wei2023neuralshadow, bokhan2022improving, wu2023supervised, pescia2023messagepassing, lou2023neural, ma2023attentionbased, kim2023neuralnetwork, zhang2023scorebased, Zhang_2023, sprague2023variational}.
Since the introduction of these states parametrized by \ac{NN} type function approximators, commonly referred to as \acp{NQS}, these have been applied in various different contexts successfully, reaching accuracies often challenging the state-of-the-art.
Concurrently, also the general understanding of the representative power of these models has progressed significantly, underlining their broad potential as a universal tool for numerical quantum studies.
It has also been shown that \ac{NN} representations, can in various scenarios, at least theoretically, represent target states of interest more efficiently than common tensor network decompositions of the state~\cite{dengQuantumEntanglementNeural2017, Borin2019, sharirNeuralTensorContractions2021, sunEntanglementFeaturesRandom2022, wuTensorNetworkQuantum2022, Chen2018, Levine2018}.

Well-established representations of states with tensor networks are, just as \acp{NN}, in principle, also able to describe a state up to essentially arbitrary accuracy.
However, these are usually particularly designed to capture states within a very specific corner of the Hilbert space efficiently, typically the ones with a low degree of entanglement~\cite{Orus2013}.
While this construction imposes some restrictions on the states that can be modelled efficiently, often it is exactly this class that is important in many physically relevant scenarios.
The specific construction of the state based on tensor decompositions also provides some very appealing practical characteristics, e.g., making it possible to evaluate many expectation values of interest for the subclass of \acp{MPS} efficiently without requiring additional approximations.
Moreover, many very powerful schemes, such as the \ac{DMRG}~\cite{Schollwoeck2011}, have been introduced to infer appropriate tensor network representations very efficiently, making such approaches the de-facto standard for obtaining the best solutions in many settings.

While states described by \acp{NN} are less restricted to targets with a low degree of entanglement, current schemes to find them typically rely on stochastic approximations of expectation values as employed in the framework of \ac{VMC} techniques~\cite{Becca2017}.
In a numerical application it is not always easy to distinguish between shortcomings of the underlying model and the method used to find the final representation (i.e., to `learn' the state).
Nonetheless, there are indications that the stochastic nature of the \ac{VMC} approaches and the applied optimization protocols sometimes hinder the practical applicability of the introduced highly flexible \ac{NQS} representations~\cite{bukovLearningGroundState2020,Westerhout2019, Szabo2020}.
With the success of the applied approach also significantly influenced by practical numerical challenges, naturally the question emerges what the best choice of method for a given problem is, i.e., how can the highest accuracy be reached  in a practical application with an affordable computational effort.
This can mean to practically choose between different representation classes and methods.
But more specifically for the case of \acp{NN} this also means that one has to find a network architecture that is suitable for the given problem of interest.
It is by no means obvious how this can be achieved in a systematic and efficient way.
In addition to the practical task of finding the model that is performing the best numerically, another interesting conceptual question is how the compressed representation of the state can be interpreted and how exactly it encodes the physical characteristics.

\section{Objectives and structure of this work}

The two challenges outlined above, systematically defining a compressed representation and interpreting the obtained solutions, are by no means unique to the modelling of quantum states.
Specifically, these are also of great interest within the general field of \ac{ML} and data science.
Various different concepts and methods have been introduced in order to understand and describe data, and \ac{NN} based representations are only a part of all the techniques that are commonly applied.
Especially the study of data within rigorous probabilistic frameworks can often provide a very clear, well-understood interpretation significantly helping with a systematic extraction of the final model.
One such approach is that of \ac{GPR} --- rigorously modelling data descriptions probabilistically~\cite{Rasmussen2006}.
Especially the large degree of interpretability provides a compelling argument for exploring such techniques for the description of many-body quantum states.

The family of quantum states emerging from this motivation, the \ac{GPS}, is the key element of interest in this work.
Building on the fundamental principles of the \ac{GPR} framework, it is a complete representation of the many-body wavefunction, which, in principle, can describe any state from the Hilbert space.
Although this is an interesting feature, also true for the \ac{NN} and the tensor network descriptions, this is ultimately of little practical relevance.
The central research question discussed in this work is rather if, and how well, the \ac{GPS} can describe states of physical interest in a computationally feasible way.
Overall, it is shown that the \ac{GPS} makes it possible to bring many-body quantum states into a compact form in various settings.
The main concepts and results around the \ac{GPS} studied within this dissertation, are (partially) similarly presented in the following publications:
\begin{itemize}
    \item \fullcite{Rath2020}, subsequently referenced as Ref.~\cite{Rath2020}
    \item \fullcite{Glielmo2020}, subsequently referenced as Ref.~\cite{Glielmo2020}
    \def\thefootnote{$^\ast$}\footnotetext{equal contribution}
    \item \fullcite{boothQuantumGaussianProcess2021}, subsequently referenced as Ref.~\cite{boothQuantumGaussianProcess2021}
    \item \fullcite{rath2023framework}, subsequently referenced as Ref.~\cite{rath2023framework}.
\end{itemize}
Some elements of the framework are also discussed in Ref.~\cite{battagliaMachineLearningWavefunction2022}.

The expressibility of the \ac{GPS} is defined by two different components, a kernel function, which can be identified as the covariance between function points in the \ac{GPR} picture, and a set of physical data points.
With these two main ingredients, the \ac{GPS} can be constructed based on a large degree of physical intuition, and various correlation properties, which explicitly underpin other physically-motivated models, can also easily be incorporated.
However, the description is not necessarily limited to such correlation properties and the underlying Bayesian framework makes it possible to select the most relevant correlation properties from some reference wavefunction data.
Ultimately, the \ac{GPS} therefore combines the idea of using a highly flexible model inspired by \ac{ML} approaches, with the more traditional paradigms to model correlation properties based on an inherent physical structure expected for the state.
Another very similar approach, also building on the idea of using kernel methods to model the many-body wavefunction, is presented in Ref.~\cite{https://doi.org/10.48550/arxiv.2303.08902}.

This work introduces the tools and concepts to use the \ac{GPS} to tackle the challenges of quantum many-body physics mainly from a point of view based on practical numerical considerations.
More specifically, this means that the central research questions are mostly discussed based on numerical results, and it is presented how the general concepts can be applied efficiently within modern computing frameworks.
The main goal of this work is to provide a holistic description of the main strengths and current challenges of using the \ac{GPS} as a tool for many-body simulations from the perspective of a numerical practitioner.

This thesis is structured into a total of seven chapters presenting different elements contributing towards this goal.

Chapter \ref{ch:theoretical_background} provides foundational background on the task of representing many-body states efficiently.
In this chapter, the many-body problem, as it is considered in this work, is formalized and the notation is established.
This part also outlines intrinsic properties that are desired for the compact representation of many-body states and a selection of different established wavefunction parametrizations is presented that significantly contribute to the intuition behind the \ac{GPS}.
These include \acp{MPS}, \acp{CPS}, Jastrow wavefunctions, as well as \acp{NQS}.
In addition to introducing the general framework of \ac{VMC} to find and study many-body ground states, the chapter moreover includes background information on the two main quantum systems used for benchmarking in this work.
These are the $J_1$-$J_2$ model of fixed spin-1/2 constituents, as well as the Fermi-Hubbard model, interpretable as a simplified model of interacting electrons moving on a lattice structure.
The introduction of a system of freely moving electrons also highlights key conceptual differences between the two types systems and motivates techniques to incorporate the required Fermionic characteristics into the descriptions.

In the subsequent chapter (chapter \ref{ch:GPS_introduction}), the concepts of Gaussian Processes for function regression and how these concepts can be used as representation of many-body states are introduced.
This leads to the general definition and introduction of the \ac{GPS} model, with the chapter particularly focussing on how to design the different building blocks defining the state.
This involves the definition of suitable, physically motivated kernel functions for the model, as well as approaches to obtain a final model based on Bayesian regression techniques from available wavefunction data.
Numerical results are presented how these statistical approaches help to obtain a compact representation of given states, and it is presented how these approaches efficiently extract the important information of a given target state in a physically interpretable way.
Lastly, the chapter also presents some further insights about the general expressibility of the model, outlining how the \ac{GPS} can be related to other models and approaches.

The central element of chapter \ref{ch:GPS_VMC} is the practical application of \acp{GPS} as a numerical tool to explore many-body systems by means of \ac{VMC} techniques.
Three different potential approaches are presented to achieve the extraction of the model.
The first is based on extrapolating the wavefunction information from small systems, allowing for an exact numerical treatment, to larger systems of interest.
Building onto this approach, a bootstrapping approach is presented approximating unknown ground states by an iterative scheme, alternating between a variational optimization of the continuous model parameters, and compression of the current state into a compact \ac{GPS} representation.
In the third scheme, the variational optimization is extended to configurations, required for the definition of the \ac{GPS}, which are parametrized as general unentangled product states.
The resulting compact form of the \ac{GPS} obtained via this approach, referred to as \ac{qGPS}, is the central element of interest for the following chapters.

Complementing the first numerical results for conceptually simple benchmarking systems presented in chapter \ref{ch:GPS_VMC}, further numerical results are presented in chapter \ref{ch:ab_initio_GPS}.
This chapter describes the application of the methodology to study the electronic structure emerging from ab-initio descriptions of molecules --- a task of very practical significance for the description of chemical properties.

In chapter \ref{ch:qGPS_learning}, specific model construction properties of the \ac{GPS} are explored relating the model to tensor decomposition approaches by identifying the \ac{GPS} as a \ac{CP} decomposition of the log wavefunction amplitudes.
An \ac{ALS} scheme is introduced, also utilizing the intrinsic connection of the \ac{GPS} to rigorous probabilistic modelling principles, to aid the process of faithfully inferring a compact state description based on limited configurational samples.
In spirit related to concepts applied in \ac{DMRG}, this scheme relies on an iterative sweeping through the system in that at each step some parameters of the model are inferred from presented wavefunction data.
Applications of the Bayesian sweeping scheme to achieve a practical compression of quantum states are discussed, including the learning of ground states for which no exact data is directly accessible~\cite{Kochkov2018}.
This also includes the extension beyond the task of describing many-body states for `standard' \ac{ML} tasks such as image recognition which is discussed briefly in section \ref{sec:classical_ML_GPS}.

Final concluding remarks, providing additional perspectives, and more general interpretations of the results and findings presented in this thesis, are given in chapter \ref{ch:conclusions}.
Additionally, this chapter also outlines further potential research directions and extensions of the \ac{GPS}.

With a strong focus on the direct numerical application of the different methods described in this work, the discussed results were mostly obtained from specific algorithmic implementations.
Several elements of the concepts are implemented in the \textit{GPSKet} library (\url{https://github.com/BoothGroup/GPSKet}), an add-on to the VMC software package \textit{NetKet}~\cite{vicentiniNetKetMachineLearning2021, Carleo2019}.
The implementations for the numerical tests presented in this work, also rely on further open-source software tool kits that greatly aided the computational execution of the ideas.
In addition to several well-established software tools with a broad application scope (including \textit{JAX}~\cite{jax2018github}, \textit{NumPy}~\cite{NumPy} and \textit{SciPy}~\cite{SciPy}), these also included the application-specific software packages \textit{Block}~\cite{Block,olivares-amayaAbinitioDensityMatrix2015}, \textit{Hyperopt}~\cite{bergstraMakingScienceModel2013, Bergstra2015}, \textit{ITensor}~\cite{2007.14822, itensor-r0.3}, \textit{mVMC}~\cite{Misawa2017}, \textit{numpy-ml}~\cite{bourginNumpyml2021}, \textit{pfapack}~\cite{wimmerEfficientNumericalComputation2012}, \textit{PySCF}~\cite{sunPySCFPythonbasedSimulations2018,sunRecentDevelopmentsPySCF2020a}, \textit{scikit-learn}~\cite{scikit-learn} with the \textit{sklearn-bayes} add-on~\cite{shaumyanAmazaspShumikSklearnbayes2022}, as well as \textit{QuSpin}~\cite{weinbergQuSpinPythonPackage2017, weinbergQuSpinPythonPackage2019}.
Additional software which was used to perform the numerical tests but is not yet included in the GPSKet package, as well as the data presented in this work, can be made available upon request.

\acbarrier

\chapter{Modelling quantum many-body wavefunctions --- the background}
\label{ch:theoretical_background}

\section{Why simulating quantum physics is hard: The many-body problem and entanglement}
\sectionmark{The many-body problem and entanglement}

\subsection{The many-body problem}
This work mostly focusses on describing the state of a quantum many-body system with discretized degrees of freedom.
Practically, these systems can take many forms and represent different physical scenarios.
However, the overall setting considered is very general, making it possible to study different physical systems with the methods introduced in this work.
Specifically, systems of $L$ different interacting quantum modes are considered, where each of the modes corresponds to a discrete local Hilbert space.
This means that the Hilbert space of the many-body states of interest, here denoted as $\mathcal{H}$, emerges from the tensor product of the local Hilbert spaces associated with the different modes, $\mathcal{H}_i$, according to
\begin{equation}
    \mathcal{H} = \bigotimes_{i=1}^L \mathcal{H}_i.
\end{equation}
Although not necessarily limited to this case, in the examples presented in this work, the local Hilbert spaces all have the same finite dimension, in the following denoted by $D$, for all the $L$ modes.
As a working example, also one of the examples studied in this work, one can consider an array of $L$ fixed spin-1/2 quantum systems that are arranged on some lattice structure.
In this case, the dimensions of the local Hilbert spaces are $D=2$.
Because a majority of this work focusses on the description of lattice models, the system modes will in the following also be referred to as lattice sites.

With the definition of the full Hilbert space, it is possible to construct a basis for this space based on tensor products of basis states of the local spaces.
In the following, the basis of the full system is denoted by states $|\mathbf{x}\rangle = \bigotimes_{i = 1}^L | x_i \rangle$, where $| x_i \rangle$ is a state from the local Hilbert space basis at mode $i$.
Practically, the label $x_i$ just works as an index identifying the local basis states.
It takes values $x_i = 1 \ldots D$.
For the example of a spin array, the local basis states $| x_i \rangle$ can for example be constructed with the two $\hat{S}_z$ eigenstates corresponding to a spin-up and spin-down realization.
In this case the two local basis states are defined as $| 1 \rangle = |\uparrow \rangle$ and $| 2 \rangle = | \downarrow\rangle$.
This is only one potential choice to construct the basis of the full Hilbert space.
The techniques outlined in the following will ultimately depend on the specific choice of the basis that will be used to represent the state of the system.
However, it will be shown that good results can be achieved with rather generic basis choices, such as the one basis on the $\hat{S}_z$ eigenstates for the studied spin systems.
Some additional investigations into the basis choice will be presented for the context of ab-initio calculations in chapter \ref{ch:ab_initio_GPS}.

With the construction of the basis as above, each basis state, $|\mathbf{x}\rangle$, can be understood as one potential many-body configuration of the system, such as a specific, experimentally observable, arrangements of the spins on the lattice.
Naturally, a general state of the system, denoted as $| \Psi \rangle$\footnote{The quantum states specified in this work will generally be defined without being explicitly normalized (i.e., the appropriate normalization of the state needs to be included for the evaluation of expectation values).} in the following, can be represented as a wavefunction over all the configurations of the computational basis according to
\begin{equation}
    | \Psi \rangle = \sum_{\mathbf{x}} \Psi(\mathbf{x}) \, | \mathbf{x} \rangle = \sum_{x_1, x_2, \ldots x_L} \Psi(x_1, x_2, \ldots x_L) \, |x_1 \rangle \otimes |x_2 \rangle \otimes \ldots \otimes |x_L \rangle.
\end{equation}
The sum runs over all basis states and $\Psi(\mathbf{x}) = \Psi(x_1, x_2, \ldots x_L)$ denotes the wavefunction amplitude for a specific configuration $| \mathbf{x} \rangle = |x_1 \rangle \otimes |x_2 \rangle \otimes \ldots \otimes |x_L \rangle$.
This representation highlights a key issue for numerical descriptions of many-body body quantum states: The number of basis states scales as $D^L$, i.e., exponentially in the size of the system $L$.
Numerical approaches working exactly with such a direct representation of the wavefunction, will therefore always encounter an exponential scaling in memory and for practical calculations also in computer time.
Consequently, such direct approaches are intrinsically limited to very small systems.

\begin{figure}[htb!]
    \centering
\begingroup%
  \makeatletter%
  \providecommand\color[2][]{%
    \errmessage{(Inkscape) Color is used for the text in Inkscape, but the package 'color.sty' is not loaded}%
    \renewcommand\color[2][]{}%
  }%
  \providecommand\transparent[1]{%
    \errmessage{(Inkscape) Transparency is used (non-zero) for the text in Inkscape, but the package 'transparent.sty' is not loaded}%
    \renewcommand\transparent[1]{}%
  }%
  \providecommand\rotatebox[2]{#2}%
  \newcommand*\fsize{\dimexpr\f@size pt\relax}%
  \newcommand*\lineheight[1]{\fontsize{\fsize}{#1\fsize}\selectfont}%
  \ifx\svgwidth\undefined%
    \setlength{\unitlength}{453.53803019bp}%
    \ifx\svgscale\undefined%
      \relax%
    \else%
      \setlength{\unitlength}{\unitlength * \real{\svgscale}}%
    \fi%
  \else%
    \setlength{\unitlength}{\svgwidth}%
  \fi%
  \global\let\svgwidth\undefined%
  \global\let\svgscale\undefined%
  \makeatother%
  \begin{picture}(1,0.36607956)%
    \lineheight{1}%
    \setlength\tabcolsep{0pt}%
    \put(0,0){\includegraphics[width=\unitlength,page=1]{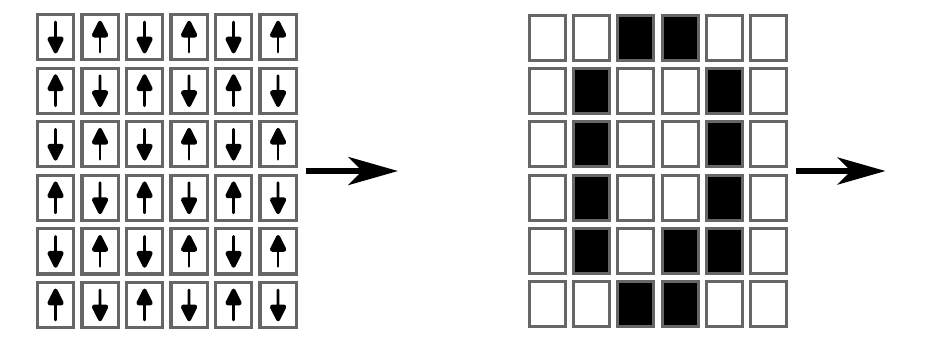}}%
    \put(0.43233883,0.16987561){\color[rgb]{0,0,0}\makebox(0,0)[lt]{\lineheight{1.25}\smash{\begin{tabular}[t]{l}{\huge$\Psi$}\end{tabular}}}}%
    \put(0.94829089,0.16987561){\color[rgb]{0,0,0}\makebox(0,0)[lt]{\lineheight{1.25}\smash{\begin{tabular}[t]{l}{\Huge0}\end{tabular}}}}%
  \end{picture}%
\endgroup%

    \caption[Analogy between the many-body problem in quantum physics and in a standard \acl{ML} tasks]{Analogy between the many-body problem in quantum physics and in a standard \ac{ML} tasks.
             The many-body wavefunction defining the system state associates amplitudes to an exponentially large set of basis configurations, e.g., it associates an amplitude $\Psi$ with each possible arrangements of spins on a two-dimensional lattice (left).
             In an exemplified task of digit recognition, a mapping from an input array of black and white pixels to a digit class label is extracted (right).}
    \label{fig:many_body_problem}
\end{figure}

The issue of an exponentially scaling state space dimensionality is commonly referred to as the `many-body problem' in the context of quantum mechanics.
The underlying problem is however not unique to numerical descriptions of quantum systems.
In fact, a direct link between the many-body problem and standard \ac{ML} tasks, e.g., image recognition, naturally emerges.
Such a task could be that of inferring a digit based on a black and white scan of some handwritten input.
In the digital representation of the scan, the image might simply be given by a two-dimensional array of pixels, each of which can either be black or white.
Based on this digital representation of the scanned image\footnote{A more detailed example of a standard digit recognition setup is presented in section \ref{sec:classical_ML_GPS}.}, the connection to many-body quantum follows naturally.
Each pixel can be understood as a local two-dimensional quantum system, similar to a spin-1/2 degree of freedom.
Images correspond to specific configurations of black or white pixels from the $2^L$ dimensional configuration space, associated with the $L$ pixels.
Whereas the central goal in the digit recognition task is to identify a digit from the exponentially large space of images, the description of the quantum state requires mapping inputs from the Hilbert space to their wavefunction amplitude.
The conceptual similarity between the two tasks is visualized in figure~\ref{fig:many_body_problem}.

The outlined analogy between the quantum many-body problem and standard \ac{ML} tasks is a central cornerstone motivating the approaches introduced in this work.
The following chapters explore how well-established \ac{ML} techniques can be transferred to the direct description of quantum states, contributing to the fast increasing applications of \ac{ML} paradigms to study quantum phenomena~\cite{Carleo2019a, dawidModernApplicationsMachine2022}.

The analogy between \ac{ML} tasks and describing quantum states can also be investigated from a reversed perspective, inspiring studies into whether the information encapsulated in many-body wavefunctions can also be used to improve current \ac{ML} algorithms.
Different novel \ac{ML} schemes have been introduced recently, explicitly building onto quantum physical concepts.
These essentially fall into one of two branches.
On the one hand it is possible to directly exploit the intrinsic complexity of many-body states and design specific quantum \ac{ML} algorithms that are designed to run (at least partially) on quantum computing hardware~\cite{ballarinEntanglementEntropyProduction2022, beerDissipativeQuantumGenerative2021, bernerQuantumBayesianNeural2021, biamonteQuantumMachineLearning2017, buffoniNewTrendsQuantum2021, caroGeneralizationQuantumMachine2021, chenEndtoendTrainableHybrid2021, chenNovelArchitectureParameterized2022, chenQuantumAlgorithmsPrediction2021, chenQuantumGaussianProcess2021, dborinMatrixProductState2021, desouzaClassicalArtificialNeural2021, desouzaQuantumWalkTrain2019, garciaSystematicLiteratureReview2022, gentinettaComplexityQuantumSupport2022, giliEvaluatingGeneralizationClassical2022, glickCovariantQuantumKernels2021, heidariTheoreticalFrameworkLearning2021, hubregtsenTrainingQuantumEmbedding2021, jagerUniversalExpressivenessVariational2022, jerbiQuantumMachineLearning2021, kobayashiOverfittingQuantumMachine2022, kwakQuantumNeuralNetworks2021, landmanQuantumAlgorithmsUnsupervised2021, liQuantumKernelsSqueezedstate2021, liQuantumNeuralNetwork2022, luQuantumConvolutionalNeural2021, manginiQuantumComputingModels2021, maroneseQuantumActivationFunctions2022, massoliLeapEntanglementNeural2021, petitzonNewQuantumNeural2022, sancho-lorenteQuantumKernelsLearn2021, schuldIntroductionQuantumMachine2015, schuldQuantumAdvantageRight2022, simeoneIntroductionQuantumMachine2022, tianRecentAdvancesQuantum2022, wossnigQuantumMachineLearning2021, wuProvableAdvantageQuantum2021, zhangQuantumAlgorithmNeural2022, zhangTrainabilityDeepQuantum2021, zhaoReviewQuantumNeural2021, zoufalGenerativeQuantumMachine2021, Biamonte2018}.
On the other hand, methods developed for the efficient description of many-body states with classical computing architectures have also been applied to help with the efficient representation of the input-output mappings within \ac{ML} tasks~\cite{baiUnsupervisedRecognitionInformative2022, barrattImprovementsGradientDescent2022, Beny2018, bhatiaMatrixProductState2019, bradleyModelingSequencesQuantum2019, chengSupervisedLearningProjected2020, convyMutualInformationScaling2021, dborinMatrixProductState2021, dymarskyTensorNetworkLearn2021, glasserExpressivePowerTensornetwork2019, gonzalezLearningDensityMatrices2021, hanUnsupervisedGenerativeModeling2018, linTensorNetworkSupervised2021, liuTensorNetworksUnsupervised2021, luTensorNetworksEfficient2021, pozas-kerstjensPhysicsSolutionsMachine2022, senguptaTensorNetworksMachine2022, stoudenmireSupervisedLearningQuantumInspired2017, strashkoGeneralizationOverfittingMatrix2022, vieijraGenerativeModelingProjected2022, Zhang2018b, zunkovicDeepTensorNetworks2022, sommerEntanglingSolidSolutions2022, https://doi.org/10.48550/arxiv.2212.14076}.
Although, the main focus of this work is to transfer the \ac{ML} techniques around the concepts of \ac{GPR} to numerically study complex quantum systems, a short exploration of how the emerging models might be useful in a standard \ac{ML} context is also discussed in section \ref{sec:classical_ML_GPS}.

\subsection{Correlation and entanglement}
The many-body problem outlined in the previous section is a significant hindrance for exactly accessing states from the underlying Hilbert space numerically.
However, just because many-body states are defined with respect to an exponentially large Hilbert space, this does not necessarily mean that states of physical importance always `explore' the full complexity of the state space.
One might, for example, consider scenarios in which no interaction takes place between the $L$ different system modes.
In this case, wavefunctions of interest (in particular eigenstates of the Hamiltonian) will separate as a product of local states $|\Psi_i\rangle$ from the local Hilbert spaces over all $L$ modes of the system.
This means that these states are given by \textit{Product States}, which are states described as follows.
\begin{ansatz}[Product States]
    Product states describe unentangled states according to
    \begin{equation*}
        | \Psi \rangle = \bigotimes_{i=1}^L | \Psi_i \rangle.
    \end{equation*}
    Its wavefunction amplitudes evaluate, in the chosen basis, to
    \begin{equation*}
        \Psi(\mathbf{x}) = \Psi(x_1) \Psi(x_2) \ldots \Psi(x_L).
    \end{equation*}
\end{ansatz}

While the factorization of a state as above can in some cases provide a reasonable description, in particular the emergence of entanglement between the system constituents gives rise to various interesting physical quantum phenomena.
Intuitively, entanglement, which is formally defined as the inability to represent the many-body state as a product state, causes correlations in the experimental outcomes of measurements of the system.
More specifically, entanglement between parts of the system causes the measurement of the state on one subsystem to be correlated with the outcome of the measurement on another.
While the ability for entanglement between subsystems to emerge is a fundamental property of quantum systems distinguishing them from classical descriptions, this also significantly complicates the numerical treatment of quantum many-body systems.
Appropriately capturing such correlations emerging in the system is therefore a key challenge in order to describe various quantum phenomena efficiently.

To go beyond the understanding of entanglement as a binary property (that is either present or not), different measures have been introduced to quantify entanglement entropy, i.e., the degree to which two subsystems $A$ and $B$ are entangled~\cite{eisertAreaLawsEntanglement2010}.
One such common measure is the von Neumann entanglement entropy.
It is defined as the negative trace over the operator $\rho_{A} \log (\rho_{A})$, where $\rho_{A}$ is the reduced density matrix for subsystem $A$, corresponding to the state for which the environment $B$ was traced out.
For pure states, this quantity can similarly be evaluated by representing the full state via the Schmidt decomposition as a linear combination of tensor products between orthonormal states from subsystem $A$ and from subsystem $B$.
That is to say, the state is decomposed according to $| \Psi \rangle = \sum_i c_i |\Psi_i\rangle_A \otimes |\Psi_i\rangle_B$, where the orthonormal set of states $|\Psi_i\rangle_A$ ($|\Psi_i\rangle_B$) are defined across the space associated with the modes in subsystem $A$ ($B$).
With this decomposition the entanglement entropy evaluates to
\begin{equation}
    \mathcal{S}(|\Psi\rangle) = -\sum_i |c_i|^2 \log(|c_i|^2),
\end{equation}
and it can directly be seen that this quantity vanishes for states that can be decomposed as a tensor product over the two subsystems.
Although it can be difficult to exactly evaluate the entanglement entropy of a state --- another manifestation of the many-body problem --- some formal results have been established that characterize the amount of entanglement expected for some states of interest~\cite{eisertAreaLawsEntanglement2010}.
Examples are area laws for the entanglement entropy.
These prove the emergence of non-vanishing entanglement for which the entropy however only grows with the size of the boundary between subsystems $A$ and $B$ (and not with their size).
Such rigorous results provide a clear intuition about correlations emerging in quantum systems therefore identifying a particular structure of the state that can potentially be exploited for efficient representations.


Whereas the entanglement between system constituents does not depend on the chosen basis representation, it can, to some degree, depend on the perspective.
In particular for systems of moving indistinguishable particles, which can occupy the different modes of the system, one might either look at the correlations between particles, or between the different modes~\cite{dingConceptOrbitalEntanglement2021,dingConceptOrbitalEntanglement2021,dingFermionicEntanglementCorrelation2022}.
These two perspectives can provide a very different picture of the correlations and give different answers to whether a state considered correlated or not.
This is exemplified by indistinguishable electrons moving between different modes, a setting also studied in this work in the context of Fermi-Hubbard models and ab-initio quantum chemistry calculations.
In a system without any interaction between the particles, the Hamiltonian eigenstates can be represented by single \acp{SD}, i.e., anti-symmetrized products of single particle wavefunctions.
Although the detailed definition of entanglement for such systems of indistinguishable particles is not necessarily obvious~\cite{benattiEntanglementIndistinguishableParticle2020}, such a product representation would typically be considered an uncorrelated state.
It does not incorporate particle-particle correlations and can be obtained from mean-field approaches.
In the picture of modes (i.e., here the lattice sites or molecular orbitals) however, this state can display a large degree of entanglement~\cite{ganahlDensityMatrixRenormalization2022}.

In this work, the term `correlation' is generally used to denote intrinsic non-trivial correlation properties that need to be extracted by the wavefunction representation.
Ultimately, the methods presented in this work aim to represent quantum states generally, especially ones that are intrinsically hard to describe due to the emerging correlation.
That is to say, these can typically not be obtained by mean-field type methods nor can these be described by simple product states in the mode picture.
These will therefore often exhibit quantum correlations, both between the modes and also between the indistinguishable particles (if the system consists of such).

From a simplified perspective, the two central elements intrinsically limiting the quantum simulation of multiple particles are therefore identified as the exponential scaling of the Hilbert space, as well as (potentially strong) entanglement and correlations building up in various states of interest.
These properties can also be seen as the central source for theoretical advantages of quantum computing algorithms over classical counterparts in some settings~\cite{nielsenQuantumComputationQuantum2002, meyerInvestigatingStudentInterpretations2022}.
On the one hand, this means that the direct simulation of relevant many-body systems might therefore provide an important application of future quantum hardware.
On the other hand, developing approaches to make the many-body problem computationally tractable with classical algorithms is thus also of great importance.
Not only can they help to identify which quantum descriptions are in fact accessible with classical simulations, despite the dimensionality of the underlying Hilbert space, but they might therefore also offer additional techniques to verify performed quantum computations.

\section{Expressing many-body states compactly}

\subsection{Product separability and size-extensivity}
\label{sec:product_sep}
The main task discussed in the following is that of representing many-body states efficiently to enable efficient numerical studies of the system of interest.
In order to design a suitable representation, it is important to incorporate some physical properties into the state that build the foundation for the success of the method.

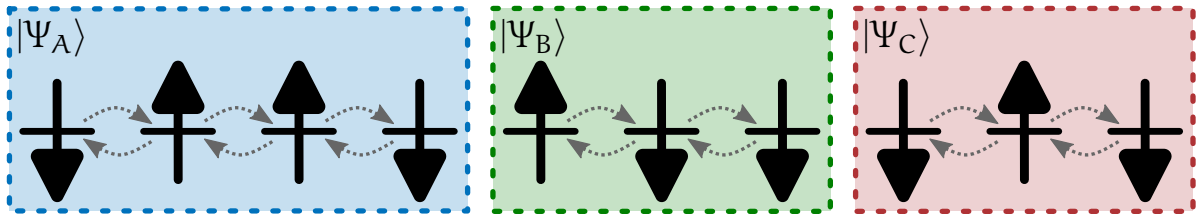
\begin{figure}[htb!]
    \centering
\begingroup%
  \makeatletter%
  \providecommand\color[2][]{%
    \errmessage{(Inkscape) Color is used for the text in Inkscape, but the package 'color.sty' is not loaded}%
    \renewcommand\color[2][]{}%
  }%
  \providecommand\transparent[1]{%
    \errmessage{(Inkscape) Transparency is used (non-zero) for the text in Inkscape, but the package 'transparent.sty' is not loaded}%
    \renewcommand\transparent[1]{}%
  }%
  \providecommand\rotatebox[2]{#2}%
  \newcommand*\fsize{\dimexpr\f@size pt\relax}%
  \newcommand*\lineheight[1]{\fontsize{\fsize}{#1\fsize}\selectfont}%
  \ifx\svgwidth\undefined%
    \setlength{\unitlength}{453.53803019bp}%
    \ifx\svgscale\undefined%
      \relax%
    \else%
      \setlength{\unitlength}{\unitlength * \real{\svgscale}}%
    \fi%
  \else%
    \setlength{\unitlength}{\svgwidth}%
  \fi%
  \global\let\svgwidth\undefined%
  \global\let\svgscale\undefined%
  \makeatother%
  \begin{picture}(1,0.17625206)%
    \lineheight{1}%
    \setlength\tabcolsep{0pt}%
    \put(0,0){\includegraphics[width=\unitlength,page=1]{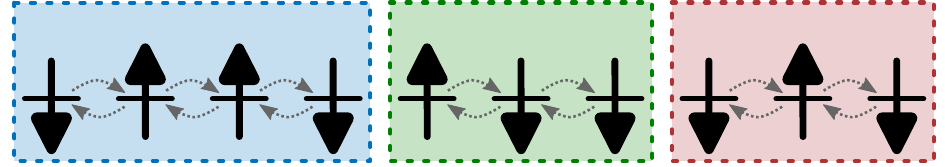}}%
    \put(0.02051958,0.14156308){\makebox(0,0)[lt]{\lineheight{1.25}\smash{\begin{tabular}[t]{l}{\large $|\Psi_A \rangle$}\end{tabular}}}}%
    \put(0.41817732,0.14156308){\makebox(0,0)[lt]{\lineheight{1.25}\smash{\begin{tabular}[t]{l}{\large $|\Psi_B \rangle$}\end{tabular}}}}%
    \put(0.71642055,0.14156308){\makebox(0,0)[lt]{\lineheight{1.25}\smash{\begin{tabular}[t]{l}{\large $|\Psi_C \rangle$}\end{tabular}}}}%
  \end{picture}%
\endgroup%

    \caption[Visualization of the product separability of states]{Visualization of the product separability of states. If the system is split into non-interacting components (indicated by the blue, green and red sectors of the displayed spin chain), then energy eigenstates factorize as a product of states only associated with individual subsystems (here $|\Psi_A\rangle$, $|\Psi_B\rangle$ and $|\Psi_C\rangle$).}
    \label{fig:size-extensivity}
\end{figure}

A main property of a state that the representation should satisfy is its product separability~\cite{Becca2017}.
This property means that a representation should be able to capture the cases of vanishing interactions between parts of the system.
Extending the concepts of product states, if a system can be decomposed into non-interacting parts then no entanglement emerges between these subsystems for the eigenstates.
Such states therefore all factorize into a product over the non-interacting parts.
Assuming for example three subsystems $A$, $B$, and $C$, the eigenstates can be represented as
\begin{equation}
    |\Psi\rangle = | \Psi \rangle_A \otimes | \Psi \rangle_B \otimes | \Psi \rangle_C \Rightarrow \Psi(\mathbf{x}) = \Psi_A(\mathbf{x}_A) \Psi_B(\mathbf{x}_B) \Psi_C(\mathbf{x}_C).
\end{equation}
The states $| \Psi \rangle_{A/B/C}$ only act on the respective subsystems, i.e., assign amplitudes to basis configurations $\mathbf{x}_{A/B/C}$, which represent the partial configuration over the subsystems.
This product factorization of a state is visualized in Fig.~\ref{fig:size-extensivity}.
Crucially, for this product factorization of an energy eigenstate, the associated energy $E$ is obtained as a sum over energies, $E = E_A + E_B + E_C$.
These subsystem energies $E_A$, $E_B$, $E_C$ denote the energies associated with the eigenstates $| \Psi \rangle_{A/B/C}$ for the corresponding contributions of the Hamiltonian.

An efficient description of quantum states should always be able to capture encapsulate a product factorization as above for any cut of the system into non-interacting parts.
This is a key requirement in order to be able to derive system properties also for larger systems efficiently.
Considering for example a translationally invariant lattice model, it can often be expected that the total energy of the system per site converges to a constant as the thermodynamic limit is approached.
Being able to represent such a scaling faithfully with a description is, especially in the quantum chemistry community, commonly referred to as a `size-extensivity' of a method~\cite{bartlettManyBodyPerturbationTheory1981}.
For a definition of an efficient quantum state representation, its size-extensivity should always be a key goal as this allows to extrapolate system properties appropriately to the thermodynamic limit (or represent large systems).

The methods introduced in the following all define efficiently treatable functional representations of the wavefunction, i.e., explicitly model the mapping between many-body configurations $\mathbf{x}$ and the wavefunction amplitude $\Psi(\mathbf{x})$.
Based on the product separability requirements for the state, it can be expected that non-trivial, size-extensive descriptions are always based on a product structure over the different system constituents.
This means that the amplitudes for the considered systems are (approximately) described as
\begin{equation}
    \Psi(\mathbf{x}) = \prod_{i=1}^L F_i(\mathbf{x}),
    \label{eq:product_separable_ansatz}
\end{equation}
where the functions $F_i$ can be seen as many-body function approximators, defining a per-mode mapping from the configuration to a scalar quantity.
The specific structure of the many-body correlators $F_i$ is directly linked to the entanglement emerging in the represented state~\cite{harneyEntanglementClassificationNeural2020} and examples for the functions $F_i$ encountered in different methods are discussed in the following sections.

Other constructions not obeying the product structure introduced above are in principle possible, e.g., using linear combinations of few product states as an ansatz.
However, these would typically not provide an efficient representation with a size-extensive increase in the representational power over a spanned product separable state (such as a single product state) as the systems get larger.
Incorporating the product structure according to \ref{eq:product_separable_ansatz} into the baseline representation therefore represents an important ingredient to the success of the method.
As will be seen for examples presented in the following, the final representation of appropriate state approximations might, nonetheless, in practice slightly deviate from this general form.
This is especially the case if additional symmetry projections are included.

\subsection{Matrix Product States}
\label{sec:MPS}

In order to find an efficient representation of a quantum state, following the general problem setup introduced in the previous section, the key question emerges what the important correlations are.
Already from a purely intuitive perspective, it can be expected that in particular local correlations are of great importance within an eigenstate of a system comprising local interactions.
Focussing on systems with local interactions is often of particular importance as typical physical interactions (e.g., the Coulomb repulsion between electrons) decay with the distance.
The expected locality of important correlations can be rigorously formalized through the analysis of the entanglement entropy scaling.
Intuitively, local correlations can be identified as ones for which the per-mode correlation functions $F_i$ are particularly governed by correlations between modes in the local vicinity around site $i$.

This can be exemplified for a one-dimensional chain of spins where the Hamiltonian only contains interactions between nearest neighbours (or potentially also next-nearest neighbours) of spins.
An example of such a system is the anti-ferromagnetic Heisenberg model introduced in section \ref{sec:J1J2_model}, but the general motivation does not rely on the specifics of the interaction.
Instead of finding the ground state of such a system with a direct specification of the $D^L$ amplitudes, one can introduce an ansatz explicitly focussing on the local correlations in the system.
This can be achieved by describing the functions $F_i(\mathbf{x})$ as a complete representation of the states across a small environment around the site with index $i$, defining a plaquette over which the correlations are modelled.
Glossing over specific details of how the representation is defined at the boundary of the system, these are therefore represented as
\begin{equation}
    F_i(\mathbf{x}) = f^{(i)}_{x_{i-J}, \ldots, x_i, \ldots, x_{i+J}}.
\end{equation}
The chain indices $j$ for which the local configuration, $x_j$ contribute to the full representation of the local correlation around mode $j$ are chosen such that these are the central mode $i$ together with the $2 J$ sites closest to it.
The full parametrization of the state space for the local environment around site $i$, comprising $P = 2J+1$ modes, then involves a total $D^P$ coefficients, here represented by the coefficient tensor $f^{(i)}$.

\begin{figure}[htb!]
    \centering
    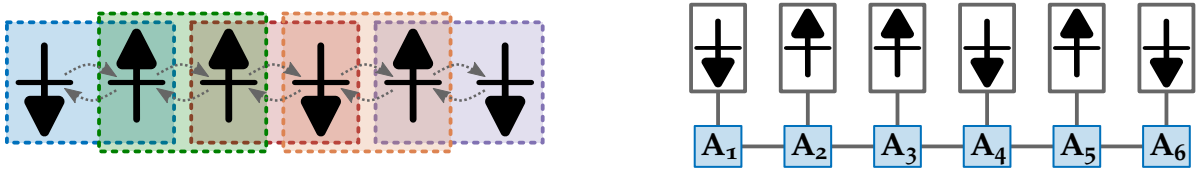
    \caption[Approximation of the wavefunction in terms of local plaquettes tiled across a one-dimensional system and a pictorial representation of the amplitude evaluation for a presented spin configuration of a \acl{MPS}]{Approximation of the wavefunction in terms of representations over local plaquettes tiled across a non-periodic one-dimensional system (left) and a pictorial representation of the amplitude evaluation for the presented spin configuration of an \ac{MPS} (right).}
    \label{fig:MPS}
\end{figure}

The construction of a state based on overlapping correlation plaquettes is visualized in the left part of Fig.~\ref{fig:MPS}.
It results in a representation of the full state according to
\begin{equation}
    \Psi(\mathbf{x}) = \prod_{i=1}^L f^{(i)}_{x_{i-J}, \ldots, x_i, \ldots, x_{i+J}}
    \label{eq:1D_correlator_product}
\end{equation}
and (assuming the same number of sites is included in each correlation plaquette) is therefore defined by $L \times D^P$ coefficients.
While this complexity scales exponentially in the size of the correlation plaquette, it only scales linearly in the size of the system.
If the initial assumption holds and only local correlations contribute significantly, it can be expected that states of interest can be approximated well with a plaquette size that is independent of the system size.
This therefore reduces the complexity of the full wavefunction parametrization significantly.

The definition according to Eq.~\ref{eq:1D_correlator_product} can be identified as a one-dimensional \ac{CPS}, introduced in the next section.
However, the one-dimensional nature of the plaquettes also makes it possible to transform this ansatz into a particularly powerful form as it is, e.g., presented in Ref.~\cite{Borin2019}.
This is achieved by introducing a set of $D^{J-1} \times D^{J-1}$ matrices $\mathbf{A}^{x_i}_i$, one for each mode $i$ and each potential local configuration of that mode $x_i$, with coefficients~\cite{Borin2019}
\begin{equation}
    (\mathbf{A}^{(x_i)}_i)_{j,k} = f^{(i)}_{x_{i-J}, \ldots, x_i, \ldots, x_{i+J}} \, \delta_{(x_{i-J}, \ldots, x_i, \ldots, x_{i+J-1}), j} \, \delta_{(x_{i-J+1}, \ldots, x_i, \ldots, x_{i+J}), k}.
\end{equation}
In this definition, the indices $j$ and $k$ are compound indices indexing all configurations over $2 J$ sites and can therefore be matched with sub-configurations of the form $(x_a, \ldots, x_{a+2J})$ over such a space with the delta function.
Assuming periodic boundary conditions of the system, the wavefunction amplitudes of the resulting state can be obtained by evaluating the trace over the matrix product of the matrices across all sites of the system.
In particular, this construction defines the ansatz class of \ac{MPS}.
\begin{ansatz}[\Aclp{MPS}]
    The \acl{MPS} defines the wavefunction amplitudes according to
    \begin{align*}
        \Psi(\mathbf{x}) = \mathrm{tr}(\mathbf{A}^{(x_1)}_1 \, \mathbf{A}^{(x_2)}_2 \, \ldots \,\mathbf{A}^{(x_L)}_L).
    \end{align*}
\end{ansatz}

Although here \acp{MPS} are, perhaps rather unconventionally, defined via a full parametrization of correlation features across fixed size plaquettes, the full class of \acp{MPS} is more general than this.
As the name suggests, the full class of \acp{MPS} is defined by the states that can be decomposed into the product of $M_b \times M_b$ dimensional matrices as above, with no particular constraints on the coefficients.
It represents probably the most widely applied form of a \ac{TNS} exploiting a tensor decomposition of the wavefunction amplitudes to capture particular states efficiently.
With many useful properties formally proven, \acp{TNS} are one of the most commonly applied and studied representations of many-body states.
The \ac{TNS} concepts are often made particularly intuitive by specific graphical representations, in which tensors are represented as nodes with legs representing the different tensor indices~\cite{penroseApplicationsNegativeDimensional1971, bridgemanHandwavingInterpretiveDance2017, MatricesTensorNetwork, TensorDiagramNotation}.
In this representation, legs connecting two nodes represent tensor contractions over the associated indices.
A standard example, visualizing the evaluation of the wavefunction amplitude of an \ac{MPS} for a basis configuration is shown in the right panel of Fig.~\ref{fig:MPS}.

The size of the matrix dimensions, $M_b$, is commonly referred to as the bond dimension of the \ac{MPS}, and the state parametrization can be made more systematically more expressive by increasing this dimension parameter.
It is well understood which specific part of the full Hilbert space can be represented efficiently with \acp{MPS}, i.e., \acp{MPS} with polynomial bond dimension~\cite{ganahlDensityMatrixRenormalization2022}.
In particular, these are exactly those states characterized by a low degree of entanglement.
This can be formalized by analysing the scaling of the entanglement entropy emerging between two different parts of the system with respect to the size of one of the two subsystems.
For one dimensional systems, \ac{MPS} represent states efficiently for which the entanglement entropy scales as the size of the boundary between the two systems, i.e., is constant~\cite{eisertAreaLawsEntanglement2010}.
Without going into the details about the proofs of such area laws, it is exactly this class of states that is of particular importance in many physically relevant settings.
This is exemplified by the result that all ground states of gapped one-dimensional systems with local interactions also fall into this class~\cite{eisertAreaLawsEntanglement2010}, underlining the usefulness of the \ac{MPS} representation of states for such systems.
Ultimately, such rigorous results describing the entanglement emerging in systems, formalize the hand-waving intuition stated above that often especially the local correlations are of particular importance.

A key benefit of the \ac{MPS} representation is that it makes it possible to evaluate expectation values for many standard operators of interest efficiently.
Specifically, this is the case if the operator can be written as a sum of polynomially many terms of operators factorizing as a tensor product over the different sites (which is, among others, also fulfilled for local Hamiltonians).
This means the operator is written as
\begin{equation}
    \hat{O} = \sum_{k=1}^K \hat{O}{(k)} \quad \mathrm{where} \quad \hat{O}{(k)} = \bigotimes_{i=1}^L \hat{O}_i^{(k)},
\end{equation}
with local operators $\hat{O}_i^{(k)}$ only acting on $\mathcal{H}_i$.
Assuming a normalized \ac{MPS}, and skipping the specifics of the derivation, its expectation value can then be evaluated as~\cite{eisertEntanglementTensorNetwork2013}
\begin{align}
    \langle \Psi | \hat{O} | \Psi \rangle &= \sum_k \sum_{x_1, \ldots, x_L} \sum_{x'_1, \ldots, x'_L} \mathrm{tr}(\mathbf{A}^{\ast(x_1)}_1 \, \mathbf{A}^{\ast(x_2)}_2 \, \ldots \,\mathbf{A}^{\ast (x_L)}_L) \mathrm{tr}(\mathbf{A}^{(x'_1)}_1 \, \mathbf{A}^{(x'_2)}_2 \, \ldots \,\mathbf{A}^{(x'_L)}_L) \langle\mathbf{x} | \hat{O}^{(k)} | \mathbf{x'} \rangle,\\
    &= \sum_k \mathrm{tr}(\mathbf{B}^{(1)}_k \mathbf{B}^{(2)}_k \ldots \mathbf{B}^{(L)}_k).
    \label{eq:MPS_exp_val}
\end{align}
Here the matrices $\mathbf{B}$ correspond to a set of $M_b^2 \times M_b^2$ matrices that can be defined by indexing the matrices with compound indices of the form $(i, i')$, with each element running from $1$ to $M_b$.
The coefficients of these matrices are given as
\begin{equation}
    (\mathbf{B}^{(i)}_k)_{(l, l'), (m, m')} = \sum_{x_i, x'_i} (\mathbf{A}^{\ast(x_i)}_i)_{l,m} (\mathbf{A}^{(x'_i)}_i)_{l',m'} \langle x_i| \hat{O}_i^{(k)} | x'_i \rangle.
\end{equation}
The evaluation of expectation values can with this formulation be achieved with a cost scaling at most as $\mathcal{O}(L K M_b^4 D^2) + \mathcal{O}(L K M_b^6)$ (with a naive assumption of an $\mathcal{O}(m^3)$ cost for the matrix multiplication of $m \times m$ matrices).
This scaling can often even be improved further with appropriate manipulations~\cite{eisertEntanglementTensorNetwork2013}.
This in particular includes the very common utilization of `open boundary' matrix product representations in which the first and the last matrix in the matrix decomposition chain are replaced by vectors, and contractions can then be performed as a sequence of matrix-vector multiplications.

In addition to being able to evaluate expectation values (such as energy expectation values) efficiently, different powerful schemes exist to optimize the parameters of \acp{MPS} in order to approximate system eigenstates.
Probably the most famous scheme is the \ac{DMRG}~\cite{Schollwoeck2011}, which represents the state-of-the-art approach for different systems of interest.
Although it is possible to apply these techniques also to the description of higher-dimensional systems, the specific construction of \ac{MPS} are particularly tailored towards one-dimensional systems exhibiting a low degree of entanglement.
While the general ideas of \acp{TNS} have also been extended to higher dimensional systems, many specific characteristics making the numerical treatment of \ac{MPS} highly efficient are typically not preserved, significantly complicating such numerical approaches.

\section{Variational Monte Carlo}
The \ac{DMRG} approach provides a powerful tool for a general task, namely that of finding an appropriate approximation of an eigenstate of a many-body Hamiltonian $\hat{H}$.
Whereas \ac{DMRG} is a method specific to \ac{MPS} representations, a more general family of approaches is given by \ac{VMC} methods.
\ac{VMC} approaches for numerical studies of many-body systems provide the main foundations for the methods outlined in this work and this section provides a brief overview of the main concepts as it can, e.g., be found in Ref.~\cite{Becca2017}.
Within the framework of \ac{VMC} an essentially arbitrary parametrization of the wavefunction amplitudes can be used as an ansatz for the state.
The only technical requirement of the model for the wavefunction amplitudes $\Psi(\mathbf{x})$ is that these can be evaluated efficiently for each configuration of the computational basis.

Obtaining the final approximation of the eigenstate of interest utilizes the variational principle of quantum mechanics.
This states that the energy expectation value of any state is bounded from below by the exact ground state of the system.
Especially focussing on the ground state of the system, an approximation can therefore be obtained by minimizing the variational energy of the chosen ansatz with respect to its free parameters.
The variational energy of a state ansatz is defined as
\begin{equation}
    E = \langle \hat{H} \rangle_\Psi = \frac{\langle \Psi | \hat{H} | \Psi \rangle}{\langle \Psi | \Psi \rangle},
\end{equation}
and approximating the system's ground state via direct minimization of this quantity is the main route taken here.

\subsection{Evaluation of expectation values}
Applying a minimization of the variational energy is the main ingredient in various approaches building on the variational principle.
One main component of \ac{VMC} techniques is that in such approaches, the state is defined via a compact functional model for the wavefunction amplitudes.
Furthermore, with the exact evaluation generally prohibited by the exponential scaling of the Hilbert space dimensionality, the expectation values are evaluated based on stochastic sampling of basis configurations.
This is exemplified by the approximation of the variational energy for a given state, $|\Psi\rangle$.
It can be expressed as
\begin{equation}
    E = \frac{\langle \Psi | \hat{H} | \Psi \rangle}{\langle \Psi | \Psi \rangle} = \sum_{\mathbf{x}} \frac{|\langle \Psi | \mathbf{x} \rangle|^2}{\langle \Psi | \Psi \rangle} \frac{\langle \mathbf{x} |\hat{H} | \Psi \rangle}{\langle \mathbf{x} | \Psi \rangle} = \sum_{\mathbf{x}} p_\mathbf{x} \, E_{loc}(\mathbf{x}),
\end{equation}
and is thus reformulated as the expectation value of so-called local energies, $E_{loc}(\mathbf{x}) = \frac{\langle \mathbf{x} |\hat{H} | \Psi \rangle}{\langle \mathbf{x} | \Psi \rangle}$, with respect to the probability distribution given by $p_\mathbf{x} = \frac{|\langle \Psi | \mathbf{x} \rangle|^2}{\langle \Psi | \Psi \rangle}$.
This quantity can be approximated by sampling configurations from the Hilbert space according to the probability distribution $p_\mathbf{x}$ and evaluating the mean of the local energies over the sampled set.
This gives the stochastic approximation
\begin{equation}
    E \approx \frac{1}{N_s} \sum_{\mathbf{x}_s} E_{loc}(\mathbf{x}_s),
\end{equation}
where the sum does not run over the full Hilbert space basis but a (typically) significantly smaller set of $N_s$ sampled configurations.
If the Hamiltonian $\hat{H}$ is sparse in the chosen basis, i.e., each row in its matrix representation only has polynomially many non-zero entries, this average can be evaluated efficiently.
Although this requirement is more restrictive than the one introduced for operators allowing for efficient evaluation of expectation values of \acp{MPS}, this constraint holds for many operators of interest, in particular local Hamiltonians.

An important property justifying this stochastic approximation of the energy is the zero variance principle.
This states that the variance over the local energies vanishes if the state corresponds to an eigenstate of the system.
Based on this, it can be expected that the error of the stochastic energy approximation decreases as the trial state becomes a better representation of the targeted ground state.
The variance over the local energy therefore also provides a figure of merit for the uncertainty, i.e., the expected error of the stochastic approximation.

\subsubsection{The Metropolis-Hastings algorithm}
An important element of the stochastic evaluation of expectation values in \ac{VMC} is the generation of configurations $\mathbf{x}$ according to the probability amplitude induced by the trial state.
While it is possible to introduce specific models allowing for a direct sampling of the configurations from $p_\mathbf{x}$~\cite{Sharir_2020}, this is not generally possible for various other sensible wavefunction parametrizations.

The common approach to generate samples for more general models, also the core backbone of the \ac{VMC} approaches discussed in this work, is to generate samples with the Metropolis-Hastings algorithm.
This does not require an explicitly normalized distribution over the configuration space and only relies on being able to evaluate the wavefunction amplitudes for configurations.
The generation of the configurational samples is achieved via an iterative scheme in which, based on a current sample at each step, a new configuration is proposed and either accepted or rejected.
Acceptance of a proposed configuration is determined stochastically based on a probability determined by the ratio of the probability amplitudes associated with the two configurations.
With this stochastic acceptance or rejection, the generated samples will, after sufficient equilibration steps, follow the underlying probability distribution.

More specifically, the algorithm generates a sequence of configurations based on a Markov chain, starting from a random initial configuration $\mathbf{x}^{(0)}$.
In the iterative generation of configurations, the $k$-th configuration in this sequence, denoted as $\mathbf{x}^{(k)}$, is thus obtained based on its predecessor $\mathbf{x}^{(k-1)}$.
From this predecessor, a proposal configuration $\mathbf{x}_{prop}$ is generated based on some underlying heuristic with a probability $P(\mathbf{x}^{(k-1)} \rightarrow \mathbf{x}_{prop})$.
This proposal configuration is accepted as the next configuration in the sequence with a probability
\begin{equation}
    Q = \mathrm{min} \left( 1, \frac{p_{\mathbf{x}_{prop}}}{p_{\mathbf{x}^{(k-1)}}} \frac{P(\mathbf{x}_{prop} \rightarrow \mathbf{x}^{(k-1)})}{P(\mathbf{x}^{(k-1)} \rightarrow \mathbf{x}_{prop})} \right).
\end{equation}
Often the algorithm is set up with equal proposal distributions for both directions, i.e., equal probability of generating $\mathbf{x}_{prop}$ from $\mathbf{x}^{(k-1)}$ as for the other way round as denoted by the equality $P(\mathbf{x}_{prop} \rightarrow \mathbf{x}^{(k-1)}) = P(\mathbf{x}^{(k-1)} \rightarrow \mathbf{x}_{prop})$.
In this case the acceptance probability only depends on the ratio of probability amplitudes,
\begin{equation}
    \frac{p_{\mathbf{x}_{prop}}}{p_{\mathbf{x}^{(k-1)}}} = \frac{|\langle \Psi | \mathbf{x}_{prop} \rangle|^2}{|\langle \Psi | \mathbf{x}^{(k-1)} \rangle|^2},
\end{equation}
which can be evaluated efficiently.

By running the Markov chains as outlined above, potentially multiple in parallel, a set of configurations sampled according to their probability amplitudes can be generated.
In order to ensure that the dependence on the randomly chosen initial configuration is removed, the first samples of the sequence are usually discarded.
Furthermore, it is also sensible to avoid correlation between the different samples by only adding configurations from the sequence that are multiple iteration steps apart to the set used for the stochastic estimation of the expectation values.

Applying the Metropolis-Hastings algorithm in practical \ac{VMC} calculations requires the specification of some algorithmic details.
This includes the specification of the number of warm-up iterations, the number of configurations from the Markov chain that are discarded between considered samples, as well as the number of chains that are run in parallel.
Furthermore, one also needs to design a mechanism to generate proposal configurations.
Such a mechanism should ideally propose configurations with a large probability amplitude (in order to avoid vanishing acceptance probabilities) while still exploring the full Hilbert space efficiently.
Depending on the system studied, different generic approaches exist that work well in many practical settings.
These, e.g., include the exchange of pairs of spins or the flip of a single spin in spin systems, or the application of a single valid particle jump from one mode to another in systems of moving particles.

While the Metropolis-Hastings algorithm is a very powerful and general approach to generate the samples, it has also been observed that sometimes problems emerge within its application.
This typically manifests in the failure to explore the full Hilbert space and emergence of correlation between the different samples and `more advanced' approaches might be required to generate appropriate uncorrelated samples~\cite{Sharir_2020, bagrovKineticSamplersNeural2020, zhangUnderstandingEliminatingSpurious2022}.

\subsection{Optimization of the parametrization}
\label{sec:VMC_optimization}
With the ability to evaluate expectation values of trial states defined by parametrized models for the wavefunction amplitude, the variational scheme can easily be applied to optimize the parameters of the ansatz.
Assuming the model is parametrized by a set of $N_{par}$ variational parameters, here described by a vector $\boldsymbol{\theta}$, the goal is to find those parameters minimizing the variational energy understood as a function of those parameters, $E = E(\boldsymbol{\theta})$.
This can in principle be achieved with various numerical schemes developed for the minimization of a target function, e.g., based on gradient descent type approaches.

If it is possible to approximate the variational energy through Markov chain sampling of configurations, also its gradient can be evaluated by the same approaches.
For general complex valued parameters, it is possible to define an energy gradient $\mathbf{G}(\boldsymbol{\theta})$ with components
\begin{equation}
    G_j(\boldsymbol{\theta}) = 2 \frac{\partial E(\boldsymbol{\theta})} {\partial \theta_j^\ast}.
\end{equation}
The expression $\frac{\partial} {\partial \theta_j^\ast} = \frac{1}{2} (\frac{\partial} {\partial \Re(\theta_j)} + i \frac{\partial} {\partial \Im(\theta_j)}) $ denotes a Wirtinger derivative, i.e., combines the derivatives with respect to the real and the imaginary part of $\theta_j$.
With this definition, the components of the energy gradient can be evaluated as
\begin{align}
    G_j &= \frac{\langle \frac{\partial}{\partial \theta_j^\ast} \Psi | \hat{H} | \Psi \rangle}{\langle \Psi | \Psi \rangle} - \frac{\langle \frac{\partial}{\partial \theta_j^\ast} \Psi | \Psi \rangle \langle \Psi | H | \Psi \rangle}{\langle \Psi | \Psi \rangle^2},\\
        &= \langle \hat{O}_j^\ast \hat{H} \rangle_\Psi - \langle \hat{O}_j^\ast \rangle_\Psi \langle \hat{H} \rangle_\Psi .
\end{align}
Here, it is assumed that the wavefunction model is holomorphic and an extension to non-holomorphic cases can easily be obtained by splitting the complex parameters into their real and imaginary part.
The introduced operators $\hat{O}_j^\ast$ represent the log wavefunction derivatives, and are defined as
\begin{equation}
    \hat{O}_j^\ast = \sum_{\mathbf{x}} O_j^\ast(\mathbf{x}) \, | \mathbf{x} \rangle \langle \mathbf{x} | = \sum_{\mathbf{x}} \frac{\partial \log (\Psi^\ast(\mathbf{x}))}{\partial \theta_j^\ast} \, | \mathbf{x} \rangle \langle \mathbf{x} |.
\end{equation}
The energy gradient can then again be estimated via the set of configurational samples, $\{\mathbf{x}_s\}$, sampled according to their wavefunction amplitudes.
This gives the approximation
\begin{align}
    G_j &= \langle \hat{O}_j^\ast \hat{H} \rangle_\Psi - \langle \hat{O}_j^\ast \rangle_\Psi \langle \hat{H} \rangle_\Psi, \\
        &= \left (\sum_{\mathbf{x}} p_{\mathbf{x}} \frac{\langle \Psi| \hat{O}_j^\ast | \mathbf{x} \rangle \langle \mathbf{x} | \hat{H} | \Psi \rangle}{\langle \Psi | \mathbf{x} \rangle \langle \mathbf{x} | \Psi \rangle} \right ) - \left (\sum_{\mathbf{x}} p_{\mathbf{x}}  \frac{\langle \Psi | \hat{O}_j^\ast | \mathbf{x} \rangle}{\langle \Psi | \mathbf{x} \rangle} \right) \left (\sum_{\mathbf{x}} p_{\mathbf{x}}  \frac{\langle \mathbf{x} | \hat{H} | \Psi \rangle}{\langle \mathbf{x} | \Psi \rangle} \right), \\
     & \approx \frac{1}{N_s} \sum_{\mathbf{x}_s} O_j^\ast(\mathbf{x}_s) (E_{loc}(\mathbf{x}_s) - E).
\end{align}

While this gradient estimation can easily be applied within standard numerical minimization techniques, the most commonly applied approaches within \ac{VMC} explicitly exploit physical specifics for finding the ground state of the system.
Probably the most widely used scheme is the \ac{SR} method to optimize the parameters.
Within the \ac{SR} method, a family of states based on a `linearization' of the trial state around small parameter variations is introduced.
Defining the infinitesimal parameter variation for parameter $\theta_j$ as $\delta_j$, the linearized state is defined as
\begin{equation}
    | \tilde{\Psi} \rangle = \delta_0 | \Psi \rangle + \sum_{j=1}^{N_{par}} \delta_j \, \hat{O}_j | \Psi \rangle.
\end{equation}
\ac{SR} then defines an iterative optimization scheme in which parameter updates are found at each step by matching this linearized state to an improved state.

Defining the state at the $k$-th iteration as $|\Psi^{(k)}\rangle$, the improved state is found by applying an imaginary-time evolution with a small time-step $\beta$, defined as
\begin{equation}
    |\Psi^{(k+1)}_{target} \rangle = e^{-\beta \hat{H}} |\Psi^{(k)}\rangle \approx (1 - \beta \hat{H}) |\Psi^{(k)}\rangle.
\end{equation}
In the approximation, a first order Taylor approximation is applied to the exponential, valid for small step sizes $\beta$.
The imaginary-time propagated state $|\Psi^{(k+1)}_{target} \rangle$ can be projected into the space spanned by the family of trial states linearized around small parameter variations~\cite{Becca2017, parkGeometryLearningNeural2020}.
This is achieved by matching the projected propagated state and the linearization of $|\Psi^{(k)}\rangle$, by equating their overlaps w.r.t. all the basis states generating the linearized family of states, $\{|\Psi^{(k)}\rangle, \, \hat{O}_1 | \Psi^{(k)} \rangle, \, \ldots, \, \hat{O}_{N_{par}} | \Psi^{(k)} \rangle\}$.
This approach leads to a system of $N_{par} + 1$ equations for the inferred parameter updates at step $k$, here denoted as $\delta^{(k)}_j$.
By solving this for the scale parameter $\delta_0$, a reduced system of $N_{par}$ equations is found that can be expressed in the compact vectorized form
\begin{equation}
    - \beta \, \boldsymbol{S} \, \boldsymbol{\delta^{(k)}} = \boldsymbol{G}(\boldsymbol{\theta}^{(k)}),
\end{equation}
where $\boldsymbol{\delta^{(k)}}$ is the vector of parameter updates with elements $\delta^{(k)}_j$, and $\boldsymbol{G}(\boldsymbol{\theta}^{(k)})$ is the energy gradient w.r.t. the variational parameters defining state $|\Psi^{(k)}\rangle$ as defined above.

The matrix $\boldsymbol{S}$, comprises elements defined according to
\begin{equation}
    S_{i,j} = \langle \hat{O}^\ast_i \hat{O}_j \rangle_{\Psi^{(k)}} - \langle \hat{O}^\ast_i \rangle_{\Psi^{(k)}} \, \langle \hat{O}_j \rangle_{\Psi^{(k)}},
\end{equation}
where the evaluation of expectation values is again approximated via stochastic sampling of configurations according to $p_\mathbf{x}$.
This matrix is commonly referred to as the Quantum Geometric Tensor~\cite{chengQuantumGeometricTensor2013} or the quantum Fisher matrix~\cite{parkGeometryLearningNeural2020}.
It can be associated with a metric over the family of states as parametrized by the variational parameters $\theta_j$, and it is the covariance of $\hat{O}$ measurements for the state $|\Psi^{(k)}\rangle$.

Solving the system of equation for the parameter updates in the \ac{SR} method, then yields updated parameter values that specify the wavefunction ansatz for the next iteration.
The updated parameter values, denoted by the vector $\boldsymbol{\theta^{(k+1)}}$, are obtained as
\begin{equation}
    \boldsymbol{\theta^{(k+1)}} = \boldsymbol{\theta^{(k)}} - \beta \, \boldsymbol{S}^{-1} \, \boldsymbol{G}(\boldsymbol{\theta}^{(k)}).
\end{equation}
This form of the parameter updates also relates the \ac{SR} approach back to gradient descent approaches.
The parameter $\beta$ can be understood as a step-size or learning rate and the matrix $\boldsymbol{S}$ defines a preconditioner, suitably modifying the update directions defined by the energy gradient.
The obtained \ac{SR} scheme, incorporating a metric over the wavefunction space to define a preconditioner, is equivalent to the ideas of Natural Gradient Descent optimization strategies for \ac{ML} parametrizations~\cite{amariNaturalGradientWorks1998}.

Updating the parameters in the \ac{SR} method requires the inversion of an $N_{par} \times N_{par}$ matrix at each step of the method.
From a computational perspective, this has two main disadvantages compared to gradient descent based methods not incorporating such a preconditioner.
Firstly, the inversion of the matrix is a computationally relatively expensive operation, with a naive implementation requiring $\mathcal{O}(N_{par}^3)$ operations.
However, the numerical complexity of the \ac{SR} can often be reduced by applying iterative schemes to solve the systems of equations, exploiting the property that matrix-vector products between $\boldsymbol{S}$ and $\boldsymbol{\delta^{(k)}}$ can be evaluated in $\mathcal{O}(N_{par} N_{s})$ operations~\cite{Misawa2017,vicentiniNetKetMachineLearning2021}.
Another potential issue of the \ac{SR} formulation is that the system of equations might be ill-conditioned, indicating redundant directions in the parametrization of the state (or a stochastic estimate with too few samples to appropriately resolve the required quantities).
This can lead to severe stability issues for the numerical the inversion of the matrix $\boldsymbol{S}$, thus resulting in potentially bad parameter updates.
Though it is not the only approach, a very common method to avoid such instabilities is to ensure that the matrix $\boldsymbol{S}$ is non-singular by adding a small constant shift $c$ to its diagonal, i.e., $S_{i,i} \rightarrow S_{i,i} + c$.
By increasing the diagonal shift, the update directions in the \ac{SR} are adjusted towards the simple gradient descent direction with a step-size of $\beta/c$.
In addition to this basic stabilization approach, further methods have been described to improve the \ac{SR} parameter updates~\cite{rothHighaccuracyVariationalMonte2022,lovatoHiddennucleonsNeuralnetworkQuantum2022,https://doi.org/10.48550/arxiv.2302.01941} that might in some instances help to improve the overall performance and stability of the method.

Although the \ac{SR} method is not always free from issues in practice, it can be considered the default approach to optimize state parametrizations in the context of \ac{VMC}.
In many scenarios, a fast convergence to appropriate solutions can be observed, and the method is typically not significantly outperformed by other approaches~\cite{frankLearningNeuralNetwork2021}.

\subsection{Functional models for Variational Monte Carlo}

The general \ac{VMC} framework makes it possible to use in principle any parametrization of the state as long as it allows for an efficient evaluation of the wavefunction amplitudes for many-body configurations from the chosen basis.
However, in order to efficiently optimize the parametrization, it should represent a compact model for the state, based on only a few variational parameters.
Choosing an appropriate model for the state is therefore vital to be able to approximate the ground state well.
The following sections introduce some  models for \ac{VMC} calculations.

\subsubsection{Correlator Product States}
An important class of parametrizations that can be used for \ac{VMC} calculations is the family of \acp{CPS}~\cite{Changlani2009}, also referred to as Entangled Plaquette States~\cite{mezzacapoGroundstatePropertiesQuantum2009}.
The key idea is to follow a similar approach as the one described as the motivation of \acp{MPS} presented in section \ref{sec:MPS}.
In particular, the general wavefunction amplitudes are obtained by tiling correlation plaquettes over which the correlations are modelled across the system.
The wavefunction amplitudes for the ansatz are then obtained based on the product over all the states across the correlation plaquettes.

\begin{figure}[htb!]
    \centering
    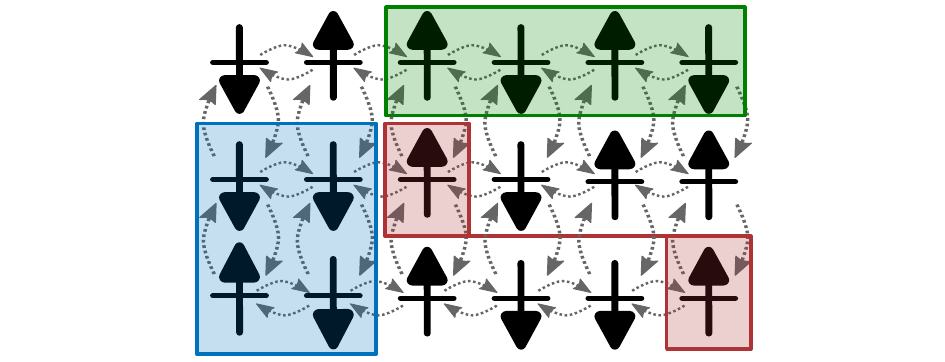
    \caption[Visualization of potential correlation plaquettes across which correlations can be extracted with \aclp{CPS} for a two-dimensional system with local interactions]{Visualization of potential correlation plaquettes across which correlations can be extracted with \acp{CPS} for a two-dimensional system with local interactions. The figure shows local plaquettes of different shapes (blue and green), as well as a longer ranged, two-body plaquette (red) also considered in Jastrow ansatzes.}
    \label{fig:2D_approximators}
\end{figure}

\begin{ansatz}[\Aclp{CPS}]
    The \acl{CPS} ansatz is defined via wavefunction amplitudes
    \begin{equation}
        \Psi(\mathbf{x}) = \prod_{i=1}^{N_p} f^{(i)}_{x_{p_1(i)}, \ldots, x_{p_P(i)}},
    \end{equation}
    where $N_p$ denotes the total number of correlation plaquettes and the indices $x_{p(i)_1}, \ldots, x_{p(i)_P}$ denote the configuration occupancies on the $P$ different modes of the $i-th$ plaquette.
\end{ansatz}
Following the product separability requirements for the states, typically one or multiple correlation plaquettes are associated with each of the modes for which the plaquette indices $p(i)_k$ comprise indices from some chosen environment around the mode.
It can directly be seen that the tiling of plaquettes containing the closest indices around the centre, as was applied in the derivation for \acp{MPS}, is an example of a \ac{CPS}.
Whereas the \ac{MPS} construction explicitly exploited a one-dimensional structure, general \ac{CPS} can be used to define entanglement plaquettes in arbitrary dimensions.
Furthermore, there is no general requirement on the shape of the different correlation plaquettes.
One can, for example, consider the correlation plaquettes comprising local environments, i.e., associate a plaquette to each mode of the system containing the $n$ modes closest to it in some sense.
Alternatively, one can also associate some longer range correlation plaquettes, such as, e.g., one-dimensional stripes, with each mode.
Both of these two examples are included in the visualization for a two-dimensional lattice of spins shown in Fig.~\ref{fig:2D_approximators}.

The standard functional form of the \ac{CPS} as introduced above explicitly relies on parametrizing the full Hilbert space across each plaquette.
It therefore results in an exponential scaling with respect to the size of the plaquettes, which practically limits the \ac{CPS} to small plaquette sizes.
However, with the explicit construction of the ansatz based on full parametrizations across specific entanglement plaquettes, such an ansatz makes it possible to very explicitly model expected correlation properties.
It is well understood what types of correlations are captured by the state.

\Acp{CPS} can be understood as a family of parametrizations defining a general framework also capturing the concepts of several other commonly employed ansatzes.
Many other ansatzes, can be understood as subclasses of some \ac{CPS} with specifically chosen correlation plaquettes and potentially relying on approximations to the functions $f^{(i)}_{x_{p_0(i)}, \ldots, x_{p_P(i)}}$~\cite{Changlani2009, Clark2018}.
These relations to other physically motivated ansatzes can help to identify what type of correlation plaquettes should be chosen to obtain a sensible parametrization.
There is however no general recipe to design entanglement plaquettes to achieve the best possible approximation for a given problem.

\subsubsection{The Jastrow ansatz}
\label{sec:jastrow}
One example of an ansatz also often used in \ac{VMC} calculations, which can also be represented easily as a compact \ac{CPS}, is the Jastrow ansatz.
While the original form of the Jastrow ansatz, as first introduced in Ref.~\cite{jastrowManyBodyProblemStrong1955}, is more specific, a Jastrow ansatz can generally be understood as an ansatz built from pairwise correlations over all possible interaction pairs in the system.
The ansatz introduced in the original work considers such pairwise correlations in the picture of particles occupying positions in real space.
However, similar ideas, explicitly building an ansatz based on pairwise correlations, have also been developed for the description in the picture of discrete modes, such as the pairwise correlations between spin-1/2 modes in the Huse-Elser ansatz~\cite{huseSimpleVariationalWave1988}.

Although Jastrow states are typically understood to be states emerging from the parametrization of pairwise correlations, the exact functional form depends somewhat on the context.
Based on the definition of \ac{CPS} in the previous section, the most general Jastrow ansatz might be defined as a \ac{CPS} based on all possible correlation plaquettes involving two system modes.
\begin{ansatz}[\ac{CPS} Jastrow ansatz]
    A generalized Jastrow ansatz, parametrized by an $L \times L \times D \times D$ tensor $J$, can be defined via wavefunction amplitudes
    \begin{equation}
        \Psi(\mathbf{x}) = \prod_{i=1}^L \prod_{j=i}^L J_{i, j, x_i, x_j}.
        \label{eq:general_Jastrow}
    \end{equation}
\end{ansatz}
In practical contexts, this general \ac{CPS} definition of a Jastrow state is restricted further by imposing additional constraints on the parametrizing tensor $J$ that depend on the specific type of system studied.

In this work, two different types of systems are considered, those of spin-1/2 modes and that of electrons moving between discrete modes.
\begin{ansatz}[Spin system Jastrow ansatz]
    The variational Jastrow ansatz used as a reference ansatz for spin systems considered in this work is defined by wavefunction amplitudes~\cite{manousakisSpinHeisenbergAntiferromagnet1991, Sorella}
    \begin{equation}
        \Psi(\mathbf{x}) = \prod_{i=1}^L \prod_{j=i}^L e^{\frac{1}{2}u_{i,j} s(x_i) s(x_j)}.
        \label{eq:spin_jastrow}
    \end{equation}
    The value $s(x_i)$ gives the $2 \hat{S}_z$ value of the configuration at spin $i$, i.e., it evaluates to $+1$ ($-1$) if $x_i$ corresponds to the up (down) state of the spin and $u_{i,j}$ denote the $L \times L$ different variational parameters.
\end{ansatz}
\begin{ansatz}[Electronic Jastrow ansatz]
    In settings of discrete electronic systems, the Jastrow ansatz used is defined as~\cite{Misawa2017}
    \begin{equation}
        \label{eq:electronic_Jastrow}
        \Psi(\mathbf{x}) = \prod_{i=1}^L e^{g_i \delta_{n_i, 2}} \prod_{j=i+1}^L e^{\frac{1}{2} u_{i,j} (n(x_i)-1) (n(x_j) - 1)},
    \end{equation}
    with variational parameters $g_i$ and $u_{i, j}$ and where $n(i)$ denotes the total number of electrons occupying mode $i$ (which can either be $0$, $1$ or $2$).
\end{ansatz}
The prefactor in the electronic Jastrow factor specified as
\begin{equation}
    \prod_{i=1}^L e^{g_i \delta_{n_i, 2}},
    \label{eq:gutzwiller}
\end{equation}
is commonly referred to as a Gutzwiller factor~\cite{gutzwillerEffectCorrelationFerromagnetism1963} and models an exponential suppression (or enhancement) of the amplitudes according to the total number of doubly occupied modes.

\subsubsection{Bridging the gap: Neural Quantum States}

While the Jastrow ansatzes introduced in the precious section can yield good approximations, the ansatz parametrization is also restricted to the pairwise correlations in the state and cannot improve upon that.
This means that the Jastrow ansatz is in general not systematically improvable.
Similar problem exists for essentially all types of \ac{CPS} types of parametrizations with fixed entanglement plaquettes.
While \ac{CPS} are, in principle, systematically improvable by increasing the size of the entanglement plaquettes considered, no general approach exists to infer which types of correlation plaquettes best to include in order to reach a certain level of accuracy.
This provides a motivating idea for \ac{NQS} which rely on the approximation power of \acp{NN} to describe the important correlations in an automated way.

The central idea of \ac{NN} function approximators is to design a functional model loosely inspired by biological neural connections.
This is achieved by modelling a network of nodes coupled with different coupling strengths.
Each node can be understood as a unit mapping an incoming signal to an output.
While different types of network architectures have been developed, typical functional mappings defined by \acp{NN} are based on two main building blocks.
The first building block are the connection strengths between the different nodes defining a weighting according to which the output of a unit is contributing to the input of the connected unit.
The second essential building of \acp{NN} are the non-linear activation functions associated with the different nodes.
This activation function describes the relationship between the input of a node to its output.

The specific network architecture of \acp{NN} can take various different shapes and forms.
This also applies to the concept of \ac{NQS} for which various different \ac{NN} type architectures have been proposed.
Based on a suitable representation of the configuration as input to the \ac{NN}, here denoted as $\tilde{\mathbf{x}}$, the general class of \ac{NQS} is therefore simply defined via the following ansatz:
\begin{ansatz}[\Aclp{NQS}]
    A \acl{NQS} associates wavefunction amplitudes with basis states via a mapping
    \begin{equation}
        \Psi(\mathbf{x}) = f(\tilde{\mathbf{x}}),
    \end{equation}
    where the function $f$ is the output of a \ac{NN} that is presented with a visible unit configuration $\tilde{\mathbf{x}}$ encoding the computational basis state $\mathbf{x}$.
\end{ansatz}
This class of states is explicitly motivated by the great success of \acp{NN} as function approximators for \ac{ML} problems.
Standard \ac{NN} architectures can easily be systematically improved in their expressiveness, typically by increasing the number of internal nodes, essentially resulting in a (formal) unlimited expressivity of \ac{NQS}.

In order to use \acp{NQS} in practical calculations, it is required to specify the architecture of the network with its variational parameters.
Most standard \acp{NN} can be represented in the form of a feed forward \ac{NN}.
In these the network comprises stacked layers of nodes where the nodes of each layer take their inputs from the previous layer and feed their output into the next layer.
Though it is not the only possible construction, in the \ac{NQS} context the input is typically encoded by $\tilde{L}$ visible nodes encoding the presented configuration in terms of floating point numbers associated with each mode.
For lattices of spin-1/2 modes, this means that each lattice site is typically associated with one unit in the input layer which takes values $+1$ or $-1$ corresponding to the $2 \hat{S}_z$ value of the associated spin in the input configuration.
The standard way to extend this representation to electronic systems is to represent each electronic mode with two different input neurons, encoding the occupancy of the two different spin levels~\cite{Choo2019a, Nomura2017}.

\begin{ansatz}[\acl{NQS} with feed-forward \ac{NN} architecture]
    Based on a chosen input representation of basis states, a general \ac{NQS} with a feed-forward \ac{NN} architecture defines the amplitudes as
    \begin{equation}
        \Psi(\mathbf{x}) = f_{N_{layer}} \circ f_{N_{layer}-1} \circ \ldots f_{1}(\tilde{\mathbf{x}}^{(1)}),
    \end{equation}
    where the functions $f_i$ describe the input-output relations for layer $i$ and $N_{layer}$ denotes the total number of layers.
    Each of the internal layers describes a mapping according to
    \begin{equation}
        f_i(\tilde{\mathbf{x}}^{(i)}) = \sigma_i(\mathbf{W}^{(i)} \tilde{\mathbf{x}}^{(i)}).
    \end{equation}
    Here, the vector $\tilde{\mathbf{x}}^{(i)}$ denotes the output vector of the previous layer, potentially together with a bias, $\mathbf{W}^{(i)}$ the matrix of weights, and $\sigma_i$ the activation function of the layer.
\end{ansatz}

\begin{figure}[htb!]
    \centering
    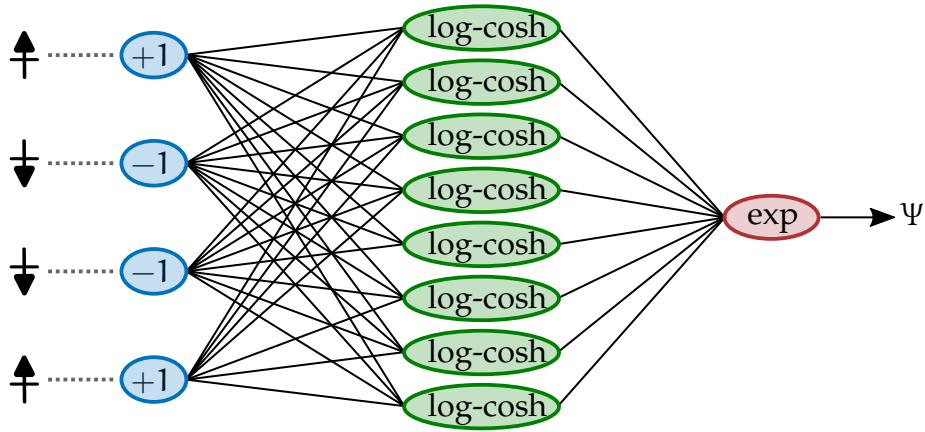
    \caption[Pictorial representation of the \ac{RBM} ansatz interpreted as a feed-forward neural network]{Pictorial representation of the \ac{RBM} ansatz interpreted as a feed-forward neural network (not incorporating network biases).}
    \label{fig:RBM_feed_forward}
\end{figure}

With the weights of the network defining the variational parameters, such feed-forward \acp{NN} can easily be applied as a model for the wavefunction in the \ac{VMC} context.
A particularly common \ac{NN} architecture, first applied as an ansatz in Ref.~\cite{Carleo2017}, is the \ac{RBM}.
The original motivation for the construction of this model is based on the statistical modelling of a probability distribution over visible units coupled to hidden units in an energy based framework.
However, the resulting model can also be understood as an exponentiated feed-forward \ac{NN} comprising a single hidden (i.e., non-input) layer with a number of nodes that is a multiple of the number of input nodes.
\begin{ansatz}[\Acl{RBM} wavefunction]
    The \acl{RBM} wavefunction amplitudes evaluate to~\cite{Carleo2017}
    \begin{equation}
        \Psi(\mathbf{x})= \prod_{i=1}^{\tilde{L}} \left ( e^{a_i \tilde{x}^{(1)}_i} \prod_{j = 1}^{\alpha_M} 2 \cosh \left(b_{i,j} + \sum_{k=1}^{\tilde{L}} W_{i,j,k} \, \tilde{x}^{(1)}_k \right) \right ).
    \end{equation}
    The parameter ${\alpha_M}$ specifies the hidden-node density of the network and the variational parameters of the model are the visible biases, $a_i$, the hidden node biases, $b_{i,j}$, as well as the network weights $W_{i,j,k}$.
\end{ansatz}
Similar to the bond dimension $M_b$ in the construction of \ac{MPS}, the number of hidden units (here defined via the density ${\alpha_M}$) controls the general complexity and therefore the expressivity of the model.
By construction, this ansatz automatically incorporates the product structure over the different modes as it was motivated for size-extensive states in section \ref{sec:product_sep}.
A pictorial representation of the \ac{RBM} ansatz, interpreted as a feed-forward architecture with a log-cosh activation function in the hidden layer followed by a final exponentiation, is presented in Fig.~\ref{fig:RBM_feed_forward}.

While the \ac{RBM} ansatz has become a widely applied \ac{NQS}, often considered the prototype for \ac{NQS}, (and also producing the state-of-the-art accuracies for some systems~\cite{nomuraHelpingRestrictedBoltzmann2020, nomuraDiractypeNodalSpin2021}), many other \ac{NN} architectures following similar constructions have been proposed as ansatzes.
These usually also incorporate an appropriate product structure of the amplitudes, which is often achieved by directly modelling the logarithm of the wavefunction amplitude with \ac{NN} architectures.

In order to achieve good numerical accuracies in practical calculations, it is often beneficial to incorporate system symmetries into the representation.
Important symmetries can include the total spin magnetization, translational symmetries and point group symmetries of the lattice.
The symmetrization of the ansatz can often be achieved in two different ways, either by a projective approach or by incorporating the symmetries directly into the design of the state (e.g., the \ac{NN} architecture).
A symmetry such as the spin magnetization can usually be enforced by including an appropriate projection into the state, which is in \ac{VMC} calculations easily achieved by restricting the sampling to configurations respecting this symmetry.

Especially translational symmetries of the system are however also often incorporated with a different paradigm, utilizing fully functional forms to extract the correlation properties.
For \ac{NQS}, this can, e.g., be formalized by application of the general concepts of convolutional \ac{NN}, as often applied in the context of image analysis.
In the context of wavefunctions, the convolutional symmetrization can be interpreted as an application of the same function to model correlations over symmetrically related environments.
Applied to the definition of \ac{RBM} this is achieved by replacing the product over all visible units by a product over the set of all symmetry operations, $\{\mathcal{S}\}$, which the amplitudes should be invariant under~\cite{Carleo2017}.
This defines the symmetrized \ac{RBM} wavefunction amplitudes as
\begin{equation}
    \Psi(\mathbf{x})= \prod_{i=1}^{|\{\mathcal{S}\}|} \left ( \prod_{j = 1}^{\alpha_M} e^{a_j \sum_{k=1}^{\tilde{L}} \mathcal{S}_i[\tilde{x}^{(1)}]_k}  2 \cosh(b_{j} + \sum_{k=1}^{\tilde{L}} W_{j,k} \, \mathcal{S}_i[\tilde{x}^{(1)}]_k) \right ),
\end{equation}
where the expression $\mathcal{S}_i[\tilde{x}^{(1)}]_k$ denotes the value of the $k$-th input neuron for an input configuration transformed under the symmetry operation $\mathcal{S}$.
This convolutional symmetrization approach often helps to reduce the total number of variational parameters of the model.
However, it was shown that using non-symmetric base ansatzes for which the symmetries are restored by projection can sometimes be more advantageous, especially when sign information needs to be described~\cite{nomuraHelpingRestrictedBoltzmann2020, nomuraDiractypeNodalSpin2021}.
More details on different symmetrization approaches will be discussed for the context of the \acp{GPS} in section \ref{sec:symmetrization}.

\section{Benchmarking systems studied in this work}
The central object of interest in this work is the \ac{GPS}, an ansatz that can be understood to combine the central ideas of \ac{CPS} and \ac{NQS} in a framework based on statistical modelling of functions.
In order to benchmark the general applicability of the \ac{GPS}, it is applied as a model to approximate the ground state of two main test systems, in addition to transferring the approaches to realistic ab-initio calculations (outlined in chapter \ref{ch:ab_initio_GPS}).

\subsection{The $J_1$-$J_2$ model}
\label{sec:J1J2_model}
The first benchmarking system studied in this work is the $J_1$-$J_2$ system of spin-1/2 modes arranged on one- and two-dimensional (square) lattice structures.
Its Hamiltonian can be defined as
\begin{equation}
    \hat{H} = J_1 \sum_{\langle i, j \rangle} \hat{\mathbf{S}}_i \cdot \hat{\mathbf{S}}_j + J_2 \sum_{\langle \langle i, j \rangle \rangle} \hat{\mathbf{S}}_i \cdot \hat{\mathbf{S}}_j.
\end{equation}
Here, $\hat{\mathbf{S}}$ denotes the vector of the three spin operators $\hat{S}^x$, $\hat{S}^y$ and $\hat{S}^z$.
The first sum in the definition includes all directly neighbouring pairs of sites and the second term includes all pairs of next-nearest neighbours within the lattice structure.
Being a system of $L$ spin-1/2 modes, this is a concrete realization of a spin system as it was used as a working example in the previous sections and the concepts can directly be applied.

The $J_1$-$J_2$ model can be understood as a general toy model capturing the main quantum effects of (frustrated) magnetism.
For vanishing next-neighbour couplings $J_2$ and positive values of $J_1$, this model is also known as the anti-ferromagnetic Heisenberg model providing a general prototype for a system exhibiting anti-ferromagnetic correlations between the different spins.
By introducing the next-nearest neighbour interactions additional quantum phenomena can be described.
As the next-nearest neighbour coupling $J_2$ is increased in a square two-dimensional system for example, the ground state transitions from one with anti-ferromagnetic N\'{e}el correlations between the spins, to one where the spins are correlated in a striped ordering.
In the transition between these phases, more intricate quantum phases have been discovered, including a spin liquid phase that cannot be associated with similar long-ranged order in the spin correlations.
Comprehensive discussions of the intricate phases emerging in such spin systems can, e.g., be found in Refs.~\cite{nomuraDiractypeNodalSpin2021, szabó_2021}.

This work does not focus on an analysis of the detailed physical properties of the $J_1$-$J_2$ model, but the model is rather understood as a generic test bed for the approaches introduced.
Due to the intrinsic challenges of modelling the state in strongly frustrated regimes, many approaches have been applied to this model providing reference results that allow for a general comparison between different methods.

A particular challenge for the description of the ground state as the systems transitions from the anti-ferromagnetic character to the striped character is a complex sign structure emerging in the target state.
Whereas the exact sign structure w.r.t. the chosen basis is known for the limit of vanishing $J_2$, a non-trivial sign structure needs to be modelled with the ansatz when $J_2$ is increased to intermediate values inducing a geometric frustration between the spins.
In the basis generated by the tensor product of $\hat{S}_z$ eigenstates, the exact sign structure for the case of vanishing $J_2$ is given by the \ac{MSR}~\cite{marshallAntiferromagnetism1955}:
\begin{ansatz}[\Acl{MSR}]
    The ground state sign structure of the anti-ferromagnetic Heisenberg model is given by the \acl{MSR} according to
    \begin{equation}
        \Psi(\mathbf{x}) = (-1)^{\sum_{i\in A} \delta_{x_i, 1}},
        \label{eq:MSR}
    \end{equation}
    where the sum in the exponent includes all spins of a sublattice $A$, which is either one of the two sublattices emerging from splitting the lattice into two sets of spins according to a chequerboard pattern (such that neither sublattice contains two spins that are coupled in the Hamiltonian).
\end{ansatz}
With an up-spin on site $i$, in the convention used here, denoted by $x_i=1$, the amplitude sign for a configuration is thus given by $+1$($-1$) if the number of up spins in the configuration on the sublattice $A$ is even(odd).

While such a sign-structure should in principle also be easily described by a chosen parametrization of the state as it is a simple product state, achieving this in practice is not always an easy task (see section \ref{sec:symmetrization}).
Instead of describing the \ac{MSR} with the chosen ansatz, it often practically helps to directly incorporate the basis transformation induced by the \ac{MSR} into the Hamiltonian.
This is easily achieved by taking all the off-diagonal matrix elements in the anti-ferromagnetic Heisenberg Hamiltonian in the chosen basis to their negative.
This gives in the representation
\begin{equation}
    \hat{H} = \sum_{\langle i, j \rangle} \hat{S}^z_i \otimes \hat{S}^z_j - \hat{S}^x_i \otimes \hat{S}^x_j - \hat{S}^y_i \otimes \hat{S}^y_j.
\end{equation}
where $\hat{S}^{x/y/z}_i$ denote standard spin operators associated with the $x/y/z$ directions acting on the spin with index $i$.
With all off-diagonal Hamiltonian matrix elements being negative in the chosen basis, the ground state wavefunction amplitudes are then guaranteed to be positive~\cite{10.5555/2011772.2011773}.

\subsection{The Fermi-Hubbard model}
\label{sec:hubbard_model}

The second benchmarking system considered in this work is the Fermi-Hubbard model.
Whereas the $J_1$-$J_2$ model comprised fixed spins, the Hubbard model describes indistinguishable Fermions that can occupy the different sites of a lattice structure.
It can be understood as a simple prototype approximating the behaviour of electrons in a crystal structure able to capture a plethora of important quantum phenomena emerging in condensed matter systems~\cite{arovasHubbardModel2021}.

Defining Fermionic creation and annihilation operators $\hat{c}^\dagger_{i,\sigma}$ and $\hat{c}_{i,\sigma}$ that create and annihilate an electron with spin $\sigma$ on site $i$, the Fermi-Hubbard Hamiltonian can be defined as
\begin{equation}
    \hat{H} = -t \sum_{\sigma \in \{\uparrow, \downarrow\}} \sum_{\langle i, j \rangle} (\hat{c}^\dagger_{i,\sigma} \hat{c}_{j,\sigma} + \hat{c}^\dagger_{j,\sigma} \hat{c}_{i,\sigma}) + U \sum_i \hat{c}^\dagger_{i,\uparrow} \hat{c}_{i,\uparrow} \hat{c}^\dagger_{i,\downarrow} \hat{c}_{i,\downarrow}.
\end{equation}
The first set of terms, weighted by the hopping parameter $t$, capture the kinetic contribution of electrons hopping between all nearest neighbour lattice sites.
The second set of terms, proportional to repulsion parameter $U$, introduces interactions between electrons by effectively increasing the local energy for configurations based on the number of doubly occupied sites.
It is exactly the inclusion of the repulsion (or attraction) between electrons that introduces correlation effects between the different electrons leading to the breakdown of mean-field approximations for non-negligible repulsion $U/t$.

The Hubbard model can represent a variety of quantum effects underpinning material properties, such as the transition from a conductor to a Mott insulator.
However, just like for the $J_1$-$J_2$ model no general analytic solutions exist to describe the physics for all parameter regimes and lattice structures, making this model another common test bed for numerical methods approximating electronic behaviour~\cite{SimonsCollaborationontheManyElectronProblem2015}.

The studies of the Fermi-Hubbard in this work are based on using a basis constructed from the occupancies of the different lattice sites, which is, e.g., outlined with more detail in Ref.~\cite{altlandCondensedMatterField2010a}.
This means that basis states $| \mathbf{x} \rangle$ are defined in a second quantized representation.
The occupancy of lattice site $i$, $x_i$, can take values $1$ to $4$ depending on whether the site is unoccupied, occupied with a single spin-up or spin-down electron, or empty.
This occupancy number representation of the basis states does not rely on a labelling the different electrons so that the indistinguishability of the particles is directly incorporated into the basis.
However, for the evaluation Fermionic expectation values, it needs to be ensured that the Fermionic commutation relations are respected.
These are defined as
\begin{align}
    \{\hat{c}^\dagger_{i,\sigma_1}, \hat{c}_{j,\sigma_2}\} &=  \hat{c}^\dagger_{i,\sigma_1}  \hat{c}_{j,\sigma_2} + \hat{c}_{j,\sigma_2} \hat{c}^\dagger_{i,\sigma_1} = \delta_{i,j} \delta_{\sigma_1, \sigma_2} \\
    \{\hat{c}^\dagger_{i,\sigma_1}, \hat{c}^\dagger_{j,\sigma_2}\} &= \{\hat{c}_{i,\sigma_1}, \hat{c}_{j,\sigma_2}\} = 0.
\end{align}

The Fermionic character can be incorporated by uniquely identifying each basis state $| \mathbf{x} \rangle$ with a configuration for which electrons are created in the occupied modes starting from the vacuum state $|0\rangle$ in a normal order.
This means that the basis states are defined as
\begin{equation}
    | \mathbf{x} \rangle = \hat{c}^\dagger_{r(1),\sigma(1)} \hat{c}^\dagger_{r(2),\sigma(2)} \ldots \hat{c}^\dagger_{r(N),\sigma(N)} |0 \rangle.
\end{equation}
Here, $r(i)$ and $\sigma(i)$ denote the lattice site and spin that electron $i$ occupies, and the total number of electrons in the configuration is $N$.
In this definition, the ordering of the creation operators is fixed to be in normal order, i.e., the labels $i$ are chosen such that the sequence of tuples $(\sigma(i), r(i))$ is sorted in ascending order.
Having defined a specific ordering of the lattice sites, the matrix elements of Fermionic operators in the chosen basis can easily be evaluated by applying the operator to the ordered string of creation operators of the basis configurations.
The evaluation of these strings then results in parity prefactors for the matrix elements that can efficiently be evaluated (see more details in section \ref{sec:ab_initio_implementation}).

The construction of the computational basis states outlined above is equivalent to the Jordan-Wigner mapping of Fermionic Hamiltonians to $2 L$ spin-1/2 (or qubit) degrees of freedom.
The parity prefactors obtained in the evaluation of operator matrix elements depend on the chosen ordering of the sites (and spins).
Therefore, also the sign structure of the modelled ground state will depend on the chosen ordering, and it can be expected that the performance of the ansatz depends on how well the emerging sign structure can be captured.
E.g., for one-dimensional Hubbard models, a canonical ordering of the sites and spins is given by ordering the site indices along the chain for each spin separately.
Under choice of appropriate boundary conditions, this results in positive ground state wavefunctions for the Hubbard chains in this basis.
In higher dimensional systems however, it is not clear how to define such a canonical ordering, which contributes to a sign structure of the target state in the chosen basis depending on the employed ordering.

It is not strictly necessary to incorporate an explicit anti-symmetrization of the state when working in the specified basis of Fock states.
Nonetheless, it is often useful to include mean-field characteristics into the ansatz that dominate for systems of weak correlation.
Including such mean-field characteristics with states explicitly anti-symmetrized with respect to exchanges of electron labels also avoids the ordering ambiguities outlined above as the spanned space of states is invariant under a change of the chosen ordering.

A key building block for this is the \ac{SD}, specifying an anti-symmetrized product of single electron wavefunctions.
The wavefunction amplitudes for this ansatz can be evaluated efficiently for configurations of the defined computational basis.
\begin{ansatz}[\Acl{SD}]
    The \acl{SD} defines an anti-symmetrized product of single-body states, giving the state
    \begin{equation}
        \label{eq:SD}
        |\Psi \rangle = \prod_{i=1}^N (\sum_{j=1}^L \sum_{\sigma \in \{\uparrow, \downarrow\}} \phi_i(j,\sigma) \, \hat{c}^\dagger_{j,\sigma}) | 0 \rangle.
    \end{equation}
    Here, the functions $\phi_i(j,\sigma)$ define the $N$ single particle wavefunctions (the orbitals) over the spin ($\sigma$) and spatial ($j$) degree of freedom, which can be fully parametrized with $2 L N$ parameters for lattice models.
    The wavefunction amplitude for a configuration $\mathbf{x}$ is given by the determinant of an $N \times N$ matrix $\mathbf{U}$, i.e.,
    \begin{equation}
        \Psi(\mathbf{x}) \sim det(\mathbf{U}).
    \end{equation}
    This matrix is constructed from the orbital values at the positions and spins occupied by the electrons in the configuration. That is to say, the coefficients of $\mathbf{U}$ are specified as
    \begin{equation}
        U_{i,j} = \phi_i(r(j), \sigma(j)).
    \end{equation}
\end{ansatz}

A description of a single \ac{SD} can easily be inferred by mean-field approaches such as the \ac{HF} method.
Because this cannot capture any correlations between the electrons, it is often useful to combine a mean-field reference state with an additional prefactor introducing appropriate descriptions of the correlations.
The general prototype for this construction is the Slater-Jastrow ansatz, in which a single \ac{SD} is multiplied with a Jastrow prefactor.
\begin{ansatz}[Slater-Jastrow]
    The amplitudes of a Slater-Jastrow ansatz are given by
    \begin{equation}
        \Psi(\mathbf{x}) = \Psi_J(\mathbf{x}) \times \Psi_{SD}(\mathbf{x}),
    \end{equation}
    where $\Psi_J$ is an electronic Jastrow ansatz according to Eq.~\ref{eq:electronic_Jastrow} and $\Psi_{SD}$ a single \ac{SD} as defined in Eq.~\ref{eq:SD}.
\end{ansatz}

The general idea of incorporating a mean-field type ansatz as a baseline reference into the state can also be extended beyond simple \acp{SD}.
Another reference state considered in this work is the Pfaffian pairing wavefunction~\cite{Bajdich2006}, which also spans the variational class of \acp{SD}.
Physically, this state is inspired by the idea of approximating a state as an anti-symmetrized product of two-electron functions (the pair orbitals), $\tilde{\phi}(r_i, \sigma_i, r_j, \sigma_j)$, over all electron pairs in the system.
By assuming an antisymmetric pair orbital (i.e., $\tilde{\phi}(r_i, \sigma_i, r_j, \sigma_j) = -\tilde{\phi}(r_j, \sigma_j, r_i, \sigma_i)$) and using the same orbital function for all electron pairs in the system, the resulting wavefunction amplitudes can be evaluated efficiently by means of Pfaffians of matrices.

The Pfaffian denotes a matrix operation that is defined for $2 n \times 2 n$ skew-symmetric matrices, i.e., ones for which the coefficients fulfil $A_{i,j} = - A_{j,i}$.
It is given by the anti-symmetrized product of the matrix coefficients evaluated w.r.t. a set of matrix index pairs.
These are the index pairs $(i,j)$ for which $i < j$ and where each index $1 \ldots 2n$ appears exactly once in an index pair of the set.
The anti-symmetrization is applied w.r.t. the different possible sets of pair indices, such that the result is antisymmetric under exchanges of a first index of one pair with a second index of another pair~\cite{bajdichPfaffianPairingBackflow2008}.
The resulting Pfaffian of a $2 n \times 2 n$ skew-symmetric matrix $\mathbf{A}$, denoted as $Pf(\mathbf{A})$, can be evaluated similarly to an evaluation of a determinant efficiently with roughly $\mathcal{O}(n^3)$ operations~\cite{wimmerEfficientNumericalComputation2012}.

\begin{ansatz}[Pfaffian wavefunction]
    The Pfaffian ansatz is, for even numbers of electrons, defined as~\cite{Misawa2017}
    \begin{equation}
        | \Psi \rangle = (\sum_{i,j = 1}^L \sum_{\sigma_i, \sigma_j \in \{\uparrow, \downarrow\}} \tilde{\phi}(r_i, \sigma_i, r_j, \sigma_j) \, \hat{c}^\dagger_{i,\sigma_i} \, \hat{c}^\dagger_{j,\sigma_j})^{N/2}| 0 \rangle.
    \end{equation}
    Its amplitudes can be evaluated as
    \begin{equation}
        \Psi(\mathbf{x}) \sim Pf(\mathbf{\tilde{U}} - \mathbf{\tilde{U}}^T),
    \end{equation}
    where the matrix $\mathbf{\tilde{U}}$ is an $N \times N$ matrix with coefficients
    \begin{equation}
        \tilde{U}_{i,j} = \tilde{\phi}(r_i, \sigma_i, r_j, \sigma_j).
    \end{equation}
\end{ansatz}

The Pfaffian reference states applied in this work are constructed to explicitly incorporate a vanishing total spin magnetization of the system.
This is achieved by fully parametrizing the part of the pair-orbitals corresponding to a pairing between a spin-up and a spin-down electron and explicitly setting the other blocks of the pair orbitals to zero.
This results in a total of $L \times L$ parameters for the Pfaffian reference state.
Further symmetries can also easily be incorporated into the reference state by including appropriate symmetry projections into the definition of the ansatz~\cite{Misawa2017}.
Due to the specific form of the mean-field reference states, it is also possible to efficiently include a projection onto an eigenstate of $\hat{\mathbf{S}}^2$ with a specified associated quantum number $S$.

The Hamiltonian of Fermi-Hubbard model in the specified basis connects each basis configuration with $\mathcal{O}(L)$ basis configurations with non-vanishing matrix elements.
A naive implementation of the local energy evaluation thus involves a number of wavefunction evaluations scaling linearly with the system size.
Neglecting marginal optimizations, each individual evaluation of a determinant or Pfaffian in the mean-field reference states scales roughly as $\mathcal{O}(L^3)$ (assuming a number of electrons proportional to $L$).
While this indicates an overall scaling of $\mathcal{O}(L^4)$ for each evaluation of the local energy, this can be reduced to a scaling of $\mathcal{O}(L^3)$ for the Hubbard models.
This can be achieved by applying efficient updating of the determinant/Pfaffian values for each connected configuration~\cite{neuscammanJastrowAntisymmetricGeminal2013, xuOptimizedImplementationCalculation2021}.
\acbarrier

\chapter{Bayesian regression techniques for wavefunctions}
\label{ch:GPS_introduction}
Based on the general framework for the description of many-body states outlined in the previous chapter, this chapter introduces the representation of \acp{GPS}.
The central motivation of the \ac{GPS} description is to define the representation based on Bayesian regression principles as emerging in the framework of \ac{GPR}.
Similar to the function approximation with \acp{NN}, such approaches can provide descriptions of input-output relationships that are extracted from presented data points.
By introducing generic probabilistic assumptions, the obtained representations are not restricted in their flexibility, but they also carry a high degree of interpretability and intuition.
These characteristics provide the main motivation for the utilization of the techniques to describe many-body quantum states, resulting in the \ac{GPS} ansatz.

\section{Bayesian learning with Gaussian Processes and linear models}
The descriptions underpinning the construction of the \acp{GPS} emerge from a probabilistic description of the presented data based on particularly suitable statistical assumptions.
In general, this means that a probability distribution over function outputs is described, thus also providing uncertainty measures for evaluated function predictions.
Basis for the utilized probabilistic description of the data is an assumed \textit{prior} probability distribution, defining the distribution over data points without any information from presented data points.
The available data points provide additional knowledge about the function that is incorporated via a \textit{likelihood}.
The key element for the Bayesian modelling of function estimations is to obtain a probability distribution that agrees with the likelihood of the data and the assumed prior.
As obtained from Bayes' theorem, this \textit{posterior} distribution is then found to be proportional to the product of likelihood and prior.

The \ac{GPR} approaches, as utilized in this work, are based on specific prior and likelihood assumptions that are intuitive and give rise to particularly compact, easily computable posteriors.
These models can be formalized via an interpretation of the target functions as \acp{GP}, or equivalently, these can also be viewed as models emerging from Bayesian regression with linear models in a (potentially high dimensional) space of features.
Both perspectives can provide valuable insight into the effectiveness and conceptual foundations of these models and are therefore briefly outlined in the following.
The formulation in terms of \acp{GP} provides foundational justifications and clear intuitions of the function approximation concepts.
The Bayesian regression with linear models, on the other hand, provide additional practical tools exploited in this work to obtain particularly compact many-body state descriptions.
A thorough description of the discussed Bayesian regression approaches can, e.g., be found in Ref.~\cite{Rasmussen2006}, where the two perspectives are denoted as the `function space' and the `weight space' view respectively.

\subsection{Gaussian Processes for function regression}
Though in a standard textbook, as the one referenced above, the \ac{GP} perspective is typically introduced as a formalization of the linear regression methods, here the \ac{GP} understanding of the applied function approximation is discussed first.

The general task achieved within the \ac{GPR} framework is that of inferring an input-output relation of an unknown function based on observed function samples (data points).
This is achieved by modelling a full probability distribution over the space of potential functions utilizing the concepts of \acp{GP}.
A \ac{GP} is formally defined as a sequence of variables for which any subset of variables taken from this sequence follows a joint normal distribution.
For the application of the \ac{GPR} framework, the function describing the wanted input-output relationship, here simply denoted as $f(\mathbf{x})$, is interpreted as a \ac{GP}.
This means that for any collection of input points $\{\mathbf{x}\}$, the probability distribution over the function values $\{f(\mathbf{x})\}$ follows a joint normal distribution.

This distribution over function outputs is specified by the mean values $\{\mu(\mathbf{x})\}$ and a covariance matrix.
The covariance matrix can be constructed by evaluating the covariances between pairs of inputs, $\mathbf{x}$ and $\mathbf{x}'$, which is defined via a kernel function $k(\mathbf{x}, \mathbf{x}')$ that is symmetric.
While a very intuitive interpretation of the kernel function is given in the next section, it can already be seen in this \ac{GP} formulation that it can be used to ensure smoothness of the approximated function.
If two inputs, $\mathbf{x}$ and $\mathbf{x}'$, are `close to each other' in some sense, then it can be expected that the function values associated with these two input points are highly correlated, and similar function values would be drawn when sampling a potential function from the \ac{GP}.
This notion of proximity between different inputs is formalized by the introduction of a suitable kernel function, which can be evaluated between different input points, in the sense that larger kernel values correspond to a larger correlation between function values.
The \ac{GP} is then fully characterized by the mean values, $\mu(\mathbf{x})$, and the kernel function defined for pairs of inputs $k(\mathbf{x}, \mathbf{x}')$.

\begin{figure}[htb!]
    \centering
    \includegraphics{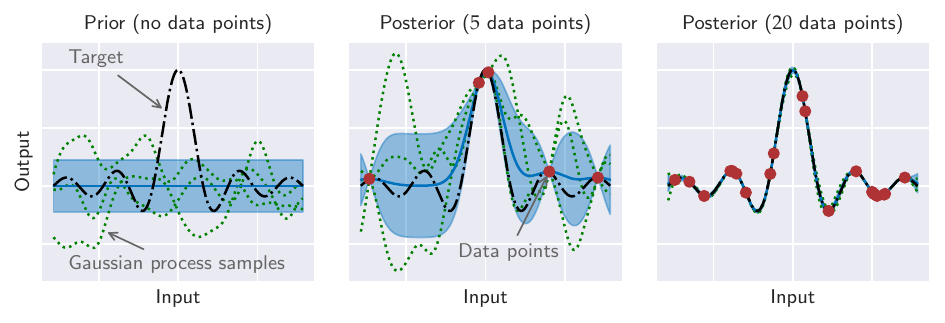}
    \caption[Exemplification of \acl{GPR} for learning the target function $sinc(x) = sin(x)/x$]{Exemplification of \ac{GPR} for learning the target function $sinc(x) = sin(x)/x$.
    The left plot shows three different functions drawn from the prior distribution (green dotted lines), defined by a constant mean of zero (solid blue line) and a covariance defined via the squared exponential kernel, also defining the fluctuation of function values around the mean (for which the standard deviation is indicated by the shaded area).
    The centre and right plot show functions drawn from the posterior distribution emerging based on $5$ (centre plot) and $20$ (right plot) samples from the target function (indicated by red points). The mean and the standard deviation of the posterior distribution are indicated by a solid blue line, respectively the shading of the surrounding area.}
    \label{fig:GPR_visualisation}
\end{figure}

Without any training data that can be used to infer the approximation, typically a vanishing mean of $\mu(\mathbf{x})=0$, is assumed.
This then defines a prior distribution over the functions, i.e., a distribution before taking any data points into account.
Specifying a concrete kernel function, it is possible to draw different realizations from this prior, which is visualized for the example of one-dimensional functions in the left panel of Fig.~\ref{fig:GPR_visualisation}.
The fluctuations about the zero mean of the function values $f(\mathbf{x})$ for the drawn realizations are described by the variance $s^2(\mathbf{x}) = k(\mathbf{x}, \mathbf{x})$.
It can also be seen that the drawn functions are smooth.
This is ensured by an appropriate choice of a kernel function that decays to zero w.r.t. the distance between any two positions on an appropriate length scale.
In this example, a squared exponential kernel is used, a very common choice for \ac{GPR}.
It can be defined as
\begin{equation}
    k(\mathbf{x}, \mathbf{x}') = s^2 e^{-\frac{|\mathbf{x} - \mathbf{x}'|}{2 l^2}},
    \label{eq:squared_exp_kern}
\end{equation}
where the parameter $s^2$ describes the variance of the fluctuations around the mean and the parameter $l$ controls the length scale of the decay.

Though it is possible to draw realizations from the prior, this does not yet incorporate any knowledge of data points based on which the final model should be constructed.
To incorporate the data points into the description, the probability distribution conditioned on the observed data points (the training set) needs to be inferred.
As a consequence of the \ac{GP} assumption, this can be evaluated easily and a closed form is obtained for this posterior distribution.
In particular, the posterior distribution over functions inferred from the available data points can again be understood as a \ac{GP} with modified mean and covariance.
Representing the set of observed function values as a vector $\mathbf{y}$, the mean of the posterior \ac{GP} evaluates to
\begin{equation}
    \mu(\mathbf{x})= \mathbf{k}(\mathbf{x})^T \, \mathbf{K}^{-1} \, \mathbf{y}.
    \label{eq:GPR_mean}
\end{equation}
In this representation, the kernel matrix $\mathbf{K}$ is obtained by the evaluation of the kernel function between all the training inputs, represented as a vector $\mathbf{X}$.
Its elements are therefore defined as
\begin{equation}
    K_{i,j} = k(\mathbf{x}_i, \mathbf{x}_j),
\end{equation}
where $\mathbf{x}_i$ denotes the $i$-th configuration from the data set.
The quantity $\mathbf{k}(\mathbf{x})^T$ represents a row vector of all the kernel values between the test input $\mathbf{x}$ and all the data set inputs with elements defined as
\begin{equation}
    k(\mathbf{x})_i = k(\mathbf{x}, \mathbf{x}_i).
\end{equation}
The covariance function for the posterior \ac{GP} can also be obtained in closed form and is given by
\begin{equation}
    \mathrm{cov}(\mathbf{x}, \mathbf{x}') = k(\mathbf{x}, \mathbf{x}') - \mathbf{k}(\mathbf{x})^T \, \mathbf{K}^{-1} \, \mathbf{k}(\mathbf{x}').
\end{equation}

This formulation of the \ac{GPR} framework underlines the commonly applied interpretation of \ac{GPR} as a parameter free function approximation scheme.
Based on the Gaussian assumptions and defining only a kernel function, a probability distribution over possible input-output relations is inferred from available data points without resorting to a functional model depending on parameters.
With the ability to evaluate the covariance matrices for the posterior \ac{GP}, the result therefore also incorporates knowledge over the variability of the inferred distribution.
This is also indicated in the centre and right panels of Fig.~\ref{fig:GPR_visualisation}.
These visualize exemplified functions drawn from the posterior distribution, which was inferred based on samples of the $sinc$ function, defined as $sinc(x) = sin(x)/x$.
Using the defined squared exponential kernel, the standard deviation in the posterior distribution, indicate by shaded areas in the figure, increases with increasing distance from the data points.
It thus represents a measure of uncertainty of the function approximation with the mean of the posterior \ac{GP}.

The results shown above assume perfect samples of the target distribution.
However, it is also possible to include an assumption of Gaussian noise of the data set into the \ac{GPR} framework.
While the details of the derivation are shown in the next sections, it is noted here that this results in a positive shift, equal to the variance of the Gaussian noise, that is added to the matrix $\mathbf{K}$ in the equations above.
It can therefore also significantly help with the numerical stability of the inference requiring a matrix inversion.

Applying the general ideas of \ac{GPR} to the description of many-body wavefunctions is the core idea leading to the definition of the \ac{GPS} outlined in this work.
While the resulting model can be motivated from different perspectives, the naming of the ansatz highlights a very important intuition: The state can be interpreted as one emerging from a \ac{GP} describing functions over the Hilbert space.
As such, it is associated with a probabilistic distribution over quantum states.

\subsection{Bayesian regression with linear models}
\label{sec:bayesian_regression}

An alternative perspective on the function estimator emerging in the context of \ac{GPR} is obtained by interpreting the framework as linear regression with Bayesian techniques in a space of features.
The function approximator, taking the form of a linear model in the space of features, can be defined as
\begin{equation}
    f_{lin}(\mathbf{x}) = \sum_{i=1}^{N_{features}} w_i \, \phi_i(\mathbf{x}).
\end{equation}
This defines a linear combination of the $N_{features}$ feature transformations, $\phi_i$, each mapping an input $\mathbf{x}$ to a scalar quantity.
The weights $w_i$ associated with the features are understood as parameters of the model, which can be obtained by fitting this model to a set of data.

\begin{figure}[htb!]
    \centering
    \includegraphics{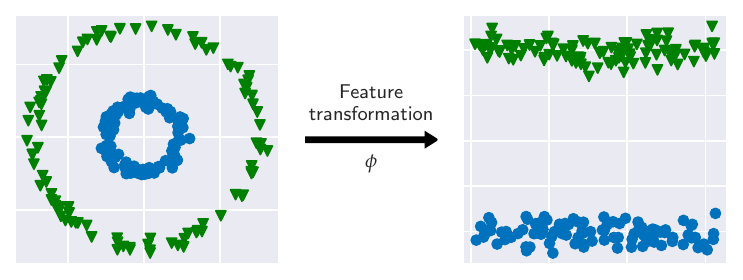}
    \caption[Exemplification of the linear separability obtained by transforming the inputs of a classification task into a specific feature space]{Exemplification of the linear separability obtained by transforming the inputs of a classification task into a specific feature space. The plots show two sets of data points associated with two different classes (blue and green scatter points). The left panel shows the exemplified data points in the original space of Cartesian coordinates, the right plot shows the same data in a feature space given by polar coordinates in which the two classes are linearly separable.}
    \label{fig:feature_space}
\end{figure}

The importance of the feature transformation is exemplified for the classification of data points as shown in Fig.~\ref{fig:feature_space}.
In the figure, two classes of data points (blue and green) are positioned in the space of Cartesian coordinates, which cannot be separated by a line in this original input space.
Due to the radial distribution of the data points however, it is possible to linearly separate the two classes in the space of polar coordinates.
Changing from Cartesian to polar coordinates is an example of a specific feature transformation of the inputs into a two-dimensional feature space in which a linear model is sufficient to separate the data points.

While, in the example of transforming the inputs into polar coordinates, the feature space is only two-dimensional, in general the feature space can be of much higher dimensionality.
Crucially, the linear model as specified above represents a universal approximator in the sense that any function can be described as the number of features is increased.
This can easily be seen by considering the feature space defined by the set of delta functions positioned at all possible positions of the input space.
This results in a one-to-one mapping between an input to an output simply given by the weight associated with this input.
It can therefore describe an arbitrary function over the input space.
While this construction shows the universal approximation property of the linear model, in practical applications it is however typically useful to introduce more generic features that are non-orthogonal.
These could, for example, be Gaussian functions placed at all positions of the input space.
This makes it possible to infer a general model from limited data that also generalizes beyond the given function samples, while it is still possible to represent any function to arbitrary accuracy.

The essential task to learning the linear model in the feature space is to find the weights $w_i$ based on a given data set.
In the following, the data set is specified by the vector of inputs $\mathbf{X}$ with an associated vector of available function values $\mathbf{y}$.
This regression task can be formulated in a rigorous Bayesian framework, which also relates the formulation back to the function estimation with \acp{GP} as introduced above.
Detailed descriptions of this Bayesian inference procedures, for which the main elements are repeated here, can, e.g., be found in Refs.~\cite{Tipping2003a} and~\cite{Rasmussen2006}.

The Bayesian inference of the weights is again achieved by bringing together a prior model with the observed data points to obtain a posterior distribution.
However, this is now achieved by performing the inference in the space of weights, rather than in the space of functions.
This means that the prior distribution, $p(\mathbf{w})$, describes a probability distribution over the model weights, here represented as a vector $\mathbf{w}$, that does not depend on any data points.
The prior is assumed to follow a multivariate Gaussian distribution with zero mean, where the covariances between the random variables are specified by a square positive semi-definite covariance matrix.
Denoting the covariance matrix as an inverse matrix $\mathbf{A}^{-1}$, the prior can therefore be defined as
\begin{equation}
    p(\mathbf{w}) = \sqrt{\det(\frac{\mathbf{A}}{2 \pi})} e^{-\frac{1}{2}\mathbf{w}^T \mathbf{A} \mathbf{w}}.
    \label{eq:prior}
\end{equation}

The data set can be taken into account by introducing a likelihood for data points describing a probability distribution over the function values $y = f(\mathbf{x})$ at the data points.
Assuming that the samples of the data set are generated around the values of the predictor $f_{lin}(\mathbf{x})$ with uncorrelated Gaussian noise, this results in a Gaussian likelihood with vanishing covariances between different data points.
The mean of the likelihood for the function value $y=f(\mathbf{x})$ is thus given by the linear predictor $f_{lin}(\mathbf{x})$ and the variance can be specified by a parameter $\sigma^2(\mathbf{x})$.
Denoting the features for a configuration $\mathbf{x}$ as a vector $\boldsymbol{\phi}(\mathbf{x})$, the likelihood over the data set of $N_{tr}$ data points can be expressed as
\begin{equation}
    p(\mathbf{y}|\mathbf{X}, \mathbf{w}) = \prod_{i=1}^{N_{tr}} \frac{1}{\sqrt{2 \pi \sigma^2(\mathbf{x}_i)}} e^{-\frac{1}{2 \sigma^2(\mathbf{x}_i)} |y_i - \mathbf{w}^T \boldsymbol{\phi}(\mathbf{x}_i) |^2}.
    \label{eq:likelihood}
\end{equation}
Inference of the weights can then be achieved by application of Bayes theorem giving a posterior probability distribution over the weights.
This combines the prior distribution with the training data.

For the discussed setup, Bayes' theorem defines the posterior distribution as
\begin{equation}
    p(\mathbf{w}|\mathbf{y}, \mathbf{X}) = \frac{p(\mathbf{y}|\mathbf{X}, \mathbf{w}) \, p(\mathbf{w})}{p(\mathbf{y}| \mathbf{X})},
\end{equation}
i.e., it is obtained as the product of likelihood and prior divided by a factor $p(\mathbf{y}| \mathbf{X})$.
This normalization factor is commonly referred to as the marginal likelihood and is obtained by integrating the numerator over the space of weights as
\begin{equation}
    p(\mathbf{y}| \mathbf{X}) = \int d \mathbf{w} \, p(\mathbf{y}|\mathbf{X}, \mathbf{w}) \, p(\mathbf{w}).
\end{equation}
One of the key consequences of the assumptions made in the Bayesian inference scheme, as introduced here, is that the posterior distribution over weights can be obtained in closed form.
As it emerges from the product of two Gaussian distributions, the resulting posterior is also a normal distribution.
The mean of the posterior, here denoted as a vector $\boldsymbol{\mu}_{mp}$, is found as
\begin{equation}
    \boldsymbol{\mu}_{mp} = \boldsymbol{\Sigma}\boldsymbol{\Phi}^T \mathbf{B} \mathbf{y}.
    \label{eq:mean_weights}
\end{equation}
The quantity $\boldsymbol{\Sigma}$ denotes the covariance matrix of the posterior distribution given by
\begin{equation}
    \boldsymbol{\Sigma} = (\boldsymbol{\Phi}^T \mathbf{B} \boldsymbol{\Phi} + \mathbf{A})^{-1}.
    \label{eq:covariance_weights}
\end{equation}
This expression utilizes the shorthand notation $\boldsymbol{\Phi}$ to denote the $N_{tr} \times N_{features}$ matrix with rows corresponding to the transposed feature vectors for all training inputs.
Similarly, $\mathbf{B}$ is a diagonal matrix for which the $i$-th diagonal is given by the inverse variance for the $i$-th data point, $1/\sigma^2(\mathbf{x}_i)$.

The derived posterior distribution describes a probability model over the weights, also factoring in the training data points.
With the specified dependence of the function outputs on the weights, this can be related back to a \ac{GP} model for the approximated function values.
Skipping some manipulations of the matrix equation, the \ac{GP} for the function values obtained in this formulation is specified by a mean function
\begin{equation}
    \mu(\mathbf{x}) = \boldsymbol{\phi}^T(\mathbf{x}) \mathbf{A}^{-1} \boldsymbol{\Phi}^T(\boldsymbol{\Phi} \mathbf{A}^{-1} \boldsymbol{\Phi}^T + \mathbf{B}^{-1})^{-1} \mathbf{y},
\end{equation}
and a covariance function
\begin{equation}
    \mathrm{cov}(\mathbf{x}, \mathbf{x}') = \boldsymbol{\phi}^T(\mathbf{x}) \mathbf{A}^{-1} \boldsymbol{\phi}(\mathbf{x}') - \boldsymbol{\phi}^T(\mathbf{x}) \mathbf{A}^{-1} \boldsymbol{\Phi} (\boldsymbol{\Phi} \mathbf{A}^{-1} \boldsymbol{\Phi}^T + \mathbf{B}^{-1})^{-1} \boldsymbol{\Phi} \mathbf{A}^{-1} \boldsymbol{\phi}(\mathbf{x}').
\end{equation}
It can be seen that these are equivalent to the \ac{GP} description (derived in the previous section for vanishing noise, i.e., $\mathbf{B}^{-1} = 0$) if one defines a kernel function
\begin{equation}
    k(\mathbf{x}, \mathbf{x}') = \boldsymbol{\phi}^T(\mathbf{x}) \mathbf{A}^{-1} \boldsymbol{\phi}(\mathbf{x}').
\end{equation}
This formulation therefore also provides a very intuitive interpretation of the kernel function as a scalar product between transformed feature vectors $\mathbf{A}^{-1/2} \boldsymbol{\phi}(\mathbf{x}')$.
These transformed features are thus simply linear combinations of the original features, and, if the covariance matrix of the prior, $\mathbf{A}^{-1}$, is chosen to be diagonal, the original features are just rescaled by the transformation.
Choosing a diagonal matrix $\mathbf{A}^{-1}$ for the covariance of the weights is a very standard choice, also commonly employed in this work.
It represents the case where no correlation between the weights is assumed a priori.

Based on the formulation above, it can be seen that the feature vectors don't need to be evaluated directly if scalar products between feature vectors can be evaluated directly.
This approach is commonly referred to as the `kernel trick', which makes it possible to use very high (or even infinite) dimensional feature spaces for which the scalar product between transformed feature vectors can be evaluated efficiently.
In practical scenarios both perspectives can be useful depending on whether the number of features is larger than the number of data points or the other way round.
Typically, the numerically most time-consuming part are the required inversions of matrices.
Whereas in the feature space formulation, the covariance matrix $\boldsymbol{\Sigma}$ needs to be calculated, which requires an inversion of a matrix of size $N_{features} \times N_{features}$, the \ac{GPR} formulation requires an inversion of the $N_{trn} \times N_{trn}$ matrix $(\mathbf{K} + \mathbf{B}^{-1})$.

Overall, the discussed Bayesian inference scheme derives a posterior probability distribution for the weights based on the assumed prior distribution and the likelihood for the presented data.
Due to the posterior distribution being Gaussian, the mean of the posterior is equal to the most probable weights, $\boldsymbol{\mu}_{mp}$.
These are therefore also exactly those parameters that maximize the product of likelihood and prior, or equivalently its logarithm, i.e.,
\begin{align}
    \boldsymbol{\mu}_{mp} &= \mathrm{argmax}_{\mathbf{w}} \left ( \log(p(\mathbf{w}) \times p(\mathbf{y}|\mathbf{X}, \mathbf{w})) \right ), \\
    &= \mathrm{argmin}_{\mathbf{w}} \left (\sum_{i=1}^{N_{trn}} \frac{|y_i - \mathbf{w}^T \boldsymbol{\phi}(\mathbf{x}_i) |^2}{\sigma^2(\mathbf{x}_i)} + \mathbf{w}^T \mathbf{A} \mathbf{w} \right ).
\end{align}
Neglecting the term $\mathbf{w}^T \mathbf{A} \mathbf{w}$, the inference of the weights based on the Bayesian approach is therefore also equivalent to a weighted least squares fit of the linear model to the training data, where the squared error terms are weighted by the inverse noise parameters $1/\sigma^2(\mathbf{x}_i)$.
The second term can be seen as an additional regularization term that favours solutions for which the norm of $||\mathbf{A}^{1/2} \mathbf{w}||^2$ is small.
Within least squares fitting approaches, this regularization approach has been discussed in different contexts and is also referred to as Tikhonov regularization~\cite{tikhonovNumericalMethodsSolution1995}.
A common choice is to apply an inverse covariance matrix proportional to the identity, $\mathbf{A} = \alpha \mathbb{1}$, resulting in a standard $L_2$ regularization favouring solutions with small weights.

While the Bayesian regression framework outlined above defines a straightforward way to obtain the solution for the weights, the results particularly depend on some design choices that need to be made to apply the method.
Firstly, suitable features need to be specified.
As it was outlined above, the specification of features is directly related to the choice of a kernel function in the \ac{GPR} framework.
It can be viewed as a transformed scalar product in the feature space (which could even be infinite dimensional).
Furthermore, the regression requires the concretization of the probabilistic models, through the choice of the prior covariance matrix, as well as the characterization of the data noise.

As another equivalence between the perspectives, the mean \ac{GPR} predictor can also be interpreted as a linear model in a feature space generated by the $N_{tr}$ kernel values evaluated for the data inputs.
This means that the mean of the \ac{GP} can be understood as a linear model
\begin{equation}
    f(\mathbf{x}) = \sum_{i=1}^{N_{tr}} w_i \, k(\mathbf{x}, \mathbf{x}_i).
\end{equation}
In the \ac{GPR} prediction, the weights would be fixed to the mean given by the expression as specified in Eq.~\ref{eq:GPR_mean} according to
\begin{equation}
    \mathbf{w} = \mathbf{K}^{-1} \, \mathbf{y}.
\end{equation}
Reinterpreting the weights instead as free model parameters, the model can be adjusted to be defined with essentially any set of generic support points.
This means that a linear model is introduced that is defined as
\begin{equation}
    f(\mathbf{x}) = \sum_{i=1}^{M} w_i \, k(\mathbf{x}, \mathbf{x}'_i),
    \label{eq:linear_kernel_model_support_points}
\end{equation}
with the key difference to the GP mean that the set of $M$ support points $\{\mathbf{x}'\}$ is allowed to be different from the set of training inputs.
For the example of the squared exponential kernel, this reformulation would therefore simply define a linear combination of Gaussian functions placed at the different support points.

The main advantage over the strict \ac{GPR} formulation is that the number of terms, $M$, can in practice be chosen to be smaller than the number of training points, therefore improving the computational efficiency of the model evaluation.
Nonetheless, the full expressivity of the \ac{GPR} model can always be restored if the training inputs are used as support points showing that such models also define universal approximators as $M$ is increased.
However, a sparser representation, giving similar overall performance, might be achieved if the model is defined with a particularly suitable set of support points.
A good choice of the support points is therefore crucial to obtain a sparse representation.
Different approaches to achieve this, exemplified for the context of \ac{GPS} representations, are discussed later in this work.

The linear combination of features, as presented in Eq.~\eqref{eq:linear_kernel_model_support_points}, does not necessarily require the functions $k(\mathbf{x}, \mathbf{x}')$ to be actual kernel functions giving rise to a valid covariance definition for the GP formulation.
This means that the functions $k(\mathbf{x}, \mathbf{x}')$, directly specifying linearly combined features, do not have to be positive semi-definite and symmetric under exchange of the arguments.
This would, however, be required for `true' kernels that can also be associated with scalar products in a feature space.
The features $k(\mathbf{x}, \mathbf{x}')$ used to construct the \ac{GPS} in this work frequently do not strictly satisfy the requirements for an actual kernel function.
Nonetheless, they are directly related to kernel functions commonly used in the \ac{GPR} framework and utilize their representational power.
Hence, these are, in the following, nonetheless often referred to as `kernels'.
This also makes the terminology equivalent with the existing literature~\cite{Glielmo2020,Rath2020,boothQuantumGaussianProcess2021}.
Based on the equivalence between a Bayesian fitting of a linear model to given data points and the \ac{GP} formulation as outlined above, it is consequently also always possible to define a new kernel function according to~\cite{Rasmussen2006}
\begin{equation}
    \tilde{k}(\mathbf{x}, \mathbf{x}') = \sum_{i,j=1}^M k(\mathbf{x}, \mathbf{x}_i') A^{-1}_{i,j} k(\mathbf{x}_j', \mathbf{x}').
\end{equation}
This function satisfies the required kernel properties and defines a covariance for an interpretation of the Bayesian fit in terms of \acp{GP}.

\section{Gaussian Process States}

\subsection{Gaussian Process models for many-body wavefunctions}
\label{sec:GPS_def}
The central idea of the \ac{GPS} representation is to construct a wavefunction ansatz based on the \ac{GPR} framework outlined in the previous section.
More specifically, an ansatz is constructed defining the wavefunction amplitudes based on the mean of a \ac{GP}.
However, the function estimation is simply interpreted as a linear combination of kernel functions as presented in Eq.~\eqref{eq:linear_kernel_model_support_points}.
In the standard \ac{GP} framework it is assumed that the model is defined directly through the data provided for the regression in a parameter free way.
Through the interpretation of the model as a weighted linear combination of kernel functions, more compact representations can be achieved by identifying particularly relevant physical configurations.
This perspective therefore offers a more flexible framework for the representation of many-body quantum states, while the Bayesian regression techniques are still applicable.
The configurations that define the linear combination of kernel functions are, in the following, denoted as the support configurations of the model.

Following the motivation outlined in section \ref{sec:product_sep} of the previous chapter, a sensible representation of quantum states should factorize as a product over the different system components according to
\begin{equation}
    \Psi(\mathbf{x}) = \prod_{i=1}^L F_i(\mathbf{x}).
\end{equation}
In this form, the different models $F_i(\mathbf{x})$ can be designed to extract the correlation of the $i$-th mode with its environment.
A natural application of the kernel models to the description of quantum states can therefore be achieved by describing each function $F_i(\mathbf{x})$ as a separate linear model of the form
\begin{equation}
    F_i(\mathbf{x}) = \sum_{j=1}^{M_i} w^{(i)}_j \, k^{(i)}(\mathbf{x}, \mathbf{x}'^{(i)}_j).
\end{equation}
Here the index $i$ is simply used to label a separate linear model associated with each of the system modes.
This model is constructed with specified kernel functions $k^{(i)}$, which can be evaluated for any computational basis state $\mathbf{x}$ and support configuration $\mathbf{x}'^{(i)}_j$.
The support configurations, are, in the following, always marked by a `prime' symbol.
These are selected quantum states, in this work, always considered to be either computational basis configurations, or, in extensions of the model discussed later, product states.
The model weights, $w^{(i)}_j$, can be interpreted as continuous variational parameters of the model.

While the product construction defines a valid wavefunction ansatz, it practically makes sense to merge the product of $L$ different linear approximators into a single one.
As it is also applied in other common \ac{VMC} wavefunction models, such as Jastrow ansatzes, this can be achieved by applying the linear model as an ansatz to model the log wavefunction amplitudes, resulting in a model
\begin{equation}
    \Psi(\mathbf{x}) = e^{\sum_{i=1}^L\sum_{j=1}^{M_i} w^{(i)}_j \, k^{(i)}(\mathbf{x}, \mathbf{x}'^{(i)}_j)}.
\end{equation}
The $L$ different linear models can be merged into a single one, which then defines the general \ac{GPS} representation.
\begin{ansatz}[\Acl{GPS}]
    The \acl{GPS} is specified by an exponentiated linear combination of kernel functions for the wavefunction amplitudes, i.e.,
    \begin{equation}
        \Psi(\mathbf{x}) = e^{\sum_{i=1}^{M} w_i \, k^{(i)}(\mathbf{x}, \mathbf{x}'_i)}.
    \end{equation}
\end{ansatz}
In this general definition of the \ac{GPS}, the kernel function carries an additional index $i$, indicating that in principle different kernel functions can be used (and the same support configuration might be included multiple times with different kernel functions).

In analogy to the bond dimension of \ac{MPS}, the total number of support configurations, $M$, is in the following denoted as the support dimension of the \ac{GPS}.
With the log wavefunction model obtained from merging $L$ linear models into one, it can be expected that this dimension should grow linearly with the size of the system to achieve appropriate size-extensivity.
However, for translationally invariant lattice systems as studied in this work, a symmetrized version of the model related to a convolutional symmetrization of \acp{NN} can be introduced.
This means, that the exact same linear model, $\sum_{j=1}^{M_i} w^{(i)}_j \, k^{(i)}(\mathbf{x}, \mathbf{x}'^{(i)}_j)$, is used for all the $L$ lattice sites (here labelled by index $i$).
That is to say, the model $F_i$ should give the same output as the model $F_j$ for an input that is shifted according to the translation from site $i$ to site $j$.

Applying the same estimator for each site, results in a model with a symmetrized kernel function.
It is symmetrized according to all translations across the system.
Further symmetries, such as point group symmetries of the lattice or spin inversion symmetries, can be incorporated equivalently, defining a feature-symmetrized (or kernel-symmetrized) \ac{GPS}.
\begin{ansatz}[Kernel-symmetrized \acl{GPS}]
    The kernel-symmetrization of the \ac{GPS} amplitudes is obtained as
    \begin{equation}
        \Psi(\mathbf{x}) = e^{\sum_{i=1}^{M} w_i \, \sum_{\{\mathcal{S}\}} k(\mathcal{S}[\mathbf{x}], \mathbf{x}'_i)}.
        \label{eq:kernel_symmetrization}
    \end{equation}
    Here, the set $\{\mathcal{S}\}$ consists of all symmetry operations the state should be invariant under (especially including translations across the lattice).
\end{ansatz}
The index $i$ for the kernel is neglected because the specific \ac{GPS} realizations considered in this work can all be specified with a single kernel function (similar to a \ac{GPR} model).
Nonetheless, it would in theory also be possible to combine multiple different kernel functions.
The kernel symmetrized \ac{GPS} defines a fully symmetric ansatz satisfying $\Psi(\mathbf{x}) = \Psi(\mathcal{S}[\mathbf{x}])$ for all included symmetry operations $\mathcal{S}$.
It is therefore not directly applicable to describe quantum states associated with non-trivial characters of the symmetry group.
Such states can either be described by including an appropriate symmetrization including the correct phase prefactors into a different reference state, or by using a projective symmetrization scheme, which is outlined in section \ref{sec:symmetrization} of chapter \ref{ch:GPS_VMC}.

Overall, the variational class of the \ac{GPS} emerges from two essential components defining the state characteristics that can be described with the model.
The definition of a \ac{GPS} ansatz class requires the definition of an appropriate kernel function, as well as a suitable set of support points for the model.
The following sections outline how physically intuitive kernel functions can be constructed and how particularly relevant features can be identified from given wavefunction data based on the Bayesian regression framework outlined above.

\subsection{Designing kernel functions for quantum many-body states}
Leaving the task of selecting a good set of support configurations aside for now, the variational expressibility of the \ac{GPS} is essentially governed by the choice of the kernel function.
The kernel function simply takes as input two configurations $\mathbf{x}$ and $\mathbf{x}'$ from the Hilbert space and maps these to a scalar quantity $k(\mathbf{x}, \mathbf{x}')$.
Following the discussion above, it can also directly be identified as a scalar product in the space of modelled correlation features if it is a symmetric positive semi-definite function.
Importantly though, this scalar product is evaluated directly through the kernel function without requiring to apply the feature transformation.
This characteristic makes it possible to capture exponentially large feature spaces and efficiently include a broad variety of different correlation properties into the \ac{GPS}.
Assuming that essentially any many-body configuration of the Hilbert space could be included into the set of support configurations, the kernel function thus specifies which exact correlation features are modelled (and what importance is attributed to them).

By modelling the log wavefunction amplitudes as a linear model, the resulting \ac{GPS} can be interpreted as a product over the weighted correlation features, in spirit very similar to the construction of \acp{CPS}.
Crucially though, based on the kernel trick, it is possible to combine multiple (potentially infinitely many) correlation features, which are only defined implicitly.
Nonetheless, the correlation features also described in \ac{CPS} representations can provide a sensible starting point to design appropriate kernel functions for the description of correlated quantum states.

As an example, the (non-symmetrized) kernel $k$ could be chosen to explicitly extract the correlation properties within a local environment around a chosen reference site.
In the non-symmetrized formulation above, which associates a separate kernel function $k^{(i)}$ with each system mode $i$, it would make sense to identify the mode label $i$ as the reference site.
However, with the translational symmetrization as above, the choice of the reference site, here denoted as $r(0)$, is arbitrary as the symmetrization ensures that the kernel comparison is applied at all sites of the system.
The correlation properties are extracted by simply comparing the configuration at the chosen environment around the reference site $r(0)$ of the test configuration $\mathbf{x}$ with that of the support configuration $\mathbf{x}'$.
Such a construction defines an $n$-body plaquette kernel, defined as
\begin{equation}
    k_n(\mathbf{x}, \mathbf{x}') = \prod_{i=0}^{n-1} \delta_{x_{r(i)}, x'_{r(i)}}.
    \label{eq:plaquette_kernel}
\end{equation}
The positions $r(i)$ denote the indices of the sites that are covered by the plaquette comprising $n$ sites around the reference site.
A central building block to construct this kernel function (and others introduced below) is the Dirac delta, $\delta_{x_{r(i)}, x'_{r(i)}}$.
This either evaluates to one if the two local states at site $r(i)$ are equal and to zero otherwise.
The plaquette kernel therefore either returns zero or one depending on whether the two configurations are the same over the specified plaquette or not.

Just as the definition of the \ac{CPS}, the plaquette kernel requires to define the correlation plaquettes over which the correlation properties are described.
In principle, it is possible to use plaquettes that span across the full lattice in which case the kernel could still be evaluated efficiently.
However, such a construction would simply give a one-to-one matching of a configuration with a specific support configuration meaning that the kernel either evaluates to one if both configurations are the same or to zero otherwise.
Nonetheless, some specific states might still be described efficiently (i.e., using a non-exponential number of support configurations) with such a construction.
Due to the direct one-to-one identification of basis configurations with features, these are exactly those wavefunctions for which the amplitudes in the chosen basis deviate from a constant only for a small set of basis states.
Examples for such states include the `W state' and the `Greenberger-Horne-Zeilinger state'~\cite{durThreeQubitsCan2000}, common prototypes for highly entangled states.

In order to use the \ac{GPS} as a tool to describe general quantum states capturing arbitrary correlation ranges across the system, it is sensible to introduce smoother kernel functions giving rise to non-orthogonal features for different basis states.
Rather than restricting the modelled correlation features to a specific range, a kernel can be introduced that defines the features as linear combinations over possible correlations of up to a specified number of constituents across the whole system.
Such a construction inspires the $p$-body kernel, which can be defined as~\cite{Glielmo2020}
\begin{equation}
    k_p(\mathbf{x}, \mathbf{x}') = \frac{\delta_{x_{r(0)}, x'_{r(0)}}}{\mathcal{N}} \left(\theta + \frac{1}{p-1} \sum_{i=1}^{L-1} \frac{\delta_{x_{r(i)}, x'_{r(i)}}}{f(r(i))}\right)^{p-1}.
    \label{eq:p_body_kern}
\end{equation}
The normalization $\mathcal{N}$ ensures that the kernel values are normalized to be between zero and one and is defined as
\begin{equation}
    \mathcal{N} = \left(\theta + \frac{1}{p-1} \sum_{i=1}^{L-1} \frac{1}{f(r(i))}\right)^{p-1}.
\end{equation}
In addition to a specified correlation order $p$, the $p$-body kernel is parametrized by an introduced order-weighting hyperparameter, $\theta$, and a displacement weighting function $f(i)$ (which can be parametrized further with additional hyperparameters).
The parameter $\theta$ controls the weighting of higher-order correlation features with respect to lower-order features, in a way that larger values of $\theta$ put a stronger weighting on those correlations that contain fewer modes.
The parametrized function $f(i)$ can be defined to additionally introduce a weighting of the features based on the positions of the sites contained in each correlation feature.

If the correlations in the target state are expected to be local (as it is the case for states with a low degree of entanglement), it is sensible to weight correlations across more local plaquettes higher in the kernel function.
This can be achieved by defining the function $f(i)$ such that it only depends on the distance from the reference site and decays with increasing distance.
A choice employed in this work is the polynomial decay modelled by the weighting
\begin{equation}
    f(i) = \left \vert r(i) - r(0) \right \vert^\gamma,
    \label{eq:polynomial_decay_weighting}
\end{equation}
depending on a new hyperparameter $\gamma$ controlling the rate of the decay.

It can be seen that an expansion of the $p$-body kernel above yields products of up to $p$ delta functions, each identifying one particular configuration pattern contributing to the extracted features.
The properties described by this kernel only ever capture correlations between up to $p$-components, i.e., these could individually all be described with $p$-body functions over corresponding plaquettes.
However, this kernel combines all possible plaquettes and does not include any restriction on the range or shape of these.
This is very similar to the construction of a Jastrow ansatz, which defines the wavefunction as a product over all possible two-body functions of pairs of sites.
The two-body kernel ($p=2$) constructed from all linear combinations of one- and two-body functions, thus defines \acp{GPS} with an equivalent variational span as the generalized Jastrow ansatz as defined in Eq.~\eqref{eq:general_Jastrow} of the previous chapter.

The parametrization of the $p$-body kernel as above highlights a key strength of the \ac{GPS} construction.
For correlation orders $p \geq L$, the emerging \ac{GPS} is a complete model.
This means that any wavefunction can be described (satisfying the incorporated symmetry) if the support configurations correspond to the full Hilbert space basis.
Crucially though, even for sparser sets of support configurations, the model combines all possible correlation features across the Hilbert space in this limit efficiently through the kernel function.
It can therefore be expected that such a model allows for a relatively smooth and systematic improvement of the model by increasing the size of the support configurations set, i.e., the support dimension.
However, the effectiveness of the model to describe target states of interest compactly practically also depends on the support configurations as well as the kernel hyperparameters that need to be specified.
Approaches to find sensible choices for these are presented in the following sections.

The $p$-body kernel can be seen as a very general sensible kernel prototype for \ac{GPS}.
It is constructed based on possible correlation plaquettes across the system, and the resulting model can still be evaluated efficiently (assuming only a small set of support configurations is chosen).
To be able to capture all possible correlations, the order $p$ should be chosen to be at least equal to the system size $L$.
Instead of scaling the correlation order $p$ with the size of the system, it is also possible to directly evaluate the $p \to \infty$ limit of the kernel~\cite{Glielmo2020}, allowing for a size independent construction.
Exploiting the identity $e^x = \lim_{n \to \infty} (1 + \frac{x}{n})^n$, the evaluation of this limit defines a kernel~\cite{Rath2020}
\begin{equation}
    k_{p \to \infty}(\mathbf{x}, \mathbf{x}') = \lim_{p \to \infty} k_p(\mathbf{x}, \mathbf{x}') = \delta_{x_{r(0)}, x'_{r(0)}} \, e^{-h(\mathbf{x}, \mathbf{x}')/\theta}.
    \label{eq:exponential_kernel}
\end{equation}
This depends on the function
\begin{equation}
    h(\mathbf{x}, \mathbf{x}') = \sum_{i=1}^{L-1} \frac{1 - \delta_{x_{r(i)}, x'_{r(i)}}}{f(i)}.
\end{equation}

\begin{figure}[htb!]
    \centering
\begingroup%
  \makeatletter%
  \providecommand\color[2][]{%
    \errmessage{(Inkscape) Color is used for the text in Inkscape, but the package 'color.sty' is not loaded}%
    \renewcommand\color[2][]{}%
  }%
  \providecommand\transparent[1]{%
    \errmessage{(Inkscape) Transparency is used (non-zero) for the text in Inkscape, but the package 'transparent.sty' is not loaded}%
    \renewcommand\transparent[1]{}%
  }%
  \providecommand\rotatebox[2]{#2}%
  \newcommand*\fsize{\dimexpr\f@size pt\relax}%
  \newcommand*\lineheight[1]{\fontsize{\fsize}{#1\fsize}\selectfont}%
  \ifx\svgwidth\undefined%
    \setlength{\unitlength}{453.53803019bp}%
    \ifx\svgscale\undefined%
      \relax%
    \else%
      \setlength{\unitlength}{\unitlength * \real{\svgscale}}%
    \fi%
  \else%
    \setlength{\unitlength}{\svgwidth}%
  \fi%
  \global\let\svgwidth\undefined%
  \global\let\svgscale\undefined%
  \makeatother%
  \begin{picture}(1,0.32000373)%
    \lineheight{1}%
    \setlength\tabcolsep{0pt}%
    \put(0,0){\includegraphics[width=\unitlength,page=1]{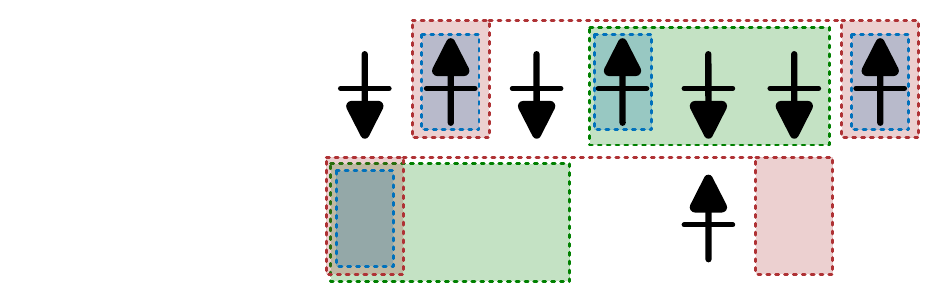}}%
    \put(0.07447265,0.22028129){\makebox(0,0)[lt]{\lineheight{1.25}\smash{\begin{tabular}[t]{l}Test configuration $\mathbf{x}$\end{tabular}}}}%
    \put(0.02470666,0.075867){\makebox(0,0)[lt]{\lineheight{1.25}\smash{\begin{tabular}[t]{l}Support configuration $\mathbf{x}'$\end{tabular}}}}%
    \put(0,0){\includegraphics[width=\unitlength,page=2]{figure05_svg-tex.pdf}}%
  \end{picture}%
\endgroup%

    \caption[Matching of sub-configuration patterns between two configurations in the $p$-body kernel and the exponential kernel]{Matching of sub-configuration patterns between two configurations in the $p$-body kernel and the exponential kernel. The final kernel value can implicitly be associated with a weighted sum of delta functions over all possible configuration patterns of up to $p$ modes including the reference site. Exactly those correlation plaquettes contribute to the final kernel value over which the two configurations are the same. The coloured plaquettes indicate such equal sub-configurations for the two example spin configurations $\mathbf{x}$ and $\mathbf{x}'$.}
    \label{fig:kernel_design}
\end{figure}

The function $h(\mathbf{x}, \mathbf{x}')$ can be understood as a weighted distance metric between the two sub-configurations of $\mathbf{x}$ and $\mathbf{x}'$ at all sites except for the reference site.
In the case of a uniform distance weighting, i.e., $f(i) \sim 1$, the function $h(\mathbf{x}, \mathbf{x}')$ is equivalent to the Hamming distance between two strings~\cite{hammingErrorDetectingError1950}, simply counting the number of positions where the two sub-configurations are not equal.
The obtained kernel is therefore directly related to generic kernel functions commonly employed in the \ac{GPR} framework based on the exponentiation of a function of a distance measure between data points.
In addition to the squared exponential kernel discussed before, these also include kernels directly exponentiating a negative (scaled) distance, commonly denoted as exponential kernels~\cite{Rasmussen2006} or Laplacian kernels~\cite{ruppMachineLearningQuantum2015}.
The $k_{p \to \infty}$ function is therefore, in the following, referred to as the exponential kernel.
Within the context of \ac{GPS} it encapsulates a large degree of intuition about the correlation properties that are described.
Representing the limit of the $p$-body kernel for infinite correlation order $p$, it can be associated with an efficient evaluation of the weighted linear combination of features including all possible correlation plaquettes across the system.
Such a kernel construction is visualized in Fig.~\ref{fig:kernel_design}, which exemplifies how the kernel value is obtained by finding matching patterns over entanglement plaquettes around a reference site within the two example configurations $\mathbf{x}$ and $\mathbf{x}'$.

The kernel construction discussed is also related to the general concept of additive \acp{GP}.
For these, the kernel functions are defined as linear combinations over different correlation orders for which the terms for the $n$-th correlation order are given by the sum over all possible products of $n$ one-dimensional base kernels~\cite{Duvenaud2011}.
For the exponential \ac{GPS} kernel, a linear combination of different correlation orders can be extracted by analysing the Taylor expansion of the kernel function.
This yields the expansion
\begin{multline}
    k_{p \to \infty}(\mathbf{x}, \mathbf{x}') = \\ \delta_{x_{r(0)}, x'_{r(0)}} \left( 1 - \sum_{i=1}^{L-1} \frac{(1 - \delta_{x_{r(i)}, x'_{r(i)}})}{\theta f(i)} + \frac{1}{2} \sum_{i=1}^{L-1} \sum_{j=1}^{L-1} \frac{(1 - \delta_{x_{r(i)}, x'_{r(i)}}) (1 - \delta_{x_{r(j)}, x'_{r(j)}})}{\theta^2 f(i)f(j)} - \ldots \right).
\end{multline}
With the prefactor $\delta_{x_{r(0)}, x'_{r(0)}}$ extracting all correlation plaquettes in which the reference site is contained, the $n$-th order of the Taylor expansion can thus be associated with all correlation plaquettes containing up to $n+1$ sites.
The weighting for the terms of the $n$-th order plaquettes involve prefactors of the form $\prod_{i=1}^L \theta \times f(j_i)$.
The combined weighting function $\theta \times f(j_i)$ thus effectively controls the weighting of the different correlation orders in this kernel.
For larger values $\theta \times f(i)$, the relative contribution of the single mode distance $(1 - \delta_{x_{r(i)}, x'_{r(i)}})$ decreases in the higher correlation orders so that a higher relative importance is associated to smaller entanglement plaquettes including modes $r(0)$ and $r(i)$.

\section{Bayesian compression of quantum states}
\label{sec:bayesian_state_learning}
The kernel functions introduced in the previous section implicitly define how correlation features are described by the \ac{GPS}.
As discussed, the introduced kernels allow for an efficient approach to capture correlation characteristics over arbitrary ranges and numbers of system components.
Together with the choice of the kernel hyperparameters, the general span of \ac{GPS} however also depends on the set of support configurations on which the representation is based.
While the kernel can be evaluated efficiently for any two configurations $\mathbf{x}$ and $\mathbf{x}'$, the \ac{GPS} representation is only an efficient representation if a compact support configuration set can be found.
Finding appropriate kernel hyperparameters and support configurations is therefore a crucial element in order to utilize the \ac{GPS} representation as a practical numerical tool.
Ideally, the most efficient \ac{GPS} representation should be extracted that either reaches some pre-defined target accuracy or for which the computational effort is fixed.

In this section, a scheme is introduced to achieve exactly this task for the \ac{GPS} in a data-driven way.
This means that the extraction of the support configurations is based on learning a representation from available wavefunction data.
This data consists of basis configurations, together with associated wavefunction amplitudes.
While this approach requires the availability of wavefunction data, it is shown that this method makes it possible to systematically obtain a highly compact representation of the data.
The introduced scheme is based on the concept of \acp{RVM} --- an approach to extract the most relevant features for a linear model based on a rigorous application of the Bayesian regression techniques as introduced in section \ref{sec:bayesian_regression}.

The kernel-symmetrized \ac{GPS}, as defined in section \ref{sec:GPS_def}, defines the log amplitudes of the wavefunction in the computational basis as a linear model according to
\begin{equation}
    \omega(\mathbf{x}) = \log \left( \Psi(\mathbf{x}) \right) = \sum_{i=1}^M w_i \sum_{\{\mathcal{S}\}} k(\mathcal{S}[\mathbf{x}], \mathbf{x}_i').
\end{equation}
The central idea presented in this section is to fit this linear model for the log amplitudes to the presented data set utilizing the Bayesian framework.
This is achieved by identifying the symmetrized kernel values as the different model features, $\Phi_{\mathbf{x}'_i} = \sum_{\{\mathcal{S}\}} k(\mathcal{S}[\mathbf{x}], \mathbf{x}_i')$, and transform the given data set of wavefunction amplitudes into the log space.
Specifically, the configurations of a data set are combined into a vector $\mathbf{X}$.
The data values are given by a vector $\boldsymbol{\omega}$, which comprises the logarithm of the given wavefunction amplitudes associated with the configurations.
As a starting point, in this section the training data is chosen to be the exact wavefunction data for ground states of systems small enough to obtain the state by exact diagonalization of the Hamiltonian.
Typically, the wavefunction amplitudes of the data set are rescaled to fix their general order of magnitude within the training procedure.
The (neglected) normalization of the state is incorporated in the evaluation of expectation values.

In addition to specifying the support configurations and the kernel hyperparameters, the regression approach requires the specification of a prior variance for the weights of the model, as well as a specification of the noise process variance, $\sigma^2(\mathbf{x})$.
In general, the exact wavefunction data, on which the model is fit, is exact, i.e., noise free.
Nonetheless, it is generally helpful to assume some degree of noise in the training data.
This allows for incorporation of the regularization through the prior specification.
Furthermore, the most relevant support configurations can be identified with the probabilistic techniques (which is the central idea of the \ac{RVM}).

The application of the Bayesian learning approach requires the definition of the prior and the likelihood variance.
As is standard, the covariance matrix of the Gaussian prior is defined as a diagonal matrix with parametrized diagonal.
The likelihood variance is chosen in a way specific to the fact that the linear model is applied in the log space and not in the space of actual amplitudes.

The variance of the likelihood essentially characterizes the variability of the log amplitudes around the mean parametrized by the linear model.
Under the assumption of a constant, configuration independent variance $\sigma^2(\mathbf{x}) = \sigma^2$, the errors of the log wavefunction amplitudes would therefore be expected to be constant.
While this is often a reasonable choice, an error on the log amplitudes that is assumed to be constant results in an error growing with the magnitude of the wavefunction amplitudes~\cite{Glielmo2020}, which is usually not desired.
The assumption of a Gaussian likelihood with variance $\sigma^2(\mathbf{x})$ for the log amplitudes, can also be transferred to the random variables associated with the actual amplitudes, $\Psi(\mathbf{x})$.
For these, a log-normal distribution is obtained.
Such a log-normal distribution of the actual amplitudes has a variance that evaluates to
\begin{equation}
    \mathrm{Var}(\Psi(\mathbf{x})) = \left ( e^{\sigma^2(\mathbf{x})} - 1 \right ) |\langle \Psi(\mathbf{x}) \rangle|^2,
    \label{eq:var_wavefunction_space}
\end{equation}
where $\langle \Psi(\mathbf{x}) \rangle$ denotes the mean of the (non-log) amplitude likelihood.
It can directly be seen, that this variance grows linearly with the mean if $\sigma^2(\mathbf{x})$ is constant.
This means that the errors of the fit would generally increase with increasing wavefunction amplitude magnitude.

In order to achieve an error of the amplitudes that is (approximately) constant, a single parameter $\tilde{\sigma}^2$ is introduced to characterize the variance $\mathrm{Var}(\Psi(\mathbf{x}))$.
Inserting this fixed variance into Eq.~\eqref{eq:var_wavefunction_space}, the equation can be solved for $\sigma^2(\mathbf{x})$.
This gives the expression
\begin{equation}
    \sigma^2(\mathbf{x}) = \log \left( \frac{\tilde{\sigma}^2} {|\langle \Psi(\mathbf{x}) \rangle|^2} +1 \right).
    \label{eq:log_space_likelihood_variance}
\end{equation}
Assuming that the presented data points can be fit well with the model, the mean of the amplitude likelihood evaluated for a data point, $\langle \Psi(\mathbf{x}_i) \rangle$, is roughly equal to the data amplitude.
This means it is assumed that
\begin{equation}
    \langle \Psi(\mathbf{x}_i) \rangle \approx e^{\omega(\mathbf{x}_i)}.
\end{equation}
Following this assumption, the fit to the log space data is then performed with the likelihood variances
\begin{equation}
    \sigma^2(\mathbf{x}_i) = \log \left( \frac{\tilde{\sigma}^2} {|e^{\omega(\mathbf{x}_i)}|^2} + 1 \right).
    \label{eq:log_space_likelihood_variance_approx}
\end{equation}

As a first test of the Bayesian learning scheme, an inference of the posterior distribution over the weights based on the given data set can be applied for randomly chosen sets of support configurations.
That is to say, a \ac{GPS} representation of the given data is obtained by adopting the mean of the posterior according to Eq.~\eqref{eq:mean_weights} as the model weights.
The value of the inferred posterior mean depends on a hyperparameter, $\tilde{\sigma}^2$, and the chosen covariance matrix $\mathbf{A}^{-1}$.

The success to `learn' the target state, $|\Psi_t\rangle$, can be gauged by evaluating the mean squared error obtained with the model defined as
\begin{equation}
    \mathcal{L} = \frac{1}{N_{states}} \sum_{\mathbf{x}} |\Psi(\mathbf{x}|\mathbf{w}) - \Psi_t(\mathbf{x})|^2.
\end{equation}
Here, the sum runs over all configurations of the Hilbert space with dimension $N_{states}$, the predicted \ac{GPS} amplitudes with the adopted weights are denoted as $\Psi(\mathbf{x}|\mathbf{w})$ and the amplitudes of the target state are denoted as $\Psi_t(\mathbf{x})$.
Overall, it can be expected that the error w.r.t. the target state decreases as the \ac{GPS} model gets more expressive, e.g., if the number of support configurations is increased.

\begin{figure}[htb!]
    \centering
    \includegraphics{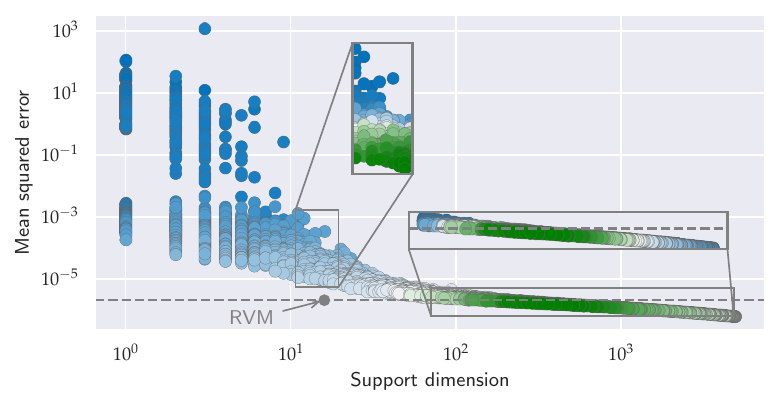}
    \caption[Mean squared error vs. the number of support configurations for a fit of the \acl{GPS} to a Fermi-Hubbard model ground state for different support configuration realizations]{Mean squared error vs. the number of support configurations for a fit of the \ac{GPS} to a Fermi-Hubbard model ground state (half-filled one-dimensional anti-periodic chain of $L=8$ sites at $U/t=8$) for different support configuration realizations. The coloured scatter points represent different random realizations of support set selections. The fits were achieved with a fixed covariance parameter $\alpha = 1$, the kernel hyperparameters were chosen as $\gamma = 1$ and $\theta = 10$, and a noise parameter of $\tilde{\sigma} = 10$ was used. The colouring of the scatter points is chosen according to the calculated marginal likelihood where larger values are represented by green, smaller values by blue colours and the scale is adjusted in the two insets, which simply magnify parts of the main plot. The dark grey scatter point indicates the result obtained with the application of the \ac{RVM} with the same noise and kernel hyperparameters. Figure (adjusted) reproduced from Ref.~\cite{Rath2020}, with the permission of AIP Publishing.}
    \label{fig:support_set_selection}
\end{figure}

This overall expected behaviour is exemplified by the results presented in figure \ref{fig:support_set_selection}.
The figure shows the error w.r.t. a target state for different sets of support configurations, chosen at random from the set of symmetrically inequivalent basis states.
In this example setup, the target state which is fitted is the exact ground state of a half-filled one-dimensional Fermi-Hubbard model comprising eight sites (with anti-periodic boundary conditions) at $U/t = 8$.
The linear features of the \ac{GPS} model are defined by the exponential kernel defined in Eq.~\eqref{eq:exponential_kernel} symmetrized according to all translations of the lattice.
This means the \ac{GPS} representation of the ground state is defined as
\begin{equation}
    \Psi(\mathbf{x}) = e^{\sum_{i=1}^{M} w_i \, k(\mathbf{x}, \mathbf{x}'_i)},
\end{equation}
with the symmetrized exponential kernel
\begin{equation}
    k(\mathbf{x}, \mathbf{x}'_i) = \sum_{\mathcal{S}} \delta_{\mathcal{S}[x]_{r(0)}, x'_{r(0)}} \, e^{-h(\mathcal{S}[\mathbf{x}], \mathbf{x}')/\theta}.
\end{equation}
The sum over symmetry operations, $\mathcal{S}$, includes all translations along the chain.
The weighting function $f$ that defines the weighting in the distance $h(\mathbf{x}, \mathbf{x}')$ is chosen to follow the polynomial decay with distance from the reference site according to Eq.~\eqref{eq:polynomial_decay_weighting}.
That is to say, the (Hamming) distance metric is defined according to
\begin{equation}
    h(\mathbf{x}, \mathbf{x}') = \sum_{i=1}^{L-1} \frac{1 - \delta_{x_{r(i)}, x'_{r(i)}}}{|r(0) - r(i)|^\gamma}.
    \label{eq:scaled_hamming_dist}
\end{equation}

Based on the specified setup, Fig.~\ref{fig:support_set_selection} includes a total of $10,000$ scatter points.
Each is associated with a Bayesian fit of the target state, based on a randomly selected support set.
The fitted data set comprises all symmetrically inequivalent configurations with associated exact amplitudes rescaled to give a vanishing mean over all log training amplitudes.
The support configurations were generated for a specified size by sampling configurations with uniform probability without replacement from the set of symmetrically inequivalent configurations.
Each scatter point in the figure represents the mean squared error of the inferred model w.r.t. the target state in relation to the number of support configurations used.
In this example, the inverse covariance matrix $\mathbf{A}$ was chosen to be proportional to the identity, i.e., $\mathbf{A} = \alpha \mathbb{1}$, with the same prior variance, $1/\alpha$, for each weight.
While different techniques to find sensible hyperparameters for the fit are described further below, the coloured scatter points in the figure all correspond to the same fixed choice for the different hyperparameters $\alpha$, $\tilde{\sigma}$, $\gamma$ and $\theta$.

It can be seen in the figure that, overall, the mean squared error of the learned representation decreases as the size of the support configuration set increases.
In the limit of large support configuration sets, a mean squared error of less than $10^{-6}$ is approached.
While the error for essentially complete support sets is ultimately dominated by the chosen noise and prior parameters, it is exactly the limit of few support configurations that is of particular interest here.
In this limit of very compact \ac{GPS} representations, a large fluctuation of the resulting errors can be observed for different random selections of the support set.
This observation highlights an important characteristic of the \ac{GPS}: Some configurations are more relevant as support configurations than others for an efficient representation of the target state.

An approach to identify the most relevant support configurations from a pool of candidates based on the \ac{RVM} is described in the next section.
For the application of the \ac{RVM}, an appropriate prior covariance is defined that gives rise to particularly compact representations of the data.
Underlining the general success of the \ac{RVM} with identifying the particularly relevant support configurations, Fig.~\ref{fig:support_set_selection} also includes a black scatter point corresponding to the fit of the data obtained with the \ac{RVM}.
It can be seen that the error achieved is significantly smaller than for all randomly selected sets of size similar to the one selected by the \ac{RVM}.

\subsection{Marginal likelihood based model selection and the Relevance Vector Machine}
\label{sec:marg_likelihood_based_model_selection}
The selection of support configurations (and kernel hyperparameters) can be achieved with strict Bayesian inference principles.
Just like a posterior distribution is inferred for the weights in a fit, one can aim to infer a full probability distribution over all possible \ac{GPS} models, specified by weights, support configurations and further hyperparameters, from the given data.

Assuming that the noise parameter $\tilde{\sigma}^2$ is fixed to specify a desired target accuracy that should be achieved by the fit, such an inference scheme therefore finds a posterior of the form
\begin{equation}
    p(\mathbf{w}, \{\mathbf{x}'\}, k| \boldsymbol{\omega}, \mathbf{X}, \tilde{\sigma}^2).
\end{equation}
This posterior is a joint probability distribution over the space of weights $\mathbf{w}$, support configuration sets $\{\mathbf{x}'\}$, and kernel functions $k$ (in the case above characterized by hyperparameters $\gamma$ and $\theta$).
It takes into account the training data and the variance hyperparameter $\tilde{\sigma}^2$.
In order to find such a joint probability distribution, a practical approach is to separate the inference of the weights from the rest of the model specifications~\cite{Tipping2003a,Rasmussen2006}.
This means the posterior is factorized into a product according to
\begin{equation}
    p(\mathbf{w}, \{\mathbf{x}'\}, k| \boldsymbol{\omega}, \mathbf{X}, \tilde{\sigma}^2, \mathbf{A}) = p(\mathbf{w} | \mathbf{A}, \{\mathbf{x}'\}, k, \boldsymbol{\omega}, \mathbf{X}, \tilde{\sigma}^2) \times p(\mathbf{A}, \{\mathbf{x}'\}, k | \boldsymbol{\omega}, \mathbf{X}, \tilde{\sigma}^2),
\end{equation}
which now also incorporates the matrix $\mathbf{A}$ characterizing the prior over the weights with further hyperparameters.

The weight posterior $p(\mathbf{w} | \mathbf{A}, \{\mathbf{x}'\}, k, \boldsymbol{\omega}, \mathbf{X}, \tilde{\sigma}^2)$ is specified by application of Bayes' theorem as before.
In order to incorporate the posterior over the remaining hyperparameters, $p(\mathbf{A}, \{\mathbf{x}'\}, k | \boldsymbol{\omega}, \mathbf{X}, \tilde{\sigma}^2)$, typically suitable approximations need to be made.
A standard approach is to model this second posterior as a distribution infinitely sharply peaked around one particular point in the parameter space (i.e., a delta function).
Under this assumption, the fit of the model can be separated into finding the peak of the hyperparameter posterior model and then applying the standard approach to infer the weights with the model specified by these most probable parameters.
Such an approach, commonly referred to as a type-II maximum likelihood method, is therefore a two-step process.
First the maximum of $p(\mathbf{A}, \{\mathbf{x}'\}, k | \boldsymbol{\omega}, \mathbf{X}, \tilde{\sigma}^2)$ is obtained.
This then defines the hyperparameters used for the inference of the weights in the second step.

Similar to the inference of the weight posterior, the maximum of $p(\mathbf{A}, \{\mathbf{x}'\}, k | \boldsymbol{\omega}, \tilde{\sigma}^2)$ can be found by application of Bayes' theorem.
In the case where the prior distribution over the hyperparameter space is assumed to be uniform, the posterior is found to be proportional to the marginal likelihood appearing in the weight inference~\cite{Tipping2003a}, i.e.,
\begin{equation}
    p(\mathbf{A}, \{\mathbf{x}'\}, k | \boldsymbol{\omega}, \mathbf{X}, \tilde{\sigma}^2) \propto p(\boldsymbol{\omega} | \mathbf{X}, \mathbf{A}, \{\mathbf{x}'\}, k, \tilde{\sigma}^2) = \int d \mathbf{w} \, p(\boldsymbol{\omega} | \mathbf{X}, \{\mathbf{x}'\}, k) \times p(\mathbf{w} | \mathbf{A}).
\end{equation}
Finding the maximum of $p(\mathbf{A}, \{\mathbf{x}'\}, k | \boldsymbol{\omega}, \tilde{\sigma}^2)$ is therefore equivalent to finding the maximum of the marginal likelihood.
This observation is a central element of the fitting procedure introduced in the following.
The marginal likelihood from the inference of the weights is used as a figure of merit for the quality of a model specification.
Hyperparameter and support set choices resulting in a larger marginal likelihood are considered to be better than ones with a smaller marginal likelihood.

Under the specified modelling assumptions for likelihood and weight prior, similar to the posterior, also the marginal likelihood can be described by a compact equation.
Its logarithm evaluates to~\cite{fletcherRelevanceVectorMachines, Rath2020}
\begin{align}
    \log \left( p_{ML} \right) &= \log \left( p(\boldsymbol{\omega} | \mathbf{X}, \mathbf{A}, \{\mathbf{x}'\}, k, \tilde{\sigma}^2) \right) \nonumber \\
    &= \frac{1}{2} \left(  \log(\det(\mathbf{A})) - \log(\det(2 \pi \mathbf{B}^{-1})) + \log(\det(\boldsymbol{\Sigma})) - \boldsymbol{\omega}^T \mathbf{B} \boldsymbol{\omega} + \boldsymbol{\mu}^T \boldsymbol{\Sigma}^{-1} \boldsymbol{\mu} \right),
    \label{eq:log_ml}
\end{align}
where $\boldsymbol{\mu}$ and $\boldsymbol{\Sigma}$ define the mean and the covariance matrix of the weight posterior as specified in Eqs.~\eqref{eq:mean_weights} and \eqref{eq:covariance_weights}.
The maximization of this property is the main approach taken in the following to specify the \ac{GPS} hyperparameters based on available data.

It is commonly observed that the marginal likelihood maximization finds balanced models in the sense that these fit the data well but are as simple as possible at the same time.
This property can already be seen in Fig.~\ref{fig:support_set_selection} showing the fit quality for different randomly selected basis sets outlined above.
The scatter points of the figure are also colour coded according to the resulting (log) marginal likelihood.
The colours transition from blue to green as the log marginal likelihood of the fit increases.

As can be seen, the log marginal likelihood typically increases with decreasing error for a fixed number of support configurations.
However, the overall maximum of the log marginal likelihood is not found in the limit of the most expressive models giving the smallest mean squared error.
Instead, the maximum lies in an intermediate regime where the model reaches a good accuracy, but it is constructed with a smaller number of support configurations.

The important observation is that the marginal likelihood is typically maximal for model choices that balance the accuracy of the fit with the complexity of the model.
This can be seen as a central inspiration for the approach to identify the most relevant configurations in the \ac{RVM}~\cite{Tipping2000, Tipping2003}.
While other approaches exist to extract the most relevant features for a linear model, the \ac{RVM} is known to extract particularly sparse sets of features for linear models.
Given a potential set of support configurations, the \ac{RVM} selects those configurations from the set that are sufficient to achieve an appropriate representation of the training data.
This is achieved by maximizing the marginal likelihood with respect to parameters specifying the prior variances of the weights.
The technical details of the marginal likelihood maximization to identify the most relevant support configurations, as, e.g., described in Refs.~\cite{fletcherRelevanceVectorMachines,Tipping2000,Tipping2003}, are summarized in appendix~\ref{ch:RVM}.
The approaches considered in the following are based on the `fast' approach to maximize the log marginal likelihood to identify relevant support configurations in the \ac{RVM} as introduced in Ref.~\cite{Tipping2003}.

\begin{figure}[htb!]
    \centering
    \includegraphics{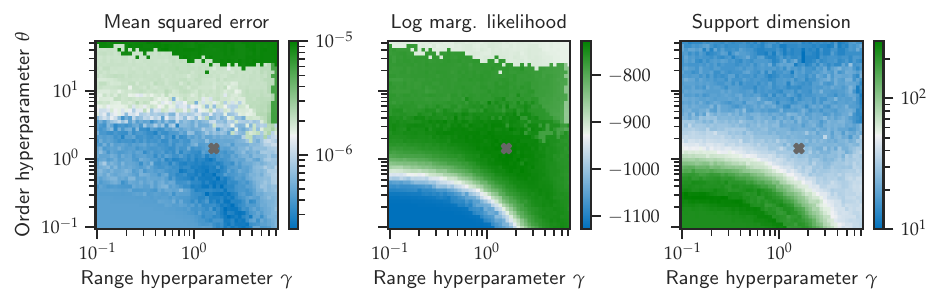}
    \caption[Heat maps representing the mean squared error, the log marginal likelihood, and the number of selected configurations for applications of the \acl{RVM} to fit a target state with different kernel hyperparameters]{Heat maps representing the mean squared error (left), the log marginal likelihood (centre), and the number of selected configurations (right) for applications of the \ac{RVM} to fit a target state with different hyperparameters $\theta$ and $\gamma$ used to define the exponential kernel.
    The target state corresponds to the ground state of an anti-periodic one-dimensional eight site Hubbard model at $U/t = 8$, and a noise parameter of $\tilde{\sigma}^2 = 10$ was chosen for the fits.
    The pair of parameter values corresponding to the maximum log marginal likelihood are indicated by a grey cross.
    Figure (adjusted) reproduced from Ref.~\cite{Rath2020}, with the permission of AIP Publishing.
    }
    \label{fig:hyperparameter_selection}
\end{figure}

The \ac{RVM} can directly be applied to the task of finding a sparse \ac{GPS} representation from given wavefunction samples in the setting discussed above.
Figure~\ref{fig:support_set_selection} already includes the single scatter point obtained for the application of the \ac{RVM}, which reaches a significantly better fit than all displayed randomly selected support sets of similar size.
A more detailed analysis of the models extracted with the \ac{RVM} (for a fit of the same target state) can be obtained from Fig.~\ref{fig:hyperparameter_selection}.
This includes heat maps for the achieved error, the log marginal likelihood, and the number of selected configurations, in relation to the chosen kernel hyperparameters $\theta$ and $\gamma$.
Such an analysis provides important details about how different choices of the kernel hyperparameters influence the \ac{GPS} model obtained by the Bayesian learning of the provided data.

Overall, it can be observed that more support configurations are selected for decreasing values of the kernel hyperparameters $\gamma$ and $\theta$.
This is in agreement with the intuition that smaller hyperparameters values correspond to more `complex' kernels putting a larger relative importance on longer-range and higher-body correlation features.
The $\theta \to 0$ and $\gamma \to 0$ limit is equivalent to the plaquette kernel defined in Eq.~\eqref{eq:plaquette_kernel} matching a single correlation plaquette extending across the full system.
On the other hand, as $\theta$ or $\gamma$ approaches infinity, only the configuration at the central reference site contributes to the kernel value, thus giving a kernel equivalent to a plaquette kernel spanning only the reference site.
Varying the values of $\gamma$ and $\theta$ interpolates between these two limits.
It can be expected that more support configurations are selected to be able to leverage the potential of the kernel in the more complex kernel limit of small hyperparameters.

Overall, the representations of the target state give relatively small mean squared errors across all hyperparameter values.
While the mean squared errors are below or around the order of $10^{-6}$ for a broad portion of the displayed hyperparameter ranges, slightly larger errors are obtained in the limit of larger order parameters $\theta$.
This can again be attributed to the fact, that larger values of $\theta$ correspond to a kernel where the low-body contributions are weighted significantly more strongly as compared to the higher order.
If the low correlation orders are too dominant, then the model with the support configurations selected by the \ac{RVM} is simply not expressive enough to capture all necessary contributions to achieve a small mean squared error.

The results presented in Fig.~\ref{fig:hyperparameter_selection} give a qualitative impression on how the \ac{GPS} obtained by application of the \ac{RVM} depends on the chosen hyperparameters.
They indicate a general understanding how choosing the right kernel parameters will typically be a balancing between sparsity and expressivity of the model.
Using very expressive kernels requires larger numbers of support configurations, more simple kernels typically lead to sparser representations.
This increased sparsity often also leads to a decrease of the fit quality due to the reduced model expressivity.
Again, it is also visualized in the figure how the log marginal likelihood provides a characterization of an ideal balancing between expressivity and simplicity of the model.
The maximum of the log marginal likelihood is found for intermediate parameter choices $\theta$ and $\gamma$, which are indicated by a grey cross on the heat maps.
The maximum marginal likelihood model corresponds to a relatively sparse representation, still reaching one of the best fit accuracies (characterized by a small mean squared error).
While the overall fit errors across the parameter space fluctuates between $\approx 2 \times 10^{-7}$ and $\approx 10^{-5}$, with the parameter values maximizing the log marginal likelihood, an error of $\approx 3.4 \times 10^{-7}$ is achieved requiring only $25$ support configurations.
This indicates that, in addition to choosing the set of support configurations, also sensible kernel hyperparameters can be extracted by marginal likelihood maximization.

Also including the optimization of the kernel hyperparameters w.r.t. the log marginal likelihood yields a fully automated method to infer a \ac{GPS} description, including the selection of support configurations, from given wavefunction samples.
The ability of the outlined scheme to compress given wavefunction data into a compact \ac{GPS} is discussed for one-dimensional Hubbard model ground states in the following.

Whereas the noise parameter $\tilde{\sigma}^2$ is used to specify a target accuracy of the fit, the kernel hyperparameters are obtained by maximization of the marginal likelihood.
The approach to optimize the kernel hyperparameters considered here is based on an iterative search through the parameter space.
For each set of parameters at a step of the parameter search, a full \ac{RVM} fit is repeated and the final log marginal likelihood is evaluated.
By comparing the different final log marginal likelihood values for different hyperparameter values, the parameter region for which the value is maximal is identified.
The search protocol that was employed to obtain the numerical results is the \textit{tree-structured Parzen Estimator Approach} outlined in Ref.~\cite{Bergstra11algorithmsfor} as it is implemented in the \textit{hyperopt} python package~\cite{bergstraMakingScienceModel2013,Bergstra2015}.
In the practical application of the scheme, the kernel parameters were optimized on logarithmic scales across suitable ranges.

The automatic scheme is exemplified here for the representation of ground states of small one-dimensional Hubbard chains at half filling as in the previous examples.
Also incorporating the optimization of the kernel hyperparameters, the approach to infer a \ac{GPS} from the given wavefunction data only requires the specification of a noise parameter $\tilde{\sigma}^2$, as well as a set of potential support configurations from which the most relevant ones are selected with the \ac{RVM}.
For the small systems considered here, for which the exact solution is directly accessible, the pool of candidates can again be chosen to comprise all symmetrically inequivalent computational basis states.

\begin{figure}[htb!]
    \centering
    \includegraphics{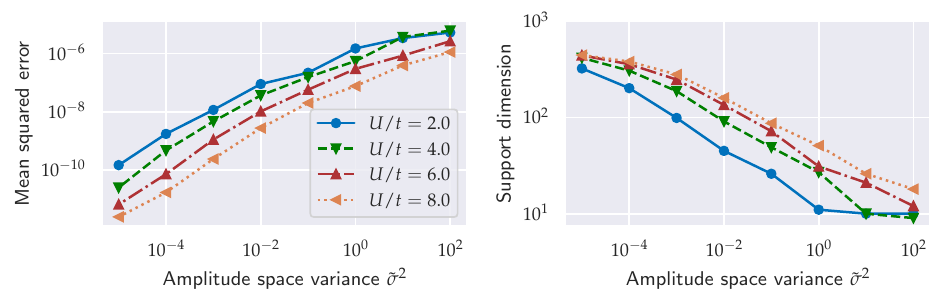}
    \caption[Mean squared error and support dimension of a learned \acl{GPS} with respect to the chosen noise parameter $\tilde{\sigma}$ for different target states]{Mean squared error (left) and support dimension (right) of a learned \ac{GPS} with respect to the chosen noise parameter $\tilde{\sigma}$ for different target states. The target states correspond to ground states of one-dimensional Hubbard models at different repulsion strengths ($U/t = 2$ (blue), $U/t = 4$ (green), $U/t = 6$ (red), $U/t = 8$ (orange)), and other Hamiltonian specifications are specified as in Figs.~\ref{fig:support_set_selection} and \ref{fig:hyperparameter_selection}.
    Figure (adjusted) reproduced from Ref.~\cite{Rath2020}, with the permission of AIP Publishing.}
    \label{fig:systematic_improveability}
\end{figure}

Figure~\ref{fig:systematic_improveability} shows the quality and complexity of \ac{GPS} obtained via the Bayesian scheme with log marginal likelihood maximization in relation to the noise parameter $\tilde{\sigma}^2$.
The left part of the figure shows the mean squared error with respect to the learned target state and the right part shows the final numbers of support configurations of the \ac{GPS} giving rise to this error.
The relationship w.r.t. the noise parameter is displayed for different target states, which are all ground states of eight site Hubbard chains at different repulsion strengths $U/t = 2, 4,6,8$.
Again, the \ac{GPS} was learned from the data set comprising the symmetrically inequivalent configurations of the computational basis with the associated ground state amplitudes.

It can be seen that, at all repulsion strengths $U/t$, a monotonic decrease is found for the incurred mean squared error of the \ac{GPS} and increased values of the parameter $\tilde{\sigma}^2$.
This highlights the intuition that the parameter $\tilde{\sigma}^2$, formally specifying a target variance for the amplitude likelihood, can intuitively be associated as a specification of a target accuracy of the \ac{GPS}.
As the accuracy of the fit improves with smaller values of $\tilde{\sigma}^2$, it can also be observed that the number of support configurations in the \ac{GPS} increases along the direction of smaller noise values.
The observed relationship between accuracy of the fit and number of support configurations specifying the \ac{GPS} can be identified as a very fundamental characteristic of the \ac{GPS}.
As the support dimension $M$ (i.e., the number of support configurations) increases, the models can become more expressive and the error w.r.t. the target state decreases systematically.

With the Bayesian compression scheme, the compactness and accuracy of the \ac{GPS} is only obtained implicitly by the choice of the parameter $\tilde{\sigma}^2$.
It can be seen in the figure that the final mean squared error as well as the number of selected support configurations for a given variance $\tilde{\sigma}^2$ also depends on the specifics of the target states.
The error at a fixed $\tilde{\sigma}^2$ overall mostly increases for larger $U/t$ values and similarly the number of selected configurations decreases.
Crucially however, a systematic improvability of the \ac{GPS} is observed in all four settings showing that the target state can be represented up to very high numerical accuracies.

\subsection{Visualization of the electronic correlation with Gaussian Process States}
The ultimate goal that is explored in this work is to apply the \ac{GPS} as a model in the framework of \ac{VMC} to approximate unknown target states efficiently.
The formulation of the Bayesian learning scheme outlined in the previous section does however rely on the availability of some wavefunction data to facilitate the extraction of a \ac{GPS}.
While this needs to be adjusted if no wavefunction data is directly available, the compression of an available state can be useful to identify particularly important correlation characteristics the \ac{GPS} extracts from the target state.

Based on the formulation of the \ac{GPS} as a linear model in the feature space with specific kernel functions and sets of support configuration, the model is easily interpretable.
The application of the \ac{RVM} together with the hyperparameter optimization can be understood to extract particularly relevant correlation features from the target state.
Analysing the support configurations and the found kernel hyperparameters of an extracted \ac{GPS} can therefore help with the understanding of the relationship between the characteristics of the wavefunction and the underlying physical regime.

\begin{figure}[htb!]
    \centering
    \includegraphics{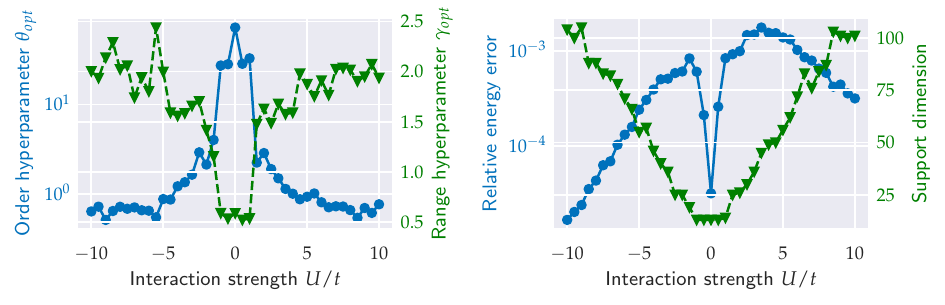}
    \caption[Optimized hyperparameters, as well as relative energy errors and number of support configurations of learned \acl{GPS} ground state representations for one-dimensional Hubbard models at different interaction strengths $U/t$.]{Optimized hyperparameters $\theta_{opt}$ and $\gamma_{opt}$ (left sub-figure), as well as relative energy errors and number of support configurations (right sub-figure) of learned \ac{GPS} ground state representations for one-dimensional Hubbard models at different interaction strengths $U/t$. The training setup is the same as for Fig.~\ref{fig:systematic_improveability}. Figures (adjusted) reproduced from Ref.~\cite{Rath2020}, with the permission of AIP Publishing.}
    \label{fig:optimised_hyperparameters}
\end{figure}

For the presented example of describing the ground states of small, one-dimensional Hubbard chains with the \ac{GPS}, one can, among other things, analyse the relationship between the optimized kernel hyperparameters and the interaction strength, $U/t$, of the Hubbard Hamiltonian.
The obtained relation is presented in Fig.~\ref{fig:optimised_hyperparameters} together with the sparsity and accuracy of the model as a reference.
The left part of the figure shows hyperparameters, $\theta_{opt}$ and $\gamma_{opt}$, which are extracted by the Bayesian learning scheme.
The right part displays the relative energy error, i.e., the fraction $\frac{|E - E_{exact}|}{E_{exact}}$, where $E$ is the variational energy of the approximated state, and $E_{exact}$ denotes the exact ground state energy.
Moreover, the right part also shows the number of selected support configurations.

The learned \ac{GPS} representations give a relative energy error of less than $2 \times 10^{-3}$ for all displayed values of $U/t$.
The achieved energy errors show maxima in intermediate regimes of intermediate interaction strengths at roughly $|U/t|\approx 3 - 4$, and decrease towards larger and smaller absolute values.
In contrast, a mostly monotonic increase in the number of selected configurations can be observed as the absolute interaction strength, $|U/t|$, is increased.
While only $13$ configurations are selected by the approach at $U/t = 0$, this number increases to approximately $M=100$ for $|U/t|=10$, showing that less sparse models are selected in the limit of stronger electronic repulsion/attraction.

Interpretable properties of the learned solution can be extracted from the left sub-figure of Fig.~\ref{fig:optimised_hyperparameters}.
As a dependency of the chosen interaction strength, $U/t$, the left axis encodes the order weighting parameter, $\theta_{opt}$, extracted by the Bayesian compression scheme.
Similarly, the right axis corresponds to the extracted distance decay rate, in the figure denoted as $\gamma_{opt}$.
Overall, the relationship between the ideal hyperparameters and the value of $U/t$ is relatively symmetric under the change from a repulsive Hubbard model ($U/t > 0$) to an attractive setting ($U/t < 0$).
As the absolute interaction strengths increases, both hyperparameters $\theta_{opt}$ and $\gamma_{opt}$ seem to converge to an approximately fixed value for increases of the absolute interaction strength.
In the regime with $|U/t| \gtrapprox 5$, $\theta_{opt}$ takes a value of  $\approx 2$ and $\gamma_{opt}$ a value of $\approx 0.8$.
While both parameters seem to converge to a fixed value, the value of $\theta_{opt}$ overall decreases and the value of $\gamma_{opt}$ increases as the value of $|U/t|$ is increased.
At $U/t=0$, a value of $\theta_{opt} \approx 70$ and a value of $\gamma_{opt} \approx 0.6$ is observed, indicating that in this limit a stronger weighting is placed on lower correlation orders, which are however longer ranged.

The general trend of the hyperparameter dependence on the value of $U/t$ follows very general physical intuitions.
For small inter-electronic interactions, the correlation plaquettes that are weighted more strongly in the expansion of the represented features are the ones spanning fewer sites that can however be further apart.
Such long-range but low-body correlations are reminiscent of the mean-field characteristics dominating in these limits.
For larger absolute interaction strengths, stronger emphasis is placed on more local correlations, which can however be of higher order.

\begin{figure}[htb!]
    \centering
    \includegraphics{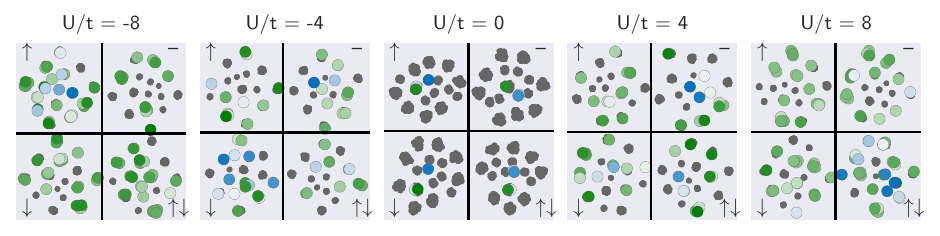}
    \caption[Two-dimensional visualization of the selected support configurations and all basis configurations of the Hilbert space extracted from Hubbard chain ground states at values $U/t = -8, -4, 0, 4, 8$]{Two-dimensional visualization of the selected support configurations and all basis configurations of the Hilbert space extracted from Hubbard chain ground states at values $U/t = -8, -4, 0, 4, 8$.
    Each scatter point is associated with one configuration of the computational basis, and configurations from the set of support configurations in the \ac{GPS} are indicated by coloured scatter points, whereas the full Hilbert space basis configurations are shown with grey scatter points.
    The configurations are grouped into four classes based on the occupancy of the central reference sites and the positioning of the configurations within each of these clusters is obtained by application of the \ac{t-SNE}~\cite{Maaten2008} as implemented in the scikit-learn library~\cite{scikit-learn}.
    The practical implementation of the \ac{t-SNE} is based on the distance metric given by the scaled Hamming distance, $h(\mathbf{x}, \mathbf{x}')/\theta$ (with $h(\mathbf{x}, \mathbf{x}')$ as defined in Eq.~\eqref{eq:scaled_hamming_dist}), as it also appears in the definition of the kernel with optimized hyperparameters.
    The specifics of target state and training setup are equivalent to the setup as in Fig.~\ref{fig:optimised_hyperparameters}.
    Figure (adjusted) reproduced from Ref.~\cite{Rath2020}, with the permission of AIP Publishing.}
    \label{fig:config_clustering}
\end{figure}

In addition to analysing the kernel hyperparameters of the kernel, additional insight about the extracted representation can be obtained by studying the set of selected support configurations in the learned model.
Figure \ref{fig:config_clustering} shows a graphical visualization of the selected support configurations in the final \ac{GPS} model as obtained in the above setting at the five different Hubbard model interaction strengths $U/t = -8, -4, 0, 4, 8$.
Each panel, associated with one particular interaction strength, shows two-dimensional scatter plots where each scatter point corresponds to a configuration from the computational basis.
Whereas those configurations contained in the set of support configurations are indicated by coloured scatter points, grey scatter points correspond to all the basis configurations.

The two-dimensional representation of the space of configurations for the visualization was obtained in two steps.
With the exponential kernel incorporating a prefactor $\delta_{x_{r(0)}, x'_{r(0)}}$, this places a specific significance on the occupancies of the support configurations at the reference site.
It is therefore sensible to group the configurations according to the reference site occupancy.
This leads to four different groups in each visualization of the figure, each associated with a single-site occupancy indicated by the four possibilities $\{\uparrow, \downarrow, -, \uparrow \downarrow \}$.
Within each group, the scatter points are displayed according to a position that was obtained by the \ac{t-SNE} approach~\cite{Maaten2008} to graphically represent the data points in a two-dimensional space.

In general, the selected support configurations appear to be mostly evenly distributed across the four different classes in the obtained representations.
The scatter points also indicate the value of the weights associated with the support configurations.
The colour follows a transition from blue to green with increases of the weight value.
A blue scatter point can be associated with a relative suppression of particular correlation characteristics attributed to the support configuration with the kernel function.
Green scatter points on the other hand, indicate a positive correlation between specific correlation features and the wavefunction amplitude.
For the graphical representation displayed in the figure, the scatter points of the support configurations are plotted such that the ones with the largest absolute weight value are plotted in the foreground.
This way, dominant contributions to the correlation characteristics can easily be identified.

For the system of vanishing interaction strength $U/t = 0$, each cluster contains at least one blue scatter point and at least one green scatter point.
However, as the system also incorporates electron-electron interaction for non-vanishing values of $U/t$, the picture changes and a less symmetric distribution of the most strongly weighted support configurations is observed.
In the limit of strong electronic repulsion, at $U/t = 8$, it can, e.g., be observed that most blue scatter points are positioned in the cluster associated with a double occupancy of the reference site.
This matches the intuition that the energetic penalty for double occupancies of sites in the repulsive regime of the Hubbard Hamiltonian, leads to a suppression of the wavefunction magnitude for configurations with doubly occupied sites.
While such a suppression is explicitly modelled in a Gutzwiller ansatz~\cite{gutzwillerEffectCorrelationFerromagnetism1963} (see Eq.~\ref{eq:gutzwiller} in chapter~\ref{ch:theoretical_background}), here it is automatically identified from the data without manually designing the ansatz around such an intuition.

In general however, it is not necessarily always easy to identify a clear physical intuition from the clustering of the support configurations.
Due to the non-orthogonality of the kernel functions, it is typically not possible to ascribe specific correlation characteristics of the \ac{GPS} to single support configurations.
This can make the two-dimensional representations of the support configuration sets as shown in Fig.~\ref{fig:config_clustering} harder to interpret.
Nonetheless, they provide an easy method to visualize the learned \ac{GPS} model vividly, exploiting the key property of the \ac{GPS} that it is constructed based on physical many-body configurations.

\acbarrier
\chapter{Numerical simulations with Gaussian Process States}
\label{ch:GPS_VMC}
In the previous chapter, it was shown how well-defined Bayesian regression approaches can be used to find a compact representation of many-body quantum states based on samples of the target state.
This Bayesian regression scheme defines the \ac{GPS} implicitly, and the extracted wavefunction ansatz depends on the data presented.
Such a construction is intrinsically different from explicitly parametrized quantum states commonly applied in \ac{VMC} approaches, as these were introduced in chapter \ref{ch:theoretical_background}.

While the presented framework can be applied for sufficiently small systems that can also be treated with exact methods, novel approaches are required to extend the \ac{GPR} framework to describe target states for which no exact data is available.
This chapter describes three different techniques to achieve the goal of using the \ac{GPS} as an ansatz for \ac{VMC} calculations, exploiting the expressibility of the constructed kernel model for the log wavefunction amplitudes.
A main challenge within this task is the extraction of suitable support configurations, which are specific states from the exponentially large Hilbert space.
Whereas other continuous model parameters (such as kernel hyperparameters or the weights) can directly be optimized with standard optimization techniques from \ac{VMC}, the selection of appropriate support configurations requires additional techniques.
The approaches discussed in this chapter include an extrapolation from a small auxiliary system, an iterative bootstrapped Bayesian extraction of configurations from the target state basis, as well as a variational parametrization of the support configurations as product states.

\section{Expressivity of Gaussian Process States}
The foundation for using the \ac{GPS} as a general \ac{VMC} model is its expressibility.
A general \ac{GPS} definition based on a kernel function $k$ and $M$ support configurations $\mathbf{x}'$ can be defined via its wavefunction amplitudes according to
\begin{equation}
    \Psi(\mathbf{x}) = e^{\sum_{i=1}^M w_i k(\mathbf{x}, \mathbf{x}'_i)}.
\end{equation}
As was outlined in the previous chapter, such an ansatz represents a universal approximator of the state, i.e., any quantum state can be represented in this form as $M$ approaches the size of the Hilbert space.
However, the usefulness of the \ac{GPS} to describe many-body states of interest will depend on whether the target state can be described in compact form.
Before outlining how to utilize the \ac{GPS} for \ac{VMC} approaches, this section gives a short description of how the model can be related to other wavefunction ansatzes.
The aim is to reach an intuition into the representational power of the model.

\subsection{Representing Correlator Product States as Gaussian Process States}
The design of the \ac{GPS} kernel functions outlined in the previous chapter is very much inspired by the construction of \acp{CPS}.
As a consequence, any \ac{CPS} emerging as a product from full parametrizations over chosen entanglement plaquettes can be represented exactly as a \ac{GPS}, if the support configurations are specified appropriately.
This requires a support dimension that is at most equal to the combined dimensionalities of the Hilbert space entanglement plaquettes.

For a single correlation plaquette, the \ac{CPS} amplitudes take the form
\begin{equation}
    \Psi(\mathbf{x}) = f_{x_{p_1}, \ldots, x_{p_P}},
\end{equation}
where the tensor $f_{x_{p_1}, \ldots, x_{p_P}}$ encodes the $D^P$ variational parameters associated with the plaquette covering the sites $p_1 \ldots p_P$.
A potential \ac{GPS} representation of this ansatz is obtained by using the $n$-body plaquette kernel as defined in Eq.~\eqref{eq:plaquette_kernel} with the same plaquette indices as for the \ac{CPS}.
This means that the kernel is defined as
\begin{equation}
    k_{n=P}(\mathbf{x}, \mathbf{x}') = \prod_{i=1}^{P} \delta_{x_{p_i}, x'_{p_i}}.
\end{equation}
With such a kernel, the relation between \ac{CPS} and \ac{GPS} follows directly by introducing any set of $D^P$ support configurations for which all configurations have a unique configuration pattern over the specified plaquette sites.
Identifying the $i$-th support configuration by a specific occupancy pattern of the plaquette modes $x_{p_1}, \ldots, x_{p_P}$, a one-to-one representation of a \ac{CPS} is obtained by setting the \ac{GPS} weights to
\begin{equation}
    w_i = \lim_{c \to \infty} \log (f_{x_{p_1}, \ldots, x_{p_P}}) + c.
\end{equation}
The constant shift $c$  is taken to infinity so that the \ac{CPS} amplitudes coincide with the \ac{GPS} amplitudes after normalization of the states.

While the description above shows a one-to-one identification between \ac{CPS} and \ac{GPS}, a key strength of the applied kernel representation is that a representation can also be achieved with less specific kernels.
If  the kernel at least captures correlations up to order $P$ across the plaquette sites, the kernel function describes an injective function w.r.t. configuration pairs on the sub Hilbert space of the plaquette sites.
In this case, the model is a universal approximator across the sub Hilbert space of the plaquette for support configurations specified as above.
Therefore, also the exponential kernel can be used to represent any \ac{CPS} exactly using a set of support configurations containing all possible sub-configuration across the plaquette exactly once.
However, with the ability to select arbitrary support configurations for a \ac{GPS}, the model is not restricted to a specific correlation plaquette but can also model correlations that go beyond a pre-specified plaquette.

Following the arguments above, also two-body correlation properties of the Jastrow representation can be captured with the \ac{GPS}.
Setting the order $p=2$ in the $p$-body kernel, defined in Eq.~\eqref{eq:p_body_kern}, gives a construction explicitly based on all two-body correlation properties similar to a generalized Jastrow ansatz.
The Jastrow ansatz can therefore always be spanned with a \ac{GPS} using a $p$-body kernel with correlation orders $p \geq 2$, again requiring the selection of corresponding support configurations.

The increased flexibility of the \ac{GPS}, not relying on pre-specified entanglement plaquettes as in the \ac{CPS}, comes at an increased evaluation cost for wavefunction amplitudes.
The cost for evaluating a kernel function defined across all lattice sites, such as the exponential kernel, typically scales at least as $\mathcal{O}(L)$.
This results in a total evaluation cost for a single \ac{GPS} amplitude (not including any symmetries) scaling as $\mathcal{O}(M \times L)$.
In contrast, the amplitude lookup in a single-plaquette \ac{CPS} can be associated with a constant cost of $\mathcal{O}(P)$.
\subsection{Gaussian Process States as Neural Network}
\label{sec:GPS_as_NN}
Choosing generic kernels such as the exponential kernel, the \ac{GPS} is able to extract the important correlation characteristics without requiring to specify specific correlation plaquettes as for the \ac{CPS}.
The general aim is in spirit very similar to what also \ac{NQS} set out to achieve: exploiting a highly flexible functional form (and the universal approximation property), ideal approximations are automatically learned without restricting the ansatz to a rigid model driven by expected physical properties.

Not only does the \ac{GPS} share this (ambitious) goal and an inspiration drawn from \ac{ML} approaches with the \ac{NQS}, it is also possible to identify the \ac{GPS} as a specific \ac{NN} architecture (with specific models for the network weights).
Formal properties have been derived relating specific \ac{NN} function approximations with infinitely wide hidden layers to \acp{GP}~\cite{nealBayesianLearningNeural1995,leeDeepNeuralNetworks2018,matthewsGaussianProcessBehaviour2018}.
However, also an explicit representation of the \ac{GPS} (with a finite number of support configurations) as a specific \ac{NN} architecture can be obtained.

To this end, again the exponential kernel is considered to define the model.
This factorizes as a product over all sites, according to
\begin{equation}
    k(\mathbf{x}, \mathbf{x}') = \prod_{i=1}^{L} k^{(i)}(x_i, x'_i),
\end{equation}
where the different single-site kernels $k^{(i)}$ only compare the local configurations at the specific sites.

A design choice difference between \ac{NQS} and \ac{GPS} is the chosen input encoding.
In the case of spin systems, common \ac{NQS} encodes the input as a vector of $L$ float elements, each proportional to the local spin eigenvalue for the configuration at that site.
This is then fed into the \ac{NN} architecture to compute the output representing the wavefunction amplitude.
On the other hand, the \ac{GPS} directly uses the indices into the local Hilbert space basis as inputs, and constructs the representation by comparing these local indices with the support configurations.
In the context of \ac{NN}, this is equivalent to a one-hot encoding of each local configuration in which $D \times L$ visible units are used in the \ac{NN}.
Each of the visible units is associated with one specific input feature per lattice site and local basis state at the lattice site.
In the case of spin systems, a one-hot encoding can be also be recovered from the standard \ac{NQS} input representation by inserting an additional first layer into the \ac{NN} with $2 L$ hidden units.
To obtain the one-hot encoding, this can, e.g., be chosen to map a visible configuration fed into the \ac{NN}, $\tilde{\mathbf{x}}$, to the $2L$ one-hot features according to
\begin{equation}
    f(\tilde{\mathbf{x}})_i = \begin{cases} \text{ReLu}(\tilde{x}_i/2) & \text{if} \  i \ \text{is even}\\ \text{ReLu}(-\tilde{x}_{(i+1)/2}) & \text{otherwise}\end{cases}.
\end{equation}
Here $f(\tilde{\mathbf{x}})_i$ denotes the output of the $i$-th unit of this additional layer and $\text{ReLu}$ denotes the rectified linear unit activation function, defined as
\begin{equation}
    \text{ReLu}(\tilde{x}_i) = \begin{cases} \tilde{x}_i & \text{if} \ \tilde{x}_i \geq 0 \\ 0 & \text{otherwise}\end{cases}.
\end{equation}

\begin{figure}[htb!]
    \centering
    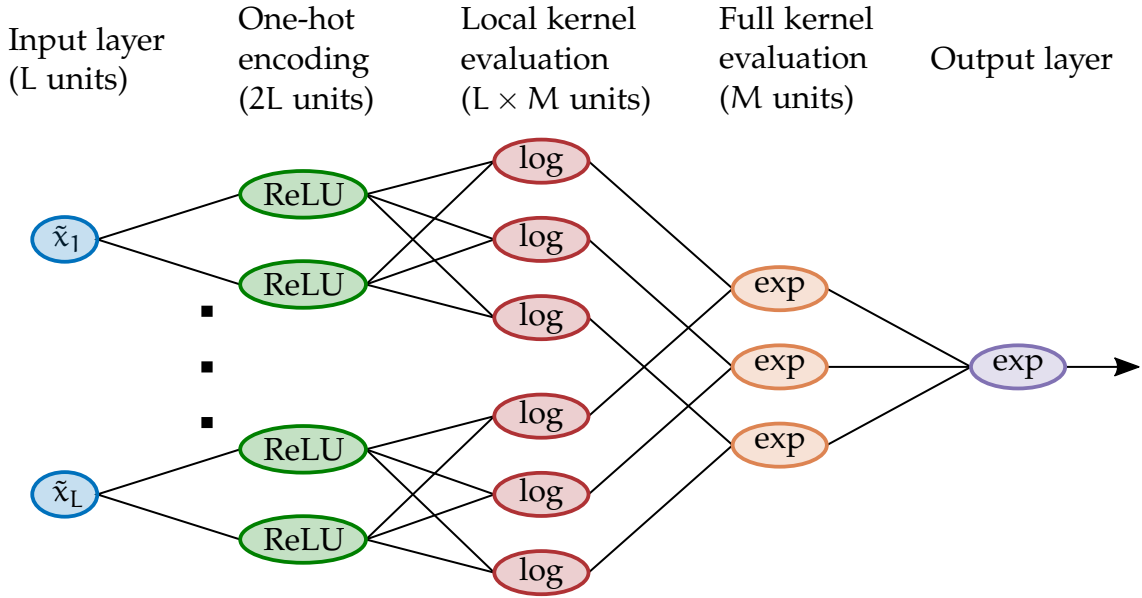
    \caption[Pictorial representation of the \acl{GPS} model formulated as an \acl{NN} for spin systems]{Pictorial representation of the \ac{GPS} model formulated as a \ac{NN} for spin systems.}
    \label{fig:GPS_NN_MPS}
\end{figure}

Based on this definition of a layer converting from the standard input representation used for \acp{NQS} to that used for \ac{GPS} and \acp{TNS}, a non-symmetrized \ac{GPS} can be represented as a particular four layer feed forward \ac{NN} according to
\begin{equation}
    \Psi(\tilde{\mathbf{x}}) = \exp \left( \sum_{i=1}^{M} w_i \left( \exp \left( \sum_{j=1}^{2L} \log \left(\tilde{W}_{i,j} f(\tilde{\mathbf{x}})_{j} \right) \right)  \right) \right).
\end{equation}
Whereas the connection weights between output unit and previous layer are simply the $M$ weights of the \ac{GPS}, $w_i$, the connections between the $2L$ units of the first hidden layer and the $M$ units of the second hidden layer are given by weights
\begin{equation}
    \tilde{W}_{i,j} = \begin{cases} k^{(j/2)}(x_{j/2}, x'^{(i)}_{j/2}) & \text{if} \  j \ \text{is even}\\ k^{((j+1)/2)}(x_{(j+1)/2}, x'^{(i)}_{(j+1)/2}) & \text{otherwise}\end{cases}.
\end{equation}
These weights are therefore specified by the values of the per-site kernel functions comparing the test configuration at site $j$ with that of the $i$-th support configuration, here denoted as $x'^{(i)}_{(j+1)/2}$.
The obtained final representation of a \ac{GPS} in the form of a \ac{NN} for spin-1/2 modes is visualized in Fig.~\ref{fig:GPS_NN_MPS}.
In addition to the hidden layer encoding the input transformation (comprising $2 L$ units), it thus contains two more hidden layers.
The layer feeding into the output is constructed from $M$ hidden units, which take their inputs from a layer consisting of $2 \times L \times M$ hidden units.

While the analogy between \ac{GPS} and \ac{NQS} does not necessarily immediately provide more insight about the representational efficiency of the \ac{GPS}, it does highlight a direct relationship between the two classes of states.
Moreover, it explicitly shows that the \ac{GPS}, as it is presented here, can be understood as a model emerging from a \ac{NN} for which the expressivity is improved by increasing the width of the network rather than its depth.
Nonetheless, the \ac{GPS} defines a very particular structure for the weights of the obtained \ac{NN} representation, significantly reducing the number of parameters as compared to a similar network that is fully connected and for which all connections are understood as optimizable model parameters.

\section{Gaussian Process States for Variational Monte Carlo}
The remainder of this chapter describes practical approaches to use the \ac{GPS} as a functional ansatz in the \ac{VMC} framework.
This section presents two different approaches to use Bayesian learning techniques to define a \ac{GPS} extended to the task of searching for ground states of one- and two-dimensional Fermi-Hubbard models.
Whereas the first approach is based on learning a representation on exact training data for a small system, which is then extrapolated, the other approach `bootstraps' the Bayesian wavefunction learning iteratively directly on the target system.
This makes it possible to uncover physical properties that only become present as the system size goes beyond what can be described with exact diagonalization methods.

\subsection{Extrapolation of Gaussian Process States}
Learning the \ac{GPS} for an unknown target state from the data set of the exact solution for a smaller system entails the extraction of a \ac{GPS} model that can be evaluated for both systems.
The main objective for the extrapolation of the state is that a \ac{GPS} appropriately inferred from the small system encodes important characteristics in a way that also provides a reasonable description for the larger system.
The feasibility of such an approach mostly relies on two main assumptions.
Firstly, the target wavefunction is assumed to be invariant w.r.t. shifts of the input representation to evaluate the amplitudes.
In the presented setting the lattice system is translationally invariant.
This invariance is explicitly built into the \ac{GPS} model by including all translation operators in the symmetrization of the kernel function as presented in Eq.~\eqref{eq:kernel_symmetrization}.
With the exponentiation, this results in a product structure of the wavefunction amplitudes, equivalent to using the same correlator estimator, previously denoted as $F_i$, for each site environment.

The second key assumption that is made for an extrapolation of a \ac{GPS} is that the correlations of the target state are sufficiently local such that the main correlation properties already emerge in the small system.
It is ultimately related to an assumption of a low degree of entanglement in the target state.
Based on this, the range of the kernel function around the reference site can be restricted so that the \ac{GPS} only extracts the local correlation properties.
More specifically, this means that a cut-off on the distance from the reference site is introduced in the (unsymmetrized) kernel function beyond which no further contributions are considered.
With the example of the $p$-body interaction kernel (and similarly the exponential kernel), this can be incorporated by restricting the sum over lattice sites to those within a specified finite range according to
\begin{equation}
    k_p(\mathbf{x}, \mathbf{x}') = \frac{\delta_{x_{r(0)}, x'_{r(0)}}}{\mathcal{N}} \left(\theta + \frac{1}{p-1} \sum_{r(i) \in P} \frac{\delta_{x_{r(i)}, x'_{r(i)}}}{f(r(i))}\right)^{p-1}.
\end{equation}
Here, the plaquette $P$ comprises all lattice sites within the chosen environment around the reference site so that the sum only involves those lattice sites that lie within the specified local environment around the reference site.
The normalization constant in the kernel, $\mathcal{N}$, is also appropriately modified to be evaluated with respect to the sum over the restricted range.

An important feature emerging through the restriction of the range is that the sizes of the systems, can be different for test configuration $\mathbf{x}$ and the support configurations $\mathbf{x}'$.
As long as the systems are large enough to accommodate the chosen environment around the reference site, the kernel value $k_p(\mathbf{x}, \mathbf{x}')$ can be evaluated.
Together with the symmetrization according to translations across the studied lattice systems, this makes it possible to obtain a \ac{GPS} by Bayesian inference for a small system, which then also defines a model for larger systems.

\begin{figure}[htb!]
    \centering
    \includegraphics{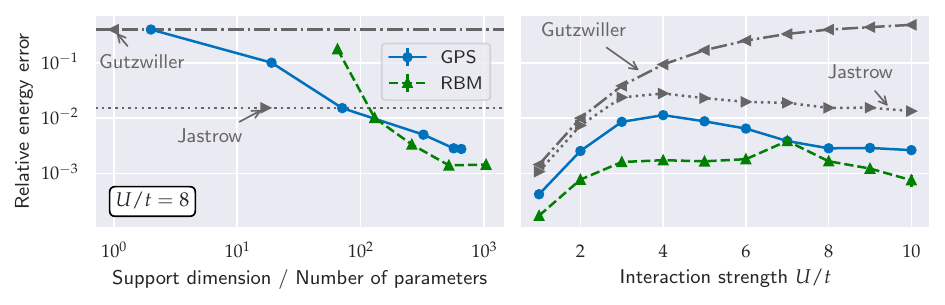}
    \caption[Relative energy errors obtained with different wavefunctions ansatzes to describe the ground states of one-dimensional Hubbard chains of $32$ sites]{Relative energy errors obtained with different wavefunctions ansatzes to describe the ground states of one-dimensional Hubbard chains of $32$ sites (with anti-periodic boundary conditions). The figures include values obtained by extrapolating a \ac{GPS} extracted from the exact data of the ground states of corresponding $12$-site anti-periodic Hubbard chains (blue), as well as variationally optimized (translationally symmetric) Gutzwiller, Jastrow and \ac{RBM} (green) states. In all cases, the ground state approximations were defined as a product with a single Slater determinant reference state, fixed to the \ac{HF} solution. The different \ac{GPS} values were obtained with the Bayesian compression scheme, using the symmetrized $p$-body kernel with $\theta = 0$, inverse distance weighting $f(r(i)) = |r(0) - r(i)|$, and an additional optimization of the noise hyperparameter $\tilde{\sigma}$. The left figure shows the relative energy error as a function either of the number of selected support configurations (for the \ac{GPS}), or as a function of the number of variationally optimized parameters (other ansatzes) at a fixed Hubbard interaction strength of $U/t = 8$. The data points for the \ac{GPS} correspond to different values of $p$ (increasing from left to right from $1$ to $6$), and the \ac{RBM} results correspond to chosen hidden unit densities of $\alpha_M = 1, 2, 4, 8, 16$. The right figure shows the relative energy errors obtained for different Hubbard model interaction strengths $U/t$. The \ac{GPS} results are associated with a correlation order of $p=5$, and for the \ac{RBM} a hidden unit density of $\alpha_M = 5$ was used. The exact reference energies were obtained with \ac{DMRG}~\cite{Block}. Figures (adjusted) taken from Ref.~\cite{Glielmo2020}.}
    \label{fig:extrapolation}
\end{figure}

The success of such an approach is shown in Fig.~\ref{fig:extrapolation}, visualizing the relative energy error of the variational energy of a \ac{GPS} w.r.t. the true ground state energy for Hubbard models of $32$ sites.
The presented data was obtained by first extracting a \ac{GPS} from the exact data obtained with exact diagonalization of $12$ site models, followed by an evaluation of the target system variational energy via Monte Carlo sampling.
Due to the one-dimensional nature of the systems, the exact reference energy references could be obtained with the \ac{DMRG} approach.
In all cases, the training of the \ac{GPS} was performed by using a kernel with a range chosen such that the central site as well as five sites closest to it are included in the considered plaquette.
With support configurations being associated with the $12$ site Hubbard models, this ensures that all possible configurational patterns over the range of $6$ sites appears in the training and in the target model, both of which are considered at half filling (i.e., with an electron number $N = L$ and vanishing total spin magnetization).

In addition to the translational symmetrization of the state according to Eq.~\eqref{eq:kernel_symmetrization}, the results presented in this section also utilize a spin-inversion symmetry.
This is obtained by extending the set of symmetry operations so that it also incorporates the operations with inversion of all the spins in the configuration.

The \ac{GPS} representation for the small system was obtained by application of the Bayesian learning scheme as described in the previous section, thus also extracting a suitable set of support configurations through the \ac{RVM} framework.
For the displayed data points, the different models were obtained with $p$-body kernels with fixed hyperparameters.
The noise parameter $\tilde{\sigma}^2$ was optimized by a hyperparameter optimization scheme to maximize the log marginal likelihood, as it was similarly applied for the optimization of kernel hyperparameters presented in the previous section~\cite{Bergstra2015,bergstraMakingScienceModel2013}.
In the discussed numerical tests, the \ac{GPS} was used to represent a multiplicative (non-signed) correction to a fixed single Slater determinant obtained via the \ac{HF} method~\cite{szaboModernQuantumChemistry2012}.
The resulting state is thus conceptually similar to a Slater-Jastrow state, as it is introduced in section~\ref{sec:hubbard_model} of chapter~\ref{ch:theoretical_background}.
The correction is, however, not just restricted to two-body correlations, but can capture arbitrary correlations over the range of $6$ sites.

The left part of the figure displays the results obtained at a fixed interaction strength in the Hubbard model of $U/t = 8$ in relation to the number of support configurations.
It also includes reference values achieved with other methods (for which the x-axis encodes the number of variational parameters).
The different data points for the \ac{GPS} are associated with different choices of the interaction order in the kernel, $p$, increasing from $p=1$ to a value of $p=6$.
The other kernel hyperparameters were kept fixed ($\theta = 0$, $f(r(i)) = |r(i)-r(0)|$).

For the displayed data, the accuracy of the \ac{GPS} overall increases systematically as more support configurations are selected.
The reductions in the relative energy error correspond to increasing values of the interaction order $p$.
Only including the single-site correlations for a kernel choice of $p=1$, a relative energy error of $\approx 0.4$ is obtained with a total of $2$ selected configurations.
This systematically improves to an error of $\approx 3 \times 10^{-3}$ for values of $p \geq 5$ in which case more than $500$ support configurations are selected.
Again, a very clear relationship between achieved accuracy and number of selected support configurations can be observed, indicating that improvements of the model accuracy are directly connected to increased support dimensions of the \ac{GPS}.
However, the overall accuracy saturates for the values of $p \geq 5$, underlining the limitations emerging due to the finite plaquette of $6$ sites over which the correlations are modelled.

Choosing a restricted correlation plaquette also ensures that finite-size effects in the state of the small systems are suppressed in the description, making it possible to achieve the high accuracies without any variational optimization of the state on the large system involved.
However, this represents a major limitation of the extrapolation approach because it does not allow for a modelling of correlations on length scales larger than the restricted plaquette.
The modelled physical properties are the ones already present in the smaller system from which the data is obtained.
Potential approaches to overcome such limitations is to include further variational optimization with \ac{VMC} techniques directly optimizing the state w.r.t. the larger system.
While it is possible to simply optimize continuous parameters of a \ac{GPS} initially extracted from a small system~\cite{Glielmo2020}, two other approaches, not relying on the compression of exact data from an auxiliary system, are presented in the next sections.

Figure~\ref{fig:extrapolation} also includes results obtained by variational optimization of other standard ansatzes for comparison, including the Gutzwiller ansatz, the Jastrow ansatz and a symmetrized \ac{RBM} with different hidden unit densities.
These were also all defined with respect to the same fixed single Slater determinant reference state as the \ac{GPS} and symmetrized w.r.t. translational symmetries.
The comparison results are plotted in relation to the number of variational parameters that are optimized.
It can be seen that the \ac{GPS} with $p=1$ essentially reaches the same accuracy of an optimized Gutzwiller ansatz, in agreement with the understanding that an order of $p=1$ extracts single site correlation characteristics.
For the two-body description in the Jastrow ansatz a relative energy error of $\approx 2 \times 10^{-2}$ was obtained.
This accuracy is approximately matched by the \ac{GPS} with $p=3$ (for which $\approx 70$ support configurations were selected) and it is surpassed for all the models with larger values of $p$.
Nonetheless, the overall best performance of the studied \ac{VMC} ansatzes can be observed for the \ac{RBM} wavefunction, reaching a relative energy error of $\approx 1.4 \times 10^{-3}$ for a hidden node density of $\alpha_M = 8$.
However, in contrast to the \ac{GPS}, which was obtained by a mostly deterministic compression scheme \footnote{The only non-deterministic elements of the algorithm are parts of the hyperparameter optimization scheme (which is not expected to significantly influence the final result).}, the optimization of the \ac{RBM} requires a variational optimization of the parameters based on stochastic estimates.

The reliability of the scheme is also highlighted in the right plot of Fig.~\ref{fig:extrapolation}, showing the relative energy errors achieved with the different ansatzes for the $32$-site Hubbard model at different (repulsive) interaction strengths $U/t$.
Whereas the fixed \ac{HF} reference state is exact at $U/t = 0$, the considered ansatzes, modelling the corrections to the single Slater determinant, capture different inter-electronic correlation structures as the value of $U/t$ is increased.
For the Gutzwiller ansatz, only modelling single-site characteristics, this results in a monotonic increase in the relative energy error as the interaction strength increases from $\approx 1.4 \times 10^{-3}$ at $U/t = 1$ to a relative error of $\approx 0.5$ at $U/t=8$.
Especially in the limit of larger interactions, the two-site contributions included in the Jastrow factor leads to an improvement over the simple Gutzwiller ansatz, giving a relative energy error of $\approx 1.3 \times 10^{-2}$ at $U/t = 10$.
At all interaction strengths, this is outperformed significantly by the \ac{GPS}, which was obtained by compression of the corresponding $12$-site ground states using the kernel as described above with an interaction order of $p=5$.

The \ac{GPS} reaches a relative energy error of $\approx 4 \times 10^{-4}$ at $U/t=1$, and a relative error of $\approx 2.6 \times 10^{-3}$ at $U/t=10$.
The maximum relative error is obtained in a regime of medium interaction strengths at around $U/t \approx 4$, where a value of $\approx 1.1 \times 10^{-2}$ is obtained.
This maximum of the error, similarly observed for the Jastrow ansatz results, points to increased difficulties with describing the state in this intermediate regime.

It can be expected that significant additional improvements of the description either require increasing the size of the system from which the data is obtained (which is ultimately limited by the exponential scaling of the Hilbert space), or additional variational optimization directly on the target system.
The ability to describe the target states to higher accuracy by variational optimization of a highly flexible functional form is indicated by the results displayed for symmetrized \ac{RBM} ansatzes with a hidden unit density of $\alpha_M = 5$.
The optimized \ac{RBM} wavefunctions mostly give rise to a consistent and significant improvement over the \ac{GPS}.
Nevertheless, an outlier in the \ac{RBM} results can be observed at $U/t=7$ for which the accuracies of \ac{RBM} and \ac{GPS} are almost identical.
Although the displayed \ac{RBM} values correspond to the best value that was obtained across $10$ different random realizations, such an outlier might still be the consequence of a non-ideal optimization of the parametrization.
The faithful optimization of such highly flexible ansatzes in the context of \ac{VMC} is not always guaranteed (and for some systems identified as a major challenge~\cite{bukovLearningGroundState2020,Westerhout2019,Szabo2020}).
This highlights a key strength of the Bayesian compression scheme underpinning the definition of the \ac{GPS}.
If only to provide a starting point for subsequent variational optimization of a state, it allows for an easy, rigorous inference of a compact form from the presented wavefunction data.

\subsection{Bootstrapping Gaussian Process States}
It is possible to extract a \ac{GPS} from a small system as a starting point for subsequent variational optimization of model parameters, such as the weights, as well as the continuous kernel hyperparameters~\cite{Glielmo2020}.
However, such an approach is ultimately always limited by the finite range over which correlations are modelled, which motivates an improved approach outlined in the following.
It is based on iterative applications of the \ac{RVM} to select support configurations directly from the Hilbert space of the target system.

The overall idea to bootstrap the approximation of a target state with the \ac{GPS} is to alternate between bringing an approximation of the target state into a sparse \ac{GPS} representation and a variational optimization of continuous model parameters.
Only a limited subset of the full Hilbert space basis is considered as potential support configurations for the model at each step.
By continuously improving the representation and identifying the most relevant support configurations out of the considered ones, the final approximation is obtained iteratively.

\begin{figure}[htb!]
    \centering
    \includegraphics{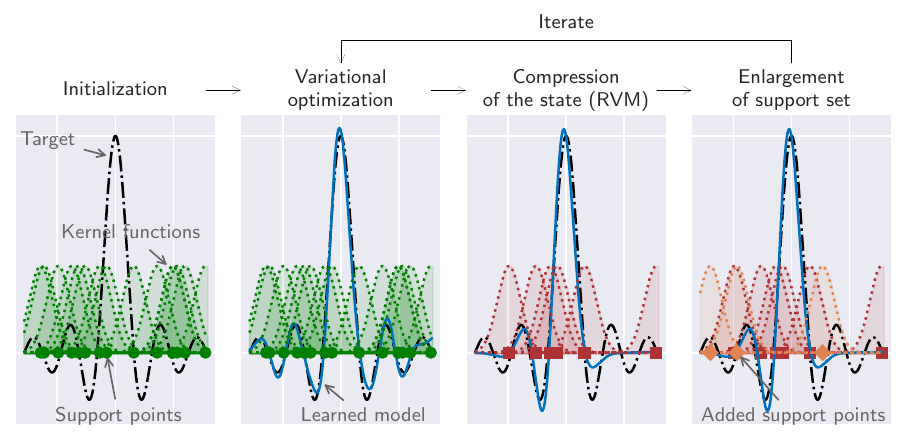}
    \caption[Schematic visualization of a bootstrapped \acl{GPS} optimization]{Schematic visualization of a bootstrapped \ac{GPS} optimization. After initialization of a \ac{GPS}, the final model is obtained by iterating a three-step process, consisting of a variational optimization of the current model, a recompression of the obtained model with the \ac{RVM}, and an enlargement of the support configuration set. The figure visualizes the approximation of the one-dimensional $sinc(x)=sin(x)/x$ function, with achieved approximations shown as the blue curves. The green scatter points indicate the positions of the initially selected support points, the red scatter points the ones selected with the \ac{RVM} and green scatter points indicate the points augmenting the support positions for subsequent optimization. Figure adapted from a figure in Ref.~\cite{Glielmo2020}.}
    \label{fig:bootstrapping_schematic}
\end{figure}

Different concrete implementations of such a protocol are imaginable.
The protocol considered here, is visualized in the schematic of Fig.~\ref{fig:bootstrapping_schematic}.
It entails a variational optimization of the weights (and potentially also further kernel hyperparameters) with the \ac{SR} approach introduced in section \ref{sec:VMC_optimization}, after initialization of a \ac{GPS}.
Having achieved an improved representation of the unknown target state via this variational optimization, the \ac{RVM} is then applied to extract the most relevant support configurations to obtain a compressed representation.
This is achieved by generating data from the constructed wavefunction (via stochastic sampling) from which a new compressed \ac{GPS} description can be obtained by application of the \ac{RVM}, removing non-relevant support configurations of the current description.
In order to increase the flexibility of the model, the \ac{GPS} extracted by the \ac{RVM} is then augmented with additional support configurations prior to the subsequent variational optimization of the model.
By iterating the process of \ac{VMC} optimization followed by an application of the \ac{RVM}, the quality of the approximation together with the choice of the support configurations can be improved until convergence is observed.

Theoretically, due to the universal approximator property of the \ac{GPS}, such an iterative process should be able to achieve an arbitrary accurate representation of the target state (albeit with a potentially exponential scaling effort).
The practical limitations of the bootstrapping approach rely on two essential components.
Firstly, the \ac{RVM} is only applied to a limited data set from the full Hilbert space.
The overall success of the approach therefore requires the \ac{RVM} to be able to extract a \ac{GPS} that reproduces the target well, also beyond the chosen data set.
In \ac{ML} terms, this means that a good generalization needs to be achieved with the fit of the data.
Secondly, the success of the bootstrapping method will also be dominated by how well the \ac{VMC} optimization of the model with the augmented support points can improve upon the previous description.
This requires the specification of suitable augmentation points (i.e., ones that appropriately improve the variational flexibility of the state), and it also requires a reliable \ac{VMC} optimization of the parameters.
The model complexities of the extracted \ac{GPS} are ultimately controlled by the number of support configurations that are selected by the \ac{RVM}.

\begin{figure}[htb!]
    \centering
    \includegraphics{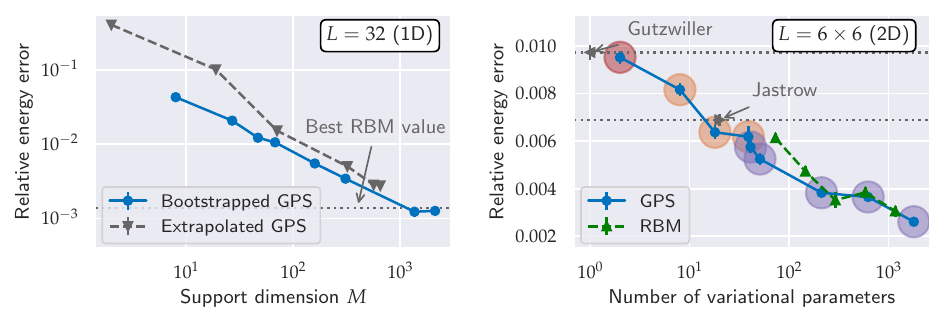}
    \caption[Relative energy error achieved with different ansatzes for strongly correlated Hubbard models at $U/t = 8$ in relation to the model complexity]{Relative energy error achieved with different ansatzes for strongly correlated Hubbard models at $U/t = 8$ in relation to the model complexities. The left plot shows the relative energy error of the bootstrapped \ac{GPS} (blue) in relation to the support dimension, as well as reference values obtained with an extrapolation approach as described in Fig.~\ref{fig:extrapolation}, for a one-dimensional anti-periodic Hubbard chain ($32$ sites) at $U/t=8$. Different noise parameters were used in the \ac{RVM} compression in the bootstrapping, decreasing from left to right as $\tilde{\sigma}^2 = 10^1, 10^0, 10^{-1}, \ldots, 10^{-7}$. The displayed results were obtained with a single Slater determinant reference state fixed to the \ac{HF} solution. For reference, the plot also includes the extrapolated GPS results, as well as the energetically lowest RBM result, from Fig.~\ref{fig:extrapolation}. The right plot shows the relative energy error w.r.t. the number of variational parameters for a Hubbard model defined on a square lattice of $6 \times 6$ sites (anti-periodic boundary conditions in one direction and periodic boundary conditions in the other direction). The results are shown for different ansatzes modelling a correction to a Pfaffian wavefunction (symmetry projected for total spin and translation symmetry quantum numbers) optimized together with the ansatzes. The displayed parameter numbers do not include the number of variational parameters of the Pfaffian state. The background shading of the \ac{GPS} data points indicates variational flexibility of the exponential kernel function used. Red indicates a fixed $\theta \to \infty$ ($M$ variational parameters, $\tilde{\sigma}^2 = 10$), orange indicates a variationally optimized value of $\theta$ and fixed distance weighting ($M+1$ variational parameters, $\tilde{\sigma}^2 = 10^{1}, 10^0, 10^{-1}$), and purple represents a full optimization of the distance weighting as for the left figure ($M+35$ variational parameters, $\tilde{\sigma}^2 = 100, 0.1, 8 \times 10^{-4}, 10^{-4}, 10^{-5}$). The displayed \ac{RBM} results correspond to hidden node densities of $\alpha_M = 1, 2, 4, 8, 16$. All ansatzes were defined to be translationally symmetric, and the \ac{GPS} was also taken to be invariant under spin inversion. The reported values denote the best out of five runs with different realizations of the random components. Figures (adapted) taken from Ref.~\cite{Glielmo2020}.}
    \label{fig:bootstrapping_results}
\end{figure}

\begin{figure}[htb!]
    \centering
    \includegraphics{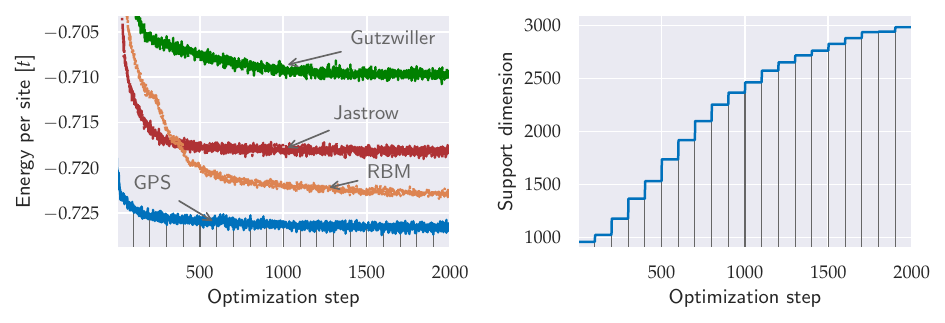}
    \caption[Variational energy per site and number of the selected support configurations in the \acl{GPS} in relation to the \acl{VMC} optimization steps for the approximation of a doped Hubbard model ground state]{Variational energy per site (left) and number of the selected support configurations in the \ac{GPS} (right) in relation to the \ac{VMC} optimization steps for the approximation of a Hubbard model ground state ($8 \times 8$ square lattice, relative filling $N/L = 0.875$, $U/t = 8$).
            The displayed data correspond to the optimization of a Gutzwiller (green), Jastrow (red), \ac{RBM} (orange) and bootstrapped \ac{GPS} (blue) ansatz, all defined as a multiplicative correction to a concurrently optimized Pfaffian reference state as defined in the setup considered in the right plot of Fig.~\ref{fig:bootstrapping_results}. The \ac{RBM} was defined with a hidden unit density of $\alpha_M = 16$, bootstrapped \ac{GPS} optimization included the optimization of the kernel hyperparameters $f(i) \times \theta$, and the optimization steps between which the \ac{RVM} were applied (here with $\tilde{\sigma}=10^{-3}$) to recompress the current state are indicated by vertical lines between the steps in the plots.
            Left plot (adapted) taken from Ref.~\cite{Glielmo2020}.}
    \label{fig:bootstrapping_results_2D_8_by_8}
\end{figure}

Results achieved with the bootstrapping algorithm are presented in Figures \ref{fig:bootstrapping_results} and \ref{fig:bootstrapping_results_2D_8_by_8}.
Both figures show results for one- and two-dimensional Hubbard models obtained with bootstrapped \ac{GPS} representations of different accuracies and model complexities.
The different model complexities for the displayed \ac{GPS} results were achieved by setting the noise parameter, $\tilde{\sigma}^2$, to different values, or by also adjusting the kernel function (either with fixed hyperparameters or with additional variational optimization of the hyperparameters).
The augmentation of the relevant support configurations extracted by the \ac{RVM} was chosen such that the number of support configurations increased by $25 \%$.
This means that the number of support configurations can (and typically will) fluctuate over the course of the optimization.
Other schemes are imaginable in which the number of support configurations is kept fixed throughout the protocol (thus also fixing the computational effort).
The data points of Fig.~\ref{fig:bootstrapping_results} show the achieved energy error in relation to the model complexity of a final ground state approximation.
The reported model complexities for the \ac{GPS} correspond to the model after a final extraction of a \ac{GPS} with the \ac{RVM} followed by a variational optimization without any added configurations.

While certainly not the only possible heuristic, in the discussed settings, the additional configurations were chosen from the Markov chain samples of the previous expectation value estimation based on their local energy.
Those configurations were added to the support configurations for which the local energy deviated the most from the mean.
This should ensure that an increased control over the associated amplitudes is possible in the next \ac{VMC} optimization step to improve the overall accuracy.
While this choice appears sensible and provided good results in example applications, more detailed studies are required to analyse the influence of different support configuration set enhancement protocols on the results.

In the numerical tests discussed here, the considered ansatzes were again used to model a correction to a mean-field type reference state (either optimized or fixed).
All discussed ansatzes were defined with real variational parameters, therefore only modelling the magnitude of the wavefunction amplitudes and not a sign structure.
If present, the sign information was therefore entirely capture by the reference state, (potentially) imposing additional constraints on the overall achievable quality.

The left part of Fig.~\ref{fig:bootstrapping_results} shows the quality of the approximation obtained for the $32$-site one-dimensional Hubbard model in the regime of strong correlation at $U/t=8$.
For reference, the figure also includes the results obtained with the extrapolation approach as already presented in the previous section.
Whereas the best result obtained with the extrapolation approach is overall limited to a relative energy error of $\approx 3 \times 10^{-3}$, with the bootstrapped optimization of the \ac{GPS}, the displayed energy errors reach $\approx 1.2 \times 10^{-3}$ for the best results.
This is roughly equal to (and even slightly improved over) the best results obtained with the \ac{RBM} ansatz displayed in Fig.~\ref{fig:extrapolation}.
Overall, the \ac{GPS} results show a mostly systematic improvement of the quality as the support dimension of the extracted model increases.
However, no further improvement could be observed over the result associated with a final support dimension of $M=1369$, thus indicating a practically achievable limit of the representation.

Comparing the data points of the extrapolated \ac{GPS} with those of the bootstrapped \ac{GPS} for similar support dimensions, it can be observed that the bootstrapped \ac{GPS} typically reaches a lower relative energy error.
This can be attributed to the fact that the bootstrapping of the \ac{GPS} makes it possible to optimize the state directly for the target system.
However, this comes at an increased computational cost incurred due to the repeated application of the \ac{RVM} and the additional variational optimization of the parameters.
In this setting, the \ac{GPS} was constructed with an exponential kernel, as defined in Eq.~\eqref{eq:exponential_kernel}, which was again symmetrized with respect to all translations and spin inversion.
The scaled distance weightings $f(i) \times \theta$, defining the exponential kernel, were treated as $L-1$ variational parameters, which were optimized together with the $M$ \ac{GPS} weights in the \ac{VMC} optimization.

Whereas essentially exact reference results for most one-dimensional systems can be obtained with \ac{MPS} approximations, typically optimized with the \ac{DMRG} approach, this is generally no longer true for two-dimensional systems.
Various two- (and higher-) dimensional systems can therefore pose significant (often unsolved) challenges for detailed numerical access to the emerging physical characteristics.
In the right panel of Fig.~\ref{fig:bootstrapping_results}, results are displayed corresponding to the ground state approximation of a two-dimensional Hubbard model, defined on a square lattice of $6 \times 6$ sites, in the strongly correlated regime at $U/t = 8$.
As the system is considered at half-filling, (numerically exact) reference energies can be obtained with the auxiliary-field quantum Monte Carlo method~\cite{Qin2016}.

The plot shows the relative energy error in relation to the total number of variational parameters for bootstrapped \ac{GPS} approximations, as well as the ones achieved with translationally symmetric Gutzwiller, Jastrow and \ac{RBM} ansatzes.
To accurately model the sign structure emerging for the target state, all ansatzes are defined with respect to a spin- and translational-symmetry projected Pfaffian reference state~\cite{Misawa2017}, optimized together with the multiplicative corrections with \ac{SR}.
With this additional optimization of the Pfaffian reference state, already the Gutzwiller ansatz (with a single variational parameter) reaches a relative energy error of approximately $10^{-2}$.
The best overall result displayed is that of a bootstrapped \ac{GPS} reaching a final relative energy error of $\approx 3 \times 10^{-3}$ with a final support dimension of $M=1759$ (therefore corresponding to $1794$ variational parameters not including the parameters of the reference state).
Instead of initializing the \ac{GPS} with random support configurations, here a starting point was obtained by compression of an optimized Jastrow ansatz (multiplied by a Pfaffian) into a \ac{GPS} (initially multiplied by the same Pfaffian reference that was optimized together with the Jastrow factor) by application of the \ac{RVM}.

Within the bootstrapped \ac{GPS} optimization, different kernel functions were considered.
For the data point highlighted with a red background shading, the kernel function was kept fixed to be the $\theta \to \infty$ limit of the exponential kernel function (thus equivalent to the $p=1$-body kernel), again matching the accuracy achieved with a simple Gutzwiller ansatz.
In this case the total number of variational parameters is equal to the support dimension.
The data points with orange background shading were obtained with the exponential kernel with a fixed distance weighting function $f(i) = |r(0) - r(i)|$\footnote{The distance is evaluated as a `Manhattan distance' on the graph, i.e., the minimum number of edges connecting sites $0$ and $i$}.
The order weighting parameter, $\theta$, was variationally optimized together with the weights for these data points, resulting in a total number of $M + 1$ free variational parameters in the \ac{GPS} part.
Lastly, a purple background shading indicates a full optimization of all the exponential kernel hyperparameters, as was also considered in the one-dimensional setting of the left plot of the figure.
These data points display a very similar relationship between the relative energy error and the number of variational parameters as the one shown for a symmetrized \ac{RBM} ansatz.

More detailed insight into the optimization characteristics of the \ac{GPS} can be obtained from Fig.~\ref{fig:bootstrapping_results_2D_8_by_8}.
This visualizes the optimization process of the ansatz for a two-dimensional Hubbard model on an $8 \times 8$ square lattice, also at $U/t = 8$, with a relative filling of $N/L =  0.875$ (where $N$ denotes the number of electrons in the system).
Such a system displays particularly intricate physical properties that are notoriously hard to capture numerically, and the exploration of the phase diagram specifics around this point has attracted a lot of attention~\cite{Zheng2017, SimonsCollaborationontheManyElectronProblem2015}.

The left part of the figure shows the variational energy per site achieved with the ground state approximation over the course of the optimization.
With specifics of the ansatz definitions as for the results for the $6 \times 6$ Hubbard model discussed above, the (stochastically estimated) variational energy is presented at each \ac{VMC} optimization step for Gutzwiller, Jastrow, \ac{RBM}, and bootstrapped \ac{GPS} ansatzes, all multiplied by a co-optimized Pfaffian reference state as above.
Based on the variational principle, smaller variational energies can be associated with more accurate approximations of the target state.

The converged energy per site reaches $\approx -0.7097(3) \, t$ for the Gutzwiller approximation, representing the largest value across the considered ansatzes.
The lowest variational energy was achieved with a bootstrapped \ac{GPS}.
This reached an energy per site of $\approx -0.7266(3) \, t$ towards the end of the optimization process, consisting of a total of $2000$ optimization steps.
In contrast to the results discussed above, this achieved final energy represents a significant improvement over an \ac{RBM} ansatz with $\alpha_M = 16$.
The \ac{RBM} energies approach a final value of $\approx-0.7229(2) \, t$ per site in the optimization, still a significant improvement over the ones obtained with a Jastrow ansatz (giving final energies of around $\approx -0.7182(3) \, t$ per site).
While the stochastic estimates are always

For the bootstrapped \ac{GPS} optimization, the kernel hyperparameters were optimized together with the model weights, as well as the parameters of the Pfaffian state, in the \ac{VMC} optimization steps.
This means a total of $M + 63 + 64 \times 64$ parameters were optimized with the \ac{SR}.
The total number of variational parameters was however not fixed across the optimization as the \ac{RVM} compression of the current state can lead to different final numbers of support configurations.
In the discussed example, the \ac{RVM} was applied after \ac{VMC} optimization steps, i.e., after every 100 energy evaluations, the most important support configurations were identified.
This is indicated by vertical lines in the left part of the figure indicating those optimization steps between which the \ac{RVM} was applied.
It can be seen, that the previous state was always matched to a high accuracy, indicated by the fact that no jumps in the variational energy are observed after application of the \ac{RVM}.
Again, the straightforward applicability of the Bayesian learning of wavefunction data with the \ac{RVM} was also exploited to obtain an improved starting point based on the optimized Jastrow (multiplied by a Pfaffian) wavefunction.

The right panel of the figure shows how the support dimension of the \ac{GPS} evolves over the course of the bootstrapped optimization.
It can be seen that the support dimension overall increases over the course of the optimization.
The compression of the Jastrow state results in an initial number of around $960$ support configurations, which grows to final support dimensions of around $M \approx 3000$.
While already reasonably good convergence in the energy values can be observed after around $1000$ optimization steps, the support dimension still increases by about 500 configurations over the last $1000$ optimization steps.
Nonetheless, convergence can also be observed for the number of support configurations for the final iterations, indicating that the maximum flexibility under the chosen noise parameter is achieved.

The stochastic exploration of the exponentially large Hilbert space in the bootstrapped \ac{GPS} optimization can never fully guarantee that the best possible set of support configurations is found.
However, the presented results overall suggest that sufficiently appropriate support configurations, achieving a relatively compact representation of the target state, can be identified with the discussed approach.


\subsection{Gaussian Process States with quantum support points}
\label{sec:qGPS}

As an alternative to the bootstrapped \ac{GPS} optimization, this section outlines a fully variational approach to identify the support configurations of the \ac{GPS}, not relying on an iterative search for particularly relevant configurations.
The general idea of this approach, presented in Ref.~\cite{boothQuantumGaussianProcess2021}, is to parametrize the support configurations of the model as general product states.
This parametrization of the product states in terms of continuous variational parameters can then be optimized with the standard \ac{VMC} approaches without any additional application of the \ac{RVM} compression scheme.
By fixing the support dimension of the model this also allows for a direct control of the computational effort required for the optimization of the parametrization.

With the product structure of the exponential kernel, the \ac{GPS} models the log wavefunction amplitudes as
\begin{equation}
    \omega(\mathbf{x}) = \sum_{i=1}^M w_i \prod_{j=1}^L k^{(j)}(x_j, x'^{(i)}_j),
\end{equation}
where again $x'^{(i)}_j$ denotes the occupancy of the $i$-th support configuration at site $j$.
This is equivalent to a linear combination of product state amplitudes, which can be seen more clearly by re-expressing this equation as
\begin{equation}
    \omega(\mathbf{x}) = \sum_{i=1}^M \prod_{j=1}^L \phi^{(x_j)}_{i,j},
    \label{eq:GPS_as_product_state_lin_combination}
\end{equation}
which is based on the definitions of local amplitudes
\begin{equation}
    \phi^{(x_j)}_{i,j} = w^{1/L}_i k^{(j)}(x_j, x'^{(i)}_j).
\end{equation}
The variational model discussed in this section is based on the idea of allowing the local amplitudes, $\phi_{i,j}^{(x_j)}$, to take arbitrary form.
This can directly be motivated by introducing product states as support configurations of the model.

For the exponential kernel, as defined in Eq.~\eqref{eq:exponential_kernel}, the local kernel values are given by
\begin{equation}
    k^{(j)}(x_j, x'_j) = e^{\frac{-(1-\delta_{x_j, x'_j})}{\theta \times f(j)}}.
\end{equation}
Instead of considering a computational basis state for the support configuration, one might also consider an arbitrary local superposition at each site, i.e., represent each local occupancy as a linear combination
\begin{equation}
    | x'_j \rangle = \sum_{k=1}^D c_{j,k} \, |k \rangle.
\end{equation}
Here, $c_{j,k}$ are the linear coefficients of the expansion in the local Hilbert space basis associated with site $j$, denoted by local states $|k \rangle$.
The kernel function can be evaluated for such quantum superpositions by replacing the delta function with the overlap between the local state of the test configuration $|x_j\rangle$ and the superposition defined for the local state of the support configuration, $|x'_j\rangle$.
The full kernel function is then obtained as
\begin{equation}
    k(\mathbf{x}, \mathbf{x}') = \prod_{j=1}^L k^{(j)}(x_j, x'_j) = \prod_{j=1}^L e^{\frac{-(1 - c_{j,x_j,x'})}{\theta \times f(j)}},
\end{equation}
where an additional index $x'$ was introduced for the tensor of coefficients $c_{j,x_j,x'}$, indicating that each support configuration is a different product state.
It can directly be seen that any \ac{GPS} with basis states as support configurations, can be recovered by appropriate choice of the expansion coefficients $c_{j,x_j,x'}$.
However, the reverse is not true, resulting in an increased expressibility of the model.

The parametrization of the kernel can be simplified by introducing new parameters $\tilde{c}_{j,x_j,x'}$, for which the index $j$ is a site label, $x_j$ the local basis state label, and $x'$ the support configuration index.
These are related to the product state expansion coefficients via the relationship
\begin{equation}
    \tilde{c}_{j,x_j,x'} = e^{\frac{-(1 - c_{j,x_j,x'})}{\theta \times f(j)}}.
\end{equation}
The kernel function can then be expressed as
\begin{equation}
    k(\mathbf{x}, \mathbf{x}') = \prod_{j=1}^L \tilde{c}_{j,x_j,x'}.
\end{equation}

Based on the reformulation, the resulting \ac{GPS} amplitudes take the form
\begin{equation}
    \Psi(\mathbf{x}) = \exp \left( \sum_{x'=1}^M w_{x'} \prod_{j=1}^L \tilde{c}_{j,x_j,x'} \right),
\end{equation}
where $w_{x'}$ denotes the weight associated with the support configuration with index $x'$.
This model defines a variational class of states extending the `classical' \ac{GPS} to products of local quantum superpositions as support configurations.
It is therefore referred to as the `\ac{qGPS}'.
\begin{ansatz}[\acl{qGPS}]
    The \acl{qGPS} ansatz class is defined by introducing parameters $\epsilon^{(x_j)}_{j, x'} = w^{1/L}_{x'} \times \tilde{c}_{j,x_j,x'}$, defining the wavefunction amplitudes as
    \begin{equation}
        \Psi(\mathbf{x}) = \exp \left( \sum_{x'=1}^M \prod_{i=1}^{L} \epsilon^{(x_i)}_{i, x'} \right).
    \end{equation}
    The ansatz is fully parametrized by $D \times L \times M$ continuous variational parameters, $\epsilon^{(x_i)}_{i, x'}$, for which the super script index is a physical index (i.e., the local occupancy of the test configuration), the first sub script index $i$ identifies lattice sites, and $x'$ is an auxiliary index enumerating the support configurations.
\end{ansatz}
For practical \ac{VMC} calculations, the $D \times L \times M$ parameters $\epsilon^{(x_i)}_{i, x'}$ can directly be optimized with standard techniques.
This avoids the necessity to explicitly identify support configurations from the computational basis, resulting in a coupled continuous-discrete optimization problem for the classical \ac{GPS}.
After an initialization of the parameters $\epsilon^{(x_i)}_{i, x'}$ for a fixed support dimension $M$, these can be updated iteratively with approaches such as the \ac{SR} to reach a final approximation of the target state.


The \ac{qGPS} extends the variational flexibility of the classical \ac{GPS}, and the full expressivity of the \ac{GPS} can, in principle, always be recovered.
Also the computational effort required to evaluate the amplitudes is the same as for the classical counterpart.
The cost of evaluating a single \ac{qGPS} amplitude (without any symmetrization of the model) scales as $\mathcal{O}(M \times L)$.

In the following, benchmark results from Ref.~\cite{boothQuantumGaussianProcess2021} are presented, which outline the practical strengths (and limitations) of the \ac{qGPS} as a variational ansatz.

\begin{figure}[htb!]
    \centering
    \includegraphics{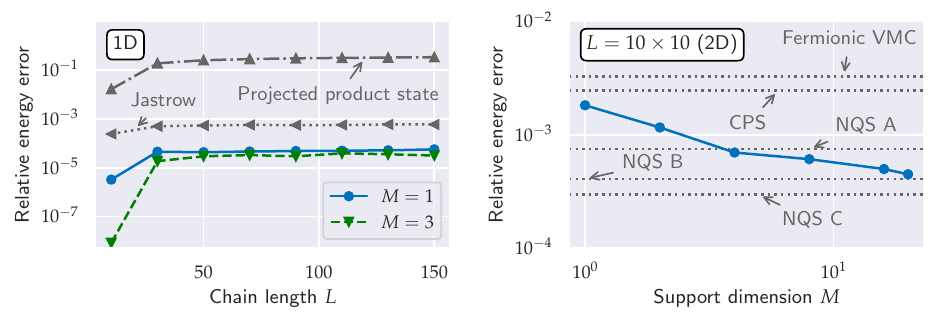}
    \caption[Relative energy errors of ground state approximations for anti-ferromagnetic Heisenberg models]{Relative energy errors of ground state approximations for anti-ferromagnetic Heisenberg models. Left figure displays the relative energy error of an optimized product state, a spin-system Jastrow ansatz (as defined in Eq.~\eqref{eq:spin_jastrow}), as well as \ac{qGPS} ansatzes with support dimensions $M=1$ and $M=3$ in relation to the chain length for one-dimensional periodic chains.
            The \ac{qGPS} and the product states were (kernel-)symmetrized according to spin-inversion, lattice translations and lattice point group symmetries. For the Jastrow ansatz, translational and point group symmetries were encoded into the structure of the parameters.
            The right plot shows the relative energy error achieved with \ac{qGPS} (equivalent symmetrization as for data points in left plot) of varying support dimension for a periodic two-dimensional square $10 \times 10$ lattice model. For comparison, the plot also indicates the best accuracies achieved with other ansatzes with results taken from the literature, including a Gutzwiller projected electronic mean-field state as described in Ref.~\cite{huDirectEvidenceGapless2013} (result taken from Ref.~\cite{Choo2019}), a \ac{CPS}~\cite{mezzacapoGroundstatePropertiesQuantum2009}, and different realizations of \ac{NN} architectures defining Neural Quantum States for the system (\ac{NQS} A [$3200$ complex variational parameters] result from Ref.~\cite{Carleo2017}, \ac{NQS} B [$5145$ real variational parameters] result from Ref.~\cite{Szabo2020}, and \ac{NQS} C [$3838$ complex variational parameters] result from  Ref.~\cite{Choo2019}). Figures (adjusted) taken from Ref.~\cite{boothQuantumGaussianProcess2021}.}
    \label{fig:qGPS_results}
\end{figure}

Figures \ref{fig:qGPS_results} and \ref{fig:qGPS_symmetrization} show accuracies achieved with a \ac{VMC} optimization of the \ac{qGPS} to describe ground states of different anti-ferromagnetic $J_1$-$J_2$ models.
Due to the availability of various benchmark values, this also allows for a direct comparison of the achieved quality with other results from the literature.

Figure~\ref{fig:qGPS_results} shows relative energy error achieved for anti-ferromagnetic Heisenberg models, i.e., $J_1$-$J_2$ models with $J_2=0$.
As discussed in section \ref{sec:J1J2_model}, in this limit, the exact sign structure in the chosen basis of $\hat{S}_z$ eigenstates is known.
It was directly incorporated by appropriate transformation of the Hamiltonian so that the \ac{qGPS} ansatz approximates a target state of positive amplitudes in the chosen basis.
Nonetheless, the parameters $\epsilon^{(x_i)}_{i, x'}$ were chosen to be complex, therefore in principle allowing for a description of signed wavefunctions.

The same kernel-symmetrization approach as for the `classical' \ac{GPS} was applied.
This gives a fully-symmetric \ac{qGPS} model with symmetrized amplitudes according to
\begin{equation}
    \Psi(\mathbf{x}) = \exp \left( \sum_{\mathcal{S}} \sum_{x'=1}^M \prod_{i=1}^{L} \epsilon^{(\mathcal{S}[x]_i)}_{i, x'} \right).
    \label{eq:kernel_symmetrized_qGPS}
\end{equation}
In the considered setting, the first sum over symmetry operations was chosen to include all combinations of translations, the point group symmetry operations of the lattice, as well as the spin-inversion.
With this symmetrization approach, particularly including all translations, general size-extensivity of the solution can therefore be expected for a fixed support dimension $M$ (i.e., one independent of the system size).

The size-extensivity of the method is indicated in the left panel of Fig.~\ref{fig:qGPS_results}.
It displays the obtained relative energy error for one-dimensional (periodic) chains of different lengths, increasing form $L=10$ to $L=150$.
Results are shown for two small support dimensions, $M=1$ and $M=3$, and it can be seen that, in both cases, relative energy errors of below $10^{-4}$ are achieved across all displayed chain lengths.
Already for the small support dimensions considered, the \ac{qGPS} achieves significantly smaller energy errors compared to a single product state (symmetrized by projection), or a symmetrized Jastrow ansatz, for which results are also displayed in the figure.
The single product state (with additional symmetrization) reaches a relative energy error of $\approx 3 \times 10^{-1}$, and the Jastrow ansatz gives a final value of $\approx 6 \times 10^{-4}$, for $L=150$.
The \ac{qGPS} values, on the other hand, converge to a smaller relative error of $\approx 6 \times 10^{-5}$ for support dimension $M=1$, therefore constructed with the same number of variational parameters as the single symmetrized product state.
Only a minor improvement over this level of accuracy is achieved with an increased support dimension of $M=3$, for which the relative energy errors converge to a value of $\approx 3 \times 10^{-5}$.

Due to the low degree of entanglement emerging for the target state of the one-dimensional local model, the target state can be represented efficiently as an \ac{MPS}, and essentially exact reference energies could be obtained with the \ac{DMRG} algorithm~\cite{2007.14822}.
To go beyond one-dimensional systems, the right plot of Fig.~\ref{fig:qGPS_results} shows the relative energy error of \ac{qGPS} ground state approximations for an anti-ferromagnetic Heisenberg model on a two-dimensional square lattice of $10 \times 10$ sites as the support dimension increases.
With the absence of a sign problem, a numerically exact reference energy can be obtained by the stochastic series expansion quantum Monte Carlo approach~\cite{sandvikFiniteSizeScalingGround1997}.

The displayed data points show a systematic decrease of the relative energy error as the support dimension $M$ is increased, resulting from the increased expressibility of the ansatz.
For the \ac{qGPS} constructed from a single product state support configuration, a relative energy error of $\approx 2 \times 10^{-3}$ is reported, which decreases to a value of $\approx 6 \times 10^{-4}$ for $M=20$.
Despite this systematic improvement of the solution, it was not possible to observe further improvements in the relative energy error for a larger support dimension.
This is an indication to difficulties with the optimization of the state, pointing to shortcomings of the \ac{VMC} optimization method.

The results achieved with the \ac{qGPS} for the two-dimensional anti-ferromagnetic Heisenberg model also compares favourably to the results achieved with other, highly flexible wavefunction ansatzes for which literature benchmark values are shown in the figure for comparison.
Already a support dimension of $M=1$ shows an improvement over the results obtained with a Gutzwiller projected Fermionic ansatz discussed in Ref.~\cite{huDirectEvidenceGapless2013} (with values taken from Ref.~\cite{Choo2019}), as well as a practical \ac{CPS} as discussed in Ref.~\cite{mezzacapoGroundstatePropertiesQuantum2009}.
Moreover, the \ac{qGPS} results are also mostly comparable with the results achieved with different proposed \ac{NQS}.
While the accuracies obtained with a generic \ac{RBM}, as discussed in Ref.~\cite{Carleo2017}, are matched and even surpassed by the \ac{qGPS} results for support dimensions $M \geq 8$, more advanced \ac{NN} architectures suggested for the problem in Refs.~\cite{Szabo2020,Choo2019} give a slightly improved final result as compared to the best shown \ac{qGPS} value.
Recently, the level of accuracy reached by neural network models for this system were improved further through the application of a novel optimization scheme that made it possible to reach relative energy errors as small as $10^{-7}$~\cite{https://doi.org/10.48550/arxiv.2302.01941}.
Nevertheless, the displayed \ac{qGPS} results indicate a great promise for general applications of the model as a universal ansatz for different scenarios, and improvements to the optimization scheme could potentially prove similarly beneficial for the optimization of the \ac{qGPS} model.

Especially due to the high simplicity of the \ac{qGPS}, for which the parametrization is specified using only a single parameter (the support dimension $M$), it can be seen as a particularly easily applicable ansatz, complementing the class of \ac{NQS}.
For practical (future) improvements of the results, it could potentially be helpful to introduce further restrictions on the structure of the parameters.
One might for example enforce some locality for the described correlations (as it was also built into the kernel function for the classical \ac{GPS}).
Such a construction, in spirit similar to finite-range convolutional filters applied in a \ac{NN}, would reduce the number of variational parameters, and could help to guide the optimization to find target states with lower degrees of entanglement more easily.

\subsubsection{Symmetrization of Gaussian Process States}
\label{sec:symmetrization}

For all results discussed so far, the \ac{GPS} was only assumed to model the magnitude of wavefunction amplitudes.
The sign structure of the target state was either approximated with a mean-field reference state of limited flexibility, or the problem was transformed so that the target state becomes non-negative in the chosen computational basis.
However, for many physically interesting target systems, a non-trivial sign structure can emerge, which also needs to be represented by the model.
As an example, in this section, $J_1$-$J_2$ Hamiltonian ground state approximations with the \ac{qGPS} are considered without employing the \ac{MSR} transformation.
In addition to testing the practical \ac{qGPS} to represent the known \ac{MSR}, also a setup associated with strong geometric frustration is considered, which is introduced through next nearest neighbour interactions.
In the regime of strong frustration at around $J_2/J_1 \approx 0.5$ of the $J_1$-$J_2$ model a highly non-trivial phase transition has been observed for the ground state~\cite{nomuraDiractypeNodalSpin2021}, and increased difficulties with the application of \ac{NQS} have been reported in different studies~\cite{Choo2019,Ferrari2019,Szabo2020,bukovLearningGroundState2020, rothGroupConvolutionalNeural2021}.

\begin{figure}[htb!]
    \centering
    \includegraphics{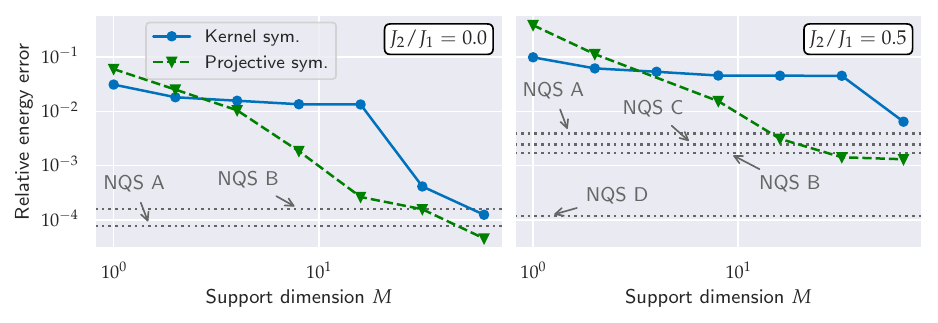}
    \caption[Relative energy error vs. the support dimension of the \acl{qGPS} for \acl{VMC} ground state approximations of two-dimensional $J_1-J_2$ models on a $6 \times 6$ square lattice structure]{Relative energy error vs. the support dimension of the \ac{qGPS} for \ac{VMC} ground state approximations of the two-dimensional $J_1-J_2$ model on a periodic $6 \times 6$ square lattice structure at coupling strengths $J_2/J_1 = 0.0$ (left) and $J_2/J_1 = 0.5$ (right). The \ac{MSR} was not imposed in either of the setups and a signed wavefunction is approximated by the \ac{qGPS}. The blue data points represent results achieved with a kernel-symmetrized \ac{qGPS} model as defined in Eq.~\eqref{eq:kernel_symmetrized_qGPS} of the main text. The green data points show the results for a projectively symmetrized \ac{qGPS} according to Eq.~\eqref{eq:projectively_symmetrized_qGPS}. The symmetrization approaches incorporate translation, point group and spin-inversion symmetries into the state. The dotted vertical lines indicate reference values for different proposed \ac{NQS} implementations from the literature, described and studied in Ref.~\cite{Choo2019} (\ac{NQS} A), Ref.~\cite{rothGroupConvolutionalNeural2021} (\ac{NQS} B), Ref.~\cite{chenNeuralNetworkEvolution2021} (\ac{NQS} C), and Ref.~\cite{nomuraHelpingRestrictedBoltzmann2020} (\ac{NQS} D). Figure (adapted) taken from Ref.~\cite{boothQuantumGaussianProcess2021}.}
    \label{fig:qGPS_symmetrization}
\end{figure}

The problem of describing the sign structure can readily be tackled with the \ac{qGPS} representation introduced in the previous section.
If the model parameters are considered to be complex, the \ac{qGPS} model can also describe the phase information of the wavefunction.
To analyse the practical ability of the model to describe the sign information, Fig.~\ref{fig:qGPS_symmetrization} displays the relative energy errors obtained with \ac{qGPS} ground state approximations in relation to the support dimension for a small rectangular lattice of $6 \times 6$ sites.
This size still allows for exact diagonalization of the $J_1$-$J_2$ system~\cite{schulzMagneticOrderDisorder1996}.
The left part of the figure shows the results for the unfrustrated case at $J_2/J_1 = 0.0$, and the right plot corresponds to a next nearest neighbour coupling of $J_2/J_1 = 0.5$.
Results obtained with a \ac{VMC} optimization of a kernel-symmetrized \ac{qGPS} as discussed before are indicated by blue data points in the figure.

Already the results for the unfrustrated case indicate practical difficulties to describe the signed target state appropriately.
Whereas the lowest relative energy error displayed reaches a value of $\approx10^{-4}$, the results are significantly worse in the limit of small support dimensions.
All data points for support dimensions $M \leq 16$ show a relative error larger than $10^{-2}$.
This is substantially worse than the value of $\approx 2 \times 10^{-3}$, which was reported for the system with explicit incorporation of the \ac{MSR} transformation in Fig.~\ref{fig:qGPS_results}.

The results for the unfrustrated setting clearly indicate a suboptimal optimization of the parametrization, finding some local minimum of the energy rather than the desired global minimum.
The only difference to the previous setup is that the \ac{MSR} needs to be described by the ansatz as well.
However, the \ac{MSR} can be represented as a \ac{qGPS} with support dimension $M = 1$ (see appendix~\ref{sec:MSR_as_qGPS}).
This means that the \ac{qGPS} should (at least) be able to reach the accuracy as for a system with \ac{MSR} imposed if the support dimension is increased by one.

The difficulty with accurately describing signed target states, also manifests for the results at the strongly frustrated point with $J_2/J_1=0.5$, shown in the right part of the figure.
With the ansatz definition as discussed before, only the \ac{qGPS} with $M=64$ reaches a relative energy error smaller than $10^{-2}$.
The achieved value of $\approx 6 \times 10^{-3}$ is only slightly larger than the reference benchmark values for \ac{NQS} presented in Refs.~\cite{Choo2019,chenNeuralNetworkEvolution2021, rothGroupConvolutionalNeural2021}, which are also included in the figure.
While the latter two works report a slight improvement over it, all reference values approximately reach a relative energy error of $\approx 4 \times 10^{-3}$.
This has been identified as a general accuracy limitation of various literature \ac{NQS} implementations in Ref.~\cite{bukovLearningGroundState2020}.

Although the overall optimization feasibility of the \ac{qGPS} description can (at least for the unfrustrated case) be improved by explicitly utilizing the \ac{MSR}, having to resort to known sign structures to improve the performance of the model is somewhat unsatisfactory.
A universally applicable method should ideally also be able to robustly find appropriate approximations of the target state, especially if the target sign structure can be represented exactly by the model.
It is clear that the results for the unfrustrated system without utilization of the \ac{MSR} are significantly influenced by practical optimization difficulties.
In general however, it is not always easy to determine the degree to which the optimization of the model limits the achieved accuracy.

While practical optimization challenges appear to contribute, at least partially, to overall bottlenecks of the approach, also tuning the definition of the model can help to improve the quality of the description.
An important component to practically achieve high quality approximations within the \ac{VMC} framework is an appropriate symmetrization of the state.
The \ac{GPS} ansatzes discussed so far, all describe a fully symmetric function for which symmetrization is achieved by an effective kernel-symmetrization, as defined for the \ac{qGPS} in Eq.~\eqref{eq:kernel_symmetrized_qGPS}.
As presented in section \ref{sec:GPS_def}, by inclusion of all translations into the set of symmetry operations, the model corresponds to a product structure of fully symmetric correlation features around each site.

As an alternative to using fully symmetric correlation features in the definition of the ansatz, an alternative approach can be considered in which the product over correlation features in the \ac{qGPS} is allowed to be non-symmetric.
Even if the target corresponds to a trivial representation of the symmetry, with the same wavefunction phase for symmetrically equivalent configurations, it might be helpful to allow the features in the product structure of the \ac{GPS} to be non-symmetric.
While this might practically require the optimization of more variational parameters, this could potentially also yield to `simpler' solutions more easily obtainable in the \ac{VMC} parameter optimization.

As an example, one can consider the representation of the \ac{MSR} as a \ac{qGPS} for which the details of the construction are shown in appendix \ref{sec:MSR_as_qGPS}.
Although the \ac{MSR} is a specific product state, representing it with a kernel-symmetrized \ac{qGPS} requires the full correlation range across the system, i.e., no cut-off can be introduced for the product across lattice sites in the exponent.
This is counter-intuitive as an unentangled product state should not require to describe the correlations across the full system.
By removing the explicit symmetrization of the features in the \ac{qGPS}, the \ac{MSR} can also be described with a \ac{qGPS} based on `local' correlation features with $M=L$.
This means that the range in the product can be restricted to a single site for each support index, giving additional points in the \ac{qGPS} parameter landscape corresponding to the \ac{MSR} (which might be found more easily in the practical ansatz optimization).

Even if the base ansatz itself is allowed to break the model symmetries (by using non-symmetric features), the system symmetries can still be exploited for the description, by including an appropriate symmetry projection of the unsymmetrized \ac{qGPS}.
The amplitudes for such a symmetry projected \ac{qGPS} can be expressed as
\begin{equation}
    \label{eq:projectively_symmetrized_qGPS}
    \Psi(\mathbf{x}) = \sum_{\{\mathcal{S}\}} \exp \left( \sum_{x'=1}^M \prod_{i=1}^L \epsilon^{(\mathcal{S}[x]_{i})}_{i, x'} \right).
\end{equation}
This projective symmetrization of the state effectively defines the amplitude as a sum over the \ac{qGPS} amplitudes over all symmetrically equivalent copies of the test configurations.
Such an approach also allows the state to be easily projected onto non-trivially symmetric solutions (i.e., ones for which the amplitudes of symmetrically equivalent configurations do not necessarily all have the same sign), by including the appropriate phase prefactors.
It should be noted however, that both symmetrization approaches for the \ac{qGPS} are, in this form, only applicable for symmetries that can be described by operations $\mathcal{S}$ that are sparse in the chosen basis.
For example, the symmetry operations for a projection onto an $\hat{S}^2$ eigenstate generally connect each computational basis state with exponentially many others, even though such a projection can be evaluated for electronic mean-field states~\cite{jimenez-hoyosVariationalApproachesMolecular2013, Misawa2017} or \ac{MPS}~\cite{Li2017,Larsson2020}.
To include a spin quantum number projection into \ac{ML} inspired \ac{VMC} ansatzes therefore requires alternative approaches~\cite{Vieijra2019}.


Figure~\ref{fig:qGPS_symmetrization} also includes results obtained with a projectively symmetrized \ac{qGPS}, indicated by green data points.
Overall, the projectively symmetrized \ac{qGPS} shows a smooth and systematic decrease of the relative energy error as the support dimension increases.
Whereas for both considered couplings $J_2/J_1$, the kernel-symmetrized \ac{qGPS} gives the higher accuracy for the smallest support dimensions $M=1$ and $M=2$, the projectively symmetrized ansatz outperforms the kernel-symmetrized \ac{qGPS} for larger support dimensions.
In the unfrustrated case, the projection approach yields the overall lowest displayed relative energy error of $\approx 5 \times 10^{-5}$ with the largest considered support dimension of $M=64$.
In the strongly frustrated limit at $J_2/J_1 = 0.5$, the projected \ac{qGPS} achieves a relative energy error of $\approx 1.3 \times 10^{-3}$, representing a slight improvement over the general accuracy barrier identified in Ref.~\cite{bukovLearningGroundState2020}.

The presented results suggest that indeed the approach of allowing the ansatz to break the symmetries, and restoring them by a projection of the state, can help to obtain accurate representations of signed target states.
The ability to overcome previously reported accuracy limitations of \ac{NQS} by fully-projective restoration of symmetries has first been described for \ac{RBM} architectures in Ref.~\cite{nomuraHelpingRestrictedBoltzmann2020}.
It was later extended via the framework of `group equivariant convolutional' architectures~\cite{cohenGroupEquivariantConvolutional2016} for other \ac{NQS} with convolutional architectures, described in Refs.~\cite{rothGroupConvolutionalNeural2021, rothHighaccuracyVariationalMonte2022}.
The first work reports an achieved relative error of $\approx 10^{-4}$ for the frustrated system of $6 \times 6$ sites~\cite{nomuraHelpingRestrictedBoltzmann2020} (also indicated in the figure as a reference).
This represents another significant improvement over the best displayed result achieved with the projectively symmetrized \ac{qGPS}.
Further confirmation of the observations contrasting the performance of different symmetrization procedures is presented for additional \ac{NN} architectures in Ref.~\cite{https://doi.org/10.48550/arxiv.2301.06788}.

While the projective symmetrization of the \ac{qGPS} seems to help with the description of signed target state, it can be expected that the support dimension needs to be scaled linearly with the system size in order to achieve general size-extensivity.
This is consistent with the observation that the projectively symmetrized \ac{qGPS}, in the limit of very small support dimensions, reaches an energy error that is larger than that of the kernel-symmetrized model.
With the total number of variational parameters in the \ac{qGPS} scaling as $M \times L \times D$, increasing the support dimension linearly with the system size results in a total number of variational parameters of $\mathcal{O}(L^2 \times D)$.
This might pose further challenges for scaling the method up to larger systems.
In addition to the increased computational cost, also a reliable \ac{VMC} optimization of the ansatz, only based on expectation value estimates with a finite number of stochastically sampled configurations, might become more difficult.
Efficient implementations together with stable optimization techniques~\cite{rothHighaccuracyVariationalMonte2022} are therefore a key requirement in order to universally scale such approaches up to larger systems, representative of the thermodynamic limit.
To improve the computational complexity, it might practically be helpful to decrease the total number of parameters in the \ac{qGPS}, e.g., by restricting the range of the modelled correlations to restricted regions around the different sites.

\acbarrier
\chapter{Ab-initio electronic structure with Gaussian Process States}
\label{ch:ab_initio_GPS}
A long-standing dream is to be able to accurately predict interesting chemical properties just from the underlying quantum mechanical laws, e.g., in order to discover novel materials for sought after technological applications.
The prototypical lattice models used as a testing ground for the methods in the previous chapters can incorporate a rich variety of quantum phenomena contributing to practically observable material characteristics.
Nonetheless, these mostly just represent toy model simplifications, only qualitatively replicating some effects emerging in real world materials, and other chemical substances.
For more practical chemical predictions, more general representations are required, in particular going beyond the explicitly local interactions described by simple lattice models.

While different quantum chemical methods allow for accurate simulations of selected systems, such as approaches explicitly improving upon a \ac{HF} description~\cite{szaboModernQuantumChemistry2012, 10.1039/9781849737289}, especially system properties emerging from strong electronic correlation remain mostly inaccessible with practical numerical techniques.
This chapter describes the extension of the introduced methodology for realistic ab-initio simulations of quantum chemical systems, similar to the presentation in Ref.~\cite{rath2023framework}.
The feasibility and applicability of the method is discussed via applications to standard benchmarking systems, and it is shown that reasonable descriptions can be achieved across different physical regimes.
However, the method is not free from practical limitations, and it is only applicable for small systems.

\section{The electronic structure Hamiltonian}
Building on the concepts explored in the previous chapter, this chapter presents a practical application of the \ac{qGPS} to describe the electronic ground state for small molecular systems.
With the description of the \ac{qGPS} for discrete degrees of freedom, the \ac{qGPS} can readily be applied for \ac{VMC} calculations in a second-quantized `linear combination of atomic orbitals' approach.
In such an approach, a discretized formulation for the electronic structure problem in the Born-Oppenheimer approximation is introduced, by considering the behaviour of electrons that occupy a finite number of fixed molecular orbitals.
These molecular orbitals correspond to single particle wavefunctions that are obtained by appropriately orthogonalizing a chosen set of atomic orbitals used as a basis to discretize the problem.

The electrons occupy the molecular orbitals similarly as the electrons occupy the lattice sites in the Hubbard model.
Representing the electronic Hamiltonian through a finite number of molecular orbitals already introduces an approximation of the many-electron behaviour across the continuous physical space.
For a systematic study into the behaviour, it would therefore typically be required to ensure that the basis set used for the calculations is actually suitable to capture the physical behaviour to the targeted accuracy.
This is different for representations directly targeting the many-electron state in real space, in which the wavefunction is directly described w.r.t. electronic configurations in the continuous physical space (corresponding to the limit of infinite atomic orbitals).
While such descriptions of the electronic wavefunction directly in the real space limit are also imaginable with a \ac{GP} inspired approach, as it has also already been tackled with \ac{NN} representations~\cite{Pfau2019,Hermann2019,spencerBetterFasterFermionic2020,gerardGoldstandardSolutionsSchr2022,hermannAbinitioQuantumChemistry2022, liInitioCalculationReal2022, vonglehnSelfAttentionAnsatzAbinitio2022, Han2018, https://doi.org/10.48550/arxiv.2302.04168, gaoAbInitioPotentialEnergy2021, https://doi.org/10.48550/arxiv.2205.14962, scherbela2023foundation, pescia2023messagepassing, lou2023neural, kim2023neuralnetwork}, the methodology discussed in the following is applied in a second quantized framework with a finite number of molecular orbitals.
It solely focusses on the task of finding an accurate description of the ground state for a specified choice of atomic orbitals, providing the basis for different practical applications.

Specifically, the problem is constructed by considering a set of $L$ electronic molecular orbitals emerging as a linear combination of chosen atomic orbitals.
These real space wavefunctions can, therefore, be described as
\begin{equation}
    \label{eq:orbital_linear_combination}
    \chi_{i}(\mathbf{r}) = \sum_{j=1}^L c_{i,j} \, \zeta_{j}(\mathbf{r}).
\end{equation}
Here, $\chi_{i}$ represents the $i$-th molecular orbital, evaluated for a position $\mathbf{r}$, $\zeta_{j}$ is the $j$-th atomic orbital of the chosen atomic orbital set, and $c_{i,j}$ denotes the linear expansion coefficient.
Crucially, the linear combination is constructed such that the molecular orbitals are orthogonal and give rise to an appropriate basis for the problem.
Here, a restricted formulation is used for which each spatial orbital can be considered as a mode in the second quantized formulation, which can be occupied by up to two electrons (with opposite spin).
This means that the $2L$ per-spin orbitals are simply defined by multiplication of the spatial orbital with one of two orthogonal spin wavefunctions (also ensuring orthogonality between the two per-spin orbitals for the same spatial orbital).

With the constructed molecular orbitals, the ab-initio Hamiltonian can be expressed for the chosen basis in a second quantized formalism, by introduction of annihilation (creation) operators $\hat{c}^{(\dagger)}_{i,\sigma}$ that annihilate (create) an electron in the $i$-th orbital with spin $\sigma$.
Neglecting the constant nuclear repulsion contribution, the resulting Hamiltonian can then be expressed as~\cite{szaboModernQuantumChemistry2012, bradbenSecondQuantizationAzure}
\begin{equation}
    \label{eq:ab_initio_hamiltonian}
    \hat{H} = \sum_{\sigma} \left( \sum_{i,j=1}^L h_{i,j} \, \hat{c}^{\dagger}_{i,\sigma} \hat{c}_{j,\sigma} + \frac{1}{2} \sum_{i,j,k,l=1}^L h_{i,j,k,l} \, (\hat{c}^{\dagger}_{i,\sigma} \hat{c}^{\dagger}_{j,\sigma} \hat{c}_{k,\sigma} \hat{c}_{l,\sigma} + \hat{c}^{\dagger}_{i,\sigma} \hat{c}^{\dagger}_{j,\bar{\sigma}} \hat{c}_{k,\bar{\sigma}} \hat{c}_{l,\sigma}) \right),
\end{equation}
where the first sum runs over both possible spin values, and a spin label $\bar{\sigma}$ denotes the inversion of spin $\sigma$.
The first terms of the Hamiltonian captures the single-particle contributions to the Hamiltonian which are specified by the one-electron integrals between pairs of orbitals, $h_{i,j}$.
These capture the electronic kinetic energy together with the static potential from the electron-nucleus interaction for the chosen molecular orbitals.
Expressed in energy units of Hartree, the one-electron integrals are given by
\begin{equation}
    h_{i,j} = \int d \mathbf{r} \, \chi^\ast_{i}(\mathbf{r}) \left( -\frac{1}{2} \nabla^2 - \sum_{k} \frac{Z_k}{|\mathbf{R}_k-\mathbf{r}|} \right) \chi_{j}(\mathbf{r}),
\end{equation}
where the inner sum runs over the static nuclei with charges $Z_k$ and positions~$\mathbf{R}_k$, $\nabla^2$~denotes the standard Laplace operator and the integration is performed across the real space.
Similarly, the two-electron integrals, $h_{i,j,k,l}$, are obtained by evaluating the contribution of the Coulomb repulsion between the electrons giving the expression
\begin{equation}
    h_{i,j,k,l} = \int d \mathbf{r} \int d \mathbf{r}' \, \frac{\chi^\ast_{i}(\mathbf{r}) \chi^\ast_{j}(\mathbf{r}') \chi_{k}(\mathbf{r}') \chi_{l}(\mathbf{r})}{|\mathbf{r} - \mathbf{r}'|}.
\end{equation}

With this definition of the Hamiltonian in a basis of molecular orbitals, the \ac{VMC} approaches outlined in the previous section can directly be applied to find the electronic ground state for the molecular system.
The computational basis of many-body configurations $\mathbf{x}$ is obtained from simply listing possible occupancies of the different molecular orbitals, equivalent to the description outlined for the Fermi-Hubbard model in section \ref{sec:hubbard_model}.
This chapter discusses the application of the \ac{qGPS} as a functional model to describe the ground state.
While the overall framework does not depend on the specifics of the \ac{qGPS} model (and other ansatzes can equally be applied), the description of ab-initio systems can be considered another test of the model's general applicability.
Similar approaches have also been discussed recently with \ac{NQS} architectures~\cite{Choo2019a, barrettAutoregressiveNeuralnetworkWavefunctions2021, zhaoScalableNeuralQuantum2022, https://doi.org/10.48550/arxiv.2301.03755, Yang2020, https://doi.org/10.48550/arxiv.2302.11588}.

\subsection{Molecular orbitals: Canonical vs. local}
\label{sec:basis_choice_ab_initio}
With the general orbital \ac{VMC} framework presented in the previous section, the \ac{VMC} approach is, in principle, directly applicable.
In general, for a given molecular geometry with a specified atomic orbital choice, suitable one- and two-electron integrals can easily be obtained from standard quantum chemistry packages, such as, e.g., the PySCF software~\cite{sunPySCFPythonbasedSimulations2018}.
However, the choice of molecular orbitals is not unique.
The only constraints for the molecular orbital functions are that these need to be orthogonal, and that they are constructed as a linear combination of the chosen atomic orbital representation.
For one valid set of molecular orbitals $\chi_i$, another set of orbitals $\tilde{\chi}_j$ can easily be obtained by applying a transformation specified by a unitary $L \times L$ matrix $U$ according to
\begin{equation}
    \tilde{\chi}_i = \sum_{j=1}^L U_{i,j} \, \chi_j.
\end{equation}

A unitary transformation of the molecular orbitals thus, in general, results in a different structure of the wavefunction amplitudes in the chosen computational basis.
The overall success of the ground state approximation depends on the ability of the \ac{qGPS} to represent the target state in the chosen basis.
It is therefore expected that the choice of molecular orbitals influences the quality of the ground state approximation that can be achieved practically.
While the perfect choice of molecular orbitals, to achieve the best possible accuracy, might in general be system-dependent (and not easily identifiable), one can consider different heuristics to construct the orbitals.

A common choice is to obtain the orbitals as eigenfunctions of a suitably chosen mean-field representation of the studied Hamiltonian.
This can be achieved by the \ac{HF} method in an iterative approach that alternates between a construction of a mean-field Hamiltonian from molecular orbitals and the diagonalization of this Hamiltonian to update the set of orbitals~\cite{szaboModernQuantumChemistry2012}.
One therefore obtains a set of molecular orbitals that can be associated with single-body eigenstates of the final mean-field Hamiltonian.
With this choice of electronic orbitals, the \ac{HF} approximation of the many-body ground state is simply obtained by taking an anti-symmetrized product (i.e., a \ac{SD}) of the energetically lowest-lying mean-field orbitals.
In the second-quantized representation, this means that the \ac{HF} wavefunction only has support on a single many-body configuration $\mathbf{x}_{HF}$ in which only the molecular orbitals with the smallest single-body energies are occupied.
Assuming that the wavefunction ansatz can describe such a peaked state, this choice of orbitals guarantees that at least the accuracy of the \ac{HF} method can be achieved in the variational approximation of the system's ground state.
With the \ac{qGPS}, for example, a support dimension of $M=1$ is sufficient to filter out the \ac{HF} configuration and obtain a state with vanishing amplitudes on the remainder of the Hilbert space.

Motivated by the demonstrated success for lattice models, one might also consider a construction of the orbitals based on a notion of locality in position space.
The idea is to obtain a set of expansion coefficients $c_{i,j}$ defining the molecular orbitals according to Eq.~\eqref{eq:orbital_linear_combination}, such that the final orbitals are as localized as possible.
Different approaches can be applied to achieve this goal in practice.
These are typically either based on analytic schemes to find an orthogonal representation of the localized atomic orbitals~\cite{lowdinNonOrthogonalityProblem1950,reedNaturalPopulationAnalysis1985,aquilanteFastNoniterativeOrbital2006}, or on a numerical minimization of a metric quantifying the locality of the molecular orbitals~\cite{fosterCanonicalConfigurationalInteraction1960,edmistonLocalizedAtomicMolecular1963,pipekFastIntrinsicLocalization1989}.
An example of localized orbitals is the basis of `Boys'-localized orbitals~\cite{kleierLocalizedMolecularOrbitals1974}, which is considered in the following.
To obtain these orbitals, a unitary rotation matrix $U$, transforming an initial set of molecular orbitals to the localized ones, is determined by numerical maximization of a locality measure defined as
\begin{equation}
    \mathcal{L}(U) = \sum_{i=1}^L \left | \int d \mathbf{r} \, \tilde{\chi}^\ast_i(\mathbf{r}) \, \mathbf{r} \, \tilde{\chi}_i(\mathbf{r}) \right |^2.
\end{equation}

When using canonical orbitals, the target state will typically have a particularly peaked structure around the \ac{HF} configuration for weakly correlated systems.
It can be expected that the Monte Carlo sampling can become problematic in such a case, since the same few configurations are sampled repeatedly when such a peaked state is approached, and the Hilbert space is not explored well.
Such difficulties were, e.g., also observed for the optimization of \ac{NQS} as discussed in Ref.~\cite{Choo2019a}\footnote{While the exact choice of orbitals was not specified in this work, the presented results strongly suggest that a canonical choice was used.}.
It was shown that a particularly large number of samples had to be generated to achieve the full potential of the model, even if a perfect sampling from the full Hilbert space was performed.

\begin{figure}[htb!]
    \centering
    \includegraphics{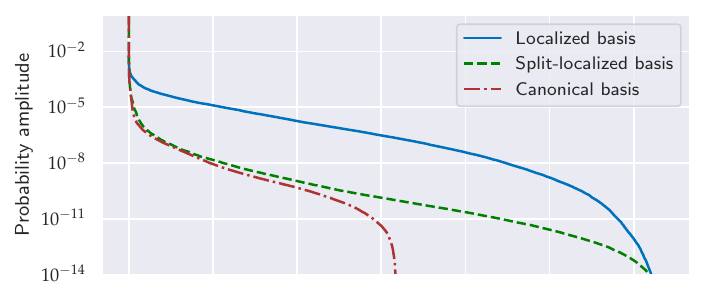}
    \caption[Distribution of the ground state probability amplitudes of a linear chain of ten hydrogen atoms w.r.t. different molecular orbital representations]{Distribution of the ground state probability amplitudes of a linear chain of ten hydrogen atoms (STO-6G atomic orbital basis, inter-atomic spacing of $1.8 \, a_0$) w.r.t. different molecular orbital representations. The distribution of probability amplitudes are shown by sorting the amplitudes by decreasing magnitude from left to right. Distributions are shown for local orbitals (blue), split-localized orbitals (green) and canonical orbitals (red). The localization of the orbitals was achieved with the Boys localization outlined in the main text~\cite{sunPySCFPythonbasedSimulations2018}. Probability amplitudes smaller than $10^{-14}$ are not displayed. Figure similarly presented in Ref.~\cite{rath2023framework}.}
    \label{fig:H10_amplitudes_distribution}
\end{figure}

In contrast to the canonical orbital choice, for a localized representation, it can generally be expected that the overall state will be less peaked and have a broader distribution across the computational basis.
The difference of the probability distributions depending on the orbital choice is exemplified in Fig.~\ref{fig:H10_amplitudes_distribution}.
It visualizes the probability distribution of the exact ground state wavefunction for a system of ten hydrogen atoms placed in a linear chain, described with a minimal basis set of atomic orbitals.
The figure displays the probability amplitudes of the target state in a basis of canonical orbitals, Boys-localized orbitals, and split-localized orbitals~\cite{olivares-amayaAbinitioDensityMatrix2015}.
In the split-localization, the Boys localization is applied separately for the orbitals occupied in the \ac{HF} solution, and the energetically higher lying orbitals (referred to as the virtual orbitals).
The most dominant probability amplitudes (sorted according to their magnitudes in descending order) show a very rapid decay in the basis defined by the canonical orbitals and the split-localized basis.
However, the probability amplitudes decay significantly less quickly for the localized orbital basis.
As the stochastic evaluation of quantities is expected to be less plagued by sampling difficulties for such a flatter distribution, this orbital choice is used in the following to benchmark the \ac{qGPS} for ab-initio quantum chemical systems.
While the following results indicate that good accuracies can be achieved with an orbital localization using the Boys heuristic, additional improvements to the orbital choice, e.g., through an additional `on-the-fly' tuning of the orbital transformation as proposed in Ref.~\cite{https://doi.org/10.48550/arxiv.2302.11588}, could potentially provide further future advancements of the methodology.

\subsection{Practical efficient implementations}
\label{sec:ab_initio_implementation}
The application of variational functions for the description of electronic states in a discrete basis of molecular orbitals is in spirit similar to the application of the models for prototypical (Fermionic) lattice models.
Using pre-calculated one- and two-electron integrals $h_{i,j}$ and $h_{i,j,k,l}$, the Hamiltonian matrix elements $\langle \tilde{\mathbf{x}} | \hat{H} | \mathbf{x} \rangle$  can directly be evaluated for two computational basis states connected by the Hamiltonian.
By creating configurational samples with the Metropolis Hastings algorithm, the energy expectation values can be evaluated and the ansatz can be optimized as discussed before.
As discussed in section \ref{sec:hubbard_model}, the Fermionic character of the state can directly be incorporated into the chosen second quantized basis.
A difference between the ab-initio Hamiltonian of Eq.~\eqref{eq:ab_initio_hamiltonian} and the Fermi-Hubbard model is the number of terms in the Hamiltonian.
Whereas the ab-initio Hamiltonian comprises $\mathcal{O}(L^4)$ terms (with $L$ being the number of orbitals), the Hubbard model only comprises $\mathcal{O}(L)$ (with $L$ being the number of lattice sites) terms in the chosen basis.
This means that the evaluation of a single local energy, $E_{loc}(\mathbf{x}) = \frac{\langle \mathbf{x} | \hat{H} |\Psi \rangle}{\langle \mathbf{x} |\Psi \rangle}$, requires the evaluation of $\mathcal{O}(L^4)$ amplitudes.
In order to scale the method up to system sizes beyond what can be achieved with exact numerical methods, special care needs to be taken to efficiently implement the evaluation of the local energy.

A practical demonstration how the evaluation of the local energy for the ab-initio Hamiltonian can be implemented efficiently, is, e.g., presented in Ref.~\cite{neuscammanJastrowAntisymmetricGeminal2013}.
In this description, the ab-initio Hamiltonian is expressed according to
\begin{equation}
    \hat{H} = \sum_{\sigma} \left( \sum_{i,j=1}^L t_{i,j} \, \hat{c}^{\dagger}_{i,\sigma} \hat{c}_{j,\sigma} + \frac{1}{2} \sum_{i,j,k,l=1}^L h_{i,l,j,k} \, (\hat{c}^{\dagger}_{i,\sigma} \hat{c}_{j,\sigma}  \hat{c}^{\dagger}_{k,\sigma} \hat{c}_{l,\sigma} + \hat{c}^{\dagger}_{i,\sigma} \hat{c}_{j,\sigma} \hat{c}^{\dagger}_{k,\bar{\sigma}} \hat{c}_{l,\bar{\sigma}} ) \right),
\end{equation}
where the coefficients $t_{i,j}$ are specified as
\begin{equation}
    t_{i,j} = h_{i,j} - \frac{1}{2} \sum_{k=1}^L h_{i,k,j,k}.
\end{equation}
The evaluation of the local energy thus requires to evaluate the ansatz amplitudes for all possible single and double electron hops, i.e., amplitudes of the form $\langle \mathbf{x} | \hat{c}^{\dagger}_{i,\sigma} \hat{c}_{j,\sigma} |\Psi \rangle$ and $\langle \mathbf{x} | \hat{c}^{\dagger}_{i,\sigma_1} \hat{c}_{j,\sigma_1}  \hat{c}^{\dagger}_{k,\sigma_2} \hat{c}_{l,\sigma_2} |\Psi \rangle$.
Such a hop will only give a non-zero amplitude if this is consistent with the configuration $|\mathbf{x}\rangle$.
This means that an amplitude $\langle \mathbf{x} | \hat{c}^{\dagger}_{i,\sigma} \hat{c}_{j,\sigma} |\Psi \rangle$ is only non-zero if the orbital $i$ ($j$) with spin $\sigma$ is occupied (unoccupied) in the configuration $|\mathbf{x}\rangle$.
By restricting the sum in the evaluation of the local energy to those single-electron terms for which $\langle \mathbf{x} | \hat{c}^{\dagger}_{i,\sigma} \hat{c}_{j,\sigma} |\Psi \rangle$ is non-zero, the number of wavefunction evaluations is reduced to $\mathcal{O}(N \times N_v)$ (where $N$ is the number of electrons, and $N_v = 2L - N$ the number of virtual spin-orbitals).
A similar restriction to the quadruple sum for the two-electron terms can be introduced, which leaves a dominant number of $\mathcal{O}(N^2 \times N_v^2)$ amplitude evaluations.

If the full wavefunction ansatz is defined as a model for second quantized basis states, each evaluation of a hop expression involves the evaluation of a parity prefactor~\cite{altlandCondensedMatterField2010a} and the amplitude evaluation for the connected configuration.
This means, e.g., for a double electron hop, the expression
\begin{equation}
    \langle \mathbf{x} | \hat{c}^{\dagger}_{i,\sigma_1} \hat{c}_{j,\sigma_1}  \hat{c}^{\dagger}_{k,\sigma_2} \hat{c}_{l,\sigma_2} |\Psi \rangle = (-1)^{N_{\mathbf{x}, \tilde{\mathbf{x}}}} \langle \tilde{\mathbf{x}} |\Psi \rangle,
\end{equation}
needs to be evaluated.
Here, $|\tilde{\mathbf{x}} \rangle$ is the connected basis state satisfying
\begin{equation}
    (-1)^{N_{\mathbf{x}, \tilde{\mathbf{x}}}} | \tilde{\mathbf{x}} \rangle = \hat{c}^{\dagger}_{l,\sigma_2} \hat{c}_{k,\sigma_2}  \hat{c}^{\dagger}_{j,\sigma_1} \hat{c}_{i,\sigma_1} |\mathbf{x} \rangle.
\end{equation}
The parity prefactor $(-1)^{N_{\mathbf{x}, \tilde{\mathbf{x}}}}$ can easily be evaluated by computing the number $N_{\mathbf{x}, \tilde{\mathbf{x}}}$ of electrons that are passed with the double electron hop w.r.t. the chosen ordering of the orbitals.
This can be achieved in constant time for each term by initially computing a cumulative orbital occupancy of each orbital (storing the total number of electrons in all previous orbitals) for the configuration $| \mathbf{x} \rangle$.

For many common wavefunction ansatzes, it is possible to compute the amplitudes of connected configurations, $\langle \tilde{\mathbf{x}} |\Psi \rangle$, more efficiently having already computed the amplitude $\langle \mathbf{x} |\Psi \rangle$ in the denominator of the local energy~\cite{neuscammanJastrowAntisymmetricGeminal2013}.
This typically involves storing intermediates in the computation of $\langle \mathbf{x} |\Psi \rangle$, so that the amplitude of the connected configuration can be obtained by appropriate update.
The `fast' updates of the amplitude can take into account that the molecular occupancies in the two configurations $|\mathbf{x}\rangle$ and $|\tilde{\mathbf{x}}\rangle$ only differ on, at most, $4$ orbitals, namely the ones from which the hop removes or adds electrons.
The fast updating can also be applied for the \ac{GPS} representations to reduce the cost for the evaluation of the connected amplitude.
It is exemplified here for \acp{GPS} with an exponential kernel and  the \ac{qGPS} form.
Similar constructions are also possible for other kernel functions (or \ac{NN} architectures).

As discussed in section \ref{sec:qGPS}, the considered \ac{GPS} functional forms can be represented as an exponentiated linear combination of $M$ product state amplitudes.
Without symmetrization, the amplitudes thus take the form
\begin{equation}
    \Psi(\mathbf{x}) = \exp \left( \sum_{x'=1}^M \prod_{i=1}^L \phi^{(x_i)}_{x', i} \right).
\end{equation}
The central element to the fast updating of amplitude values is to store the $M$ product state amplitudes, $\psi_{x'}(\mathbf{x}) = \prod_{i=1}^L \phi^{(x_i)}_{x', i}$, for the evaluation of the configuration $|\mathbf{x}\rangle$.
These $M$ product state amplitudes can then be updated efficiently for the connected configuration $|\tilde{\mathbf{x}}\rangle$ according to
\begin{equation}
    \psi_{x'}(\tilde{\mathbf{x}}) = \prod_{i} \psi_{x'}(\mathbf{x}) \frac{\phi^{(\tilde{x}_{i})}_{x', {i}}}{\phi^{(x_{i})}_{x', {i}}}.
\end{equation}
Crucially, here the product only includes those indices $i$ identifying orbitals for which the occupancy has changed (which are at most $4$ in the discussed setting).
This means that each of the $M$ product state amplitudes can be evaluated in constant time.
For the evaluation of the amplitude, these are then summed together giving an overall cost to evaluate the connected configuration of $\mathcal{O}(M)$, representing an $\mathcal{O}(L)$ improvement over the direct evaluation of the amplitude.

The fast updating scheme for updating a (q)\ac{GPS} amplitude under the change of the configuration on few modes is equally applicable in the context of other models where the Hamiltonian connects configurations differing only on few sites.
Examples also include the prototypical lattice models discussed before.
Furthermore, it is also applicable to gain speed improvements in the generation of samples with the Metropolis-Hastings algorithm, assuming that proposals of new configurations are based on small numbers of updates to the current sample (which would be the case for standard choices).
To what extent the fast updating improves the overall computational effort of the method is however ultimately determined by the contribution of the local energy evaluation and the sample generation to the full runtime.
This can differ depending on the specifics of the approach and system.
Especially if a model that can be evaluated relatively efficiently comprises many variationally optimized parameters, then an optimization of the parameters with the \ac{SR} can also become a significant contribution to the computational cost.
If however, the ansatz has fewer optimized parameters for a similar amplitude evaluation cost (such as the classical \acp{GPS}), or an increased evaluation cost (e.g., for \acp{SD}), then fast update mechanisms likely have a greater impact on the overall computational efficiency.

With the increased computational cost of the local energy evaluation in the ab-initio calculations discussed in this section, the fast update scheme will typically help to decrease the overall runtime, especially for larger numbers of orbitals.
For a fixed support dimension $M$ of the \ac{GPS}, the dominating cost to evaluate a single local energy for the full ab-initio Hamiltonian, using the efficient formulation and the fast updating, scales as $\mathcal{O}(M \times N^2 \times N_v^2)$.
An additional improvement of this scaling can be achieved (asymptotically) by exploiting (approximate) sparsity of the Hamiltonian.
If localized orbitals are chosen, the overlap between different molecular orbitals often vanishes due to their separation in position space.
This means that the values of the two-electron integrals $h_{i,j,k,l}$ vanishes for many pairs of orbital indices $(i,l)$ and $(j,k)$, and asymptotically only $\mathcal{O}(L^2)$ non-vanishing terms contribute to the Hamiltonian.
Different schemes have been proposed to efficiently prune non-significant terms from the Hamiltonian, decreasing the overall computational effort to evaluate the local energies in the method~\cite{hachmannMultireferenceCorrelationLong2006, weiReducedScalingHilbert2018, sabzevariFasterLowerScaling2018, doi:10.1063/5.0025055}.
The use of localized orbitals therefore also enables potential routes to push the practical applicability of second quantized wavefunction parametrizations to larger systems with computational cost that is in-line with descriptions in real-space~\cite{weiReducedScalingHilbert2018, foulkesQuantumMonteCarlo2001}.

The results presented in the next section benchmark the overall applicability of the \ac{qGPS} to the description of ab-initio quantum chemistry setups in the discussed framework.
With an efficient implementation of the overall method, also including fast updating of the \ac{qGPS} amplitudes, and directly usable with modern graphics processing unit (GPU) computing architectures, results were obtained for molecular systems with up to $64$ electrons in $64$ spatial orbitals (i.e., $128$ per-spin orbitals).
While such systems sizes can in principle also be treated with other standard post-\ac{HF} quantum chemistry methods, these are larger than the systems previously studied with comparable \ac{NQS} approaches (going up to $N=52$ and $L=38$~\cite{zhaoScalableNeuralQuantum2022}).
Perhaps more importantly however, it is shown in the next section that the discussed \ac{VMC} approaches are applicable in regimes of particularly strong correlation, in contrast to other standard quantum chemistry methods, such as coupled-cluster approaches.

\section{Results}
As a first benchmark test of the ab-initio system description with the \ac{qGPS}, a linearly arranged sequence of hydrogen atoms with a minimal basis-set description (i.e., one spatial orbital per hydrogen atom) is considered.
Such a system is very similar to a simple Hubbard model.
However, the strict locality of the lattice model (restricting the electronic interactions to the same site and the kinetic movement to nearest-neighbours), is replaced by general interactions and movement across the full system.
Such systems are a very common testing ground to benchmark different ab-initio methods~\cite{mottaSolutionManyelectronProblem2017}.

\begin{figure}[htb!]
    \centering
    \includegraphics{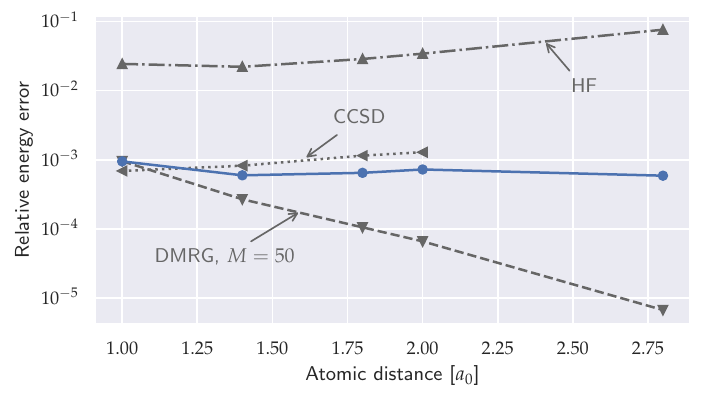}
    \caption[Relative ground state energy error obtained with a \acl{qGPS} ($M=50$) for a linear $50$-atom hydrogen chain at different inter-atomic separations]{Relative ground state energy error obtained with a \ac{qGPS} ($M=50$) for a linear $50$-atom hydrogen chain at different inter-atomic separations in a minimal basis set (STO-6G). Comparison values as obtained with the \ac{HF} method, \ac{CCSD} and \ac{DMRG} with an \ac{MPS} bond dimension of $M_b=50$ are also displayed. Relative energy errors were computed w.r.t. a reference energy obtained with \ac{DMRG} (with \ac{MPS} bond dimension $M_b=500$). Reference energy and comparison values taken from Ref.~\cite{hachmannMultireferenceCorrelationLong2006}. Figure similarly presented in Ref.~\cite{rath2023framework}.}
    \label{fig:H50_results}
\end{figure}

Figure~\ref{fig:H50_results} displays the relative energy error obtained by \ac{qGPS} ground state approximations, using a fixed support dimension of $M=50$ for different inter-atomic separations.
Due to the one-dimensional structure, reference values obtained with \ac{DMRG}~\cite{hachmannMultireferenceCorrelationLong2006} with a large bond dimension in the \ac{MPS} can be considered exact and give a reference to evaluate the relative energy error.
The data points showing the accuracy achieved with the \ac{qGPS} (blue dots) indicate a high accuracy across the full range of considered inter-atomic distances, consistently reaching a relative energy error of slightly less than $10^{-3}$.
Also shown in the figure are reference values obtained with \ac{HF}, \ac{CCSD} (where available) and \ac{DMRG} (with \ac{MPS} bond dimension of $M_b=50$) calculations, all taken from Ref.~\cite{hachmannMultireferenceCorrelationLong2006}.
It can be seen that a significant degree of correlation emerges for these systems resulting in a large error for the mean-field \ac{HF} description.
The level of accuracy achieved with the \ac{qGPS} is comparable to that obtained with the \ac{CCSD} calculation for the data points going up to the separation of $2 \, a_0$ (with Bohr radius $a_0$).
The \ac{CCSD} calculations typically fail as stronger electronic correlation emerges for larger separation between the atoms, and no \ac{CCSD} comparison values are available for the inter-atomic separation of $2.8 \, a_0$.
That the \ac{qGPS} gives a similar level of accuracy across different atomic separations, is a promising indication of a universal applicability of the method for different ab-initio systems beyond the reach of standard ab-initio methods.

Interestingly, the relationship between the achieved energy error and the atomic separation appears to be different for an \ac{MPS} (optimized with \ac{DMRG}) than for the \ac{qGPS}.
For a compressed geometry with a separation of $1 \, a_0$, the \ac{qGPS} with support dimension $M = 50$ also matches the accuracy of an \ac{MPS} with bond dimension $M_b=50$.
As the distance between atoms is increased however, the relative energy error of the \ac{MPS} with $M_b = 50$ decreases.
This points to a decay of the amount of entanglement between the orbitals present in the target ground state.

\begin{figure}[htb!]
    \centering
    \includegraphics{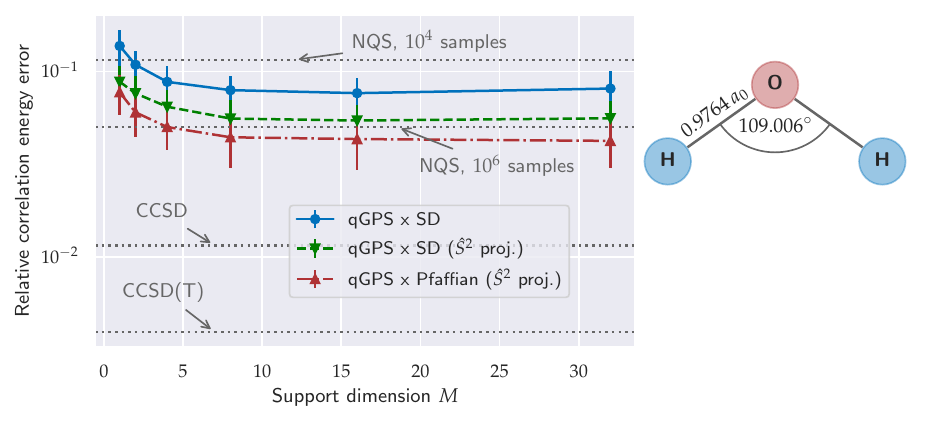}
    \caption[Relative correlation energy error for a ground state approximation of a water molecule in a 6-31G atomic orbital basis]{Relative correlation energy error for a ground state approximation of a water molecule in a 6-31G atomic orbital basis in the geometry as chosen in Ref.~\cite{Choo2019a}, which is also depicted in the right part of the figure. Values are displayed for an optimized mean-field reference state augmented \ac{qGPS} with different support dimensions (optimized using $\approx 10^4$ Monte Carlo samples). The results correspond to different reference state ansatzes (optimized together with the \ac{qGPS}). These include a single \ac{SD} with fixed spin-magnetization but allowed to break the $\hat{S}^2$ symmetry (blue), the same reference ansatz also including a spin projection (green), as well as a spin projected Pfaffian reference state with fixed spin magnetization (red). Reference results achieved with an \ac{RBM} ansatz from Ref.~\cite{Choo2019a}, obtained with $10^4$ and $10^6$ samples respectively, together with \ac{CCSD} and \ac{CCSD(T)} results~\cite{sunPySCFPythonbasedSimulations2018}, are also displayed for comparison. Figure similarly presented in Ref.~\cite{rath2023framework}.}
    \label{fig:H2O_results}
\end{figure}

Due to the one-dimensional structure of the linear hydrogen chains, the ground state can be approximated efficiently to very high accuracy with \acp{MPS}.
For more general molecules however, this will not necessarily be true, and no general method exists to accurately describe the electronic state appropriately in all correlation regimes.
Figure~\ref{fig:H2O_results} presents results obtained for the ground state approximation of a single water molecule in the $6$-$31$G atomic orbital basis in an arrangement near the equilibrium geometry (depicted in the right part of the plot), thus going beyond one-dimensional molecular geometries.
This setup has also been considered in Ref.~\cite{Choo2019a}, discussing the application of \ac{NQS} to the task of modelling ab-initio wavefunctions, apparently in a basis of canonical molecular orbitals.
Depending on the number of samples used, in that work relative correlation energy errors for the ground state approximation with a \ac{RBM} ansatz ranging from $5 \times 10^{-2}$ to $\approx 7 \times 10^{-1}$ are reported.

It was not possible to practically achieve similar energy errors solely using a \ac{qGPS} in a basis constructed of localized molecular orbitals.
Whereas the \ac{HF} state is easily expressed as a \ac{qGPS} with the canonical orbitals, this is no longer true in the basis constructed from localized orbitals.
Moreover, the ordering of the molecular orbitals, which needs to be defined to evaluate Hamiltonian matrix elements, becomes ambiguous.
As discussed in appendix \ref{sec:ferm_ordering_GPS}, modelling the sign structure in the target wavefunction due to different orderings potentially comes at the cost of requiring a support dimension that needs to be scaled quadratically with the number of orbitals.
Such large support dimensions would thus significantly increase the complexity of the model evaluation and number of variational parameters, likely hindering a practical optimization of the parametrization.

To circumvent the two conceptual issues emerging when using a pure \ac{qGPS} as a variational ansatz in a second quantized basis of localized orbitals, the \ac{qGPS} was again augmented by multiplication of a mean-field type reference state.
Multiplying the (theoretically) systematically improvable \ac{qGPS} with such reference states means that the \ac{HF} state can always be spanned.
The ansatz is therefore expected to improve upon this baseline for all support dimensions.
Moreover, the span of the considered parametrization is independent of the chosen orbital ordering since the reference state represents an explicitly antisymmetric wavefunction w.r.t. electronic configurations represented in first quantization.

Figure~\ref{fig:H2O_results} shows the obtained relative error of the correlation energy, defined as the improvement over the variational energy of the \ac{HF} description, as the \ac{qGPS} support dimension $M$ is increased.
Firstly, a single \ac{SD} with fixed total spin magnetization, but which can break the $\hat{S}^2$ symmetry, was optimized together with the \ac{qGPS}.
This gives a relative correlation energy of $\approx 1.4 \times 10^{-1}$ using a support dimension of $M=1$ in the \ac{qGPS}.
This is only slightly larger than the error that was obtained with the \ac{RBM} architecture discussed in Ref.~\cite{Choo2019a} with an optimization of the state using a total of $10^{4}$ samples.
This is also roughly equal to the numbers of samples that was used to obtain the \ac{GPS} results of this chapter.

In the chosen basis of localized orbitals with a mean-field augmented reference state, further improvements to the description manifest by increasing the support dimension.
Whereas the relative error of the correlation energy shows a systematic decrease as the support dimension is increased for small values of $M$, the improvement flattens off, and the accuracy converges to a relative correlation energy error of $\approx 8 \times 10^{-2}$ for support dimensions $M \gtrapprox 8$.
However, not being able to improve upon this value indicates some shortcomings of the approach.
While simple \acp{SD} do not introduce a principled restriction of the model~\cite{morenoFermionicWaveFunctions2021}, slight improvements to the accuracies could be obtained by adjusting the chosen reference state.
As shown in the figure, a magnetization conserving \ac{SD} with spin projection (green data points) gives slightly smaller errors than the non-spin-projected \ac{SD} reference, for all considered support dimensions.
An additional small improvement could be achieved with a magnetization conserving Pfaffian reference state, also including a spin projection, for which the data points are shown as red triangles in the figure.
With this reference state, the relative correlation energy error approaches a value of $\approx 4 \times 10^{-2}$ for larger support dimensions of $M \gtrapprox 8$.
Nonetheless, the overall accuracy still appears to remain limited to this value and no further improvements could be observed.

Whereas great accuracies can be achieved for specific systems with standard methods systematically building upon the \ac{HF} description, such as coupled cluster approaches, these are often not well suited for describing particularly strong electronic correlations.
The high accuracy achieved with \ac{CCSD} and \ac{CCSD(T)} as reported for the water molecule, will therefore not necessarily always be reached for other molecular arrangements.
Especially in the limit where these approaches break down, the ability of the \ac{GPS} to describe correlation in the state could be a valuable asset to improve upon existing descriptions.

\begin{figure}[htb!]
    \centering
    \includegraphics{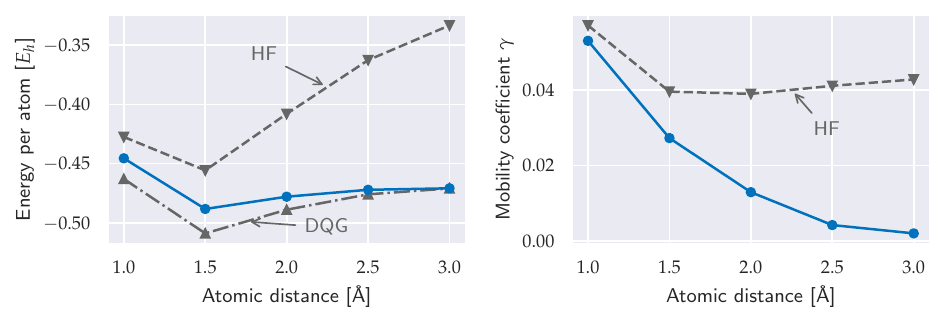}
    \caption[Ground state energy per atom and mobility coefficient $\gamma$ obtained for a three-dimensional cubic arrangement of $4 \times 4 \times 4$ hydrogen atoms at different distances between the nearest neighbour atoms]{Ground state energy per atom (left) and mobility coefficient $\gamma$ (right) obtained for a three-dimensional cubic arrangement of $4 \times 4 \times 4$ hydrogen atoms in a minimal atomic orbital basis set (STO-6G) at different distances between the nearest neighbour atoms. The blue data points represent the results achieved with a \ac{qGPS} ansatz (with support dimension $M=96$) multiplied by a single (spin breaking) \ac{SD} with fixed spin magnetization. The figure also includes comparison values obtained with the \ac{HF} method, as well as the energies obtained with the \ac{DQG} approach taken from Ref.~\cite{sinitskiyStrongCorrelationHydrogen2010}. The coefficient $\gamma$, shown in the right plot, captures the harmonic average of the off-diagonal elements of the \ac{1RDM} in the atomic orbital basis as defined in Eq.~\eqref{eq:gamma_metal_insulator} of the main text. Figure similarly presented in Ref.~\cite{rath2023framework}.}
    \label{fig:H_cube_results}
\end{figure}

Figure~\ref{fig:H_cube_results} shows results for a three-dimensional cubic arrangement of $4 \times 4 \times 4$ hydrogen atoms represented in a minimal basis set (constructed from STO-6G atomic orbitals).
The results were obtained with a variationally optimized \ac{qGPS}, with a support dimension of $M=96$, augmented by a single \ac{SD} reference state.
No general numerically exact methods exist to obtain exact reference energies for this system, making this a showcase application of the method, introducing a novel benchmark.

The left plot of the figure shows the final variational energy per atom, in relationship to the inter-atomic separation between the hydrogen atoms.
The \ac{qGPS} ansatz results in a similar equilibrium geometry as the one obtained from the \ac{HF} method, with a local energy minimum at a separation of around $1.5 \, \text{\r{A}}$.
However, as can be seen in the figure, the \ac{qGPS} ansatz shows a significant improvement over the \ac{HF} level of accuracy across the full range of considered atomic distances.
Crucially, the \ac{qGPS} energies seem to converge to a constant energy per atom for large atomic separations, with a value of $\approx -0.47 \, E_h$ obtained for a separation of $3 \, \text{\r{A}}$.
A single \ac{SD} optimized in the \ac{HF} method is not able to capture the convergence of the energy per atom as the structure is dissociated with large inter-atomic separations.
This also leads to increasing energy difference between the \ac{HF} baseline energy and the one reported for the \ac{qGPS}.
At an atomic distance of $1 \, \text{\r{A}}$, the \ac{qGPS} result improves upon the \ac{HF} energy per atom by $\approx 2 \times 10^{-2} \, E_h$, and this value increases to a difference of $\approx 1.4 \times 10^{-1} \, E_h$ at $3 \, \text{\r{A}}$.

The plot also includes energies obtained with the \ac{DQG} approach, discussed in Ref.~\cite{sinitskiyStrongCorrelationHydrogen2010}.
While the \ac{VMC} calculations give approximations to the energy that present an upper bound to the true ground state energy, the \ac{DQG} approach provides a lower energy bound.
Neglecting statistical errors in the sampling, the actual ground state energy will therefore lie between the \ac{DQG} and the \ac{qGPS} energy.
As it can be seen, the value obtained for the energy per atom, as the atomic distance approaches the largest displayed separation of $3 \, \text{\r{A}}$, is in agreement between the two methods, confirming a high accuracy of both methods.
For smaller separations however, a non-negligible difference between the energies can be observed.
At a distance of $1 \, \text{\r{A}}$ between the atoms, the difference between \ac{VMC} energy and \ac{DQG} energy is $\approx 2 \times 10^{-2} \, E_h$, indicating a less accurate description of at least one of the approaches.
Based on the results for the water molecule discussed above, it can be expected that the \ac{VMC} ansatz is likely not perfectly accurate in the less-strongly correlated limit.

Although the \ac{qGPS} (augmented with a mean-field reference state) model is not giving the perfect ground state approximation in all limits, the improvement over the \ac{HF} mean-field description is significant.
An exemplification how this improvement helps to capture physical effects that are not observable with a mean-field level of theory can be obtained from the right panel of Fig.~\ref{fig:H_cube_results}.
This plot shows how the cube of hydrogen atoms undergoes a decrease of an `electronic mobility', leading a transition from a metal to an insulator, as the distance between atoms is increased.
This metal-to-insulator transition can be quantified by a decay in the coherences of the \ac{1RDM}, i.e., the $2 L \times 2 L$ matrix comprising expectation values of the form $D_{(i,\sigma_1),(j, \sigma_2)} = \langle \Psi|\hat{c}^\dagger_{i,\sigma_1} \hat{c}_{j, \sigma_2} |\Psi \rangle$ as elements.
Here, the \ac{1RDM} is considered in the basis of the original atomic orbitals.
In spirit similar to Ref.~\cite{sinitskiyStrongCorrelationHydrogen2010}, a root mean square over the \ac{1RDM} off-diagonal elements is taken to quantify the decay of spontaneous electronic transitions between atomic orbitals in the ground state.
Its value, displayed in the right panel of Fig.~\ref{fig:H_cube_results} in relation to the inter-atomic separation for the cube of hydrogen atoms, is defined as
\begin{equation}
    \gamma = \sqrt{\frac{\sum_{i \neq j} \sum_{\sigma} |D_{(i,\sigma),(j, \sigma)}|^2}{2 L \times (2 L-1)}},
    \label{eq:gamma_metal_insulator}
\end{equation}
where the indices $i$ and $j$ label the different atomic orbitals.

As is displayed in the figure, such a metal-to-insulator transition cannot be described by the \ac{HF} description, giving a value of $\gamma$ that is mostly independent of the atomic separation.
This is different for the obtained \ac{qGPS} results, which reproduce the expected decay of $\gamma$ as the cube of hydrogen atoms is dissociated.
These results give a value of $\gamma \approx 0.05$ at an atomic distance of $1 \, \text{\r{A}}$, which monotonically decreases to $\gamma \approx 0$ for the largest considered separation of $3 \, \text{\r{A}}$.
This decay also qualitatively matches the behaviour obtained with the \ac{DQG} approach as discussed in Ref.~\cite{sinitskiyStrongCorrelationHydrogen2010}.

While the discussed results indicate an applicability of the \ac{GPS} for ab-initio quantum chemical simulations, they also show that systematic improvability of the approximation can not always be achieved in practice.
Again, it is not entirely clear whether this is a limitation of the ansatz, or a shortcoming emerging from the stochastic optimization of the parametrization.
To improve the general quality of the model, different additional extensions of the ansatz are possible.
One might, e.g., use the \ac{qGPS} as a general model to define configuration-dependent many-body functions for the evaluation of \acp{SD}.
This `backflow' construction forms part of the standard approach to represent electronic wavefunctions with \acp{NN} in the real space~\cite{gerardGoldstandardSolutionsSchr2022, Pfau2019,Hermann2019, Han2018, hermannAbinitioQuantumChemistry2022, liInitioCalculationReal2022} and can also be applied in a discretized picture of orbitals~\cite{Luo2019}.
However, this would likely significantly increase the computational cost to evaluate and optimize the model, posing additional challenges to apply the methodology also for larger system sizes.

\acbarrier

\chapter{Tensor network perspectives on Gaussian Process States}
\label{ch:qGPS_learning}
The previous two chapters discussed the application of \ac{GPS} models within standard \ac{VMC} approaches.
In this framework, essentially any model can be used to define a mapping associating wavefunction amplitudes to computational basis states (as long as it can be evaluated efficiently).
In that sense, the application of the \ac{GPS} is very similar to that of the \ac{NQS}.
While universal approximation properties guarantee to be able to describe states theoretically to essentially arbitrary accuracy (albeit potentially with an exponential effort), the results that can be achieved in practical calculations are dominated by different factors.
In addition to the efficiency of the model to represent relevant target states compactly, also the ability to robustly optimize the parametrization contributes to the practical usability of these states.

Especially with the introduction of quantum support points, the previous approximation of target states with \ac{GPS} in a \ac{VMC} context follows essentially the same approaches as one would apply with \ac{NQS}.
Indeed, it was shown that the \ac{GPS} can not only be represented as a specific \ac{NN} architecture (see section~\ref{sec:GPS_as_NN}), but also numerical evidence was presented that the representation practically reaches mostly comparable results to other \ac{NQS}.
Overall, the results seem to suggest that for some applications and comparison results, the \ac{GPS} model improves upon \ac{NQS} realizations, for others it does not.
But all in all, no universal advantage (or disadvantage) manifested for the \ac{GPS} over other discussed \ac{NQS}.
Especially with a full parametrization of the support configuration with continuous parameters, one might ask whether there is any fundamental difference between \acp{GPS} and \acp{NQS}.

In this chapter, a different perspective on the \ac{qGPS} model is discussed, contrasting the model from general \ac{NN} architectures.
This explores more explicit connections between the \ac{GPS} and tensor decompositions of quantum states.
By bringing this perspective together with the Bayesian regression framework outlined in chapter \ref{ch:GPS_introduction}, novel tools are discussed to find a state based on rigorous supervised learning mechanisms.

\section{Gaussian Process States as exponentiated tensor decompositions}
As outlined in section \ref{sec:qGPS}, the \ac{GPS}, with exponential kernel, models the log wavefunction amplitudes, $\omega(\mathbf{x})$, as a linear combination of product states.
Following Eq.~\eqref{eq:GPS_as_product_state_lin_combination}, the log amplitudes are specified as
\begin{equation}
    \omega(\mathbf{x}) = \sum_{x'=1}^M \prod_{j=1}^L \phi^{(x_j)}_{x',j}.
\end{equation}
With parametrized product states as support configurations, the amplitudes $\phi^{(x_j)}_{i,j}$ can also directly be associated with the variational parameters of the \ac{qGPS}, i.e.,
\begin{equation}
    \phi^{(x_j)}_{x',j} = \epsilon^{(x_j)}_{j, x'}.
\end{equation}
In mathematical terms, this form is equivalent to a tensor decomposition of the full $D^L$ tensor of the log wavefunction amplitudes, in terms of a sum over $M$ (tensor) products of $L$ one-dimensional tensors (where the one-dimensional tensors are indexed by the physical index $x_j$).
Such a decomposition is known by various names and in recent literature typically simply denoted as a \ac{CP} decomposition~\cite{kiersStandardizedNotationTerminology2000, koldaTensorDecompositionsApplications2009}.
The linear combination of $M$ product states is also equivalent to the amplitudes associated with \acp{MPS} of bond dimension $M$, where all the matrices are constrained to be diagonal.
While such a construction for the actual wavefunction amplitudes would in general not fulfil the product separability requirements, the exponentiation of the \ac{CP} decomposition ensures that a product of correlation features is obtained.

It is tempting to leverage the connection between \ac{GPS} and tensor networks to define an alternative scheme to contract the representation in the evaluation of expectation values, similar to the evaluation of expectation values for \ac{MPS} (c.f. Eq.~\ref{eq:MPS_exp_val}).
Considering the Taylor expansion of the exponential around zero, the \ac{GPS} amplitudes can be represented as a Taylor series
\begin{equation}
    \Psi(\mathbf{x}) = \frac{\left( \sum_{i=1}^M \prod_{j=1}^L \phi^{(x_j)}_{i,j} \right)^k}{k!}.
\end{equation}
Each term in this Taylor expansion can then be rephrased as a specific linear combination of product state amplitudes by application of the multinomial theorem.
This gives the expression
\begin{equation}
    \frac{\left( \sum_{i=1}^M \prod_{j=1}^L \phi^{(x_j)}_{i,j} \right)^k}{k!} = \sum_{\mathcal{P}(k_1, k_2, \ldots, k_M)}\frac{1}{k_1! k_2! \ldots k_M!} \prod_{l=1}^M \left( \prod_{j=1}^L \phi^{(x_j)}_{l,j} \right)^{k_l}.
\end{equation}
Here, the sum includes all possible choices of $M$ non-negative integers $(k_1 \ldots k_M)$ that sum to the expansion order $k$.
With this reformulation of the \ac{GPS} amplitudes, the $k$-th expansion order is therefore associated with a linear combination of $|\mathcal{P}|(k)$ product states.
The total number of terms arising from the different choices for the integers $(k_1, \ldots, k_M)$ is given by
\begin{equation}
    |\mathcal{P}|(k) = {{k+M-1} \choose {M-1}}.
\end{equation}
This number thus grows exponentially with the expansion order, $k$, which limits the evaluation of the expansion terms to small expansion orders.

The Taylor expansion of the \ac{GPS} amplitudes gives a representation in terms of linear combinations of product states.
If the Taylor series can be truncated at small orders (independently of the system size), the \ac{GPS} thus describes a vanishing degree of entanglement, as it emerges for a linear combination of few product states.
While the contribution of higher orders in the Taylor series generally prevent exact contractions of the state, it thus also enables to describe states exhibiting stronger entanglement.
Though the specifics of the entanglement that can be described have not yet been investigated for the \ac{GPS} model, the potential to capture a volume-law type scaling of the entanglement entropy (i.e., an entanglement between two sub-systems growing with the size of the sub-systems) has been shown for other \ac{NQS}~\cite{dengQuantumEntanglementNeural2017, sunEntanglementFeaturesRandom2022}, and similar characteristics are expected to hold for the \ac{GPS}.

\section{Iterative Bayesian sweeping}
Whereas the deterministic evaluation of expectation values is intrinsically limited to states represented efficiently by linear combinations of product states, the specifics of the \ac{CP} decomposition can be exploited for data-driven approaches to `learn' a \ac{qGPS}.
Building on the relations between the \ac{qGPS} and tensor network representations, this section introduces an iterative approach to compress given wavefunction data points into the form of a \ac{qGPS}.
This is achieved by iteratively updating the \ac{qGPS} parameters for one extracted reference site at a time with standard Bayesian approaches.
By moving the choice of reference sites across the $L$ modes, iteratively a \ac{qGPS} representation is learned based on the presented data.
This sweeping through the physical space is conceptually directly related to iterative \ac{MPS} optimization techniques, such as \ac{DMRG}~\cite{Schollwoeck2011} and the time evolving block decimation~\cite{paeckelTimeevolutionMethodsMatrixproduct2019}.
It corresponds to an \ac{ALS} approach~\cite{faberRecentDevelopmentsCANDECOMP2003, koldaTensorDecompositionsApplications2009, minsterCPDecompositionTensors2021} applied in the log wavefunction space to compress given wavefunction data into a \ac{qGPS} in a supervised learning setup.
The Bayesian regression principles can be used to appropriately regularize the optimization, giving a fully automated approach directly applicable in various settings.


In section \ref{sec:bayesian_sweeping_j1_j2} it is shown that this iterative procedure helps with the stability of the compression, and it utilizes the principles from the Bayesian regression framework to introduce appropriate regularization to learn a state from small numbers of finite samples.
Such a task can directly be related back and applied to the \ac{VMC} optimization of a state by tracking the imaginary time evolution as a method to approximate the (unknown) target state~\cite{kochkovLearningGroundStates2021}.
This is discussed and exemplified in section \ref{sec:ground_state_search_sweeping}.
Lastly, section \ref{sec:classical_ML_GPS} shows how the method can be extended for different tasks of supervised \ac{ML}, here exemplified by a simple image recognition experiment.

Central property that is exploited for the iterative data compression into a \ac{qGPS} is the observation that the \ac{qGPS} represents a multilinear model for the log wavefunction amplitudes according to
\begin{equation}
    \omega(\mathbf{x}) = \log \left( \Psi(\mathbf{x}) \right) = \sum_{x'=1}^M \prod_{i=1}^L \epsilon^{(x_i)}_{i, x'}.
\end{equation}
Due to the multilinearity, it is possible to extract one parameter per support index and local occupancy index as linear prefactors in a weighted sum of features defined by the other parameters.
In this construction, the log wavefunction amplitude is re-expressed as
\begin{equation}
    \omega(\mathbf{x}) = \sum_{x'=1}^M \sum_{l=1}^D \epsilon^{l}_{I, x'} \, \delta_{x_{I}, l} \prod_{i \neq I} \epsilon^{(x_i)}_{i, x'},
\end{equation}
where $I$ is a chosen reference site.
While not considered in the numerical tests discussed in the following, the reference site can, in principle, be different for different support indices $x'$ and local occupancies $l$, i.e., $I = I(x', l)$.
With the identification of weights and features, the equation above can be written more compactly as a linear combination of features according to
\begin{equation}
    \omega(\mathbf{x}) = \sum_{i=1}^{M \times D} w_i \, \phi_i(\mathbf{x}).
    \label{eq:linear_model_reference_site}
\end{equation}
The weights are given by the parameters associated with the picked reference site, $w_i = \epsilon^{l}_{I, x'}$, and the other parameters define the $D \times M$ features $\phi_i(\mathbf{x}) = \delta_{x_{I}, l} \prod_{j \neq I} \epsilon^{(x_j)}_{j, x'}$  (where in both cases $i$ is seen as a compound index of $x'$ and $l$, and, as before, $D$ denotes the local Hilbert space dimension).

\begin{figure}[htb!]
    \centering
    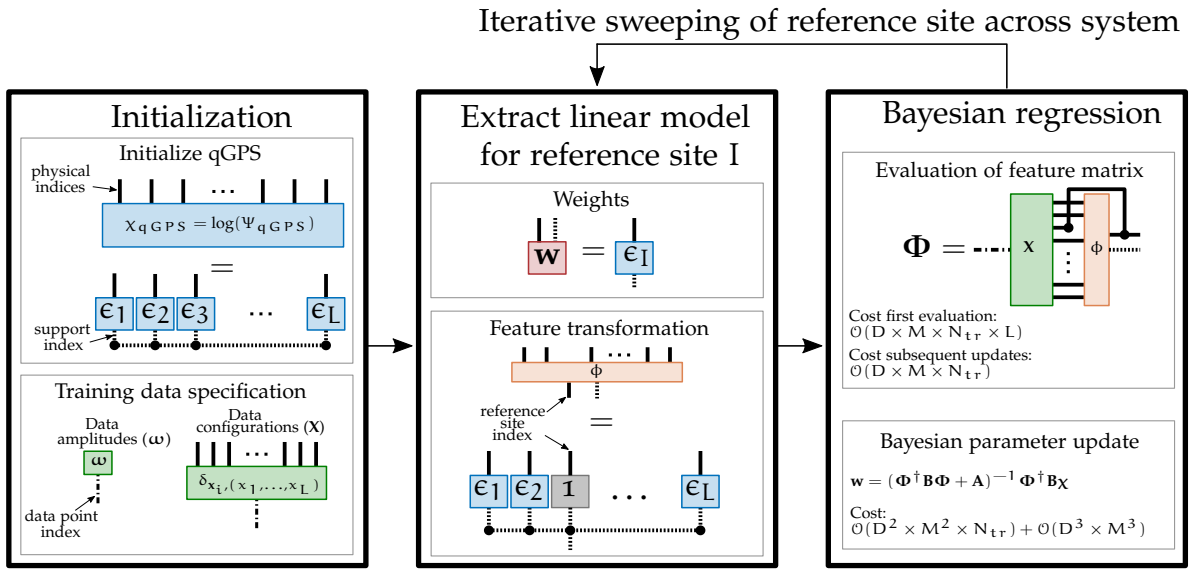
    \caption[Schematic overview of the Bayesian sweeping algorithm for a supervised learning of a \acl{qGPS}]{Schematic overview of the Bayesian sweeping algorithm for a supervised learning of a \ac{qGPS}. The visualization of tensorial quantities is inspired by the standard diagrammatic notation for tensor network diagrams, e.g., summarized in figure 1 of Ref.~\cite{millerProbabilisticGraphicalModels2021}. The representation includes the use of black dots to represent `COPY' tensors, i.e., multidimensional extensions of the Kronecker delta~\cite{biamonteCategoricalTensorNetwork2011, dennyAlgebraicallyContractibleTopological2012, peiCompactNeuralnetworkQuantum2021}.}
    \label{fig:sweeping_schematic}
\end{figure}

With the reformulation of the \ac{qGPS} model according to Eq.~\eqref{eq:linear_model_reference_site}, the Bayesian regression techniques discussed in section~\ref{sec:bayesian_regression} are directly applicable to obtain the weights $w_i$ in a well-defined, statistically meaningful approach from given wavefunction data.
Applying the Bayesian regression as before, a closed expression for a Gaussian posterior distribution over the weights can be obtained from which the most probable weights can be adapted to update the corresponding model parameters $\epsilon^{l}_{I, x'}$.
The local regression can be iterated by repeatedly sweeping the choice of the reference site across the different modes of the system, updating $D \times M$ parameters of the \ac{qGPS} at a time.
A pictorial schematic, outlining the different steps of the approach to compress data into a \ac{qGPS}, is presented in Fig.~\ref{fig:sweeping_schematic}.

The inference of the statistical distribution over the weights at each inner regression step can be achieved with the same concepts as outlined in section \ref{sec:bayesian_regression}.
This means that the log wavefunction amplitude likelihoods are modelled as independent normal distribution around the log predictions of the \ac{qGPS}.
Furthermore, also the prior distribution of the weights is modelled as a Gaussian distribution centred at zero.
If also sign information is described by the model, the normal distributions, modelling the likelihood of the log wavefunction amplitudes and the weight prior, can be extended to be normal distributions of complex random variables~\cite{halliwellComplexRandomVariables2015, hankinComplexMultivariateGaussian2015}.
Here, the normal distributions are specified with real-valued, diagonal covariance matrices and vanishing pseudo-covariance matrices.

As discussed in section \ref{sec:bayesian_regression}, (now extended to the considered case of complex variables), the modelling assumptions lead to closed form expressions for the weight posterior.
Its mean is given by
\begin{equation}
    \boldsymbol{\mu}_{mp} = \left( \boldsymbol{\Phi}^\dagger \mathbf{B} \boldsymbol{\Phi} + \mathbf{A} \right)^{-1} \boldsymbol{\Phi}^\dagger \mathbf{B} \boldsymbol{\omega}.
    \label{eq:weight_update_posterior}
\end{equation}
In this formulation, the vector $\boldsymbol{\omega}$ comprises the different data log-amplitudes, $\boldsymbol{\Phi}$ represents the corresponding $N_{tr} \times (D M)$ matrix of features, and the `$^\dagger$' symbol denotes the hermitian-conjugation of a matrix.
The matrices $\mathbf{A}^{-1}$ and $\mathbf{B}^{-1}$, (taken to be diagonal) encode the prior and (log space) likelihood variances on their diagonal.
As it was also discussed in the learning setup with the `classical' \ac{GPS}, it is sensible to tune additional hyperparameters, defining the variances, by maximization of the (log) marginal likelihood.
This provides fully automated ways to achieve appropriate regularization of the fit, particularly useful to achieve good generalization of the model if only limited data is presented.

\subsection{Supervised learning of signed target states}
\label{sec:bayesian_sweeping_j1_j2}

Being able to exploit the closed analytic expressions to update the weights in the Bayesian regression framework, is a key element setting the iterative sweeping approach apart from other approaches more generally applicable to fit function parametrizations to data.
In this section, the sweeping is compared to a fitting via a direct numerical minimization of a squared error loss function with a generic gradient descent type approach.
The squared error loss for \ac{qGPS} wavefunction model, $\Psi(\mathbf{x}_i)$, fitted to training data is given by
\begin{equation}
    \mathcal{L} = \sum_{i=1}^{N_{tr}} |e^{\omega_i} - \Psi(\mathbf{x}_i)|^2,
\end{equation}
where the sum is taken over all elements of the training set for which $\omega_i$ denotes the log training amplitude associated with training configuration $\mathbf{x}_i$.

It is exemplified in the following that the iterative sweeping scheme provides a particularly robust approach for learning a \ac{qGPS} model from a limited set of training data.
In particular, it is shown that the algorithm, underpinned by rigorous Bayesian principles, yields solutions typically generalizing well across the full Hilbert space, without requiring separate validation.

\subsubsection{Experiment setup}
For a concrete realization of the sweeping learning, here, it is discussed for the ground state approximation for $J_1$-$J_2$ models defined on a square lattice of $6 \times 4$ sites, without explicit incorporation of the \ac{MSR}.
Some small fraction of the full Hilbert space data is presented as training data, from which a \ac{qGPS} wavefunction model should be learned, ideally capturing the target state well across the full Hilbert space.
The learning setup is directly inspired by one previously considered for \ac{NN} architectures discussed in Ref.~\cite{Westerhout2019}.
In that work, it was described that particular generalization difficulties emerge for the learning of the \ac{NQS} representation, especially when describing the ground state in the frustrated regime.
This observation was also related to the increased difficulty with achieving good accuracies for \ac{VMC} optimizations of such ansatzes.
Relying on stochastic sampling of expectation values, the optimization of the state in the \ac{VMC} context intrinsically relies on the ability to optimize a model based on few samples from the Hilbert space.
Even if the expressivity of the model allows for a highly accurate description of the target state, the ability to learn a representation generalizing across the full Hilbert space is also crucial to the success of the method.


In the practical application of the sweeping approach discussed in this section, the learning of the state was performed with a kernel-symmetrized \ac{qGPS} (ensuring translational symmetries, lattice point group symmetries, and spin-inversion symmetry).
With this symmetrization of the model, the sweeping scheme is directly applicable by defining the linear model for the log amplitudes at each inner step with symmetrized features according to
\begin{equation}
    \omega(\mathbf{x}) = \sum_{i=1}^{M \times D} w_i \, \left(  \sum_{\mathcal{S}} \phi_i(\mathcal{S}[\mathbf{x}])\right).
\end{equation}
This expression includes an inner sum over all included symmetry operations $\mathcal{S}$.

At each step of the sweeping, the particular features and weights for the chosen reference site are specified by the variational parameters of the model.
While not the only possibility, the schedule to pick the reference site employed in the numerical tests, was based on using a global reference site $I$, and deterministically moving this reference site across the lattice in a sweep.

Based on the chosen sweep protocol, the variational parameters associated with the reference site $\epsilon^{l}_{I, x'}$ are updated according to Eq.~\ref{eq:weight_update_posterior} at each step.
The prior at each local fit was specified with a diagonal inverse covariance matrix $\mathbf{A} = \alpha_I \mathbb{1}$, with a single real parameter $\alpha_I$, however allowed to be different for different reference sites.
Furthermore, a site-independent variance parameter $\tilde{\sigma}^2$ was introduced to specify the variances of the log space likelihood.
Following the approaches outlined in section~\ref{sec:bayesian_state_learning}, this parameter was chosen to approximate the variance of the non-log wavefunction amplitudes to achieve an (approximately) magnitude-independent error of the wavefunction amplitudes with the fit.
As given in Eq.~\eqref{eq:log_space_likelihood_variance} this means that the diagonal $N_{tr} \times N_{tr}$ matrix $\mathbf{B}$ was defined according to
\begin{equation}
    B_{i,i} = \frac{1}{\sigma^2(\mathbf{x}_i)} = \frac{1}{\log \left( \frac{\tilde{\sigma}^2} {|\langle e^{\omega_i} \rangle|^2} +1 \right)}.
\end{equation}
The additional hyperparameters $\alpha_I$ and $\tilde{\sigma}^2$, regularizing the fit in a probabilistic framework, were updated during the sweeping to maximize the log marginal likelihood at each local fit.
With the marginal likelihood maximization, the regularization hyperparameters are automatically obtained from the data, allowing to fit the model on all available data without any additional validation~\cite{Tipping2003a}.

In the practical implementation, the optimization of $\alpha_I$ and $\tilde{\sigma}^2$ were separated because the parameters $\alpha_I$ can be updated without requiring a recalculation of the matrix-matrix products of the form $\boldsymbol{\Phi}^\dagger \mathbf{B} \boldsymbol{\Phi}$.
This means that, at each step, first the parameter $\alpha_I$ was updated by repeated updates according to
\begin{equation}
    \label{eq:alpha_update}
    \alpha_I \rightarrow \frac{\sum_i (1 - \alpha_I \Sigma_{i,i})}{|\boldsymbol{\mu}_{mp}|^2}.
\end{equation}
This update is based on a standard update formula to maximize the marginal likelihood, commonly employed in the \ac{RVM} (see appendix \ref{ch:RVM}).
After the optimization of the parameter $\alpha_I$, a single gradient ascent update to the parameter $\tilde{\sigma}^2$ was applied in the log parameter space, updating the parameter according to
\begin{equation}
    \tilde{\sigma}^2 \rightarrow \exp \left( \log (\tilde{\sigma}^2) + \eta \frac{d \log \left(p^{(I)}_{ML} \right)}{d \tilde{\sigma}^2} \tilde{\sigma}^2 \right),
    \label{eq:gradient_ascent_noise}
\end{equation}
with a small learning rate $\eta = 10^{-5}$.
The derivative of the log marginal likelihood with respect to the parameter $\tilde{\sigma}^2$ is stated in appendix~\ref{ch:RVM}.
To keep the value of $\tilde{\sigma}^2$ within reasonable bounds (especially important during the initial stages of the sweeping), the gradient ascent updates were however capped so that the value of $\tilde{\sigma}^2$ never exceeded its initial value, chosen as the mean squared error across the training set.
The sweeping was iterated until convergence in the log marginal likelihood values, averaged over a full sweep, was observed.



\subsubsection{Results}
\begin{figure}[htb!]
    \centering
    \includegraphics{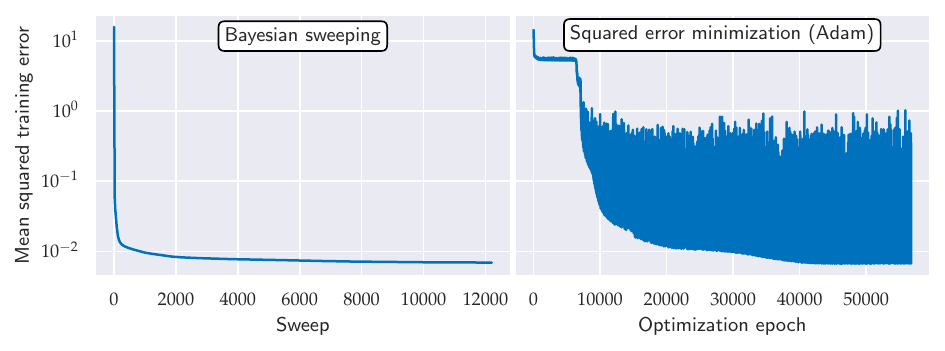}
    \caption[Mean squared training error achieved in the fit of a \acl{qGPS} to a data set from the antiferromagnetic Heisenberg model ground state for a $6 \times 4$ square lattice]{Mean squared training error achieved in the fit of a \ac{qGPS} to a data set from the antiferromagnetic Heisenberg model ground state for a $6 \times 4$ square lattice. The left panel shows the evolution of the error against the number of sweeps in the Bayesian sweeping approach as outlined in the main text. The right panel shows the error in relation to the number of epochs for the minimization of the squared error loss with Adam optimizer~\cite{kingmaAdamMethodStochastic2017}, using a learning rate of $10^{-4}$, other hyperparameters set to standard values~\cite{bourginNumpyml2021}, using mini-batches of $\approx 64$ training configurations, and holding back $20 \%$ of the training data for validation.}
    \label{fig:sweeping_example}
\end{figure}

Figure~\ref{fig:sweeping_example} shows the example evolution of the mean squared error between the \ac{qGPS} and the training data for a fit of the ground state in the unfrustrated limit at $J_2/J_1=0$.
The training data consisted of a randomly selected set of configurations with associated ground state amplitudes, corresponding to $\approx 1 \%$ of the full Hilbert space size.
The right plot shows the mean squared error in relation to the number of sweeps that were applied in the Bayesian sweeping approach described above.
As can be seen, the mean squared training error over the training amplitudes, which were rescaled to achieve a vanishing mean of the log training amplitudes, shows a rapid decay.
It decreases from an initial value of $\approx 16$ to a value of less than $10^{-2}$, achieved after $\approx 850$ sweeps.

The results achieved by the sweeping can be compared to the results achieved with a fitting of the \ac{qGPS} to the same data set by a direct minimization of the squared error loss function with a gradient descent based optimizer.
The right panel of fig.~\ref{fig:sweeping_example} reports the mean squared error vs. the number of optimization epochs in which the parameter updates were applied based on the Adam optimizer ~\cite{kingmaAdamMethodStochastic2017}.
Following standard approaches, each epoch of the optimization comprised multiple parameter updates based on mini-batches of $\approx 64$ training data points at a time, and $\approx 20 \%$ of the training data was held back to be used for subsequent validation of the model.
Each training data point from the remaining $\approx 80 \%$ was considered exactly once in the randomly generated mini-batches of an epoch (so that the last mini-batch of an epoch might have contained less than $64$ training samples).

Whereas an overall decrease in the error is generally also observed for the direct minimization of the squared error with the Adam optimizer, in the presented example, the optimization got stuck at an early stage of the minimization.
Multiple optimization epochs were required to escape from the apparent local minimum in the squared error to reach errors comparable to the ones already obtained after few sweeps with the Bayesian sweeping approach.
Furthermore, the presented training error shows significant fluctuations between different epochs.
The displayed non-monotonic behaviour already indicates significant difficulties to learn a \ac{qGPS} from the signed target wavefunction with generic approaches, here exemplified by the squared error minimization with the Adam optimizer.
While different adjustments to the optimization hyperparameters and protocol might help to improve the stability, the Bayesian sweeping learning does not require manual hyperparameter tuning.

The displayed training error decay with the \ac{qGPS}-specific sweeping algorithm already indicates an advantageous applicability of the sweeping approach to the learning of quantum states from small numbers of samples.
Ultimately however, the key challenge is to learn a model that generalizes well beyond the training data.
The model will only be sensible representation of the target state if the wavefunction amplitudes of the target are captured well for all configurations of the computational basis.
The utilized Bayesian regression framework provides an essentially fully automated approach to balance the accuracy of the fit with an appropriate level of regularization through the maximization of the marginal likelihood~\cite{Tipping2003a}.
It is exactly this probabilistic interpretation of the \ac{GPS} that could prove to be helpful to improve the reliability of the optimization of highly flexible wavefunction parametrizations.

\begin{figure}[htb!]
    \centering
    \includegraphics{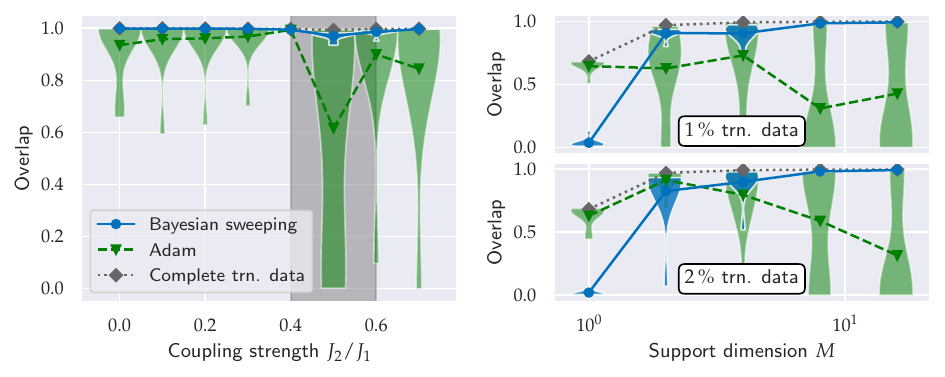}
    \caption[Overlap between a \acl{qGPS} representation trained on a small data set and the exact ground state for $J_1$-$J_2$ systems on a square lattice of $6 \times 4$ sites]{Overlap between a \ac{qGPS} representation trained on a small data set and the exact ground state for $J_1$-$J_2$ systems on a square lattice of $6 \times 4$ sites. The results were obtained with a direct minimization of the squared error with Adam validated on $20 \%$ of the training set (green), and with the Bayesian sweeping (blue). The solid data points indicate the mean overlap, and violin plots indicate the spread, across ten different random realizations. The grey data points indicate the model expressiveness as obtained from training the model on the complete data set of all configurations and associated wavefunction amplitudes. The left panel shows results for different values of $J_2/J_1$ (where the regime of strong frustration is indicated by a dark grey shading~\cite{Westerhout2019}) using a fixed support dimension of $M=5$ trained on $27042$ configurations of the Hilbert space.
            The right panels show the results at fixed $J_2/J_1=0.5$ against the support dimension of the model.
            The top (bottom) panel of the right figure shows results for a training set corresponding to $27042$ ($54083$) configurations of the full Hilbert space comprising $\approx 2.7 \times 10^6$ basis configurations.
            Figure (adjusted) taken from Ref.~\cite{boothQuantumGaussianProcess2021}.}
    \label{fig:J1J2_sweeping_results}
\end{figure}

Inspired by the setup discussed in Ref.~\cite{Westerhout2019}, the generalization properties of the \ac{qGPS} fitting procedures can directly be analysed for the setup discussed above by evaluating the overlap between the target state and the \ac{qGPS} trained on a limited data set.
Figure~\ref{fig:J1J2_sweeping_results} shows the obtained overlap evaluated between learned model and target state obtained for different parameter regimes of the system.
As a baseline, the figure also reports the accuracy obtained by a fit on the full wavefunction data as grey data points\footnote{The fit on the complete wavefunction data was practically achieved by a quasi-Newton method minimization of the full squared error after suitable initialization~\cite{boothQuantumGaussianProcess2021}.}.
The main goal is to be able to match this expressivity limit as well as possible, solely based on the information from the presented training data.
For each setup, the fit to the restricted data set was repeated ten times with different realizations of the random elements (in particular including the training data selection).
The violin plots displayed in the figure visualize the distribution around the mean of the outcomes represented by the solid data points.
The outcomes are presented for the Bayesian fitting approach (blue circles), as well as the squared error minimization with Adam (green triangles).
For the latter approach, the fit was validated by analysing the mean squared error for a validation subset comprising $20 \%$ of the training data not used to optimize the parametrization.
The parameter values that gave the smallest validation error across different optimization epochs (also determining an early-stopping type regularization~\cite{Westerhout2019}), and over two different learning rates ($10^{-3}$ and $10^{-4}$), were used as the final parameters of the learned model.

The left sub-plot of the figure reports the results obtained by fitting the $J_1$-$J_2$ model target state with a \ac{qGPS} to randomly selected training data as the next-nearest neighbour interaction strength $J_2/J_1$ is varied.
The \ac{qGPS} was defined with a fixed support dimension of $M=5$, and $1 \%$ of the full space was considered as training data.
Whereas a significant spread of the results across different random seeds can be observed with the direct least squares minimization, the Bayesian sweeping approach gives a much less varying approximation of the target state.
Importantly, for most displayed interaction strengths, the Bayesian learning approximately reaches the maximum overlap of one between compressed representation and target state.
Only for the displayed data points from the frustrated regime ($ 0.4 \lessapprox J_2/J_1 \lessapprox 0.6$), a slight increase in the spread across seeds and a deviation from the model expressivity limit becomes apparent.
This observation is in agreement with the ones reported for the supervised \ac{NQS} fitting~\cite{Westerhout2019}.
The described decrease of the generalization quality in the regime of strong frustration, gives the largest deviation between mean fitted state overlap and the model expressivity limit at $J_2/J_1 = 0.5$ with a mean overlap of $\approx 0.97$.
Crucially however, the overlap values obtained with the Bayesian sweeping algorithm show a much greater consistency as compared to the validated least squares fit of the training data with Adam.
This results in a mean overlap of $\approx 0.62$ at the highly frustrated parameter point $J_2/J_1 = 0.5$.

This greater reliability also persists for larger support dimensions as indicated in the right part of the figure.
It reports the overlap in the strongly frustrated parameter regime at $J_2/J_1 = 0.5$ as the support dimension is increased from $M=1$ to $M=16$.
Perhaps more importantly however, the Bayesian sweeping approach yields a convergence to the target state with overlap approximately equal to one (and almost vanishing spread) for support dimensions $M=8$ and $M=16$.
This convergence is observed for both considered training set sizes, corresponding to $\approx 1 \%$ (top panel) or $\approx 2 \%$ (bottom panel) of the full Hilbert space size.

Without appropriate regularization of the fit, an increase of model expressiveness (achieved by the increase of the support dimension) might lead to an increased susceptibility to overfitting the training data.
A very simple model on the other hand will likely not be sufficiently expressive to describe the target state well enough.
These characteristics can directly be observed for the data points corresponding to the direct minimization of the mean squared error with Adam.
Whereas the spread across different seeds is smaller and the mean overlap almost reaches the limit of the full model expressivity for very simple models, significant accuracy fluctuations across different random seeds emerge with larger support dimensions.
As the support dimension increases, the deviation between the potential expressivity and the results achieved with the Adam minimization get larger.
While the results obtained with the Bayesian sweeping match the maximum expressivity limit less well for $M=1,2,4$, good agreement is obtained for the larger support dimensions with $M=8,16$.
Nonetheless, the overall highest overlap for a \ac{qGPS} trained on a limited data set is achieved with the direct mean squared error minimization for a run with the \ac{qGPS} with $M=16$ reaching an overlap of $\approx 0.9975$.

The presented results suggest that one sacrifices some degree of accuracy for a more consistent generalization of the model with the applied the Bayesian sweeping algorithm to infer a model from the presented data.
Ultimately, a more consistent approach to learn a representation of the quantum state is likely often more desirable than one potentially reaching a more compact representation, albeit less reliably.
Applying the fully automated approach to learn a \ac{qGPS} representation from presented training data could therefore provide a valuable tool in different scenarios where especially the generalization of the model is of great importance.

\subsection{Iterative imaginary time tracking for ground state search}
\label{sec:ground_state_search_sweeping}

The Bayesian sweeping method is readily applicable to compress a presented data set to a \ac{qGPS} of given support dimension for various different applications.
The support dimension is the only input parameter of the method that controls the model complexity (and thus its maximum expressivity as well as the computational effort associated with the evaluation of wavefunction amplitudes).

In order to apply the supervised learning to the task of approximating many-body ground states from which no exact data is available, the method can also directly be utilized for the \ac{SWO} presented in Ref.~\cite{Kochkov2018}.
Originally described for the optimization of \acp{NQS}, a similar application was also discussed recently for kernel models in Ref.~\cite{https://doi.org/10.48550/arxiv.2303.08902}.
In this approach, the ground state approximation is obtained by iteratively generating wavefunction data from an improved version of a current ansatz, which is then used to recompress the improved state back into a compact functional form.
This section discusses the application of the Bayesian sweeping as a technique to achieve the compression of the target state into the form of \ac{qGPS} at each iteration.

To iteratively learn a \ac{qGPS} ground state description, here essentially the same approach is considered as was discussed in the original work introducing the \ac{SWO} framework~\cite{Kochkov2018}.
Given a \ac{qGPS} $| \Psi_{qGPS}^{(k)} \rangle$ at the $k$-th macro-iteration step, the updated target state is defined as
\begin{equation}
    | \Psi_{target}^{(k)} \rangle = (1 - \tau \hat{H}) | \Psi_{qGPS}^{(k)} \rangle,
    \label{eq:target_step_SWO}
\end{equation}
with an appropriately chosen parameter $\tau$, and where $\hat{H}$ is the system Hamiltonian.
This target state is equivalent to the first order approximation of the imaginary time evolved state $e^{-\tau \hat{H}} | \Psi_{qGPS}^{(k)} \rangle$.
By iteratively learning the target state $| \Psi_{target}^{(k)} \rangle$ as a new \ac{qGPS} model, for sufficiently small values of $\tau$, the method thus tracks the imaginary time evolution of the initial ansatz as a \ac{qGPS}.
This is also directly related to the \ac{SR} approach to optimize the parametrization, which utilizes closed form expressions to recompress the target state for small parameter variations (see section \ref{sec:VMC_optimization}).
Whereas the updates in the \ac{SR} approach are explicitly based on small model parameter changes, in the \ac{SWO} approach, there are no such restrictions for the improved target state.
This means, that often also larger values of the time step, $\tau$, corresponding to a power method type scheme~\cite{Kochkov2018}, can be chosen.

For standard \ac{VMC} ansatzes not allowing for an efficient contraction of expectation values, the approximation of the target state $| \Psi_{target}^{(k)} \rangle$ typically needs to be achieved based on stochastic sampling from the Hilbert space.
In the \ac{SWO} formulation essentially any supervised learning scheme can be applied to fit the variational model to $| \Psi_{target}^{(k)} \rangle$.
Because for states that cannot be contracted efficiently, the fit of the target state is typically based on finite numbers of samples, avoiding overfitting is key to achieve a good approximation of the targeted ground state.
Due to the high degree of reliability demonstrated for the supervised learning of a \ac{qGPS} with the Bayesian sweeping approach, its application to the \ac{SWO} seems sensible.
An exemplified application of the scheme is discussed in the following.

To apply the Bayesian sweeping, for each compression of the target state, a corresponding data set needs to be generated.
This requires the sampling of configurations for which the target state amplitudes are evaluated to define the training data.
While different realizations are imaginable, the approach considered here involves the generation of samples according to two different probability distributions.
To ensure that large wavefunction amplitudes are described appropriately, samples according to the probability distribution of the target state, $|\langle \mathbf{x} | \Psi_{target}^{(k)} \rangle|^2$, are generated via Markov chain sampling.
Especially if a peaked target state distribution needs to be fitted, it can be expected that only choosing such data points can become problematic to learn an appropriate representation.
For a successful application, the model also needs to represent the target amplitudes well for configurations with vanishing amplitudes, which would not be contained in the sampled set.
Hence, it is sensible to augment the data set, e.g., by sampling an additional set of configurations with uniform probability from the computational basis.

The sampled configurations $\{\mathbf{x}\}$, together with the associated wavefunction amplitudes, $\{\langle \mathbf{x} | \Psi_{target}^{(k)} \rangle\}$ define the training set used for the supervised compression of the state.
To achieve an unbiased error for all amplitudes, the sampling probability for the inclusion of configurations into the training set needs to be taken into account by appropriate modification of the loss function~\cite{Kochkov2018}.
For the Bayesian regression approaches, it is sensible to follow the approaches discussed before to achieve an approximately constant likelihood variance for the actual wavefunction amplitudes by setting the log-space likelihood variances to
\begin{equation}
    \sigma_\mathbf{x}^2 =\log \left( \frac{\tilde{\sigma}^2 \, p(\mathbf{x})} {|\langle \mathbf{x} | \Psi_{target}^{(k)} \rangle|^2} +1 \right).
\end{equation}
Here $p(x)$ is the sampling probability according to which the samples are generated (which can also be non-normalized).
As before, $\tilde{\sigma}^2$ is a noise hyperparameter, characterizing the approximate variance of the amplitude likelihood (following a log-normal distribution).

\subsubsection{Practical application}
Based on the \ac{SWO} protocol specified above, the Bayesian sweeping learning can straightforwardly be used as a tool to iteratively approximate many-body ground states.
At each macro-iteration, a training set is generated from the target state defined according to Eq.~\eqref{eq:target_step_SWO}.
This is then used to learn a \ac{qGPS} with the Bayesian sweeping protocol.
In this section, a toy-model set up of this approach is discussed, only meant as a first indication of the feasibility of the approach.

\begin{figure}[htb!]
    \centering
    \includegraphics{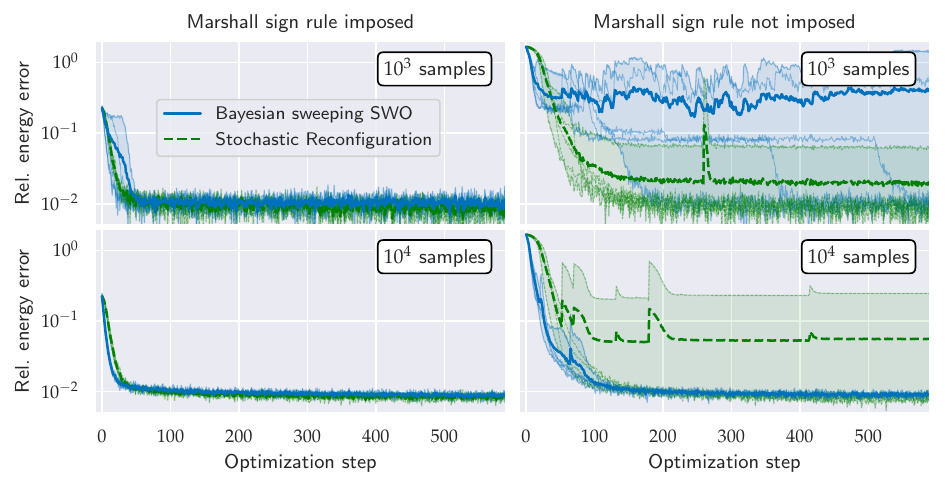}
    \caption[Convergences of the relative energy error for the approximation of the antiferromagnetic Heisenberg model ground state (square lattice with $4 \times 4$ sites) utilizing a Bayesian sweeping for \acl{SWO} and \acl{SR}]{Convergences of the relative energy error for the approximation of the antiferromagnetic Heisenberg model ground state (square lattice with $4 \times 4$ sites) utilizing a Bayesian sweeping for \ac{SWO} (blue) and \ac{SR} (green). The left (right) panels show results with \ac{MSR} (not) incorporated into the Hamiltonian, the top (bottom) panels correspond to the optimization of the ansatz with $10^3$ ($10^4$) configurational training samples. For the \ac{SWO} with Bayesian sweeping, a time step of $\tau=0.1$ was used, for the \ac{SR} it was chosen to be $\tau=0.02$.}
    \label{fig:imag_time_sweeping_example}
\end{figure}

Exemplified evolutions of the state approximation over the different macro-iterations are visualized in Fig.~\ref{fig:imag_time_sweeping_example}.
The plot displays the relative energy error between a \ac{qGPS} trained with the \ac{SWO} approach and the ground state of a simple antiferromagnetic Heisenberg model on a $4 \times 4$ square lattice as a function of the macro-iteration steps.

The key question that is discussed with the simple test setup is whether simple sign structures of the target state, can practically be learned with the iterative \ac{SWO} approach.
Rather than utilizing complex parameters in the \ac{qGPS}, the sign structures are modelled by introducing separate models for magnitude and phase component of the model.
The \ac{qGPS} parametrization is specified as
\begin{equation}
    \Psi(\mathbf{x}) = \exp \left({\sum_{x'=1}^M (\prod_{j=1}^L \epsilon^{(x_j)}_{j, x'} + i \prod_{j=1}^L \tilde{\epsilon}^{(x_j)}_{j, x'})}\right),
\end{equation}
with two different sets of real-valued variational parameters $\epsilon^{(x_j)}_{j, x'}$ and $\tilde{\epsilon}^{(x_j)}_{j, x'}$.
Each of the two-different parameter sets can be trained by separately fitting the real and the imaginary part of the sampled log data amplitudes with the sweeping approach.

In the setup discussed here, a single sweep through the lattice was applied with the Bayesian learning to compress the target state $|\Psi_{target}^{(k)}\rangle$ at each iteration, before generating data from an updated target state for the next iteration.
Because the sign structure only needs to be described accurately for non-vanishing parameters, the phase component of the ansatz was solely trained on $N_{tr}$ samples generated according to $p(\mathbf{x}) = |\langle \mathbf{x} | \Psi_{target}^{(k)} \rangle|^2$.
This data set was associated with a constant log space variance for each data point, $\sigma^2_{sign} = \log \left(\tilde{\sigma}^2_{sign} +1 \right)$.
The magnitude part was trained on a combined data set comprising $N_{tr}/2$ configurations sampled according to $p(\mathbf{x})$, augmented by another $N_{tr}/2$ configurations sampled according to a uniform distribution.
To assign a similar weighting for the fit to both components of the training data set for the magnitude part, the log space variances were chosen to be $\log \left( \frac{\tilde{\sigma}^2_{abs}} {|\langle \mathbf{x} | \Psi_{target}^{(k)} \rangle|^2} +1 \right)$ for the uniformly sampled data points, and $\log \left( \frac{\tilde{\sigma}^2_{abs}} {\langle |\langle \mathbf{x} | \Psi_{target}^{(k)} \rangle|^2 \rangle} +1 \right)$ for the Markov chain sampled configurations.
Here, $\langle |\langle \mathbf{x} | \Psi_{target}^{(k)} \rangle|^2 \rangle$ denotes the mean of $|\langle \mathbf{x} | \Psi_{target}^{(k)} \rangle|^2$ taken across the configurations selected according to this distribution.
The noise hyperparameters, $\tilde{\sigma}^2_{sign}$ and $\tilde{\sigma}^2_{abs}$, were again updated during the sweeping to maximize the log marginal likelihood.
The log data amplitudes were rescaled at each step such that a vanishing mean over the log amplitudes for the component of the data sets sampled according to $|\langle \mathbf{x} | \Psi_{target}^{(k)} \rangle|^2$ was obtained.
This way, the overall order of magnitude of the training amplitudes is approximately kept fixed across the iterations, and it is sensible to use (hyper)parameters from the previous iteration as start values for the next iteration.

With the specific set up of the model and the training data set as specified above, each sweep across the lattice essentially followed the protocol as discussed in section \ref{sec:bayesian_sweeping_j1_j2}.
In one sweep, the reference site was moved in a zigzag pattern across the lattice, and variance hyperparameters were continuously optimized by marginal likelihood maximization for each local fit.
At each local fit, first the local $\alpha_I$ parameters were optimized with the standard update formula (Eq.~\eqref{eq:alpha_update}).
This is followed by a single update to the global noise parameters $\tilde{\sigma}^2_{sign}$ and $\tilde{\sigma}^2_{abs}$.
The magnitude parameter $\tilde{\sigma}^2_{abs}$ was again updated by taking a single gradient ascent step in the log space.
As the likelihood variance is data-independent for the sign part of the model, the standard update formula commonly employed for \acp{RVM} can be applied to update the value of $\tilde{\sigma}^2_{sign}$.
As presented in appendix~\ref{ch:RVM}, this defines the update to the variance parameter $\sigma^2_{sign}$ according to~\cite{Tipping2000,fletcherRelevanceVectorMachines}
\begin{equation}
    \sigma^2_{sign} \rightarrow \frac{|\boldsymbol{\omega}- \boldsymbol{\Phi} \boldsymbol{\mu}_{mp}|^2}{N_{trn} - tr(\mathbb{1}- \mathbf{A}\boldsymbol{\Sigma})}.
\end{equation}

The regularization, automatically employed through the application of the Bayesian principles, effectively biases the inference of the model parameter updates according to the prior.
In the previous example applications of the Bayesian sweeping, the updates to the model parameters were always biased towards zero (as dictated by the Gaussian weight prior centred at zero).
Updating the \ac{qGPS} model parameters, $\epsilon^{(x_i)}_{i, x'}$, with such priors can, however, be problematic since the log amplitudes incorporate a product over all lattice sites, i.e.,
\begin{equation}
    \omega(\mathbf{x}) = \sum_{x'=1}^M \prod_{i=1}^L \epsilon^{(x_i)}_{i, x'}.
\end{equation}
With a product over parameters that all fluctuate around zero, this can lead to vanishingly small or heavily fluctuating products, $\prod_{i=1}^L \epsilon^{(x_i)}_{i, x'}$.
To avoid instabilities emerging due to this, a different biasing of the model parameters was applied for the practical \ac{SWO} realization.
In particular, the weight inference was, for all but one site, biased towards one.
For the other site (for simplicity taken to be the site with index $i=1$) the inference was biased towards zero.
Practically, this can be achieved by re-expressing the functional form of the \ac{qGPS} ansatz in the form
\begin{equation}
    \omega(\mathbf{x}) = \sum_{x'=1}^M (\bar{\epsilon}^{(x_1)}_{1, x'} \prod_{j=2}^L (\bar{\epsilon}^{(x_j)}_{j, x'} + 1)),
\end{equation}
with transformed model parameters $\bar{\epsilon}^{(x_i)}_{i, x'}$, and applying the sweeping to update the transformed parameters using Gaussian priors centred at zero.

The results displayed in Fig.~\ref{fig:imag_time_sweeping_example}, correspond to four different experiment setups in which a \ac{qGPS} with $M=L=16$ was learned with the protocol as outlined above.
The left two plots show the relative energy error that is obtained with explicit incorporation of the \ac{MSR} into the Hamiltonian, so that the amplitudes of the learned target state are known to be non-negative.
For the results presented in the right two plots however, the \ac{MSR} was not imposed so that the \ac{qGPS} is approximating a signed target state.
With the split of the \ac{qGPS} into separate parts for phase and magnitude of the wavefunction amplitudes, both target phase structures approximations can be represented exactly by the model (see appendix \ref{sec:MSR_as_qGPS}).
This means that theoretically the same level of accuracy can be achieved in both setups, and emerging differences between the results can directly be attributed to shortcomings of the optimization protocol.
Whether the sign structure is incorporated into the Hamiltonian or not can equivalently be understood to correspond different initializations of the model.
In the practical setup discussed here, the initialization of the models was obtained by a random initialization of the parameters $\bar{\epsilon}$, drawn from a narrow normal distribution centred at zero.
This effectively initializes the state `similar' to one with a uniform distribution of amplitudes in the chosen basis.
Imposing the \ac{MSR} thus results an initial sign structure more similar to the exact one than in the case where the \ac{MSR} is not imposed.

Figure~\ref{fig:imag_time_sweeping_example} displays the evolution of the relative energy error between the variational energy and the exact ground state energy across the optimization steps of the protocol.
Each sub-plot summarizes the results of five different random realizations with thick lines indicating the mean relative energy error across the different runs.
The individual relative errors for the different runs are displayed by faint lines and the shaded areas indicate the range of the relative energy errors.
The top row of the figure represents the results achieved using a total of $N_{trn} =10^3$ data points at each step, and the bottom row shows results for an increased number of $N_{trn} =10^4$ training samples.

With the utilization of the \ac{MSR} in the Hamiltonian setup, a final accuracy with a relative energy error of approximately $10^{-2}$ was reached for all different runs, likely indicating the maximum accuracy that can be reached with this model.
This value is also in agreement with the values obtained with an optimization of the state with \ac{SR} (for which results are also shown in the figure).

Whereas the achieved energy errors are, especially after multiple optimization steps, mostly consistent across the different protocols, runs, and number of samples with \ac{MSR} imposed, this is not the case for the realizations not utilizing the sign transformation.
While some runs converged to a relative energy error of $10^{-2}$, not all runs achieved this value within the displayed range of $590$ optimization steps for an optimization of the state with $10^{3}$ samples.
For the iterative Bayesian optimization of the state, the expected final value is only approached in three out of the $5$ different runs.
The other two runs failed to converge to the expected level of accuracy.

While significant instabilities are apparent in the setup with $10^3$ samples, with the increased number of $10^4$ training data points, the reliability of the state optimization appears to improve and the (assumed) expressivity limit is reached in all instances.
In agreement with the discussions in Ref.~\cite{Westerhout2019}, this observation indicates that more samples are required for a reliable optimization of states if a sign structure needs to be learned.
The increased difficulty for the learning of the sign information also becomes apparent within the \ac{SR} approach to optimize the state.
As shown in the figure, for both considered numbers of samples, in one of the five random realizations, the \ac{SR} method failed to achieve a final energy error of $\approx 10^{-2}$ within the displayed range of optimization steps.

The discussed experiment setup does not allow for a detailed comparison between the \ac{SR} and the \ac{SWO} approach.
At this stage, no general advantage of the sweeping \ac{SWO} compared to \ac{SR} could be observed, and similar difficulties emerged in the practical application of both techniques.
Different observations, and the results presented here, suggest that problems especially emerge for the learning of intermediate states.
Assuming that the target state can be described accurately by the chosen ansatz, it appears to be required to explicitly steer the optimization along trajectories of states that can be learned well.
In addition to projecting the states to respect appropriate system symmetries, it could also be sensible to initially attribute a greater importance to the learning of the sign information and delay the full optimization of the magnitude part.
Similar approaches have also been considered to stabilize the optimization of \ac{NQS} representations with \ac{SR}~\cite{rothHighaccuracyVariationalMonte2022,Szabo2020}.

All in all, to efficiently learn target states with \ac{VMC} approaches, it is of particular importance to find ways to learn the state with as few samples as possible, be it with \ac{SR} or the Bayesian \ac{SWO} protocol.
While the Bayesian \ac{SWO} protocol does not provide any general advantages over standard \ac{SR} at this stage, it offers a new perspective for the iterative ground state approximation.
The sweeping could also easily be extended to utilize the \ac{RVM} to dynamically adjust the support dimension during the optimization.
It is sensible to expect that different intermediate states require a different degree of flexibility of the model.
Therefore, an automatic adjustment of the model complexity according to the presented data could offer automated ways to utilize available computational resources most efficiently.

Ultimately, the Bayesian sweeping protocol to compress given data samples of a state into a \ac{qGPS} is general and can be applied in various different contexts.
The application for a \ac{SWO} style ground state approximation is only an exemplification of the applicability of the iterative sweeping approach.
It can similarly be extended to other scenarios, including other tasks tackled recently with \ac{NQS} ansatzes, such as the description of the (real) time evolution of quantum states~\cite{Carleo2017, hofmannRoleStochasticNoise2021, linScalingNeuralnetworkQuantum2021, donatellaDynamicsAutoregressiveNeural2022, vargas-calderonEmpiricalStudyQuantum2022}, the extraction of dynamical system properties~\cite{Hendry2019, hendryChebyshevExpansionSpectral2021, https://doi.org/10.48550/arxiv.2303.08184}, or the simulation of quantum circuits~\cite{jonssonNeuralnetworkStatesClassical2018, medvidovicClassicalVariationalSimulation2021}.

\subsection{Classical Machine Learning with Gaussian Process States}
\label{sec:classical_ML_GPS}
The (q)\ac{GPS} was in this work introduced as a specific ansatz to model particular many-body quantum states.
While the approaches and schemes particularly focussed on making phenomena of many-body quantum system numerically accessible, the ansatz itself can also be seen as a more general functional form associating a scalar value to vectorial inputs.
Such representations can therefore equally be used as a representation to model input-output relationships in various other contexts.
Just like the concepts around \acp{TNS} have been applied beyond the description of quantum states~\cite{stoudenmireSupervisedLearningQuantumInspired2017, hanUnsupervisedGenerativeModeling2018, bradleyModelingSequencesQuantum2019, glasserExpressivePowerTensornetwork2019, bhatiaMatrixProductState2019, chengSupervisedLearningProjected2020, dymarskyTensorNetworkLearn2021, convyMutualInformationScaling2021, liuTensorNetworksUnsupervised2021, luTensorNetworksEfficient2021, linTensorNetworkSupervised2021, barrattImprovementsGradientDescent2022, baiUnsupervisedRecognitionInformative2022, senguptaTensorNetworksMachine2022, vieijraGenerativeModelingProjected2022, strashkoGeneralizationOverfittingMatrix2022, sommerEntanglingSolidSolutions2022, zunkovicDeepTensorNetworks2022, helmsDynamicalPhaseBehavior2019, banulsUsingMatrixProduct2019, causerOptimalSamplingDynamical2022, jahromiVariationalTensorNeural2022, Stokes_2019, Liu_2019, https://doi.org/10.48550/arxiv.2212.14076, millerProbabilisticGraphicalModels2021, Otgonbaatar2023, Sun_2020, chen2023machine}, similarly the \ac{GPS} framework can be extended to other scenarios offering a novel set of tools for numerical studies and applications.

As a first indication of the different possibilities, in this section a rather naive application of the Bayesian sweeping learning approach to a common benchmark image recognition task is presented.
The setup is inspired by the one discussed in Ref.~\cite{stoudenmireSupervisedLearningQuantumInspired2017}, exploiting the representational power of \acp{MPS} for the task.
Due to relation between the \ac{GPS} model and \ac{CP} decompositions, supervised learning approaches with the model are related to similar techniques leveraging the compression ability of this decomposition for different \ac{ML} applications, commonly used in conjunction with \acp{NN}~\cite{lebedevSpeedingupConvolutionalNeural2015, caoTensorRegressionNetworks2018, jiFastCPcompressionLayer2022}.
The discussed setup represents an exemplified indication how the different perspectives brought together in the \ac{GPS} model, namely Bayesian regression principles, fundamentals of many-body wavefunction modelling, and tensor decompositions, can provide universal tools for such tasks.

A very standard testing ground for image recognition, which is also discussed here, is the identification of scanned handwritten digits from the MNIST data set~\cite{lecunGradientbasedLearningApplied1998}.
The MNIST data set comprises digital representations of the scans as $28 \times 28$ greyscale pixels, appropriately pre-processed to identify the digits with \ac{ML} techniques.
The digit recognition task thus represents a simple supervised learning task of classification.
A set of training examples is used to train the method in order to associate presented images to one of the ten different digit classes.
Being a very prototypical setup for a practically relevant classification task, learning from and testing methods on the MNIST data set has become a standard benchmarking setup for different methods, including ones inspired by Tensor Network representations~\cite{stoudenmireSupervisedLearningQuantumInspired2017, hanUnsupervisedGenerativeModeling2018, chengSupervisedLearningProjected2020, convyMutualInformationScaling2021, dymarskyTensorNetworkLearn2021, liuTensorNetworksUnsupervised2021, baiUnsupervisedRecognitionInformative2022, strashkoGeneralizationOverfittingMatrix2022, zunkovicDeepTensorNetworks2022, Liu_2019, Sun_2020}.

\begin{figure}[htb!]
    \centering
    \includegraphics{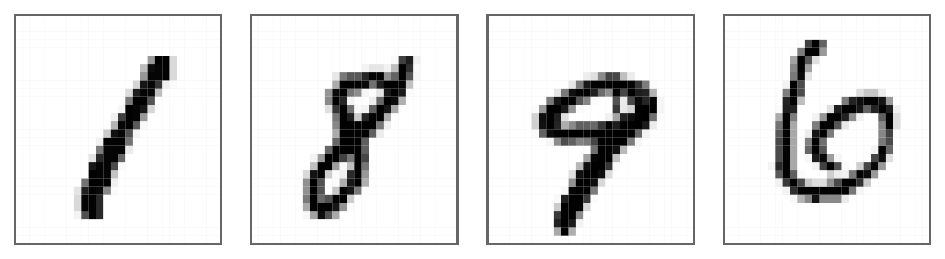}
    \caption[Example images from the MNIST data set]{Example images from the MNIST data set.}
    \label{fig:MNIST_examples}
\end{figure}

For the exemplified application of the \ac{qGPS} to the handwritten digit recognition, the MNIST dataset comprises 60,000 training images, and 10,000 further images to test the method.
Each image of the dataset represents a $28 \times 28$ array of greyscale values.
In analogy to the many-body configurations, these are, in the following, represented as a flattened vector $\mathbf{x}$ for which the element $x_i$ captures the value of the $i$-th pixel.
Different example inputs from the data set are visualized in Fig.~\ref{fig:MNIST_examples}.

To apply the \ac{qGPS} to the classification of MNIST inputs, a `one vs. rest' approach~\cite[section 7.6]{scholkopfLearningKernelsSupport2001} is followed.
In this, a separate model is introduced for each of the digit classes to identify whether an input is part of the class or not.
To this end, ten different \ac{qGPS} models, $\Psi^{(d)}_{qGPS}(\mathbf{x})$, are introduced, one for each of the ten digit classes $d = 1 \ldots 10$.
Here, $\Psi^{(d)}_{qGPS}(\mathbf{x})$ represents a \ac{qGPS}-style mapping from the input to an (unnormalized) probability determining whether the input is considered to be element of that class or not.

In the discussed applications of the \ac{GPS} for quantum systems of discretized degrees of freedom, the elements of the input vectors $\mathbf{x}$ took one out of $D$ values (with $D$ being the dimensionality of the local Hilbert space).
For the considered digit classification however, the vector elements are continuous greyscale values (for the MNIST dataset, represented with a precision of eight bits).
The core element of the \ac{qGPS} is to construct the functional estimator as an (exponentiated) linear combination of $M$ support points described as product states.
To extend this for continuous local degrees of freedom, in principle, different approaches are possible to parametrize the different local amplitudes $f_{d, i, x'}(x_i)$.
Each of these associates an amplitude with a local greyscale value $x_i$ for pixel $i$, support point $x'$, and digit class $d$.
While further investigations are required to assess the influence of different encodings on the final results, here, a simple linear model is assumed for the local state.
This parametrizes the state in the fashion of a visible unit of a \ac{NN} as
\begin{equation}
    f_{i, x'}(x_i) = \epsilon^{(0)}_{d, i, x'} + \epsilon^{(1)}_{d, i, x'} x_i,
\end{equation}
where $\epsilon^{(0)}_{d, i, x'}$ and $\epsilon^{(1)}_{d, i, x'} x_i$ are the variational parameters associated with the \ac{qGPS} for digit class $d$.
Alternative choices for the input encoding could, e.g., be obtained by discretizing the greyscale value, or by encoding the greyscale value as a local spin rotation, a construction used in Ref.~\cite{stoudenmireSupervisedLearningQuantumInspired2017}.

Based on the chosen encoding, the \ac{qGPS} functional model for the classification is defined as
\begin{equation}
    \Psi^{(d)}_{qGPS}(\mathbf{x}) = \exp \left(  {\sum_\mathcal{S} \sum_{x'=1}^M \prod_{i=1}^L (\epsilon^{(0)}_{d, i, x'} + \epsilon^{(1)}_{d, i, x'} \mathcal{S}[x]_i)} \right),
\end{equation}
where $L$ is the total number of pixels.
This functional model includes an additional (generally optional) sum over symmetry operations, $\mathcal{S}$, which can be included to symmetrize the model according to a 'kernel-symmetrization' approach.
In the discussed setup of image classification, the symmetry operations were given by all shifts of the image by up-to two pixels in any direction, therefore giving a total of $25$ considered symmetry operations.
For the translational shifts of the image data, white pixels were added at the opposite side of pixels shifted across the boundary (of which the values are discarded).
This symmetrization w.r.t. to short distance translations of the images is only an exemplification of how symmetries can be incorporated easily.
Based on the model as specified above, other symmetrization approaches are directly applicable.
It could, e.g., also prove beneficial to employ a full symmetrization of the model according to all translations with assumed periodic boundary conditions~\cite{luTensorNetworksEfficient2021}, or to incorporate other symmetry operations such as rotations of the image~\cite{byerlyNoRoutingNeeded2021, anEnsembleSimpleConvolutional2020}.

Because the probabilities $\Psi^{(d)}_{qGPS}(\mathbf{x})$, which are not explicitly normalized, are still exponentiated multilinear models, the Bayesian sweeping discussed in the previous sections is readily applicable to learn the probability models from the available training data.
In the discussed setup, each of the ten different models is trained on the full set of training data, consisting of images and classification labels.
To directly apply the regression of the models, the \ac{qGPS} models are fit on probability amplitudes, either vanishing if the training configuration is not associated with the class, or giving a value of one if they are.
Each of the ten different \ac{qGPS} models can then be fit with the Bayesian sweeping approach on a separate data set consisting of the set of training images, $\{\mathbf{x}\}_{tr}$, with a set of log training amplitudes $\{\omega^{(d)}\}$.
To facilitate the fitting in the log space of the probabilities, the vanishing amplitudes are set to a small value, which was chosen to the (approximate) variance of the amplitude likelihood, $\tilde{\sigma}^2$, for the practical implementation.
The log training amplitude for a training image $\mathbf{x}$ for class label $d$ was thus defined as
\begin{equation}
    \omega^{(d)}(\mathbf{x}) = \begin{cases} 0 \quad &\text{ if the label of }\mathbf{x} \text{ is } d  \\ \log (\tilde{\sigma}^2) \quad &\text{ otherwise} \end{cases}.
\end{equation}

To learn the \ac{qGPS} models, the sweeping protocol was applied as before, however involving the fit of ten different models associated with the different classes instead of a single one.
As motivated in the previous section, it is sensible to stabilize the iterative Bayesian fitting by appropriate biasing of the parameter updates.
Again, this was achieved by appropriately re-expressing the \ac{qGPS} with respected to shifted parameters, here defined as $\bar{\epsilon}^{(k)}_{d, i, x'} = \bar{\epsilon}^{(k)}_{d, i, x'} - \delta_{k,0} (1 - \delta_{i,1})$.
The shifted parameters were optimized with Bayesian priors biasing these shifted parameters towards zero.

As before, prior and likelihood were specified by variance parameters $\alpha$ and $\tilde{\sigma}^2$, updated to maximize the log marginal likelihood (where however the dependence of the training amplitudes on the parameter $\tilde{\sigma}^2$ is not taken into account).
Whereas a single noise parameter $\tilde{\sigma}^2$ was used (initialized with a value $\tilde{\sigma}^2=0.1$), simultaneously specifying the likelihood variance for all classes, each pixel and label class was assigned a separate $\alpha$ value that was optimized when fitting the corresponding model at that pixel.
The value of $\tilde{\sigma}^2$ was again optimized by applying a single gradient ascent step in the log space after the update of the $\alpha$ parameters, using a gradient averaged over all classes with a learning rate of $\eta=10^{-3}$.

Having optimized the models with the Bayesian sweeping protocol, digit labels can easily be obtained for other image inputs.
To classify an image $\mathbf{x}$ (potentially one not included in the training data), the probability amplitudes $\Psi^{(d)}_{qGPS}$ are evaluated for all classes, and a label is predicted according to the class for which the evaluated amplitude is the largest.
For such a prediction approach, the exponentiation applied in the \ac{qGPS} model appears to be irrelevant as it represents a monotonic transformation of the multilinear log model output.
With fixed model parameters, the predicted label for any input is therefore the same, irrespective of whether the exponentiation is included or not.
However, the exponentiation ensures that the amplitudes are appropriate (unsigned), albeit unnormalized, probability amplitudes, which could also easily be normalized across all image classes, although this is not applied here.
In addition to having influences on the training procedure (and thus the practically achieved results), this also provides the foundation for further extensions to obtain more probabilistic information in the label prediction with standard approaches~\cite{Rasmussen2006}.

\begin{figure}[htb!]
    \centering
    \includegraphics{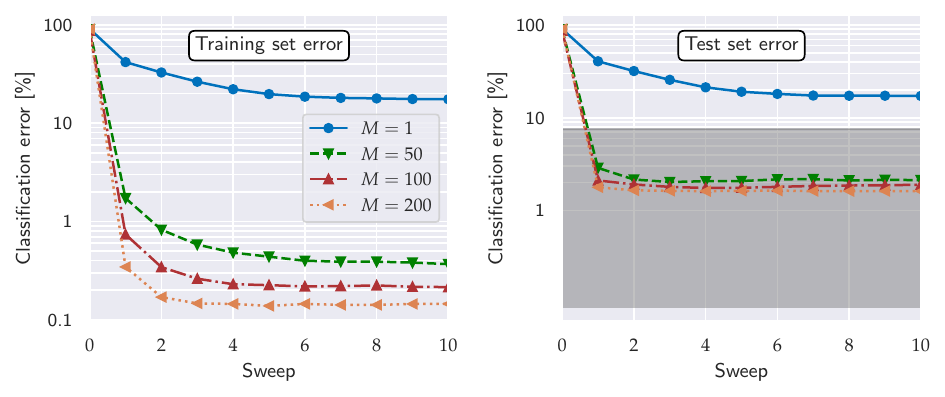}
    \caption[Percentage of incorrectly classified images from the MNIST data set in relation to the number of sweeps applied to train \acl{qGPS} models]{Percentage of incorrectly classified images from the MNIST data set in relation to the number of sweeps applied to train \ac{qGPS} models with support dimensions $M=1$ (blue data points), $M=50$ (green data points), $M=100$ (red data points), and $M=200$ (orange data points). The left plot shows the classification error of the training inputs, the right plot shows the error for the classification of the test images. The shaded area in the right plot denotes the range between the overall lowest test error of $0.09 \%$ misclassified images achieved with the approach from Ref.~\cite{anEnsembleSimpleConvolutional2020}, and the highest test error rate ($7.53 \%$~\cite{shi2022personalized}) from the comparison of state-of-the-art approaches listed on \url{https://paperswithcode.com/sota/image-classification-on-mnist} (last accessed on 4/6/2023).}
    \label{fig:MNIST_results}
\end{figure}

Following the method outlined above, figure~\ref{fig:MNIST_results} shows the percentage of misclassified images from the MNIST data set in relation to the number of sweeps applied in the training using different \ac{qGPS} support dimensions $M=1,50,100,200$.
The left plot reports the classification error obtained for the prediction of the labels from the training set.
It can be seen that for all displayed support dimensions the training set error decreases rapidly and convergence is observed after few sweeps with the applied sweeping protocol.
Furthermore, the approached value shows systematic improvement with respect to increases in the support dimension of the models.
With a support dimension of $M=1$, slightly less than $20 \, \%$ of the images from the training set are not correctly classified after ten sweeps.
This error decreases to a value of $\approx 0.4 \, \%$ for the model with $M=50$, and a value of $\approx 0.1 \, \%$ for the model with $M=200$.

While the quick decrease to small errors on the training set indicates that the training data is fit appropriately, the key quantity of interest is how well the representations generalize to data not presented in the training.
To gauge the ability of the model to distinguish between classes for unseen images, the right plot of the figure shows the achieved error across the $10,000$ test images of the MNIST data set.
The percentage of incorrectly classified test images also overall shows a rapid decrease during the first few sweeps for all considered support dimensions $M$.
Whereas the test set error is approximately equal to the training set error for the simplest model with $M=1$, slight discrepancies of about $2 \%$ between training and test errors can be observed for the more expressive models.
This indicates a small degree of overfitting of the models with $M=50,100,200$ to the training data.
Nonetheless, the overall test accuracies approached still show marginal improvements for the considered increases of the support dimension, and test error rates of $\approx 1.6 - 1.7 \, \%$ are achieved with $M=200$ after three sweeps.
Although some slight increases of the error rate can be observed as the sweeping is continued, again indicating further overfitting of the training data, these increases are relatively small (giving values of $\approx 1.8 \, \%$ after $\approx 50$ sweeps with the $M=200$ model).
Based on the observed results, it appears to be sensible to stop the sweeping as early as possible based on simple heuristics, e.g., when no more sufficient improvements of the training classification error are observed.

Overall, the discussed application of the \ac{qGPS} to the task of digit classification from scanned images provides an exemplary presentation of how the \ac{GPS} specifics can be extended beyond the representation of quantum many-body states.
The MNIST data set can be considered a rather simple data set, and comprehensive benchmarks of the method comparing the performance against other approaches will generally also require the extension of the techniques to more advanced tasks.
Nonetheless, the introduced scheme is relatively general and the presented results for the MNIST data set already provide a first indication of a general applicability of the method.
The classification accuracies achieved for the MNIST dataset do not reach the overall highest accuracies achieved with some \ac{ML} approaches often especially fine-tuned for the task at hand and potentially including further augmentation of the training set~\cite{byerlyNoRoutingNeeded2021}.
For the MNIST dataset, test accuracies as small as $0.09 \%$ have been reported~\cite{anEnsembleSimpleConvolutional2020}.
Nonetheless, the accuracies obtained with the discussed Bayesian sweeping method are within the range of other state-of-the-art results.
For comparison, the performance benchmarks listed on the website \url{https://paperswithcode.com/sota/image-classification-on-mnist} (last accessed on 4/6/2023) include test set classification error rates of state-of-the-art methods introduced between 2013 and 2023 ranging from $0.13 \%$~\cite{byerlyNoRoutingNeeded2021} up to $7.53 \%$~\cite{shi2022personalized}.
The range of test accuracies achieved with different state-of-the-art methods is indicated in the right plot of fig.~\ref{fig:MNIST_results} where the region between the lowest test error rate of $0.09 \%$~\cite{anEnsembleSimpleConvolutional2020} and the test error rate of $7.53 \%$~\cite{shi2022personalized} is shaded.

In particular the rigorous Bayesian regression framework underpinning the discussed concepts is a particularly useful component helping to find a general applicable scheme to automatically introduce appropriate regularization and learn a compact representation for the problem essentially fully deterministically.
Whereas the \ac{GPS} model is based on exponentiated \ac{CP} decompositions, extensions of the Bayesian sweeping framework might also prove useful to aid the model optimization for other tensor network type models~\cite{strashkoGeneralizationOverfittingMatrix2022}.
\acbarrier

\chapter{Conclusions and outlook}
\label{ch:conclusions}

The central question discussed in this work is how quantum many-body wavefunctions can be modelled efficiently utilizing models from Bayesian regression frameworks.
The emerging wavefunction ansatz, the \ac{GPS}, was applied in different contexts to model the ground state of quantum many-body systems for which intricate correlation characteristics emerge, making the accurate numerical simulation of such systems inherently hard.
It was shown that the compressed representation can provide a valuable tool aiding such simulations, and that the \ac{GPS} offers a numerically efficient (approximate) representation of target states in different scenarios.
The success can be understood to be rooted in two fundamental modelling assumptions incorporated into the construction of the state.
Firstly, the state features a strict product-separability of the modelled correlation features, allowing for (non-trivial) compact representations of the state as system sizes go beyond what can be captured with exact numerical treatment.
Secondly, a careful design of the kernel function, describing an assumed co-variance between function points in the \ac{GPR} picture, makes it possible to incorporate explicit physicality into the model, without imposing an a-priori restriction of the overall state expressivity.

The ultimate goal of the discussion in this work is to contribute towards the aim of developing a detailed understanding of how available computational resources can be utilized efficiently to achieve accurate insight about intricate quantum systems.
A key building block of this is to design methods that do not only provide accurate results, but are also as universally applicable as possible.
Exploiting the power of system-agnostic \ac{ML} approaches might therefore provide a viable route forward to improve upon well-established methodology and harness the increasing power of computational hardware.

Also building upon this motivation, the \ac{GPS} discussed in this work is in spirit very similar to applying \ac{NN} architectures to the description of quantum states.
The numerical application of these states also follows similar constructions in order to apply these in numerical contexts for practical calculations.
These functional forms employed as wavefunction ansatz typically do not allow for exact and efficient evaluations of expectation values.
This leads to the requirement to apply stochastic sampling approaches to derive physical quantities from the description in the framework of \ac{VMC}.
Numerical approaches are thus inherently influenced by noise of the estimation procedures, potentially limiting the practical feasibility of the method.
Overall, the results presented in this work show that the application of \ac{GPS} in \ac{VMC} contexts enables treatments of systems to high accuracy in different settings.
However, these also indicate that this is not always true, and a degradation of the quality that can be achieved practically is observed in other cases.
A very fundamental question is whether quantum states of interest can be learned efficiently based on a limited and practically feasible numbers of samples.
Only for the instances where this is the case, the \ac{VMC} framework offers a viable route to tackle the many-body problem.

At this stage, no general answer can be provided if, and how, it is generally possible to approximate states of interest in all regimes of interest reliably with \ac{VMC} techniques.
Especially for the description of signed target states, it was observed that the optimization of parametrizations can become problematic, and specific care needs to be taken to achieve reliable numerical schemes not troubled by instabilities and limited performance.
The observed difficulties are not necessarily specific to the \ac{GPS} ansatz, and similar difficulties have been reported for other highly flexible representations, particularly \ac{NQS}.

In general, practically achieved results are often greatly improved by explicitly incorporating additional physical structure into the method.
Mostly independent of the particular functional model used, this can, e.g., be the explicit incorporation of system symmetries into the ansatz~\cite{nomuraHelpingRestrictedBoltzmann2020, rothGroupConvolutionalNeural2021, rothHighaccuracyVariationalMonte2022}.
Moreover, also other physically-motivated design choices can be made to improve the practicability of the method.
One such approach is to explicitly include an appropriate transformation into the choice of the computational basis, respectively the definition of the Hamiltonian.
This is, e.g., achieved when explicitly imposing the \ac{MSR} for Heisenberg type spin systems, or by working in a basis associated with canonical \ac{HF} orbitals for Fermionic systems.
An alternative route, also frequently pursued in this work, is to enhance the model with mean-field type reference states to ensure that the mean-field accuracy can, theoretically, be reproduced.

Nevertheless, all the approaches to improve the practically achievable results are, from a conceptual point of view, slightly unsatisfactory as these are, to a large degree, explicitly system dependent and not truly universal.
The symmetrization techniques require appropriate symmetries to be present, which is not always the case.
Furthermore, the incorporation of the mean-field structure generally only shifts the problem of appropriate state learnability to a different accuracy level.
The correction to mean-field physics still needs to be learned, which often carries a non-trivial sign structure.
Reliably learning the representations can be identified as a major challenge.
Additional improvements appear to be necessary to further advance the techniques, so that a systematic improvability to high accuracies, even for very challenging systems, are achievable.
As demonstrated in this work, the results practically obtained with \ac{GPS} are mostly in-line with different considered \ac{NN} ansatzes, and similar learnability problems appear to emerge for both types of ansatzes.

A strong focus of this work is to discuss supervised learning approaches to obtain a compact quantum state representation based on available wavefunction data.
This supervised learning of states based on wavefunction data can also be seen as a simplified prototype for general \ac{VMC} calculations, requiring to learn a state based on (vanishingly small) proportions of data from the full Hilbert space.
Discussing and evaluating different approaches to such compression tasks is therefore a key contribution to understanding and circumventing the observed optimization difficulties emerging in \ac{VMC} calculations~\cite{Westerhout2019}.
It is shown that Bayesian regression approaches provide methods to extract particularly compact \ac{GPS} forms for which the correlation features driving the model definition are automatically selected.
In addition to appropriately guiding the model selection process, the Bayesian regression frameworks were also found to be able to automatically introduce appropriate regularization for the data-driven compression of quantum states.
Although so far no clear advantages could be observed for the learning of generally unknown target states with standard \ac{VMC} techniques, these perspectives offer new ways to interpret the practical optimization problems in a probabilistically data-driven way.
Such supervised learning perspectives, e.g., allow for an automatic adaption of the model based on the amount (and quality) of data presented for the training, target accuracy specifications, or specific properties deemed particularly important in the state.

The task of compressing a state into a compact representation reliably is of great importance well beyond the discussed tasks of deriving the ground state properties of systems.
Various other applications of such schemes, leveraging the described efficiency of the \ac{GPS} to represent many-body states efficiently, can be imagined (as have also been studied with other parametrizations).
The \ac{GPS} learning frameworks discussed are generally applicable, and can easily be extended to different tasks relevant for the understanding of quantum simulations, in particular including recently discussed applications of \ac{NQS}.

Data-driven wavefunction representations could, amongst others, be utilized for the reconstruction of quantum states from experimental measurements~\cite{Torlai2018a, carrasquillaReconstructingQuantumStates2019,beachQuCumberWavefunctionReconstruction2019, czischekDataEnhancedVariationalMonte2022, iouchtchenkoNeuralNetworkEnhanced2022, koutnyNeuralnetworkQuantumState2022, gomezReconstructingQuantumStates2022, zhuFlexibleLearningQuantum2022,golubevaPruningRestrictedBoltzmann2021, schmaleScalableQuantumState2021, carrasquillaNeuralNetworksQuantum2021, Neugebauer2020, Torlai2019, zhongQuantumStateTomography2022, wangPredictingPropertiesQuantum2022, lohaniDemonstrationMachinelearningenhancedBayesian2022, Bennewitz_2022, wei2023neuralshadow, ma2023attentionbased}, or even the classical simulation of quantum circuits~\cite{jonssonNeuralnetworkStatesClassical2018, medvidovicClassicalVariationalSimulation2021, knitterNeuralNetworkSimulation2022, Pehle2018}, e.g., as a verification mechanism for quantum hardware.
All such approaches are based on the requirement to learn a representation of the quantum state based on limited samples, either observed experimentally or artificially generated, therefore explicitly requiring to avoid the overfitting of the data.

In addition to applying the \ac{GPS} ansatz in its presented form to different quantum simulation tasks, different extensions of the model might help to improve practically achievable results.
One can, e.g., extend the \ac{GPS} to fulfil the properties of autoregressive \ac{NN} constructions~\cite{Sharir_2020, HibatAllah2020, barrettAutoregressiveNeuralnetworkWavefunctions2021, donatellaDynamicsAutoregressiveNeural2022, humeniukAutoregressiveNeuralSlaterJastrow2022, zhaoScalableNeuralQuantum2022,sharirNeuralVariationalMonte, hibatallah2023investigating, PhysRevResearch.5.013216}.
This general construction, exemplified with the GPS in Ref.~\cite{bortone2023impact}, explicitly normalizes the state across the Hilbert space and allows for an efficient generation of configurational samples from the associated probability distribution without having to resort to Markov chain algorithms.
It is therefore particularly appealing for tasks in which the explicit normalization of the state can help to avoid overfitting, or where the generation of uncorrelated samples according to the distribution is challenging.
A very central question is, how such a construction practically affects the ability of the model to represent target states efficiently.
While the improved ability to sample the state is generally expected to be an advantageous feature, it might come at the cost of reduced (practically observable) expressibility required for accurate representations.

Especially for the descriptions of Fermionic systems, further extensions, also explored with \ac{NQS}, are possible for the \ac{GPS} model.
These could include the extension of the \ac{GPS} framework to real space descriptions~\cite{Pfau2019,Hermann2019,spencerBetterFasterFermionic2020,gerardGoldstandardSolutionsSchr2022,hermannAbinitioQuantumChemistry2022, liInitioCalculationReal2022, vonglehnSelfAttentionAnsatzAbinitio2022, pesciaNeuralNetworkQuantumStates2021}, or using it as extended, explicitly anti-symmetrized Fermionic ansatzes in backflow~\cite{Rios, Luo2019, lamiMatrixProductStates2022}, or hidden-fermion model type parametrizations~\cite{morenoFermionicWaveFunctions2021}.

While such extensions inspired by recent developments of the \ac{NQS} methodology provide novel tools to improve upon the quality of the descriptions, one might ask if there is any property that really sets the \ac{GPS} apart from \ac{NN} function estimators.
Although no numerical evidence has been found so far that the \ac{GPS} carries an intrinsic practical advantage over state representations based on \acp{NN}, a particularly appealing property of the model is its simplicity.
As discussed in this work, the \ac{GPS} ansatz can be identified as an exponentiated linear combination of product state amplitudes, equivalent to a \ac{CP} decomposition, also equivalent to \ac{MPS} amplitudes with diagonal matrices.
The span of the model is essentially controlled by a single hyperparameter, the support dimension $M$, similar to the bond dimension of \acp{MPS}, or the number of hidden nodes in \acp{RBM}.
Especially through the introduction of unentangled quantum states as support points of the model, the ansatz class is defined without requiring further specification of network architectures, or other specifics.

It might in fact be considered striking that such a simple parametrization was, in the exemplified applications, able to reach accuracies competitive with other state-of-the-art parametrizations, despite the fact that only very moderate support dimensions were considered.
With current \ac{DMRG} calculations going up to \ac{MPS} bond dimensions of $M=2^{16}$~\cite{ganahlDensityMatrixRenormalization2022}, it would be interesting to investigate the performance of the model for significantly increased support dimensions.
However, with the requirement to evaluate expectation values via stochastic sampling and large prefactors incurred in \ac{VMC} optimizations, going to such support dimensions with the (q)\ac{GPS} seems to be out of reach at this stage.
From a practical perspective it is indeed of great interest whether (respectively for what systems) it is beneficial to employ methods such as \ac{DMRG} over \ac{VMC} approaches in practical like-for-like comparisons.
Whereas the former benefits from easy applicability and fully deterministic contractions of expectation values, \ac{VMC} calculations are generally plagued by stochastic noise.
However, these allow for a utilization of models alleviating the foundational shortcomings of states such as \ac{MPS}, and also offer a great parallelizability.

It appears to be intrinsically challenging to transfer the fully deterministic contraction techniques for the evaluation of expectation values from tensor network approaches to representations such as the \ac{GPS} ansatz.
Nonetheless, the Bayesian sweeping approach builds upon the specific form of the \ac{CP} decomposition for the log wavefunction amplitudes and ultimately brings together different quantum state modelling perspectives discussed in this work.
The applied quantum state inference relies on learning input-output relationship from stochastically generated data points, formulated as a standard \ac{ML} task of supervised learning.
The explicitly data-driven supervised compression of the state itself is directly based on the tensorial form of the model, and constitutes an \ac{ALS} approach with iterative sweeps through the space, in spirit similar to \ac{DMRG} type methods.
Appropriate regularization of the description is achieved by considering the model in a rigorous Bayesian regression framework, foundationally providing the inspiration for the \ac{GPS} construction in the first place.

All in all, the main contribution of this work is the exploration of these three different perspectives, namely data-driven \ac{ML}, Bayesian modelling frameworks, and tensor network decompositions, to design a physically motivated many-body ansatz for simulations of quantum systems.
To what extent the discussed \ac{GP} representation of quantum many-body states really provides a useful tool for practical numerical calculations mostly remains an open question at this stage.
The methods have, so far, not offered any contributions to advance the physical understanding about challenging quantum systems, ultimately representing the main objective.
Nonetheless, exploring novel routes is an important element in order to discover appropriate numerical techniques with the ability to settle long-standing quantum physical questions.

\acbarrier

\begin{appendices}
    \chapter{The Relevance Vector Machine}
\label{ch:RVM}

The \ac{RVM} represents a rigorous application of the Bayesian learning techniques outlined in chapter \ref{ch:GPS_introduction}, first introduced in Ref.~\cite{Tipping2000}.
By introduction of an appropriate prior for the weights of a linear model, and application of the marginal likelihood maximization, the \ac{RVM} enables the identification of the most relevant features.
In the considered settings, the features are appropriately defined kernel functions evaluated w.r.t. specific support configurations $\{\mathbf{x}'\}$.
The \ac{RVM} thus extracts a set of these support configurations from a pool of potential candidates (e.g., all the data configurations the model is trained on).
This section gives a brief outline of the \ac{RVM} as applied in this work.
More technical details can, e.g., be found in Ref.~\cite{fletcherRelevanceVectorMachines}.

Following the modelling assumptions outlined in the main text, the linear combination of kernel functions evaluated w.r.t. the support configurations $\{\mathbf{x}'\}$, defines a mapping $f(\mathbf{x}) = \sum_{\{\mathbf{x}'\}} w_{\mathbf{x}'} k(\mathbf{x}, \mathbf{x}')$.
This linear combination specifies the mean of a Gaussian likelihood.
The likelihood can be evaluated for the data points the model is trained on, specified by a vector of $N_{tr}$ amplitudes $\mathbf{y}$ (in this work typically the log wavefunction amplitudes $\boldsymbol{\omega}$) associated with corresponding data configurations, here specified as a matrix $\mathbf{X}$.
No correlations are assumed in the likelihood between different data points, resulting in a diagonal covariance matrix of the likelihood, $\mathbf{B}^{-1}$.
Reiterating the result presented in Eq.~\eqref{eq:likelihood}, the likelihood is thus specified by a normal distribution according to
\begin{equation}
    p(\mathbf{y}|\mathbf{X}, \mathbf{w}) = \prod_{i=1}^{N_{tr}} \frac{1}{\sqrt{2 \pi \sigma^2(\mathbf{x}_i)}} \exp \left(  {-\frac{1}{2 \sigma^2(\mathbf{x}_i)} |y_i - (\sum_{\{\mathbf{x}'\}} w_{\mathbf{x}'} k(\mathbf{x}, \mathbf{x}')) |^2}\right).
\end{equation}
Here, $\mathbf{x}_i$ is the $i$-th data configuration with associated amplitude $y_i$ (here assumed to be real), and likelihood variance $\sigma^2(\mathbf{x}_i) = 1/B_{i,i}$.
The vector $\mathbf{w}$ denotes the vector of weights, also assumed to be real, for which the posterior distribution can be inferred by application of Bayes' theorem.

As in all the examples discussed in this work, for the inference of the posterior via Bayes' theorem, a Gaussian prior with diagonal covariance matrix is introduced for the weights.
Crucial element for the extraction of the support configurations in the \ac{RVM} is to allow the prior variances to be different for different features of the model.
That is to say, the inverse prior covariance matrix $\mathbf{A}$ is a diagonal matrix with diagonal elements $A_{i,i} = \alpha_i$, where the parameter $\alpha_i$ specifies the inverse variance for weight $w_i$.
The prior probability distribution therefore takes the form (see Eq.~\eqref{eq:prior})
\begin{equation}
    p(\mathbf{w}) = \prod_{i=1}^{N_{features}} \sqrt{\frac{\alpha_i}{2 \pi}} \exp \left(  -\frac{\alpha_i}{2} |w_i|^2 \right),
\end{equation}
where $N_{features}$ is the total number of considered support configurations.

Bayes' theorem gives the posterior as
\begin{equation}
    p(\mathbf{x}|\mathbf{X}, \mathbf{y}) = \frac{p(\mathbf{y}|\mathbf{X}, \mathbf{w}) \times p(\mathbf{w})}{p(\mathbf{y}| \mathbf{X})},
\end{equation}
with posterior distribution $p(\mathbf{x}|\mathbf{X}, \mathbf{y})$, and marginal likelihood $p(\mathbf{y}| \mathbf{X})$.
As outlined in the main text, the posterior, as well as the marginal likelihood, can then be evaluated in closed form and are also found to be Gaussian.
Equation \ref{eq:mean_weights} specifies the mean of the Gaussian posterior distribution over the space of weights as
\begin{equation}
    \boldsymbol{\mu} = \boldsymbol{\Sigma}\boldsymbol{\Phi}^T \mathbf{B} \mathbf{y}.
\end{equation}
Here, $\boldsymbol{\Phi}$ denotes the $N_{tr} \times N_{features}$ matrix of kernel function values evaluated between the considered support configurations and the configurations of the data set, i.e., its elements are given as $\Phi_{i,j}= k(\mathbf{x}_i, \mathbf{x}'_j)$.
The covariance matrix of the posterior, $\boldsymbol{\Sigma}$, is, as specified in Eq.~\eqref{eq:covariance_weights}, given by the expression
\begin{equation}
    \boldsymbol{\Sigma} = (\boldsymbol{\Phi}^T \mathbf{B} \boldsymbol{\Phi} + \mathbf{A})^{-1}.
\end{equation}
The logarithm of the marginal likelihood, is specified in Eq.~\eqref{eq:log_ml}, giving the expression~\cite{fletcherRelevanceVectorMachines, Rath2020}
\begin{equation}
    \log \left(p(\mathbf{y}| \mathbf{X})  \right) = \frac{1}{2} \left(  \log(\det(\mathbf{A})) - \log(\det(2 \pi \mathbf{B}^{-1})) + \log(\det(\boldsymbol{\Sigma})) - \boldsymbol{y}^T \mathbf{B} \boldsymbol{y} + \boldsymbol{\mu}^T \boldsymbol{\Sigma}^{-1} \boldsymbol{\mu} \right).
\end{equation}

As outlined in section \ref{sec:marg_likelihood_based_model_selection}, it is common to optimize hyperparameters, here in particular the variance parameters, $\alpha_i$, by maximization of the (log) marginal likelihood.
The key property that is exploited in the application of the \ac{RVM} is that the marginal likelihood is often found to be maximal with many of the $\alpha_i$ parameters tending to infinity.
For an infinite $\alpha$ parameter, the prior distribution of the associated weight therefore becomes a delta function at zero, so that the inferred most probable posterior weight goes to zero.
This also means, that the corresponding support configuration does not contribute to the output of the inferred model, and it can be pruned from the set of support configurations.
By maximizing the log marginal likelihood, the relevant support configurations are therefore identified (i.e., the ones with finite $\alpha$ values), and a sparse model can be extracted with the probabilistic techniques.

Different approaches are possible to find the maximum of the marginal likelihood w.r.t. the sparsity parameters $\alpha$ with iterative approaches.
In the original formulation of the \ac{RVM}~\cite{Tipping2000}, the values are initialized to some finite value, giving an initial model incorporating all potential support configurations.
The $\alpha$ are then iteratively updated, so that support configurations can be pruned from the initial set of candidates if corresponding sparsity parameters are updated to infinity (or to large values greater than some sensible threshold).
With a more detailed derivation shown in Ref.~\cite{fletcherRelevanceVectorMachines}, update equations for the sparsity parameters can, e.g., be obtained by equating the gradient of the log marginal likelihood w.r.t. these to zero.
This leads to the update equation at each iterative parameter update according to~\cite{Tipping2000,fletcherRelevanceVectorMachines}
\begin{equation}
    \alpha_i \rightarrow \frac{1- \alpha_i \Sigma_{i,i}}{|\mu_i|^2}.
\end{equation}
Each sparsity parameter update therefore requires the re-evaluation of the posterior mean $\boldsymbol{\mu}$ and covariance matrix $\boldsymbol{\Sigma}$ with the updated parameters.
By iterating these updates until convergence is observed, the final set of relevant support configurations (which is often significantly sparser than the initial set of considered candidates) is obtained.

Similar to the optimization of the sparsity parameters, it can also be sensible to optimize the noise hyperparameter(s) characterizing the likelihood variances by maximization of the log marginal likelihood.
In the standard formulation of the RVM, the likelihood variances are assumed to be data point independent, i.e., the matrix $\mathbf{B}$ is proportional to the identity, $\mathbb{B} = \beta \mathbb{1}$.
Again, equating the derivative of the log marginal likelihood w.r.t. the beta parameter to zero~\cite{Tipping2000,fletcherRelevanceVectorMachines} gives an iterative update equation.
This can be employed to update the value of $\beta$ alongside the sparsity parameters according to
\begin{equation}
    \beta \rightarrow \frac{N_{tr} - \text{tr}(\bm{1} - \mathbf{A} \boldsymbol{\Sigma})}{|\mathbf{y} - \boldsymbol{\Phi} \boldsymbol{\mu}|^2}.
\end{equation}

Going beyond the standard assumption of a data-independent noise specification, various applications of the Bayesian regression approaches in this work utilize a likelihood variance that depends on the magnitude of the fitted wavefunction amplitudes.
In particular, as outlined in section \ref{sec:bayesian_state_learning}, to compensate the fitting in the log space of the wavefunction amplitudes, the diagonal elements of the matrix $\mathbf{B}$ are often specified as (see Eq.~\eqref{eq:log_space_likelihood_variance_approx})
\begin{equation}
    B_{i,i} = \frac{1}{\sigma^2(\mathbf{x}_i)} = \frac{1}{\log \left( \frac{\tilde{\sigma}^2} {|e^{\omega(\mathbf{x}_i)}|^2} + 1 \right)}.
\end{equation}
This utilizes a parameter $\tilde{\sigma}^2$, denoting an approximate variance of the wavefunction amplitude likelihood, which are fit with the Bayesian approaches in the space of log amplitudes $\omega(\mathbf{x}_i)$ (in which the likelihood variances are $\sigma^2(\mathbf{x}_i)$).
If the hyperparameter $\tilde{\sigma}^2$ should also be adapted to maximize the log marginal likelihood, one approach considered is to repeat the full \ac{RVM} training with different $\tilde{\sigma}^2$ values to find a suitable value maximizing the log marginal likelihood.
Alternatively, it is also possible to employ gradient ascent type updates of the parameter $\tilde{\sigma}^2$ (usually employed in the log space) alongside the $\alpha$ parameters.
While this was only considered in the iterative Bayesian sweeping context discussed in chapter \ref{ch:qGPS_learning} (in which the same prior variance was assumed for all weights at a particular reference site), this is also applicable in combination with the sparse prior specification of the \ac{RVM}.
The derivative of the log marginal likelihood w.r.t to the variance parameter evaluates to~\cite{boothQuantumGaussianProcess2021}
\begin{equation}
    \frac{d (\log \left(p(\mathbf{y}| \mathbf{X})  \right))}{d \tilde{\sigma}^2} = \frac{1}{2} \left(
    \text{tr} \left( \mathbf{B}'\mathbf{B}^{-1} - \boldsymbol{\Phi}^T \mathbf{B}' \boldsymbol{\Phi} \boldsymbol{\Sigma} \right) - \mathbf{y}^T \mathbf{B}' \mathbf{y} - \boldsymbol{\mu}^T \boldsymbol{\Phi}^T \mathbf{B}' \boldsymbol{\Phi} \boldsymbol{\mu} + 2 \mathbf{y}^T \mathbf{B}' \boldsymbol{\Phi} \boldsymbol{\mu}  \right).
\end{equation}
The diagonal matrix $\mathbf{B}'$, giving the derivative of the matrix $\mathbf{B}$ w.r.t. the parameter $\tilde{\sigma}^2$, can be expressed via its diagonal elements according to~\cite{boothQuantumGaussianProcess2021}
\begin{equation}
    B'_{i,i} = \frac{d (1/\sigma^2(\mathbf{x}_i))}{d \tilde{\sigma}^2} = - \frac{(B_{i,i})^2}{|e^{\omega(\mathbf{x}_i)}|^2 + \tilde{\sigma}^2}.
\end{equation}
In a Bayesian regression setup with complex-valued model outputs as outlined in the main text, the noise derivative can be obtained by replacing transposed with hermitian conjugated quantities, and ignoring the prefactor of $\frac{1}{2}$ in the expression above.

\section{Fast marginal likelihood maximization}
The iterative maximization of the marginal likelihood in the \ac{RVM} requires the inversion of $N_{active} \times N_{active}$ matrices at each step, where $N_{active}$ is the number of `active' support configurations with finite $\alpha$ value.
This results in a computational cost of roughly $\mathcal{O}(N_{active}^3)$ at each update step.
In the original description of the \ac{RVM} from Ref.~\cite{Tipping2000}, as it is outlined above, the iterative scheme starts with all potential support configurations considered as active before configurations are removed from the active set to obtain a sparse model.
Assuming that the finally extracted model only contains a fraction of the potential candidates as relevant support configurations, especially the initial steps from the selection of the relevant configurations are particularly expensive.
Furthermore, the (approximate) cubic scaling, also restricts the approach to relatively small sets of initial support configurations from which the relevant set is extracted.

As an alternative to the pruning of non-relevant configurations, the applications of the \ac{RVM} in this work are based on the fast scheme introduced in Ref.~\cite{Tipping2003}.
Rather than removing non-relevant configurations from the active set, this scheme is based on the idea of starting from a small initial set of support configurations, potentially only containing a single point,
Further configurations can then be added from the set of candidates, which are identified by a marginal likelihood maximization.
This means that initially most of the $\alpha$ parameters are set to infinity and then at most a single configuration is added to or removed from the active set at each iteration.
If the number of relevant support configurations is small compared to the total number of candidates (and it can be kept small throughout), this can therefore result in a significant speed up as compared to the scheme outlined above.

In the fast iterative scheme, only one of the sparsity parameters is optimized at a time, and the update is determined by extracting the associated contribution to the full log marginal likelihood.
To obtain this update, the log marginal likelihood is, for each potential support configuration $\mathbf{x}_i$, separated into a component depending on the corresponding sparsity parameter $\alpha_i$ and one depending on all other $\alpha$ values.
This means that, as presented in more detail in Refs.~\cite{NIPS2001_02b1be0d,Tipping2003}, the log marginal likelihood is expressed according to
\begin{equation}
    \log \left(p(\mathbf{y}| \mathbf{X})  \right) = \log [p_{ML}](\alpha_i) + \log [p_{ML}](\alpha_{j \neq i}).
\end{equation}
The function $\log [p_{ML}](\alpha_i)$ denotes the extracted contribution of the parameter $\alpha_i$ to the log marginal likelihood, and the $\log [p_{ML}](\alpha_{j \neq i})$ captures the remaining part not depending on this parameter.
The important contribution depending on the updated parameter, i.e., $\log [p_{ML}](\alpha_i)$, evaluates to~\cite{NIPS2001_02b1be0d, Tipping2003}
\begin{equation}
    \log [p_{ML}](\alpha_i) = \frac{1}{2} \left( \log (\alpha_i) - \log(\alpha_i + s_i) + \frac{q_i^2}{\alpha_i + s_i} \right),
\end{equation}
which utilizes additional quantities $s_i$ and $q_i$.
These can practically be expressed as~\cite{Tipping2003}
\begin{align}
    s_i &= \frac{\alpha_i \mathbf{S}_i^T \boldsymbol{\Phi}_i}{\alpha_i - \mathbf{S}_i^T \boldsymbol{\Phi}_i}, \\
    q_i &= \frac{\alpha_i \mathbf{S}_i^T \mathbf{y}}{\alpha_i - \mathbf{S}_i^T \boldsymbol{\Phi}_i},
\end{align}
with
\begin{equation}
    \mathbf{S}_i^T = \boldsymbol{\Phi}_i^T \mathbf{B} - \boldsymbol{\Phi}_i^T \mathbf{B} \boldsymbol{\Phi} \boldsymbol{\Sigma} \boldsymbol{\Phi}^T \mathbf{B},
\end{equation}
and where in both equations $\boldsymbol{\Phi}_i$ denotes the $i$-th column of the matrix $\boldsymbol{\Phi}$ interpreted as a column vector.

At each step of the fast marginal likelihood maximization, a single $\alpha_i$ associated with one identified (potential) support configuration is optimized by maximizing the corresponding contribution $\log [p_{ML}](\alpha_i)$.
Crucially, a sensible update can be computed not requiring the considered configuration to be in the active set.
As presented in Ref.~\cite{Tipping2003}, three different cases can be distinguished for the parameter update.
If the value of $q_i^2$ is greater than $s_i$, then the maximum of value of $\log [p_{ML}](\alpha_i)$ is obtained by updating the value of $\alpha_i$ according to
\begin{equation}
    \alpha_i = \frac{s_i^2}{q_i^2 - s_i}.
\end{equation}
This either updates the sparsity parameter for an active support configuration, or it adds a new configuration to the active set if it was not active before.
If, however, the quantity $s_i$ is greater than or equal to $q_i^2$, then the maximum of the log marginal likelihood is obtained by pruning the corresponding support configuration from the active set.

The updates of a single $\alpha$ parameter at a time can be iterated until convergence is observed.
To single out a configuration for which the parameter is updated, it is, e.g., possible to identify the configuration for which the update results in the largest increase in the log marginal likelihood~\cite{Tipping2003}.

\acbarrier
\chapter{Representing specific sign structures as Gaussian Process State}

\section{Representing the Marshall Sign Rule}
\label{sec:MSR_as_qGPS}
As discussed in section \ref{sec:J1J2_model}, the \ac{MSR}~\cite{marshallAntiferromagnetism1955} represents the exact sign structure of anti-ferromagnetic Heisenberg models.
In the considered basis constructed from tensor products of the $\hat{S}_z$ eigenstates, it can be defined by the wavefunction amplitudes according to Eq.~\eqref{eq:MSR}, specifying the signs as
\begin{equation}
    \Psi(\mathbf{x}) = (-1)^{\sum_{i\in A} \delta_{x_i, 1}}.
\end{equation}
The indices of a sublattice $A$ comprise one of two disjoint chequerboard sublattices of the system.
This means that the \ac{MSR} counts the number of spin-up occupancies on the specified sublattice and gives a positive sign for an even number, and a negative sign otherwise.
The \ac{MSR} can be represented efficiently as a \ac{GPS} with different approaches.

By utilizing the formulation of the \ac{GPS} with unentangled product states as support configurations for the model, the \ac{MSR} can be represented exactly up to a global phase, requiring a support dimension of $M=1$.
The resulting (non-symmetrized) \ac{qGPS} defines the wavefunction amplitudes as
\begin{equation}
    \Psi(\mathbf{x}) = \exp \left( \prod_{j=1}^L \epsilon^{(x_j)}_{j} \right),
\end{equation}
with specifically chosen parameters $\epsilon^{(x_i)}_{i}$.
It can directly be verified that a choice of the parameters as
\begin{equation}
    \epsilon^{(x_j)}_{j} = \begin{cases} -(i \pi/2)^{1/L} \quad &\text{if } j \in A \text{ and } x_j = 1 \\ (i \pi/2)^{1/L} \quad &\text{otherwise}\end{cases},
\end{equation}
gives the sign structure as for the \ac{MSR} with a global phase prefactor $e^{-i * \pi/2}$.

If the \ac{MSR} for a system is entirely symmetric with respect to symmetry operations (i.e., two symmetrically equivalent configurations always give the same sign value), then a similar construction as above can be defined for a kernel-symmetrized \ac{qGPS} model.
The resulting \ac{qGPS} with a support dimension of $M=1$ defines the wavefunction amplitudes according to
\begin{equation}
    \Psi(\mathbf{x}) = \exp \left( \sum_{\{\mathcal{S}\}} \prod_{j=1}^L \epsilon^{(\mathcal{S}[x]_j)}_{j} \right),
\end{equation}
where the sum includes the included symmetry operations.
The \ac{MSR} amplitudes are obtained for this kernel-symmetrized \ac{qGPS}, by setting the parameters to
\begin{equation}
    \epsilon^{(x_j)}_{j} = \begin{cases} - \frac{(i \pi/2)^{1/L}}{|\{\mathcal{S}\}|} \quad &\text{if } j \in A \text{ and } x_j = 1 \\ \frac{(i \pi/2)^{1/L}}{|\{\mathcal{S}\}|}  \quad &\text{otherwise}\end{cases},
\end{equation}
where $|\{\mathcal{S}\}|$ denotes the total number of symmetry operations considered.

The two \ac{qGPS} representations of the \ac{MSR} above explicitly rely on the ability to construct the log amplitude by a product over $L$ factors for which the signs can be different.
These constructions are therefore not local, in the sense that the product is taken across all $L$ sites and cannot be truncated at short ranges (nor is a weighting of shorter ranged correlation sensible).
It is also possible to represent the \ac{MSR} with an ansatz resembling a classical \ac{GPS} extracting local correlations and computational basis states as support configurations (therefore giving positive kernel values).
This is achieved by expressing the \ac{MSR} as
\begin{equation}
    \Psi(\mathbf{x}) = \exp \left( \sum_{j=1}^L \epsilon^{(x_j)}_{j} \right),
\end{equation}
with parameters
\begin{equation}
    \epsilon^{(x_j)}_{j} = \begin{cases} i \pi \quad &\text{if } j \in A \text{ and } x_j = 1 \\ 0  \quad &\text{otherwise}\end{cases}.
\end{equation}
This reformulation thus represents the \ac{MSR} as a \ac{GPS} with support dimension $M=L$.

\section{Representing Fermionic orbital reordering}
\label{sec:ferm_ordering_GPS}
Due to the anti-symmetry, the representation of Fermionic systems in a second-quantized basis of discrete modes results in a sign structure of the wavefunction amplitudes that depends on the ordering chosen for these modes.
Assuming spinless Fermions, this section briefly outlines how an additional sign structure in the wavefunction amplitudes, emerging from a change of this ordering, can be represented as a \ac{GPS}.
This representation can directly be extended to the case of spinful Fermions.

As outlined in section \ref{sec:ab_initio_implementation}, the evaluation of matrix elements for Fermionic operators include specific sign prefactors.
These depend on the number of occupied modes that are passed in the application of the creation and annihilation operators at a specific site with respect to the chosen ordering~\cite{altlandCondensedMatterField2010a}.
Two different choices to order the system modes can (and typically will) therefore also result in different sign structures of the wavefunction amplitudes for an electronic state $| \Psi \rangle$.
More concretely, encoding the same electronic occupancies with two different orderings of the modes, denoted by configurations $|\mathbf{x}\rangle$ and $|\bar{\mathbf{x}}\rangle$, the associated wavefunction amplitudes can be related as
\begin{equation}
    \langle \mathbf{x} | \Psi \rangle = (-1)^{\mathcal{N}(\mathbf{x}, \bar{\mathbf{x}})} \langle \bar{\mathbf{x}} | \Psi \rangle.
\end{equation}
The sign prefactor relating the two ordering choices, $(-1)^{\mathcal{N}(\mathbf{x}, \bar{\mathbf{x}})}$, corresponds to the parity of the permutation of occupied modes to bring the occupied modes from one order into the other.
This can be obtained by evaluating the number of exchanges of occupied modes, $\mathcal{N}(\mathbf{x}, \bar{\mathbf{x}})$, to bring one representation into the order of the other.

To evaluate the number of exchanges, $|\mathbf{x}\rangle$ can be assumed to correspond to a default ordering of the modes in ascending order, i.e., $1, \ldots, L$.
The ordering for $|\bar{\mathbf{x}}\rangle$ is assumed to be a permutation $\mathcal{P}$ of this ordering, i.e., modes are ordered according to $\mathcal{P}(1), \ldots, \mathcal{P}(L)$.
The number of exchanges can then directly be evaluated as
\begin{equation}
    \mathcal{N}(\mathbf{x}, \bar{\mathbf{x}}) = \sum_{i=1}^{L-1} \sum_{j=i+1}^L \delta_{x_{\mathcal{P}(i)}, 1} \delta_{x_{\mathcal{P}(j)}, 1} \times \begin{cases} 1 & \quad \text{if } \mathcal{P}(i) > \mathcal{P}(j) \\ 0 & \quad \text{otherwise} \end{cases},
\end{equation}
and therefore contains at most $\mathcal{O}(N^2)$ terms (with $N$ being the number of electrons).
Based on the standard approach to encode Fermionic occupancies of the modes, this representation utilizes local occupancies, $x_k$, which are either zero or one, depending on whether the mode is occupied or not.

Following the representation of the relative sign between wavefunction amplitudes for different mode orderings via the number of exchanges, this relative sign can directly be expressed efficiently in the form of a \ac{GPS}.
Utilizing a support dimension $M= \frac{L \times (L-1)}{2}$, the \ac{GPS} representation can, e.g., be specified as,
\begin{equation}
    (-1)^{\mathcal{N}(\mathbf{x}, \bar{\mathbf{x}})} = \exp \left( \sum_{i=1}^{L-1} \sum_{j=i+1}^L \prod_{k=1}^L \epsilon^{(x_k)}_{k,(i,j)} \right),
\end{equation}
where $(i,j)$ denote the compound indices for the support configuration index.
In this representation, a potential choice of parameters is
\begin{align}
    \epsilon^{(x_k=1)}_{k,(i,j)} &= (i \pi)^{1/L} \times \begin{cases} 1 & \quad \text{if } \mathcal{P}(i) > \mathcal{P}(j) \\ 0 & \quad \text{otherwise} \end{cases}, \\
    \epsilon^{(x_k=0)}_{k,(i,j)} &= (i \pi)^{1/L} \times \begin{cases} 0 & \quad \text{if } k = \mathcal{P}(i) \text{ or }  k = \mathcal{P}(j) \\ 1 & \quad \text{otherwise} \end{cases}.
\end{align}

The accuracy of a \ac{GPS} for one choice to order the Fermionic modes, can therefore always be recovered for any other ordering by a polynomial increase of the support dimension, in particular by increasing it by at most $\mathcal{O}(L^2)$.
Within current numerical setups, such a quadratic scaling of the support dimensions could however likely quickly become prohibitive for practical applications, and the inclusion of other approaches to alleviate potential ordering ambiguities is sensible.
\end{appendices}

\renewcommand*{\inintro}[1]{\relax}

\backmatter

\printbibliography[heading=bibintoc]
\end{document}